\documentclass[12pt,a4paper]{article}
\usepackage{cite}
\usepackage{latexsym,amsfonts,amsmath,amssymb,graphics}
\usepackage{dsfont}
\usepackage[dvips]{graphicx}
\usepackage{mathrsfs}  %\mathscr{...} - czcionki jak \mathcal ale faniejsze
\usepackage{bbm} %\mathbbm{..} - czcionki jak \mathbb ale te? dla malych liter
\usepackage{slashed}   %\slashed{a} - Dirac slash
\DeclareMathAlphabet{\mathpzc}{OT1}{pzc}{m}{it}

\def\bea{\begin{eqnarray}}
\def\eea{\end{eqnarray}}
\def\bc{\begin{center}}
\def\ec{\end{center}}

\newcommand{\eq}[1]{\begin{equation}#1\end{equation}}
\newcommand{\eqs}[1]{\setlength\arraycolsep{2pt}
\begin{eqnarray}#1\end{eqnarray}}
\newcommand{\refer}[1]{(\ref{#1})}
\newcommand{\nothing}[1]{}    %PUSTE POLECENIE
\newcommand{\ds}[1]{\slashed{#1}}

\def\nn{\nonumber}

\newcommand{\tr}[0]{{\rm tr}}

\newcommand{\KR}[0]{\mathcal{R}}
\newcommand{\SR}[0]{\ds{\mathcal{R}}}

\newcommand{\FeMa}[1]{{\mathscr{S}}(#1)}

\newcommand{\FeMaR}[1]{{\mathscr{S}}_\Lambda(#1)}
\newcommand{\YM}[0]{y}     %Yukawy Majorany
\newcommand{\YF}[0]{Y}     %Yukawy Weyla
\newcommand{\TS}[0]{\mathcal{T}}       %Reprezentacja na polach Skalarnych
\newcommand{\TV}[0]{e}%Reprezentacja na polach Wektorowych
\newcommand{\TF}[0]{{ \mathfrak{f}}}%Reprezentacja na polach Weyla
\newcommand{\TM}[0]{t}%Reprezentacja na polach Majorany
\newcommand{\eSC}[3]{{e}^{#1}_{\phantom{#1}#2 #3}} %Sta?e struktury
\newcommand{\tig}[0]{{\widetilde{\Gamma}}}
\newcommand{\gt}[0]{\tilde{g}}
\newcommand{\OLB}[0]{{(1B)}}
\newcommand{\OL}[0]{{(1)}}
\newcommand{\anti}[1]{\overline{#1}}
\newcommand{\derp}[2]{\frac{\partial #1}{\partial #2}} %Cz?stkowa
\newcommand{\deru}[2]{\frac{{\rm d} #1}{{\rm d} #2}}     %Zwyk?a
\newcommand{\derf}[2]{\frac{\delta #1}{\delta #2}}     %Funkcjonalna
\newcommand{\volel}[1]{\!\! {{\rm d}^4 #1 }}
\newcommand{\vol}[1]{\!\! \frac{{\rm d}^d #1}{\left(2\pi\right)^d}}
\newcommand{\volfour}[1]{\!\! \frac{{\rm d}^4 #1}{\left(2\pi\right)^4}}
\newcommand{\intt}[3]{\int{{\rm d}^{#1} #2 \text{ } #3}}

\newcommand{\bracket}[1]{\left(#1\right)}
\newcommand{\matri}[1]{\left[#1\right]}
\newcommand{\VEV}[1]{\left<#1\right>}
\newcommand{\VEVOPI}[1]{\left<#1\right>_{{\text{\tiny1PI}}}}
\newcommand{\VEVOPITIL}[1]{\left<#1\right>_{{\widetilde{\text{\tiny1PI}}}}}

\newcommand{\matrixind}[3]{{\matri{#1}}^{#2}_{\phantom{#2}{#3}}}
\newcommand{\mind}[3]{{{#1}}^{#2}_{\phantom{#2}{#3}}}
\newcommand{\SI}[0]{\mathcal{S}_{{I}_0}}
\newcommand{\dd}[0]{{\rm d}}
\newcommand{\rd}{{\rm d}}
\newcommand{\pa}[0]{{\partial}}

\newcommand{\bE}[0]{{\mathbb{E}}}

\newcommand{\fB}[0]{{\mathfrak{B}}}

\newcommand{\fW}[0]{{\mathfrak{W}}}

\newcommand{\cC}[0]{{\mathcal{C}}}

\newcommand{\cG}[0]{{\mathcal{G}}}

\newcommand{\cJ}[0]{{\mathcal{J}}}

\newcommand{\cN}[0]{{\mathcal{N}}}
\newcommand{\cO}[0]{{\mathcal{O}}}

\newcommand{\cS}[0]{{\mathcal{S}}}

\newcommand{\cV}[0]{{\mathcal{V}}}
\newcommand{\cW}[0]{{\mathcal{W}}}

\newcommand{\sC}[0]{{\mathscr{C}}}

\newcommand{\al}{\alpha}
\newcommand{\be}{\beta}
\newcommand{\de}{\delta}

\newcommand{\ga}{\gamma}

\newcommand{\ka}{\kappa}
\newcommand{\la}{\lambda}
\newcommand{\La}{\Lambda}
\def\th{\theta} %nadpisuje symbol \th juz zdefiniowany

\newcommand{\ep}{\epsilon}
\newcommand{\vp}{\varphi}
\newcommand{\vep}{\varepsilon}
\newcommand{\om}{\omega}

\makeatletter
\def\underbracket{%
  \@ifnextchar [ %
    {\@underbracket}%
    {\@underbracket [\@bracketheight]}}
\def\@underbracket[#1]{%
  \@ifnextchar [ %
    {\@under@bracket[#1]}%
    {\@under@bracket[#1][0.4em]}}
\def\@under@bracket[#1][#2]#3{%\message {Underbracket: #1,#2,#3}
  \mathop {%
    \vtop {%
      \m@th \ialign {%
        ##\crcr $\hfil \displaystyle {#3}\hfil $%
       \crcr \noalign %
       {\kern 3\p@ \nointerlineskip }%
        \upbracketfill {#1}{#2}
       \crcr \noalign %
       {\kern 3\p@ }%
     }%
   }%
  }%
  \limits%
}
\def\upbracketfill#1#2{%
  $\m@th \setbox \z@ \hbox {$\braceld$}
  \edef\@bracketheight{\the\ht\z@}\bracketend{#1}{#2}
  \leaders \vrule \@height #1 \@depth \z@ \hfill
  \leaders \vrule \@height #1 \@depth \z@ \hfill%
  \bracketend{#1}{#2}$%
}
\def\bracketend#1#2{\vrule height #2 width #1\relax}

\begin{document}
\pagestyle{empty}
\begin{flushright}
%IFT-?/2015\\
%{\tt hep-ph/yymmnnn}\\
{\bf \today}
\end{flushright}
\vspace*{5mm}
\begin{center}

{\large {\bf Two-loop RGE of a general renormalizable Yang-Mills theory
in a renormalization scheme with an explicit UV cutoff}}
\vspace*{1cm}

{\bf Piotr H. Chankowski$^1$}, {\bf Adrian Lewandowski$^{2,1}$}\\
 and {\bf Krzysztof A. Meissner$^1$}\\
\vspace{1cm}
{{\it
$^1$Institute of Theoretical Physics, Faculty of Physics,
University of Warsaw\\
Pasteura 5, 02-093 Warsaw, Poland\\
\vspace{0.5cm}
$^2$ Max-Planck-Institut f\"ur Gravitationsphysik
(Albert-Einstein-Institut)\\
M\"uhlenberg 1, D-14476 Potsdam, Germany\\
}}

\vspace*{1.7cm}
{\bf Abstract}
\end{center}
\vspace*{5mm}
\noindent
{We perform a systematic one-loop renormalization of a general
renormalizable Yang-Mills theory coupled to scalars and fermions using
a regularization scheme with a smooth momentum cutoff $\Lambda$ (implemented through an exponential damping factor). We
construct the necessary finite
counterterms restoring the BRST invariance of the effective action by analyzing
the relevant Slavnov-Taylor identities. We find the relation
between the renormalized parameters in our scheme and in the conventional
$\overline{\rm MS}$ scheme which allow us to obtain the explicit two-loop
renormalization group equations in our scheme from the known two-loop ones
in the  $\overline{\rm MS}$ scheme. 
We calculate in our scheme the divergences of two-loop
vacuum graphs in the presence of a constant scalar background field which allow us to rederive the two-loop beta functions for parameters of the scalar potential. 
We also prove that consistent
application of the proposed regularization leads to counterterms which,
together with the original action, combine to a bare action expressed in
terms of bare parameters. This, together with treating $\Lambda$ as an
intrinsic scale of a hypothetical underlying finite theory of all interactions,
offers a possibility of an unconventional solution to the hierarchy problem
if no intermediate scales between the electroweak scale and the Planck scale
exist.

}
\vspace*{1.0cm}
\date{\today}

%\noindent PACS numbers: \ldots

%\rule[.1in]{16.5cm}{.002in}
\vspace*{0.2cm}

\vfill\eject
\newpage

\setcounter{page}{1}
\pagestyle{plain}

\section{Introduction}

\renewcommand{\thesection}{\arabic{section}}
\renewcommand{\theequation}{\arabic{section}.\arabic{equation}}
\renewcommand{\thefigure}{\arabic{section}.\arabic{figure}}

\setcounter{equation}{0}
\setcounter{figure}{0}

Renormalization is in quantum field theory a standard procedure. It not only
renders calculated quantities finite but also, when the freedom in implementing
it is judiciously exploited, allows to analyze the behavior of the computed
Green's functions and observables when the characteristic energy scale
changes. The first step in this procedure is usually the introduction of an
ultraviolet (UV) regularization (an UV cutoff). The second one is performing
appropriate subtractions (usually interpreted as an effect of taking into
account contributions of suitable counterterms) after which the UV cutoff
can be removed leaving finite amplitudes. The freedom in the subtractions
(in the choice of the renormalization scheme) can be used either to directly
parametrize the computed quantities in terms of a selected set of measured
observables or to introduce an arbitrary scale $\mu$ and parametrize the
theory predictions with a set of finite, $\mu$ dependent parameters (hybrid
schemes are also possible).
The requirement that physical results be independent of $\mu$
gives then rise to the renormalization group (RG) which in turn allows
for the mentioned possibility of analyzing  the dependence of predictions
on the energy scale. The most frequently used scheme of this second type is
the (modified) minimal subtraction $\overline{\rm MS}$ applied to
dimensionally regularized amplitudes which automatically introduces an
arbitrary scale $\mu$. Renormalization of Yang Mills (YM) theories is usually
studied using this scheme \cite{'tHooft:1972fi} the main reason being that
the dimensional regularization (DimReg), unlike other more physical
UV cutoffs, automatically preserves (in theories like QCD, without
fermions in chiral representations) the BRST symmetry. This
greatly facilitates the construction of the finite (renormalized) effective
action which must be  BRST-symmetric.
This property of the effective action is indispensable to ensure decoupling
of unphysical degrees of freedom (Faddeev-Popov ghosts and antighost, scalar
components of vector bosons, would-be Goldstone modes in the case of broken
gauge symmetries or longitudinal vector bosons of unbroken gauge symmetries)
and unitarity of the $S$-matrix in the physical subspace of the full
(pseudo-)Hilbert space.

However, DimReg, while being elegant and convenient as a technical tool,
has some rather unphysical features. In particular it sets (by definition) to
zero the whole class of contributions to the effective action which are due
to real fluctuations of quantum fields but which happen to be quadratically
divergent with an explicit momentum ultra-violet cutoff $\Lambda$ (however
introduced). It is also hard to interpret physically the departure from the
integer dimension of the space-time. These drawbacks do not, of course, create
any problem for practical calculations aiming at expressing low energy
observables in terms of a selected set of other low energy observables (or
in terms of another set of finite parameters), in which, after performing
subtractions, the cutoff is completely removed, but certainly obscure
understanding of the problem of stability of the electroweak scale
$G_{\rm F}^{-1/2}$ versus the Planck scale $M_{\rm Pl}$.
\vskip0.2cm

In this paper we would like to adopt a more fundamental point of view on
renormalization (close in spirit to the one taken in applications of field
theory to statistical physics problems), proposed in \cite{CHLEMENI} (see
also \cite{Latosinski}), which
we motivate (in Section \ref{sec:HP}) by its possible connection with the
hierarchy problem. This view precludes using unphysical regularizations like
DimReg and requires treating the momentum space cutoff $\Lambda$ as a
{\it bona fide} physical scale which in our approach is viewed as an
intrinsic scale of a fundamental theory of physics at the Planck scale
(and, therefore, the limit $\Lambda\rightarrow\infty$ is not taken). This
leads us to study renormalization of a general YM theory coupled to scalars
and fermions using an explicit momentum cutoff $\Lambda$. The use of the
momentum cutoff as the regulator in YM theories immediately brings in the
problem that the regulated Green's functions do not satisfy the requisite
Slavnov-Taylor (ST) identities following from the BRST invariance. This calls
for a special form of subtractions which must restore these
identities.\footnote{An alternative approach is to device a
cutoff regularization which preserves an appropriately modified version of
the BRST symmetry \cite{KleWoo, KleWooEv}.}
We recall in this connection the general procedure for achieving this, which
is based on the Quantum Action Principle (QAP) \cite{Lam,Low}, and implement
it in the explicit one-loop calculations. We point out, however, that
strict BRST invariance is recovered with the help of this procedure
only in the limit of infinite $\Lambda$; for
finite  $\Lambda$ the ST identities remain broken by terms suppressed by
inverse powers of $\Lambda^2$ and one has to assume that other effects of the
underlying fundamental theory act so that effectively all potential problems 
associated with this breaking are cured.

To our knowledge, renormalization of YM theories in the regularization based
on an explicit momentum cutoff has never been studied systematically. In this
paper we provide the necessary technical tools for developing the approach
sketched in \cite{CHLEMENI} and perform the systematic one-loop renormalization
of a general renormalizable YM theory coupled to scalars and fermions in
arbitrary (but non-anomalous) representations using the explicit UV cutoff
proposed there. The paper is organized as follows. In Section
\ref{Sec:Dzialanie_S_I_itp} we explain our notation and conventions and recall
basic facts concerning the BRST symmetry. In Section \ref{Sec:UV-cutoff} we
specify our choice of the UV cutoff which introduces a scale $\Lambda$ and
present some technicalities concerning practical evaluation of Feynman
diagrams. Section \ref{sec:Subtractions} is devoted to the general procedure
of making subtractions restoring the BRST invariance. Here we also specify our
renormalization scheme which, similarly as the ordinary $\overline{\rm MS}$
scheme, introduces an arbitrary scale $\mu$. Explicit determination of the
one-loop counterterms and of the relation between renormalized parameters in
our scheme and in DimReg occupy Sections \ref{sec:cterms} and \ref{Sec:RelSch}.
The results of Section \ref{Sec:RelSch} can be also read as an extension
to the most general case of the results of \cite{Martin:1999cc}, namely as a
proof of equivalence at one-loop of the $\overline{\rm MS}$ scheme with 
anticommuting $\gamma^5$ matrix with a fully consistent
renormalization prescription.
In Section \ref{sec:rg} we introduce the RG equation. We argue that the standard
reasoning justifying it is not directly applicable to regularizations which
break the BRST invariance and, therefore, $\mu$ independence of the results
requires a separate proof (which we offer). The renormalization group allows
for the use the concept of bare action whose structure in the case of our
regularization is elucidated. In the same section using the
relation of our subtraction
scheme with the standard $\overline{\rm MS}$ scheme we derive two-loop
renormalization group equations satisfied by parameters (couplings and mass
parameters) of a general YM theory. In Section \ref{Vac:Diag} we apply our
regularization prescription to the two-loop computation of the scalar fields
effective potential focusing, however, only on its divergences. We determine
in this way the two-loop coefficient proportional to $\Lambda^2$ of the
counterterm to the effective potential which turns out to be different than
that found using the dimensional reduction (DimRed) \cite{Jones,JonesSTARY}
which has been recently reproduced in \cite{HAKAOD} using a cutoff
regularization superficially similar to ours. We explain the difference
between our result and that in \cite{HAKAOD}.
We also determine the one-loop coefficient of $\Lambda^2$ in the counterterm
to vector boson masses squared which is not present in DimReg (or DimRed)
but is unavoidable in the regularization by a physical momentum cutoff.

The possibility to formulate the theory in terms of the bare action and
treating the introduced momentum cutoff scale $\Lambda$ as a physical
(finite) scale allows to discuss the hierarchy problem and to propose its
possible solution along the lines of ref. \cite{CHLEMENI}. In Section
\ref{sec:HP} we recall the basic idea of this solution (which owing to the
results presented in this paper gain more solid foundations) and use the
derived two-loop RG equations and the coefficient of $\Lambda^2$ divergence
of the effective potential to discuss (non)viability of this solution in
the SM. Section \ref{sec:Final} contains our conclusions.

\section{Lagrangian and conventions}
\label{Sec:Dzialanie_S_I_itp}

\renewcommand{\thesection}{\arabic{section}}
\renewcommand{\theequation}{\arabic{section}.\arabic{equation}}
\renewcommand{\thefigure}{\arabic{section}.\arabic{figure}}

\setcounter{equation}{0}
\setcounter{figure}{0}

As the starting point of our approach we consider a general renormalizable
Yang-Mills theory with the gauge group which is a direct product of an
arbitrary number of compact simple Lie groups and $U(1)$ groups coupled to
scalar and fermionic fields in arbitrary representations of the gauge group.
We work with real scalars $\phi^i$ and represent all fermionic fields as
four-component Majorana spinors  $\psi^a$  built out of fundamental
two-component Weyl spinors.\footnote{Although calculations with the
Majorana fields involve the charge conjugation matrix $C$ defined by
the relation $\anti{\psi}=\psi^{\rm T}C$, they are more convenient as
they lead to a smaller number of diagrams.
}
It is also convenient to consider the theory in the presence of an arbitrary
constant scalar background $\varphi^i$ which can eventually be identified
with the vacuum expectation value (VEV) of ``the symmetric phase'' field
$\Phi^i=\phi^i+\varphi^i$. (This identification, however, will not be used
in what follows). Thus the classical gauge-invariant action $I_0^{GI}$
(prior to regularization) is given by the integral of the Lagrangian density
\begin{eqnarray}
\mathcal{L}^{GI}_0=-{1\over4}
\delta_{\alpha \beta}F^\alpha_{\phantom{a}\mu\nu}F^{\beta~\!\mu\nu}+
{1\over2}\delta_{ij}(D_\mu\phi)^i(D^\mu\phi)^j\!
-\mathcal{V}(\phi+\varphi)\nonumber\\
+{1\over2}~\!\anti{\psi}^a\!\left\{\delta_{ab}\,
i(\gamma^\mu D_\mu\psi)^b
-[\widehat{\mathcal{M}}_F(\phi+\varphi)]_{ab}{\psi}^b
\right\}.\phantom{aa}
\label{Eq:LagrTreeGI}
\end{eqnarray}
The potential $\mathcal{V}(\Phi)$ is a fourth order polynomial. It is
parametrized by the following coupling constants and mass parameters:
\begin{eqnarray}
\lambda_{ijkl} = \mathcal{V}^{(4)}_{ijkl}(\varphi),\phantom{aa}~
[\rho_i]_{jk} = \mathcal{V}^{\prime\prime\prime}_{ijk}(\varphi),\phantom{aa}~
m_{S\,ij}^2=\mathcal{M}_S^2(\varphi)_{ij}=\mathcal{V}^{\prime\prime}_{ij}(\varphi),
\phantom{aa} \mathcal{V}^\prime_{i}(\varphi)\neq0.\phantom{a}
\label{Eq:RegFey:LamRhoMs}
\end{eqnarray}
which, with the exception of $\lambda_{ijkl}$, are $\varphi$-dependent.
The generalized fermion mass matrix, which is a first order polynomial
in $\Phi^i$, includes also the Yukawa couplings
\begin{eqnarray}
\widehat{\mathcal{M}}_F(\Phi)=
\widehat{\mathcal{M}}_F(0)+\YM_i\,\Phi^i.\label{eqn:FermionMassMatrixDef}
\end{eqnarray}
Different kinds of indices are lowered/raised  with the aid of the
appropriate metrics: $\delta_{ij}$, $\delta_{\alpha\beta}$, $\delta_{ab}$ for
internal indices and $\eta_{\mu\nu}={\rm diag}(+1,-1,-1,-1)$ for
Lorentz indices.

The explicit form of $F^\alpha_{\phantom{a}\mu\nu}$ is
\begin{eqnarray}\nn
F^\alpha_{\text{ }\mu\nu}=\partial_\mu A^\alpha_\nu\!
-\partial_\nu A^\alpha_\mu\!+e^\alpha_{\phantom{a}\beta\gamma}A^\beta_\mu A^\gamma_\nu
\end{eqnarray}
and the covariant derivatives read
\begin{eqnarray}
D_\mu\phi =\partial_\mu \phi
+ A^\alpha_\mu[{\TS}_\alpha(\phi+\varphi)+\bar{P}_\alpha],\phantom{aaa}
D_\mu\psi =\partial_\mu \psi+ A^\alpha_\mu\,{\TM}_\alpha\psi.
\label{Eq:CovDer}
\end{eqnarray}
${\TS}_\alpha$ are real antisymmetric generators of the
gauge group in the representation formed by the scalars $\phi^i$; they
satisfy the commutation relations
$[{\TS}_\alpha,~\!{\TS}_\beta]={\TS}_\gamma~\! e^\gamma_{\phantom{a}\alpha\beta}$
with the real structure constants $e^\gamma_{\phantom{a}\alpha\beta}$. Obviously,
$e^\gamma_{\phantom{a}\alpha\beta}$, which themselves are matrix elements
of the generators $e_\alpha$ in the adjoint representation
($[e_\alpha]^\gamma_{\phantom{a}\beta}=e^\gamma_{\phantom{a}\alpha\beta}$), are, similarly
as ${\TS}_\alpha$ and ${\TM}_\alpha$ (and ${\TF}_\alpha$ - see below),
proportional to the gauge coupling constants. We work in a natural basis of
the gauge Lie algebra, so that the indices $\alpha$ split into Abelian
ones ($\alpha_A$) and semisimple ones ($\alpha_S$). Coefficients $\bar P_\alpha$
obeying $\TS_\beta \bar{P}_\alpha=0$ must vanish for non-Abelian indices
$\alpha=\alpha_S$. If $\bar{P}_{\alpha_A}\neq0$ for some Abelian indices
${\alpha_A}$, Stueckelberg fields are present (see e.g. \cite{Stue} and
references therein) among components of the scalar fields $\phi$  as
explained in Appendix \ref{App:Ren-Det:Part1-AuxCond}.
In the generators ${\TM}_\alpha$, similarly
as in the generalized fermion mass matrix (\ref{eqn:FermionMassMatrixDef}),
the chiral projectors $P_{L,R}={1\over2}(1\mp\gamma^5)$ are included:
\begin{eqnarray}\label{Eq:Majorana}
\TM_\alpha = \TF_\alpha P_L+\TF^\ast_\alpha P_R,
\phantom{aaa}
\widehat{\mathcal{M}}_F(\Phi)=
\mathcal{M}_F(\Phi)P_L+\mathcal{M}^\ast_F(\Phi) P_R,
\end{eqnarray}
(likewise $\YM_i\equiv \YF_i P_L + \YF^\ast_i P_R$). Here $\TF_\alpha$
are ordinary antihermitian matrix generators (satisfying the relation
$[{\TF}_\alpha,~\!{\TF}_\beta]={\TF}_\gamma~\! e^\gamma_{\phantom{a}\alpha\beta}$)
of the gauge group representation realized by the Weyl fields. The
background-dependent mass matrix $m_F$ of the Weyl fermions has the structure
\begin{eqnarray}\label{Eq:m_F-def}
m_F\equiv\mathcal{M}_F(\varphi)=\mathcal{M}_F(0)+\YF_i \varphi^i~\!,
%\qquad\quad \widehat{m}_F\equiv\widehat{\mathcal{M}}_F(\varphi)\,.
\end{eqnarray}
We also write $\widehat{m}_F\equiv\widehat{\mathcal{M}}_F(\varphi)$ for
its Majorana counterpart.
The mass matrix of the vector bosons is given by
\begin{eqnarray}
\matri{m_V^2}_{\alpha\beta}=[\mathcal{M}_V^2(\varphi)]_{\alpha\beta}\equiv
-{1\over2}\varphi^{\rm T}\!\{\TS_\alpha,\ \TS_\beta\}\varphi
+ \bar{P}_\alpha^{\rm T}\bar{P}_\beta~\!.\label{Eq:RelA12}
\end{eqnarray}
If Stueckelberg fields are absent, $m_V^2$ vanishes unless the background
$\varphi$ has some nonzero components breaking (at least partly) the gauge
group. Gauge invariance of $\mathcal{L}^{GI}_0$ implies also various important
relations between parameters, like e.g.
\begin{eqnarray}
\bracket{m_S^2 \TS_\alpha \varphi}_k
=-\mathcal{V}^\prime_l(\varphi)[\TS_\alpha]^l_{\phantom{a}k}
=(\TS_\alpha \mathcal{V}^\prime(\varphi))_k~\!.
\end{eqnarray}
\vskip0.2cm

To generate Green's functions of the quantum theory, the classical action
$I_0^{GI}$ must be supplemented with a gauge fixing term  and with the ghost
fields action. The structure of divergences arising in the perturbative
expansion can be then controlled by working with the BRST invariant
tree-level action
\begin{eqnarray}
I_0=I_0^{GI}+I_0^{Rest}=\int\!{\rm d}^4 x~\! (\mathcal{L}^{GI}_0
+\mathcal{L}^{Rest}_0)~\!,\label{Eq:I_0=I_0^GI+I_0^Rest}
\end{eqnarray}
where ${\cal L}_0^{Rest}$ depends on the Nakanishi-Lautrup fields $h_\beta$, the
ghost fields $\omega^\alpha$ and $\overline{\omega}_\alpha$ and the so-called
antifields $K_i$, $\bar{K}_a$, $K^\mu_\alpha$ and $L_\alpha$:
\begin{eqnarray}
\mathcal{L}^{Rest}_0=
s\!\left(\!\overline{\omega}_\alpha \mathfrak{F}^\alpha+
{1\over2}~\!\overline{\omega}_\alpha\xi^{\alpha\beta}h_\beta\!\right)\!
+\!L_\alpha\, s({\omega}^\alpha)+K_i\, s(\phi^i)
+\bar{K}_a\, s(\psi^a)+K^\mu_\alpha\, s(A^\alpha_\mu)~\!.\phantom{aa}
\label{Eq:LagrTreeRest}
\end{eqnarray}
Here $\xi^{\alpha\beta}$ are arbitrary gauge fixing parameters. In what
follows we will work in the Landau gauge
\begin{eqnarray}
\mathfrak{F}^\alpha\equiv-\partial_\mu A^{\alpha\mu},
\phantom{aaaaaaaaa} \xi^{\alpha\beta}\equiv 0~\!,
\label{Eq:Gauge_Fix_Funct_falpha_Gen_Gauge}
\end{eqnarray}
which leads to some simplifications due to the presence in this gauge of
additional symmetries of $I_0^{Rest}$ (see
Appendix \ref{App:Ren-Det:Part1-AuxCond}).

The action on fields of the BRST ``differential'' $s(\cdot)$ is
given by \cite{BRS}
\begin{eqnarray}\label{Eq:BRS}
s(\phi^i)=\omega^{\alpha}
\left[\TS_\alpha(\phi\!+\!\varphi)\!+\!\bar{P}_\alpha\right]^i,\phantom{aaa}
s(\psi^a) = \omega^{\alpha}(\TM_\alpha \psi)^a,
\phantom{aaa}
s(A^\gamma_\mu)=-\partial_\mu\omega^{\gamma}\!+\!
e^{\gamma}_{\phantom{a}\alpha \beta}~\!\omega^{\alpha}A^\beta_\mu,
\phantom{a}
\!\!\!\!\!\!\!\!\!\!\!\!\!
\nonumber\\
s(\omega^\alpha)={1\over2}~\!
e^{\alpha}_{\phantom{a}\beta \gamma}~\!\omega^\beta\omega^\gamma,
\phantom{aaaaaaaaaa}
s(\overline{\omega}_\alpha) =h_\alpha,
\phantom{aaaaaaaa}
s(h_\alpha)=0.\phantom{aaaaaaaaaaaa}
\end{eqnarray}

The antifields $K_i$, $\bar{K}_a$, $K^\mu_\alpha$ and $L_\alpha$, treated as
external sources, control the renormalization of the composite operators
$s(\phi^i)$, $s(\psi^a)$, $s(A^\alpha_\mu)$ and $s(\omega^\alpha)$.
Setting $s(K_i)=s(\bar{K}_a)=s(K^\mu_\alpha)=s(L_\alpha)=0$ makes the action
$I_0^{Rest}$ a BRST-exact functional: $I_0^{Rest}=s(W)$. Nilpotency $s^2=0$ of
the $s(\cdot)$ operation ensures then the BRST invariance of the complete
action  (\ref{Eq:I_0=I_0^GI+I_0^Rest}): $s(I_0)=0$.
\vskip0.2cm

In writing identities expressing the BRST invariance of the effective
action we will work in the momentum space representing fields by
their Fourier images according to the formulae
\begin{eqnarray}
\label{Eq:Mom-Pos-Rel}
A^{\alpha}_{\mu}(x)=\int\,\volel{l} ~\! e^{-i l x} \tilde{A}^{\alpha}_{\mu}(l),
\qquad
\derf{}{A^{\alpha}_{\mu}(x)}=\int\volfour{l} ~\! e^{i l x}~\!
\derf{}{\tilde{A}^{\alpha}_{\mu}(l)}\,.
\end{eqnarray}
Momentum space one-particle irreducible (1PI) Green's functions are then
given by (all momenta are incoming into the 1PI vertices)
\begin{eqnarray}
\VEVOPI{\tilde{\psi}^b(p')  \tilde{\psi}^a(p) \tilde{A}^\alpha_\mu(l)}
\equiv
\derf{}{\tilde{A}^\alpha_\mu(l)} \left.\derf{}{\tilde{\psi}^a(p)}
\derf{}{\tilde{\psi}^b(p')}~\!\Gamma\matri{\phi,\psi,A,\ldots}\right|_{0}\nn\\
=(2\pi)^4 \delta^{(4)}(p'+p+l)
\widetilde{\Gamma}^{\phantom{ba}\mu}_{ba\alpha} (p',p,l)~\!.
\phantom{aaaaa}\label{Eq:Notacja1}
\end{eqnarray}
The functional derivatives (which act always from
the left) in \refer{Eq:Notacja1} are taken at
the ``point'' at which \emph{all} fields vanish. Notice also the order of
the fermionic variables and the ``wrong'' height of indices inside the bracket
$\VEVOPI{\cdot}$. For the 1PI functions we will also use the notation
\eq{\label{Eq:Notacja2}
\VEVOPI{\tilde{\psi}^b(p')  \tilde{\psi}^a(p) \tilde{A}^\alpha_\mu(l)}
=(2\pi)^4 \delta^{(4)}(p'+p+l)
\VEVOPITIL{\tilde{\psi}^b(p')  \tilde{\psi}^a(p) \tilde{A}^\alpha_\mu(l)}.
}
Green's functions like \refer{Eq:Notacja1} become ``physical'' when the 
background $\varphi$ is chosen so that the following condition is satisfied 
\eq{\nn\label{Eq:Tadpole}
\left.\derf{\Gamma\matri{\phi,\psi,A,\ldots}}{\phi^i(z)}\right|_{0}=0.
}
As we have already said, in studying renormalization we do {\emph{not}}
impose the above relation, treating $\varphi^i$ as arbitrary external
parameters. Contributions of order $\hbar^n$ to the 1PI function are
denoted $\tig(\cdots)^{(n)}$, e.g.:
\eq{\nn
\tig^{\phantom{ba}\mu}_{ba\alpha}(p',p,l)=\sum_{n=0}^\infty \hbar^n\,
\tig^{\phantom{ba}\mu}_{ba\alpha}(p',p,l)^{(n)}.
}
In what follows it will be convenient to further split
$\tig^{\phantom{ba}\mu}_{ba\alpha}(p',p,l)^{(1)}$ into the contribution of the
counterterm diagrams and the sum of genuine one-loop diagram contributions.
The latter will be denoted $\tig^{\phantom{ba}\mu}_{ba\alpha} (p',p,l) ^{(1B)}$.
If a given function is convergent \emph{by power-counting}, the superscripts
${}^{(1B)}$ and ${}^{(1)}$ are used interchangeably.

\section{The UV regularization}
\label{Sec:UV-cutoff}

\renewcommand{\thesection}{\arabic{section}}
\renewcommand{\theequation}{\arabic{section}.\arabic{equation}}
\renewcommand{\thefigure}{\arabic{section}.\arabic{figure}}

\setcounter{equation}{0}
\setcounter{figure}{0}

As the UV regularization in our study of the renormalization of a general YM theory 
we choose (out of many other possibilities) the prescription which consists of
modifying {\it every} derivative in the Lagrangian according to the rule
\begin{eqnarray}
\partial_\mu\rightarrow
\exp\!\left\{\frac{\partial^2}{2\Lambda^2}\right\}\partial_\mu~\!.
\label{eqn:PrescriptionDef}
\end{eqnarray}
The replacement \refer{eqn:PrescriptionDef} is to be done at the
level of the Lagrangian densities \refer{Eq:LagrTreeGI} and
\refer{Eq:LagrTreeRest}; in the latter the BRST operations $s(\cdot)$
have to be carried out first (this should be considered a part of
the  regularization definition).

In the momentum space the above prescription is
equivalent to the replacement
\begin{eqnarray}
k_\mu\rightarrow{\cal R}_\mu(k)\equiv
\exp\!\left\{-\frac{k^2}{2\Lambda^2}\right\}k_\mu~\!.
\label{eqn:MomSpacePrescription}
\end{eqnarray}
Strictly speaking, the rule \refer{eqn:MomSpacePrescription} should be applied to the Euclidean counterpart of the action \refer{Eq:I_0=I_0^GI+I_0^Rest}, in the form 
$k_{E\mu}\rightarrow
\exp\!\left\{+k^2_E/(2\Lambda^2)\right\}k_{E\mu}~\!$. Indeed, if \refer{eqn:MomSpacePrescription} is applied literally to, say, the massless one-loop one-point function in the Minkowski space-time, the integral w.r.t. the time-like component of the momentum is badly divergent. By contrast, the corresponding Euclidean integral is undoubtedly convergent owing to the exponential damping factor (see below for consideration of an arbitrary diagram), which effectively restricts the integration region to Euclidean momenta obeying 
$k_E\lesssim \Lambda$; therefore we will call $\La$ in the following the UV cutoff. The resulting amplitudes computed perturbatively in the Euclidean space are easily continued to the Minkowski space-time. 
(Such a treatement of the regularization does not 
preclude investigating non-perturbative effects, e.g. bound states, by 
summing infinite series of subtracted and continued to the Minkowski 
space Feynman diagrams.) 
In actual calculations we prefer to work with the Minkowski space-time  Feynman rules.
 Therefore, instead of explicitly reformulating the theory in the Euclidean space, we work with the action \refer{Eq:I_0=I_0^GI+I_0^Rest} and the prescription \refer{eqn:PrescriptionDef}, but  perform in Feynman diagrams a \emph{formal} Wick rotation, that is neglect  contributions arising from (divergent) integrals over contours at infinity (in other words, all integrals over time-like components of loop momenta are in practice taken over the imaginary axis).   
In the perturbative expansion this procedure just implements the analytic continuation of the corresponding (convergent) integrals of the Euclidean version of theory. 
We also stress that in principle one could try to find a similar regularization acting directly in the Minkowski space-time by replacing the exponential in \refer{eqn:PrescriptionDef} with a polynomial, what gives a variant of the higher derivative regularization, see e.g. \cite{Fujikawa,Kataev:2013csa,Aleshin:2016rrr} -- however, we prefer to work with the exponential form for the sake of calculational simplicity.

In the more fundamental perspective (see Section \ref{sec:HP}) 
we would like to treat the Euclidean version of the Lagrangian density modified 
according to the prescription \refer{eqn:PrescriptionDef} 
as a part of the complete Lagrangian density of an
effective field theory 
for some fundamental finite theory of all interactions. 
The scale $\Lambda$
should be therefore identified with an intrinsic physical scale of the
putative fundamental theory rather than with the scale introduced by the
Wilsonian procedure of integrating out some high energy degrees of freedom, and the limit $\Lambda\to\infty$ should not be taken. 
%%%%%%%
    Consistency of such an interpretation requires probably the fundamental theory 
    to be formulated in the Euclidean space. The question then arises 
    whether the prescription (3.1) in the effective theory can have a 
    meaning also outside the
    perturbative expansion. Since the action has then a nonlocal character,
    standard arguments (appealing to the Osterwalder-Schrader theorem,
    whose status in YM theories remains, however, unclear) in favor of
    uniqueness of the analytical continuation to the Minkowski space-time
    of non-perturbatively determined Green's functions may not apply.
    Moreover with the exponential factors (3.3) not expanded, the 
    propagators can, after continuation, develop unphysical poles, signaling 
    potential problems. However, as will be seen (see the end of the next section),
     if the limit 
    $\Lambda\to\infty$ is \emph{not} taken, the (Euclidean) action (2.1) and (2.10)
    with the substitution (3.1) cannot be considered a complete action
    of the effective theory: further terms suppressed by inverse powers of $\La$ must be added to it to restore the BRST symmetry for finite values of $\La$.  In the spirit of our further considerations
    we can therefore speculate that the complete Euclidean effective theory 
    action is not sick when treated non-perturbatively and does allow for a 
    unique continuation to the Minkowski space of the non-perturbative
    amplitudes.

\vskip0.2cm

The important virtue of the proposed
prescription \refer{eqn:PrescriptionDef} is that it preserves the formal invariance of the path
integral with respect to shifting fields by constant backgrounds,
leading to the 1PI effective action
$\Gamma$ satisfying the ``translational Ward identity" \cite{Pilaf}
\begin{eqnarray}
\Gamma[A,\psi,\phi,\ldots,\varphi]=\Gamma[A,\psi,\phi+\varphi,\ldots,0]~\!.
\label{Eq:shift-sym}
\end{eqnarray}
It is therefore applicable without modifications also to theories with
spontaneous symmetry breaking by nonzero VEVs of scalar fields.
On the practical side, the prescription (\ref{eqn:PrescriptionDef})
allows for an easy extraction of finite and divergent parts of
amplitudes which can be automatized using standard computer packages for
symbolic manipulations.

With the prescription (\ref{eqn:PrescriptionDef}) the propagators
of vector bosons (in the Landau gauge), fermions, scalars and ghosts take
respectively the forms:
\begin{eqnarray}
iD^{\alpha_1\alpha_2}_{\mu_1\mu_2}(k;\,\La)
=\left[{-i\over\KR^2(k)-m_V^2}\right]^{\alpha_1 \alpha_2}
\!\!\bracket{\eta_{\mu_1 \mu_2}-{k_{\mu_1} k_{\mu_2}\over k^2}},\nn\\ \nn\\
iS^{a_1a_2}(p;\,\La)=
\left[{i\over\SR(p)-\widehat{m}_F}~\!C^{-1}\right]^{a_1a_2},
\phantom{aaaaaaaaaaaaa}   \label{Eq:FermProp}\!\\   \nn\\
i\Delta^{i_1i_2}(q;\,\La)=
\left[{i\over{\cal R}^2(q)-m_S^2}\right]^{i_1i_2},
\phantom{aaaaaaaaaaaaaaaaa}~\!\nonumber\\ \nn\\
iD_{gh~\!\beta}^\alpha(k;\,\La)={i\over {\cal R}^2(k)}~\!
\delta^\alpha_{\phantom{a}\beta}~\!.\phantom{aaaaaaaaaaaaaaaaaaaaaaa}~\!\nonumber
\end{eqnarray}
The mixed scalar-vector propagator vanishes owing to the choice of the Landau
gauge (\ref{Eq:Gauge_Fix_Funct_falpha_Gen_Gauge}). We also list the vertices
which get modified by the prescription (\ref{eqn:PrescriptionDef}):
\begin{eqnarray}
\tilde{\mathcal{L}}_{AAA}(\{k\})
\!\!&=&\!{i\over3!}~\!e_{\alpha_1 \alpha_2 \alpha_3}
\{[\KR(k_3)\!-\!\KR(k_2)]^{\mu_1}\eta^{\mu_3\mu_2}
 +[\KR(k_1)\!-\!\KR(k_3)]^{\mu_2}\eta^{\mu_1\mu_3}
\phantom{aaa}\!\nonumber\\
&{}&\phantom{a}
 +[\KR(k_2)\!-\!\KR(k_1)]^{\mu_3}\eta^{\mu_2\mu_1}\}
\tilde A^{\alpha_1}_{\mu_1}(k_1)~\!
\tilde A^{\alpha_2}_{\mu_2}(k_2)~\!
\tilde A^{\alpha_3}_{\mu_3}(k_3)~\!,\label{Eq:RegFey:AAA-vertex}\\
\tilde{\mathcal{L}}_{A\phi\phi}(\{k\})
\!\!&=&\!
{i\over2!}[\KR(k_2)\!-\!\KR(k_1)]^{\mu} (\TS_\alpha)_{i_1 i_2}
\tilde A^{\alpha}_{\mu}(k_3)~\!\tilde\phi^{i_1}(k_1)~\!\tilde
\phi^{i_2}(k_2)~\!,
\nn\\
\tilde{\mathcal{L}}_{A\anti{\omega}{\omega}}(\{k\})
\!\!&=&\!
-i\KR^\mu(k_1)~\!e^{\alpha_1}_{\phantom{a}\beta\alpha_2}
\tilde A^\beta_\mu(k_3)~\!\tilde{\anti{\omega}}_{\alpha_1}(k_1)~\!
\tilde{\omega}^{\alpha_2}(k_2)~\!.\nonumber
\end{eqnarray}
We have used here the notation
\begin{eqnarray}
I^\Lambda_0=\sum_{\{\Phi_1,..., \Phi_n\}} \int\!\prod_{i=1}^{n}{{\rm d}^4 k_i}
\ \tilde{\mathcal{L}}_{\Phi_1\cdots\Phi_n}(k_1,..., k_n)~\!
(2\pi)^4\delta^{(4)}\!\bracket{k_1 +\cdots + k_n}.
\end{eqnarray}
The remaining vertices having $n\geq3$ are not modified.\footnote{The
two-point vertex $\tilde{\mathcal{L}}_{K^\mu_\al \omega^\be}$  is omitted here as
it does not contribute to \emph{loop} 1PI diagrams. For the same reason 
propagators involving the Nakanishi-Lautrup multipliers $h_\al$ are omitted.
}
\vskip0.2cm

To see that indeed all relevant diagrams are regularized by the prescription
(\ref{eqn:PrescriptionDef}), consider a 1PI diagram $\gamma$
consisting of $V_i$ vertices of type $i$ involving (prior to regularization)
$d_i$ derivatives and to which $n_{i\Phi}$ lines of fields of type $\Phi$ are
attached, $I_\Phi$ internal  and $E_\Phi$ external lines of type $\Phi$.
The corresponding integrand (after \emph{formal} Wick rotation) acquires the factor
$\exp(\bar\omega(\gamma)k^2_E/2\Lambda^2)$, where
\begin{eqnarray}
\bar\omega(\gamma)\equiv\sum_\Phi I_\Phi(2s_\Phi-2)+\sum_iV_id_i  \,,
\end{eqnarray}
(the factor $s_\Phi$ characterizes the $\Phi$ line propagator which
behaves as $k_E^{2s_\Phi-2}$ as $k_E\rightarrow\infty$).
Obviously, a diagram $\gamma$ gets regularized if $\bar\omega(\gamma)<0$.
Moreover, since $\omega(\gamma)=4L+\bar\omega(\gamma)$, where $L\geq0$ is
the number of loops and $\omega(\gamma)$ is the textbook degree of
superficial divergence \cite{WeinbergVol1}, it follows that
superficially convergent diagrams (of $\omega(\gamma)<0$) necessarily have
$\bar\omega(\gamma)<0$. Using the standard identities one gets that
\begin{eqnarray}
\bar\omega(\gamma)=-4(L-1)-\sum_\Phi E_\Phi(1+s_\Phi)-\sum_iV_i\Delta_i~\!,
\end{eqnarray}
where $\Delta_i=4-d_i-\sum_\Phi n_{i\Phi}(1+s_\Phi)$. This shows that in
renormalizable theories, in which all vertices have
$\Delta_i\geq0$, unregulated by the prescription (\ref{eqn:PrescriptionDef})
remain only one-loop ($L=1$) vacuum ($E_\Phi=0$) diagrams which cannot
appear in physically interesting amplitudes as divergent subdiagrams.
All other diagrams arising in renormalizable theories get regularized.
\vskip0.2cm

Computation of diagrams regularized with the help of the prescription
(\ref{eqn:MomSpacePrescription}) is based on the following expansion
\begin{eqnarray}
{i\over{\cal R}^2(k)-m^2}=e^{k^2/\Lambda^2}{i\over k^2-m^2}\sum_{n=0}^\infty
\left[{m^2\over m^2-k^2}\left(1-e^{k^2/\Lambda^2}\right)\right]^n,
\label{eqn:PropagatorExpansion}
\end{eqnarray}
($k$ may stand for a sum of several loop and external line momenta).
It is clear that in the Euclidean space, for $k^2\rightarrow-k^2_E$, the
expansion (\ref{eqn:PropagatorExpansion}) would be absolutely convergent. In
particular, owing to the growing inverse powers of $m^2-k^2$ in successive
terms, for a given one-loop diagram only a finite number of terms yield
integrals that are divergent when the factors
$e^{k^2/\Lambda^2}(1-e^{k^2/\Lambda^2})^n$ are
omitted. The remaining terms are integrable without these factors which
implies that their contributions vanish in the limit
$\Lambda\rightarrow\infty$. Thus the practical recipe for computing diagrams
regularized with the help of (\ref{eqn:MomSpacePrescription}) consists of
the following steps (see also \ref{App:Diag}):
{\it i)} expanding all regularized propagators as in
(\ref{eqn:PropagatorExpansion}), {\it ii)} combining denominators using
the standard trick introducing integrals over Feynman parameters $\alpha_i$,
{\it iii)} shifting and formally Wick-rotating the momenta, {\it iv)} expanding
the exponential factors in powers of external momenta,  {\it v)} performing
integrals over angular variables. After these steps every one-loop diagram
gets represented in the form of the confluent hypergeometric function
\begin{equation}\label{eqn:ConflHypGeom}
U\!\left(a,b,z\right)=\frac{1}{\Gamma(a)}\int_0^\infty \!{\rm d}t \
t^{a-1}(1+t)^{b-a-1}\exp(-z t),
\end{equation}
in which $a$ and $b$ are some real numbers, $t\propto k_E^2$ and $z$ is the
ratio of a linear combination of masses squared and external momenta squared
weighed by the Feynman parameters $\alpha_i$ and of $\Lambda^2$. One is
therefore led to study the limit of $z\rightarrow0$ of $U(a,~\!b,~\!z)$ which
can be extracted using the well known formulae \cite{ABRSTEG}. In this way
one-loop diagrams get represented in the standard form of integrals over
Feynman parameters.

Although this is not necessary for one-loop calculations, we note that in
general extraction of the $\Lambda\rightarrow\infty$ asymptotics can be
efficiently done by exploiting a theorem by Handelsman and Lew \cite{HANLEW}
which relates the requisite coefficients in the asymptotics
of the Laplace transform
of the general form
\begin{eqnarray}
L[f,z]=\int_0^\infty\!{\rm d}t~\!f(t)~\!e^{-zt}~\!,\nonumber
\end{eqnarray}
directly to the coefficients of the $t\rightarrow\infty$ asymptotics
of the function $f(t)$ and to constant terms in the Laurent expansions
of (the analytic continuation of) the Mellin transform
\begin{eqnarray}
M[f,z]=\int_0^\infty\!{\rm d}t~\!f(t)~\!t^{z-1}~\!.\nonumber
\end{eqnarray}
around its poles. Thus, the Handelsman-Lew theorem is crucial for finding
the asymptotic form of multi-loop diagrams, which cannot be expressed in
terms of the function \refer{eqn:ConflHypGeom}.

\section{The subtraction procedure}
\label{sec:Subtractions}

\renewcommand{\thesection}{\arabic{section}}
\renewcommand{\theequation}{\arabic{section}.\arabic{equation}}
\renewcommand{\thefigure}{\arabic{section}.\arabic{figure}}

\setcounter{equation}{0}
\setcounter{figure}{0}

The UV cutoff introduced in Section \ref{Sec:UV-cutoff} explicitly breaks
the BRST symmetry -  $s(I^\Lambda_0)\neq0$, where $I^\Lambda_0$ is the action
(\ref{Eq:I_0=I_0^GI+I_0^Rest}) modified according to the prescription
(\ref{eqn:PrescriptionDef}). Consistency of the quantized gauge theory
does not require, however, BRST invariance of $I^\Lambda_0$, but only BRST
invariance of the 1PI effective action $\Gamma$ - the functional generating
one-particle irreducible (1PI) Green's functions. This can be restored by
using  the  general methodology based on the Quantum Action Principle
\cite{Lam,Low} (see also \cite{Bon,PiguetSorella} for reviews). In practical
terms it consists of starting with the local BRST invariant action expressed
in terms of renormalized fields and parameters and in making in the computed
Green's functions (or the effective action) order by order in the loop
expansion appropriate subtractions in such a way, that the Zinn-Justin (ZJ)
identity \cite{Z-J:74}
\begin{eqnarray}
{\cal S}(\Gamma)=0~\!,
\label{eqn:ZJidBasic}
\end{eqnarray}
in which $\mathcal{S}(\cdot)$ is the differential operator whose action
on an arbitrary functional $F$ of fields and antifields
is given by\footnote{We use the notation
${k}\cdotp\!{g}\equiv\intt{4}{x}{k(x)~\!g(x)}$.}
\begin{eqnarray}
\mathcal{S}(F)\equiv
\derf{F}{K^\mu_\alpha}\!\cdot\!\derf{F}{A^\alpha_\mu}+
\derf{F}{K_i}\!\cdot\!\derf{F}{\phi^i}+
\derf{F}{\bar K_a}\!\cdot\!\derf{F}{\psi^a}+
\derf{F}{L_\alpha}\!\cdot\!\derf{F}{\omega^\alpha}+
h_\alpha\!\cdot\!\derf{F}{\anti{\omega}_{\alpha}}.\label{Eq:ZinnJustinOperQuad}
\end{eqnarray}
is satisfied (up to higher order terms) by the subtracted effective action
$\Gamma$. Within the general framework the possibility to restore BRST
invariance of the effective action (in non-anomalous theories) in this way
was first demonstrated in \cite{BRS} using the BPHZ scheme \cite{BPHZ} in
which subtractions are made directly in integrands of the integrals
corresponding to Feynman diagrams and thus no explicit regulator is introduced.
This approach is usually used in formal proofs of existence (within the
perturbation theory) of unitary gauge theories for which no symmetry preserving
regularization is available \cite{BBBC1,BBBC2,Kraus:1997bi,Hollik:2002mv};
some practical calculations within the Standard Model (SM) based on this
approach can be found in
\cite{Grassi:1999tp,Grassi:2001zz,Grassi:2001kw,Grassi:2000kp}.

The general QAP methodology can obviously be applied also in conjunction
with any explicit BRST symmetry violating regulator. In such an approach one
constructs order by order in the perturbative expansion the counterterms:
the divergent (as the regulator is removed) ones, which in our scheme
will be uniquely determined by the regularization and the adopted
``minimal'' subtraction prescription,
and the additional finite counterterms restoring the ZJ identity. This
approach
has been used in particular to renormalize YM theories with chiral fermions
using DimReg and the original 't Hooft-Veltman definition of the $\gamma^5$
matrix which avoids inconsistencies \cite{BreiMei} but breaks the BRST
symmetry already at one-loop. The full set of one-loop
counterterms was determined in specific models \cite{SanchezRuiz:2002xc},
including supersymmetric ones \cite{Fischer:2003cb} as well as in
an arbitrary renormalizable gauge theory without scalars \cite{Martin:1999cc}.

In this paper we apply this approach to the regularization of a general renormalizable YM
theory by the explicit UV momentum cutoff defined in Section
\ref{Sec:UV-cutoff} (see
\cite{Becchi:1996an,Bonini:1994kp,D'Attanasio:1996jd,Bonini:1994dz,Bonini:1993sj,Frob:2015uqy,Frob:2016mzv} for partially related applications in the context of the Wilson-Polchinski
renormalization group).
Below we recall the general procedure based on the QAP and specify
our way of fixing its arbitrariness (our renormalization conditions).
\vskip0.2cm

As said, the starting point is the regularized action $I_0^\Lambda$ obtained
by applying the prescription (\ref{eqn:PrescriptionDef}) to the local BRST
symmetric action $I_0$ defined by (\ref{Eq:I_0=I_0^GI+I_0^Rest}).
All fields and parameters of $I_0$ have the interpretation of renormalized
quantities. The action $I_0$ is such that ${\cal S}(I_0)=0$ and satisfies
a number of additional conditions listed in
Appendix \ref{App:Ren-Det:Part1-AuxCond}. Since
\begin{eqnarray}
\Gamma[I_0^\Lambda]\equiv \Gamma_0^\Lambda=I_0^\Lambda+{\cal O}(\hbar)~\!,
\nonumber
\end{eqnarray}
the ``asymptotic part'' of $\Gamma_0^\Lambda$ (denoted $\Gamma_0$) obtained
by neglecting all terms which vanish in the limit
$\Lambda\rightarrow\infty$ satisfies the ZJ identity (\ref{eqn:ZJidBasic})
up to terms of order $\hbar$.

We now show that having a local action $I_n$ (with all counterterms up
to the order $\hbar^n$ included) satisfying the conditions of
Appendix \ref{App:Ren-Det:Part1-AuxCond} and such that in the asymptotic part
$\Gamma_n$ of $\Gamma^\Lambda_n\equiv\Gamma[I_n^\Lambda]$
\begin{eqnarray}\label{Eq:Gamma_n-expansion}
\Gamma_n=I_0+\sum_{k=1}^\infty\hbar^k\Gamma_n^{(k)}
\end{eqnarray}
the functionals $\Gamma_n^{(k)}$ are already $\Lambda$-independent for
$k\leq n$ and
\begin{eqnarray}
{\cal S}(\Gamma_n)=\hbar^{n+1}\Omega_n+{\cal O}(\hbar^{n+2})~\!,
\label{eqn:Omega_n}
\end{eqnarray}
it is possible to construct $I_{n+1}$ extending these results to the next
order in $n$. Useful in this, in addition to the operator
(\ref{Eq:ZinnJustinOperQuad}), is also its linearized version
$\mathcal{S}_F$ \cite{PiguetSorella} defined by
$\mathcal{S}(F+\varepsilon~\!G)=\mathcal{S}(F)+\varepsilon~\!\mathcal{S}_F(G)
+\mathcal{O}(\varepsilon^2)$, whose explicit form reads
\begin{eqnarray}
\mathcal{S}_F=
\derf{F}{K^\mu_\alpha}\!\cdot\!\derf{}{A^\alpha_\mu}+
\derf{F}{K_i}\!\cdot\!\derf{}{\phi^i}+
\derf{F}{\bar K_a}\!\cdot\!\derf{}{\psi^a}+
\derf{F}{L_\alpha}\!\cdot\!\derf{}{\omega^\alpha}+
h_\alpha\!\cdot\!\derf{}{\anti{\omega}_{\alpha}}\nonumber\\
+\derf{F}{A^\alpha_\mu}\!\cdot\!\derf{}{K^\mu_\alpha}+
\derf{F}{\phi^i}\!\cdot\!\derf{}{K_i}+
\derf{F}{\psi^a}\!\cdot\!\derf{}{\bar K_a}+
\derf{F}{\omega^\alpha} \!\cdot\! \derf{}{L_\alpha}.\phantom{aaaaa}
\label{Eq:ZinnJustinOperLin}
\end{eqnarray}
The operations $\mathcal{S}(\cdot)$ and $\mathcal{S}_F$ have two important
properties \cite{PiguetSorella}. Firstly,
\begin{eqnarray}
\mathcal{S}_F\mathcal{S}(F)=0~\!.\label{Eq:LinZJoper:wlasnosc1}
\end{eqnarray}
for any functional $F$. Secondly, if $\mathcal{S}(F)=0$, then
\begin{eqnarray}
\mathcal{S}_F^2=0~\!.\label{Eq:LinZJoper:wlasnosc2}
\end{eqnarray}
In particular, $\cS^2_{I_0}=0$.

It is the well known property of the ordinary renormalization procedure
that the lowest order divergent (in the infinite cutoff limit) part of
$\Gamma_n$, that is $\Gamma_n^{(n+1)\rm div}$, is an integral of a local
operator which can be removed by adding to $I_n$ appropriate counterterms.
Similarly, the QAP guarantees
\cite{Lam,Low}, that $\Omega_n$ in (\ref{eqn:Omega_n}) is an integral of a
local operator (of ghost number 1 and dimension $\leq5$). Moreover, using
the identity \refer{Eq:LinZJoper:wlasnosc1} applied to
$F=\Gamma_n$ in conjunction with the expansion
$\cS_{\Gamma_n}=\cS_{I_0}+\cO(\hbar)$ one learns that $\Omega_n$
satisfies the Wess-Zumino consistency condition (WZCC)
\begin{eqnarray}
\cS_{I_0}\Omega_n=0~\!.
\label{Eq:WZC}
\end{eqnarray}
Restoring the BRST invariance of $\Gamma$ in the order $\hbar^{n+1}$ relies on
the possibility of representing $\Omega_n$ in a cohomologically trivial form
\begin{eqnarray}
\Omega_n =\mathcal{S}_{I_0}\sC_n~\!,
\label{Eq:Om0=}
\end{eqnarray}
with $\sC_n$ being the integral of some local operator (of ghost number 0 and
dimension $\leq 4$), which can therefore be used as an additional (symmetry
restoring) counterterm. This is so if the representation of the gauge group
realized on fermionic fields fulfills (cf. Eq. \refer{Eq:Majorana})
\begin{eqnarray}
\tr(\TF_\alpha\{\TF_\beta,\ \TF_\gamma\})=0~\!,
\label{Eq:Non-anomaly-cond}
\end{eqnarray}
for all triplets $(\alpha,\beta,\gamma)$ of the gauge indices.\footnote{For
semisimple gauge groups the only cohomologically non-trivial solution to the
WZCC \refer{Eq:WZC} is the Adler-Bell-Jackiw anomaly, which vanishes to all
orders if \refer{Eq:Non-anomaly-cond} holds (see e.g. \cite{PiguetSorella}).
Additional (Abelian) anomalies that could potentially appear in the case of
non-semisimple gauge groups \cite{Barnich:1994ve} are excluded if the
Abelian antighost equation \refer{Eq:AbelAntiGhost} is imposed as one of the
conditions  defining the theory (see \cite{Grassi} and references therein).}

In the ``algebraic renormalization" framework usually explicit renormalization
conditions are used to fix the counterterm $\sC_n$ \cite{PiguetSorella}. Here,
aiming at constructing a mass-independent renormalization scheme, we
adopt a two-step procedure instead. In the first step a local action
\eq{\label{Eq:tilde-I-n}
\tilde I_n=I_n-\hbar^{n+1}\Gamma_n^{(n+1){\rm div}},
}
is constructed with the divergent part $\Gamma_n^{(n+1){\rm div}}$ defined in
the spirit of (the modified) minimal subtraction as the ``pure divergence",
i.e. by imposing the condition
\begin{eqnarray}\label{Eq:RenCond}
\Gamma_n^{(n+1){\rm div}}\big|_{\delta_\Lambda=0}\bigg|_{\Lambda^2=0}=0\,,
\end{eqnarray}
in which $\delta_\Lambda$ is the ``basic logarithmic divergence"
\begin{eqnarray}\label{Eq:delta_Lambda}
\delta_\Lambda\equiv\ln{\Lambda^2\over\mu^2}-1-\gamma_{\rm E}-\ln2~\!
=\ln{\bar\Lambda^2\over\mu^2}.
\end{eqnarray}
The arbitrary scale $\mu$ is introduced on dimensional ground to render
the subtraction procedure mass-independent. The ``asymptotic'' (in the
sense explained above) part $\tilde\Gamma_n$ of the effective action
$\tilde\Gamma^\Lambda_n\equiv\Gamma[\tilde I_n^\Lambda]$ obtained from the
regularized version $\tilde I^\Lambda_n$ of $\tilde I_n$ has then the
form\footnote{The form \refer{Eq:TiGaN} is correct, because quadratic
divergences are independent of external momenta. For this reason, terms
of the form
\eq{\nn
\Lambda^2 \exp\left\{ - \frac{\ell^2}{\Lambda^2}  \right\}
=\Lambda^2 - \ell^2+\cO(\Lambda^{-2})~\!,
}
will not be produced by the prescription \refer{eqn:PrescriptionDef}.
}
\begin{eqnarray}
\tilde{\Gamma}_n={\Gamma}_n-\hbar^{n+1}\Gamma_n^{(n+1){\rm div}}
+\mathcal{O}(\hbar^{n+2})~\!,\label{Eq:TiGaN}
\end{eqnarray}
and it is easy to see that
\eq{\label{Eq:Def:tilde-Delta}
\mathcal{S}(\tilde{\Gamma}_n)=\hbar^{n+1}\tilde{\Omega}_n
+\mathcal{O}(\hbar^{n+2})~\!,
}
where $\tilde{\Omega}_n$ is related to $\Omega_n$ in (\ref{eqn:Omega_n}) by
\begin{eqnarray}
\tilde{\Omega}_n\equiv{\Omega}_n
-\mathcal{S}_{{I}_0}\Gamma_n^{(n+1){\rm div}}~\!.\label{Eq:WZCC-mod}
\end{eqnarray}
As all $\Lambda$-dependent terms in $\tilde{\Gamma}_n$ are at least of  order
$\hbar^{n+2}$, Eq. \refer{Eq:Def:tilde-Delta} means that $\tilde{\Omega}_n$ is
$\Lambda$-independent. Furthermore, \refer{Eq:Om0=} (if true) implies that
\begin{eqnarray}
\tilde{\Omega}_n=-\mathcal{S}_{{I}_0}\delta_\flat\!\Gamma_n^{(n+1)}~\!.
\label{Eq:DeltaGammaCond}
\end{eqnarray}
with $\delta_\flat\!\Gamma_n^{(n+1)}$ being the integral of a
(cutoff-independent) local operator (of ghost number 0 and dimension
$\leq4$). Regularized version $I^\Lambda_{n+1}$ of the next order local action
\begin{eqnarray}
I_{n+1}=\tilde I_n+\hbar^{n+1}\ \! \delta_\flat\!\Gamma_n^{(n+1)}=I_n+
\hbar^{n+1}\left\{\delta_\flat\!\Gamma_n^{(n+1)} -\Gamma^{(n+1){\rm div}}_n\right\},
\label{Eq:Def:In+1}
\end{eqnarray}
leads then to  $\Gamma_{n+1}^\Lambda\equiv\Gamma[I_{n+1}^\Lambda]$ whose
asymptotic part $\Gamma_{n+1}$ reads
\begin{eqnarray}
{\Gamma}_{n+1}={\Gamma}_n+\hbar^{n+1}\left\{\delta_\flat\!\Gamma_n^{(n+1)}
-\Gamma_n^{(n+1){\rm div}}\right\}+\mathcal{O}(\hbar^{n+2}),\label{Eq:GaN+1}
\end{eqnarray}
and breaks the ZJ identity (\ref{eqn:ZJidBasic}) only at the $\hbar^{n+2}$ order:
\eq{
\mathcal{S}({\Gamma}_{n+1})=\hbar^{n+1}\left\{\Omega_n-\mathcal{S}_{I_0}\!
\left[\Gamma_n^{(n+1){\rm div}}\!-\!\delta_\flat\!\Gamma_n^{(n+1)}\right]
\right\}+\mathcal{O}(\hbar^{n+2})=0+\mathcal{O}(\hbar^{n+2}).
}
To complete the inductive step it is still necessary to show that $I_{n+1}$
satisfies also all the auxiliary conditions
\refer{Eq:TransWT}-\refer{Eq:Stue-2}. This is done in Appendix
\ref{App:Ren-Det:IndStep}.
\vskip0.2cm

Due to the non-triviality of ker$\cS_{I_0}$,  the counterterm
$\delta_\flat\!\Gamma_n^{(n+1)}$ is not uniquely determined by the condition
\refer{Eq:DeltaGammaCond} - any functional $\mathpzc{v}_{{}_0}$ belonging to
$\mathpzc{V}\cap\ker{}\cS_{I_0}$ can be added to it. Here $\mathpzc{V}$
denotes the vector space of integrals of local operators of dimension
$\leq4$ and zero ghost number satisfying the homogeneous versions of the
conditions \refer{Eq:TransWT}-\refer{Eq:Stue-2} and having other symmetries of
$I_0$. It is easy to check that any $\mathpzc{v}_{{}_0}\in \mathpzc{V}\cap \ker{}\cS_{I_0}$ has the form
\eqs{\label{Eq:K-ker-0}
\mathpzc{v}_{{}_0} &=&\int{{\rm d}^4 x\,}\left\{\bE^{\ka_S}_{\ \ \be_S}
\left[(K^\mu_{\ka_S}-\pa^\mu\anti{\om}_{\ka_S}) \pa_\mu\om^{\be_S}
+\derf{I_0^{GI}}{A^{\ka_S}_\mu} A_\mu^{\be_S}\right]\right.+\\
&{}&\phantom{aaaaaaa}\left.-\frac{1}{2}
z^A_{\alpha \beta}F^\alpha_{\text{ }\mu\nu}F^{\beta\text{}\mu\nu}+
z^\phi_{ij}\bracket{D_\mu\phi}^i\!\bracket{D^\mu\phi}^j
+i\!\bracket{\anti{\psi}\gamma^\mu}^a\! z^{\psi}_{ab}\bracket{D_\mu\psi}^b
-\mathpzc{w}\!\bracket{\phi,\psi}\right\},
\nn}
with the matrices $\bE$, ${z}$ (of course, $z^{\psi}={z}^F P_L+{z}^{F*} P_R$)
and the polynomial $\mathpzc{w}\!\bracket{\phi,\psi}$ constrained by the
global symmetries of $I_0$ (including those which belong to the gauge group);
moreover $z^\phi_{ij}=0$ if either $i$ or $j$ corresponds to the Stueckelberg
scalar and $\mathpzc{w}\!\bracket{\phi,\psi}$ is independent of the
Stueckelberg fields.

Remembering that the tree-level action $I_0$ is (up to a rescaling of field)
the most general functional consistent with the power-counting and a given
set of symmetries, it is easy to check that Eq. \refer{Eq:K-ker-0} can be
can be rewritten in the form (here $g^C$ denotes collectively all parameters
of $I_0$ except for components of the background $\varphi$, i.e. couplings
constants and explicit mass parameters)
\eq{\label{Eq:K-ker=Der-I-0}
\mathpzc{v}_{{}_0}=-\left\{{\fB}^C \derp{}{g^C}
-\cN_\phi(z^{\phi})-\cN_\psi(z^{\psi})-\cN_\omega(z^A)-\cN_A(\anti\bE+z^A)
\right\} I_0\,  ,
}
where
\begin{eqnarray}\label{Eq:cN-operators}
\!\!\!\!&&{\cal N}_\phi(z^\phi)=(z^\phi)^i_{\phantom{a}j}\left\{
(\phi+\varphi)^j\cdot{\delta\over\delta\phi^i}
-K_i\cdot{\delta\over\delta K_j}\right\},\nonumber\\
\!\!\!\!&&{\cal N}_\psi(z^\psi)=(z^\psi)^a_{\phantom{a}b}\left\{
\psi^b\cdot{\delta\over\delta\psi^a}
-\bar K_a\cdot{\delta\over\delta \bar K_b}\right\},\nonumber\\
\!\!\!\!&&{\cal N}_\omega(\mathpzc{z})=\mathpzc{z}^\alpha_{\phantom{a}\beta}
\left\{\omega^\beta\cdot{\delta\over\delta \omega^\alpha}-L_\alpha
\cdot{\delta\over\delta L_\beta}\right\},\\
\!\!\!\!&&{\cal N}_A(\mathpzc{z})=\mathpzc{z}^\alpha_{\phantom{a}\beta}
\left\{A_\mu^\beta\cdot{\delta\over\delta A^\alpha_\mu}
-K^\mu_\alpha\cdot{\delta\over\delta K^\mu_\beta}
-\overline\omega_\alpha\cdot{\delta\over\delta \overline\omega_\beta}
-h_\alpha\cdot{\delta\over\delta h_\beta}\right\}\,,\nonumber
\end{eqnarray}
are the ``counting operators'' \cite{PiguetSorella}, $\anti\bE$ equals
$\bE$ for non-Abelian indices and vanishes otherwise
\eq{\label{Eq:bE}
\anti\bE^{\al}_{\ \be}
=\delta^{\al}_{\ \ka_S}\bE^{\ka_S}_{\ \ \th_S}~\!\delta^{\th_S}_{\ \ \beta}\,,
}
while the coefficients ${\fB}^C$ satisfy the relations
\eq{\label{Eq:zeta-A-rel-1-000}
{\fB}^C \derp{}{g^C} T_\alpha=[z^A]^\gamma_{\ \al}\, T_\gamma
\,,\qquad\quad \text{for}
\qquad\quad T_\gamma=\TS_\gamma,\,\TM_\gamma,\,\TV_\gamma\,,
}
and
\eq{\label{Eq:zeta-A-rel-2-000}
{\fB}^C \derp{}{g^C} \bar{P}_{\alpha_A}=
[z^A]^{\de_A}_{\ \ {\al_A}}\, \bar P_{\de_A}\,.
}
Of course, the coefficients ${\fB}^C$ corresponding to non-gauge couplings
$g^C$ (which parametrize the $\mathpzc{w}(\phi,\psi)$ polynomial) are not
constrained by the relations
\refer{Eq:zeta-A-rel-1-000}-\refer{Eq:zeta-A-rel-2-000}.

The form  \refer{Eq:K-ker=Der-I-0} of $\mathpzc{v}_{{}_0}$ implies that this
functional can be obtained from $I_0$ by an infinitesimal
``finite renormalization" of its fields and couplings $g^C$. This shows that the
necessity of fixing the freedom in the form of the counterterms
$\delta_\flat\!\Gamma_n^{(n+1)}$ is equivalent to the usual necessity
of specifying the renormalization conditions.

In our approach we impose the \emph{implicit} renormalization conditions
by requiring that the counterterms $\delta_\flat\Gamma_n^{(n+1)}$ belong to
a subspace $\mathpzc{W}\subset \mathpzc{V}$ which is complementary to the
subspace $\mathpzc{V}\cap \ker{}\cS_{I_0}$, that is such that
$\mathpzc{V}=[\mathpzc{V}\cap\ker\mathcal{S}_{\mathcal I_0}]\oplus \mathpzc{W}$.
Different choices of $\mathpzc{W}$ correspond to different mass-independent
renormalization schemes. Since a generic element $\mathpzc{v}$ of
$\mathpzc{V}$ is of the form
\eqs{\nn %\label{Eq:K-ker-1}
\mathpzc{v}&=&\cJ[A,\phi,\psi]+
\tilde{\bE}^{\ka_S}_{\ \ \be_S}\!\int{\rd^4 x}\,
[K^\mu_{\ka_S}-\pa^\mu\anti{\om}_{\ka_S}]~\!\pa_\mu\om^{\be_S}\,,
}
where the functional $\cJ$ is independent of the Stueckelberg fields and
constrained by power-counting and global (gauge and other)
symmetries of $I_0$, it is easy to see that one (particularly natural)
choice is the subspace $\mathpzc{W}$ spanned by the following integrated
operators (in the symbolic form)
\eqs{\label{Eq:DodKontr:PostacSymb}
\delta_\flat\!\Gamma_n^{(n+1)}
&\in&\int
(\partial^\mu\! A_\mu)(\partial^\nu\! A_\nu)\oplus A_\mu A^\mu\oplus
A_\mu\anti{\psi}\gamma^\mu P_L\psi\oplus
A_\mu\anti{\psi}\gamma^\mu P_R\psi\oplus
\phi\phi A_\mu A^\mu\oplus
\nn\\
&{}&\ \ \ \oplus A_\mu\partial^\mu\!\phi\oplus
\phi A_\mu\partial^\mu\!\phi\oplus
\phi A_\mu A^\mu\oplus A A\,\partial\! A\oplus AAAA~\!,\quad
}
in which each component represents a set of operators with all possible
assignments of the ``color'' (and ``flavor'') indices. In the last
two terms suppressed Lorentz indices have to be contracted in a
Lorentz-invariant way. The counterterm \refer{Eq:DodKontr:PostacSymb}
vanishes for $A=0$ and
does \emph{not} involve the $A_\mu \partial_\nu \partial^\nu\! A^\mu$
operator nor the Stueckelberg fields. We will call this choice
the  $\La$-$\anti{\rm MS}$ scheme.
\vskip0.2cm

As a result of the procedure outlined above the action $I^\Lambda_\infty$ is
constructed which, modulo exponents introduced according to the prescription
(\ref{eqn:PrescriptionDef}), has a renormalizable form but is obviously not 
BRST symmetric. In typical
applications of the procedure, mentioned at the beginning of this section,
the structure of the resulting BRST symmetry violating counterterms is not
very interesting in itself - the counterterms serve only as a technical
mean to consistently calculate finite amplitudes satisfying the appropriate
identities (which embody the requirements of the BRST invariance).
Therefore one usually does not exploit the fact that, as will be shown in
Section \ref{sec:rg}, the action $I_\infty^\Lambda$, can be given the
interpretation of the ``bare'' action $I_{\rm B}$ expressed in terms of the
``bare'' parameters. This fact, however, will be crucial in discussing our
view on the hierarchy problem in Section \ref{sec:HP}.

Before closing this section an important comment must be made. From the
above description of the procedure for constructing counterterms it is
clear that the full BRST invariance of the effective
action $\Gamma$ (i.e. the ZJ identity) is recovered only in the strict
limit $\Lambda\rightarrow\infty$. This is perfectly fine if one does not
ask about the origin of the low energy field theory model and is interested
only in obtaining renormalized (finite) amplitudes satisfying the
requirements of the BRST symmetry. On the other hand, if the bare action
and the cutoff $\Lambda$ are to be given a physical meaning (and the limit
$\Lambda\rightarrow\infty$ is not to be taken), one has to assume that the
complete bare action $I_{\rm B}$ has additional terms, suppressed by inverse
powers of $\Lambda$, which are not obtained with the help of the outlined
procedure applied to the regularized renormalizable action
(\ref{Eq:I_0=I_0^GI+I_0^Rest}), and which conspire to restore the full BRST
invariance of the amplitudes. Indeed, the experimental limit on the photon
mass $M_\gamma<10^{-18}$ eV \cite{PDG} does not leave room for BRST (gauge)
symmetry breaking at order $M^4_{\rm top}/\Lambda^2$ (or $M_W^4/\Lambda^2$),
even for $\Lambda$ as high as the Planck scale. %
As pointed out in the discussion of the regularization prescription \refer{eqn:PrescriptionDef}, in the complete Euclidean action additional terms postulated here may be also important in the problem of the non-perturbative continuation to the Minkowski space-time.
We do not attempt here to
determine the form of these terms. We only point out that such a situation
can be somewhat analogous to the one encountered in superstring theory: while
the anomaly is shown to cancel out exactly at the string theory level, the
minimal supergravity - the effective low energy theory of massless string
excitations derived from string tree-level amplitudes is anomalous. Making
it anomaly-free requires modifying the field strength $H=dB$ by adding
a term which originates from one-loop string amplitudes; this correction
taken alone breaks supersymmetry; restoring  supersymmetry
reintroduces, in turn, the anomaly and so on.

\section{Determination of the BRST symmetry restoring counterterms}
\label{sec:cterms}

\renewcommand{\thesection}{\arabic{section}}
\renewcommand{\theequation}{\arabic{section}.\arabic{equation}}
\renewcommand{\thefigure}{\arabic{section}.\arabic{figure}}

\setcounter{equation}{0}
\setcounter{figure}{0}

At the one-loop order the ZJ identity (\ref{eqn:ZJidBasic}) is equivalent
to the condition
\begin{eqnarray}\label{Eq:ZJ-one-loop}
\mathcal{S}_{I_0}\, \Gamma^{(1)} =0,
\end{eqnarray}
where $\Gamma^{(1)}\equiv\Gamma^{(1)}_{1}$ is the one-loop contribution to
the renormalized 1PI effective action (for the notation, see
\refer{Eq:Gamma_n-expansion}). In our renormalization scheme (see Section
\ref{sec:Subtractions}) the BRST symmetry restoring
counterterm $\delta_\flat\!\Gamma^{(1)}_0$ must be of the form
\refer{Eq:DodKontr:PostacSymb}. In order to determine the coefficients of its
individual terms it is sufficient to consider the derivative of
\refer{Eq:ZJ-one-loop} with respect to the ghost field restricted to
the ``physical submanifold''
\begin{eqnarray}
\omega^\alpha=\anti{\omega}_\alpha=K^\mu_\alpha=K_i=\bar{K}_a=L_\alpha=h_\alpha=0.
\end{eqnarray}
In the momentum space, cf. Eq. \refer{Eq:Mom-Pos-Rel}, the resulting
identity reads\footnote{To simplify the notation we write
$\Gamma^{(1)}_{ ph}=\left.\Gamma^{(1)}\right|_{ph}$.}
\begin{eqnarray}\label{Eq:WTgenJaw}
il_\mu \derf{{\Gamma}^{(1)}_{ph}}{\tilde{A}^{\gamma}_{\mu}(l)}
&+&
\int\!{\rm d}^4 p\ \! \mind{e}{\alpha}{\!\gamma\beta}
\tilde{A}^{\beta}_{\mu}(p-l) \derf{{\Gamma}^{(1)}_{ph}}{\tilde{A}^{\alpha}_{\mu}(p)}
+\int\!{\rm d}^4 p\ \!
\matrixind{\TM_\gamma\tilde{\psi}(p-l)}{a}{\!\!}\derf{{\Gamma}^{(1)}_{ph}}
{\tilde{\psi}^{a}(p)}+\nn\\&{}&\hspace*{-60 pt}
+\int\!{\rm d}^4 p\ \!
\matrixind{\TS_\gamma\tilde{\phi}(p-l)}{i}{\!\!}
\derf{{\Gamma}^{(1)}_{ph}}{\tilde{\phi}^{i}(p)}+
\matrixind{\TS_\gamma\varphi}{i}{\!\!}
\derf{{\Gamma}^{(1)}_{ph}}{\tilde{\phi}^{i}(l)}+\nn\\
&{}&\hspace*{-60 pt}+
\int\volfour{p}\derf{{I}^{GI}_{0}}{\tilde{A}^{\alpha}_{\mu}(p)}
\left[\derf{}{\tilde{\omega}^{\gamma}(l)}
\derf{{\Gamma}^{(1)}}{\tilde{K}_{\alpha}^{\mu}(-p)}\right]_{ph}
\!\!\!\!-\!
\int\volfour{p}\derf{{I}^{GI}_{0}}{\tilde{\psi}^{a}(p)}
\left[\derf{}{\tilde{\omega}^{\gamma}(l)}
\derf{{\Gamma}^{(1)}}{\tilde{\bar{K}}_{a}(-p)}\right]_{ph}
\!\!\!\!+\nn\\
&{}&\hspace*{-60 pt}+
\int\volfour{p}\derf{{I}^{GI}_{0}}{\tilde{\phi}^{i}(p)}
\left[\derf{}{\tilde{\omega}^{\gamma}(l)}
\derf{{\Gamma}^{(1)}}{\tilde{{K}}_{i}(-p)}\right]_{ph}=0,
\end{eqnarray}
(the sum $\TS_\gamma\varphi+\bar{P}_\gamma$ appearing in \refer{Eq:BRS} has
been replaced here by $\TS_\gamma\varphi$, because the Stueckelberg fields,
if present, are free in the Landau gauge -- see Eq. \refer{Eq:Stue-1}).

As it is easy to realize (by looking at the Feynman rules), the last
three terms of the left hand side of \refer{Eq:WTgenJaw} vanish if the
index $\gamma$ corresponds to an Abelian generator. The identity
\refer{Eq:WTgenJaw} takes then the form of the  standard QED-like
Ward-Takahashi (WT) identity.\footnote{Thus, Abelian ideals do not have to
be considered separately -- relevant constraints are already contained in
the identity \refer{Eq:WTgenJaw}. This statement generalizes
to higher orders, because the regularization (\ref{eqn:PrescriptionDef})
automatically preserves the Abelian antighost equation \cite{Grassi},
see also Eq. \refer{Eq:AbelAntiGhost}, in the infinite cutoff limit.
In particular, Abelian WT identities follow from the ZJ identity
(\ref{eqn:ZJidBasic}) as a consequence of the algebraic relation
\refer{Eq:RelKom-ZJ--abel-AntiGh}.}

Taking functional derivatives of \refer{Eq:WTgenJaw} w.r.t. ``physical''
fields and setting all fields to zero one obtains various Slavnov-Taylor
(ST) identities. If the first term on the LHS, obtained as a result of
differentiation of \refer{Eq:WTgenJaw}, is a 1PI function $X$, we call
the resulting relation ``the identity involving the $X$ function''. At
the one-loop order the 1PI functions related by a given ST identity receive
contributions from bare one-loop diagrams,\footnote{As there is no one-loop
contribution to the function $\VEV{K_{i} \omega^\gamma}$, the last term of
\refer{Eq:WTgenJaw} does not contribute if all  differentiations act on the
$I_0^{GI}$ factor. For this reason and because non-minimal counterterms are
not allowed for this function (cf. \refer{Eq:DodKontr:PostacSymb}), all
terms with $\VEV{K_{i}\omega^\gamma}$ are omitted in the formulae below.}
from minimal counterterms and from non-minimal ones. The strategy which we
follow below is to take a ST identity and compute first the contributions
(marked by the superscript $\OLB$) of regularized bare one-loop diagrams.
Because the regularization (\ref{eqn:PrescriptionDef}) (``$\La$Reg") breaks the BRST
invariance, these contributions to the ST identity do not sum up to zero,
but according to the QAP their sum, denoted $\Omega$ with appropriate
indices, should be local in the infinite cutoff limit. This can be verified by doing more or less
standard manipulations on regularized integrals. Since the calculations
are rather lengthy, we do not show their details except for one case:
in \ref{Sec:ApB-AAAA} we outline the steps necessary to work out
the contribution of bare one-loop fermionic diagrams to the identity
involving the $\VEV{AAAA}$ function. The functions $\Omega$ obtained in
this way represent one-loop breakings of the respective ST identities and
have the obvious interpretation of appropriate derivatives w.r.t. to fields
of the functional $\Omega_0$ defined in \refer{eqn:Omega_n}. 

The next step
is to take into account minimal counterterms specified by the prescription
\refer{Eq:RenCond}. The resulting one-loop breaking factors $\tilde\Omega$
with appropriate indices are just the appropriate
derivatives w.r.t. to fields of the functional $\tilde\Omega_0$ defined in
\refer{Eq:Def:tilde-Delta}. From \refer{Eq:WZCC-mod} and
\refer{Eq:Def:tilde-Delta} it follows that obtaining $\tilde\Omega$'s
reduces to setting to zero in the corresponding $\Omega$'s all factors
$\delta_\Lambda$
defined in (\ref{Eq:delta_Lambda}) and all terms proportional to $\Lambda^2$.
(In fact, the universality of one-loop logarithmic divergences
makes it clear that factors $\delta_\Lambda$ cannot appear in
$\Omega$'s and to obtain $\tilde\Omega$'s it is enough to set quadratic
divergences to zero in the corresponding $\Omega$'s).

The last step is the determination of the non-minimal counterterms,
which in principle means
solving Eq. \refer{Eq:DeltaGammaCond} with the auxiliary
condition \refer{Eq:DodKontr:PostacSymb}.
Before presenting the systematic of this procedure, we remark that there
is an alternative way of obtaining the necessary one-loop breaking factors
$\Omega$. It relies on the fact that the bare 1PI functions
$\tig^{\OLB}_{{\rm Dim}}$ calculated using DimReg do satisfy the ST
identities, provided the naive definition of $\gamma^5$ is
employed.\footnote{Terms which are ambiguous due to using the
anticommuting  $\gamma^5$ in $d$-dimensions vanish if the condition
\refer{Eq:Non-anomaly-cond} is fulfilled.}
Thus, replacing in the ST identities each bare one-loop 1PI function
$\tig^{\OLB}\equiv\tig^{\OLB}_{\Lambda}$ calculated in our regularization
\refer{eqn:PrescriptionDef} by the difference
\begin{eqnarray}
\Delta\tig^\OLB \equiv\tig^{\OLB}-\tig^{\OLB}_{\rm Dim},\label{Eq:DeltaReg:def}
\end{eqnarray}
must produce the same factors $\Omega$. The necessary differences
$\Delta\tig^\OLB $, which will also be used in Sections \ref{Sec:RelSch}
and \ref{sec:rg} to derive the two-loop RGE satisfied by the
renormalized parameters in our scheme are calculated in \ref{App:Diag}.
This approach is obviously much simpler than the direct calculations in
$\La$Reg, firstly, because the differences (\ref{Eq:DeltaReg:def}) are
already local expressions\footnote{
Strictly speaking, functions on the RHS of \refer{Eq:DeltaReg:def} depend on two different sets of couplings, say, $\{g^{C}\}$ and $\{\check{g}^{C}\}$. However, 
as will be shown in the next section, $g^{C}-\check{g}^{C}=\cO(\hbar)$ and thus the resulting non-localities are of $\cO(\hbar^2)$ order. Similarly, we assume here that non-local terms of order of $\cO(\La^{-1})$ (or $\cO(d-4)$) are neglected. 
}  and, secondly, because in this method the only
1PI function with antifields that contributes to the factors $\Omega$ is
$\VEV{K^\mu_\al\om^\be}$ (the corresponding difference is given in
\refer{Eq:DeltaZ:KmuOmega}); the remaining functions with antifields are
the same in DimReg and $\La$Reg (even though the degree of divergence may
indicate otherwise) due to the additional ``symmetry" \refer{Eq:AntiGhost}
of ${\cal L}_0^{Rest}$ \refer{Eq:LagrTreeRest} in the
Landau gauge, which is preserved by both regularizations.

We stress however that, except for the bosonic
contribution to the identity involving the $\langle AAAA\rangle$
function, all factors $\Omega$ have been computed directly in $\La$Reg (along the lines described in \ref{Sec:ApB-AAAA}) and 
the results are, therefore, unaffected by ambiguities of DimReg with the naive prescription for $\ga^5$.

Systematic determination of non-minimal counterterms restoring the BRST
symmetry consists of considering first those ST identities in which only
one 1PI function can have such a counterterm (this is established by
inspection of the allowed set \refer{Eq:DodKontr:PostacSymb} of non-minimal
counterterms) and moving successively
to those in which more functions can have non-minimal counterterms but
only one such counterterm which has not been determined yet.
We have divided these steps into separate subsections.\footnote{Since minimal counterterms can be immediately obtained
from divergent parts of formulae listed in \ref{App:Diag} we do not give
them explicitly here; those needed for the calculation of the
$\cO(\hbar^2)$ vacuum graphs are given in  Section \ref{Vac:Diag}.}

\subsection{Identity involving the $\VEV{\psi\psi A}$ function }
\label{Sec:WT:psipsiA}
Functionally differentiating \refer{Eq:WTgenJaw} twice w.r.t the Majorana
fields one obtains the identity (we use the notation explained in
\refer{Eq:Notacja1}-\refer{Eq:Notacja2})
\begin{eqnarray}\label{Eq:WT-psi-psi-A}
{\phantom{a}}&{\phantom{=}}&\hspace*{-20 pt}
i l_\mu \tig^{\phantom{b_1b_2}\mu}_{b_1b_2\gamma}(k_1,k_2,l)^\OL
+\TM^a_{\gamma\,  b_1} \tig_{ab_2}(k_1+l,k_2)^\OL
+\tig_{b_1a}(k_1,k_2+l)^\OL \TM^a_{\gamma\, b_2}
+\nn\\
{}&{}&\!\!\hspace*{-10 pt}
+(\TS_\gamma \varphi)^n \tig^{\phantom{b_1b_2}}_{b_1b_2 n}(k_1,k_2,l)^\OL
-\mathcal{V}^\prime_n(\varphi)
\VEVOPITIL{\tilde{K_n}(0)\tilde{\psi}^{b_1}(k_1)\tilde{\psi}^{b_2}(k_2)
\tilde{\omega}^\gamma(l)}^\OL
+\nn\\
{}&{}&\!\!\hspace*{-10 pt}
+i (C\gamma^\mu  \TM_\alpha)_{b_1 b_2}
\VEVOPITIL{\tilde{K}^\mu_\alpha(k_1\!+\!k_2) \tilde{\omega}^\gamma(l)}^\OL
%+
%\nn\\{}&{}&\!\!\hspace*{-10 pt}
-[C(\ds{k}_1\!-m_F)]_{ab_1}\!
\VEVOPITIL{\tilde{\bar{K}}_a(k_1)
\tilde{\omega}^\gamma(l)\tilde{\psi}^{b_2}(k_2)}^\OL+\nn\\
{}&{}&\!\!\hspace*{-10 pt}
+[C(\ds{k}_2\!-m_F)]_{ab_2}\!
\VEVOPITIL{\tilde{\bar{K}}_a(k_2)
\tilde{\omega}^\gamma(l)\tilde{\psi}^{b_1}(k_1)}^\OL=0.\qquad
\end{eqnarray}
The contribution $\Omega^{\phantom{b_1b_2}\phantom{\mu}}_{b_1b_2\gamma}(k_1,k_2,l)$
of the bare one-loop diagrams to the LHS of \refer{Eq:WT-psi-psi-A}
(in the limit of infinite cutoff)  is
%\footnote{This result can be obtained from Eqs.
%\refer{Eq:DeltaZ:PsiPsiA:Tot}, \refer{Eq:DeltaZ:PsiPsi:AiB},
%\refer{Eq:DeltaZ:PsiPsiPhi:Tot} and \refer{Eq:DeltaZ:KmuOmega}.}
\begin{eqnarray}
\Omega^{\phantom{b_1b_2}\phantom{\mu}}_{b_1b_2\gamma}(k_1,k_2,l)
&=&\frac{1}{2(4\pi)^2}\Big\{C\,l_\mu \gamma^\mu
\left[\TM_\epsilon\TM^\epsilon\TM_\gamma+\frac{3}{2}\left(1-\ln\frac{3}{4}\right)
\mind{e}{\kappa}{\!\epsilon\beta}\mind{e}{\epsilon\beta}{\!\gamma}\TM_\kappa
+\YM_i^* \YM_j \TS_\gamma^{\,ij}+\nn\right.\\
&{}&\left.%\hspace*{-15 pt}
\qquad\qquad\qquad\ \
-\left(\frac{1}{3}+\ln\frac{3}{4}\right)
\YM_i^* \YM^i \TM_\gamma\right]\Big\}_{b_1 b_2}.\nn
\end{eqnarray}
Since $\Omega^{\phantom{b_1b_2}\phantom{\mu}}_{b_1b_2\gamma}(k_1,k_2,l)$ turns
out to be $\Lambda$-independent, it is just equal
$\tilde\Omega^{\phantom{b_1b_2}\phantom{\mu}}_{b_1b_2\gamma}(k_1,k_2,l)$
(notice that none of the 1PI functions involved is quadratically divergent).
Inspection of \refer{Eq:DodKontr:PostacSymb} reveals that only the vertex
$\VEV{\psi\psi A}$ can have a non-minimal counterterm, of the general form
\begin{eqnarray}
\label{Eq:FunDeltaFlat_PsiPsiA}
\delta\tig^{\phantom{b_1 b_2}\mu}_{b_1 b_2\gamma}(k_1,k_2,l)
=i\left(C\gamma^\mu\delta^\flat\! \widehat{\mathscr{X}}_{F\gamma}\right)_{b_1 b_2},
\end{eqnarray}
(for simplicity we write $\delta\Gamma$
rather than $\delta_\flat\!\Gamma^{(1)}_0$ from now on). The equation
\begin{eqnarray}
\tilde\Omega^{\phantom{b_1b_2}\phantom{\mu}}_{b_1b_2\gamma}(k_1,k_2,l)
+i\,l_\mu \delta\tig^{\phantom{b_1 b_2}\mu}_{b_1 b_2\gamma}(k_1,k_2,l)=0,
\end{eqnarray}
necessary for fulfilling the ST identity \refer{Eq:WT-psi-psi-A}
has the unique solution
\begin{eqnarray}
\delta^\flat\!\widehat{\mathscr{X}}_{F\gamma}
&=&\frac{1}{2(4\pi)^2}\left[\TM_\alpha\TM^\alpha\TM_\gamma
+\frac{3}{2}\left(1-\ln\frac{3}{4}\right)
\mind{e}{\kappa}{\!\alpha\beta}\mind{e}{\alpha\beta}{\!\gamma}\TM_\kappa
+\YM_i^* \YM_j \TS_\gamma^{\,ij}+\nn\right.\\&{}&\left.
%\hspace*{-15 pt}
\qquad\ \ \ \ \
-\left(\frac{1}{3}+\ln\frac{3}{4}\right)\YM_i^* \YM^i \TM_\gamma\right].
\label{Eq:KontrFlat:X_F}
\end{eqnarray}

%%%%%%%%%%%%%%%%%%%%%%%%%%%%%%%%%%%%%%%%%%%%%%%%%%%%%%%%%%%%%%%%%%%%%%%%%%%%

\subsection{Identity involving the $\VEV{\phi A}$ function}
\label{Sec-SubSec-Phi-A}
%\vspace*{-0.5 cm}
Functional differentiation of \refer{Eq:WTgenJaw} w.r.t the scalar field
yields the ST identity
\begin{eqnarray}\label{Eq:WT-phi-A}
{\phantom{a}}&{\phantom{=}}& \!\!\!\!\!\!
i l_\mu \tig^{\phantom{i}\mu}_{i\alpha}(p,l)^\OL+
\TS^j_{\alpha i} \tig_{j}(p+l)^\OL+
(\TS_\alpha \varphi)^j \tig_{ij}(p,l)^\OL+\nn\\
{}&{}&\!\!-\mathcal{V}^\prime_j(\varphi)
\VEVOPITIL{\tilde{K_j}(0)\tilde{\phi}^i(p) \tilde{\omega}^\alpha(l)}^\OL
-i p^\mu (\TS_\gamma \varphi)_i
\VEVOPITIL{\tilde{K}^\mu_\gamma(-l) \tilde{\omega}^\alpha(l)}^\OL=0 \,.\qquad
\end{eqnarray}
The contribution $\Omega_{i\alpha}(p,l)$ of bare one-loop diagrams reads
\begin{eqnarray}\label{Eq:WT-br-phi-A--jaw}
\Omega_{i\alpha}(p,l)
&=&
\frac{l^2}{(4\pi)^2}\left\{
\left(\frac{7}{12}+\ln\frac{3}{2}\right)
\tr\big[\YF_i\TF_\alpha m_F^*-\YF^*_i m_F \TF_\alpha\big]
+
\right.\\&{}&\left.
\qquad\quad
+
\frac{3}{4}\ln\frac{3}{4}
\left(\varphi^{\rm T}\TS^\kappa\TS_\alpha\TS_\kappa\right)_i
-
\frac{3}{4}\ln\frac{3}{2}
\left(\varphi^{\rm T}\TS_\alpha\TS^\kappa\TS_\kappa\right)_i
\right\}.
\nn
\end{eqnarray}
Again, since $\Omega^{\rm div}_{i\alpha}(p,l)=0$,
$\tilde\Omega_{i\alpha}(p,l)=\Omega_{i\alpha}(p,l)$. Only the $\VEV{\phi A}$
function can have a non-minimal counterterm of the form
\begin{eqnarray}\label{Eq:FunDeltaFlat_PhiA}
\delta\tig^{\phantom{i}\mu}_{i\alpha}(p,-p)
=i\ \! p^\mu \delta^\flat\!c_{i\alpha},
\end{eqnarray}
in which $\delta^\flat\!c_{i\alpha}$ is a constant matrix. The Slavnov-Taylor
identity \refer{Eq:WT-phi-A} requires
\begin{eqnarray}\label{Eq:Eq-for-del-phi-A}
%\left\{\Omega_{i\alpha}(p,-p)-\Omega^{\rm div}_{i\alpha}(p,-p)\right\}
\tilde\Omega_{i\alpha}(p,-p)-i\ \! p_\mu \delta\tig^{\phantom{i}\mu}_{i\alpha}(p,-p)=0,
\end{eqnarray}
whose unique solution is
\begin{eqnarray}
\label{Eq:KontrFlat:c}
\delta^\flat\!\! c_{i\alpha}
&=&
-\frac{1}{(4\pi)^2}\left\{
\left(\frac{7}{12}+\ln\frac{3}{2}\right)
\tr\big[\YF_i\TF_\alpha m_F^*-\YF^*_i m_F \TF_\alpha\big]
+
\right.\\\nn&{}&\left.
\qquad\qquad
+
\frac{3}{4}\ln\frac{3}{4}
\left(\varphi^{\rm T}\TS^\kappa\TS_\alpha\TS_\kappa\right)_i
-
\frac{3}{4}\ln\frac{3}{2}
\left(\varphi^{\rm T}\TS_\alpha\TS^\kappa\TS_\kappa\right)_i
\right\}.
\end{eqnarray}

%%%%%%%%%%%%%%%%%%%%%%%%%%%%%%%%%%%%%%%%%%%%%%%%%%%%%%%%%%%%%%%%%%%%%%%%%%%%%

\subsection{Identity involving the $\VEV{A A}$ function}

%\vspace*{-0.3 cm}
The identity involving the vacuum polarization tensor reads
\begin{eqnarray}\label{Eq:WT-A-A}
{\phantom{a}}&{\phantom{=}}& \!\!\!\!\!\!
i l_\mu \tig^{\mu\nu}_{\alpha\beta}(l,p)^\OL
+(\TS_\alpha \varphi)^i \tig^{\phantom{i}\nu}_{i\beta}(l,p)^\OL
-\mathcal{V}^\prime_i(\varphi)
\VEVOPITIL{\tilde{K_i}(0) \tilde{\omega}^\alpha(l) \tilde{A}^\beta_\nu(p)}^\OL
+\nn\\
{}&{}&\!\!+\left\{
{m_V^2}_{\beta\kappa}\eta^{\nu\sigma}
+{\delta}_{\beta\kappa}\bracket{p^\nu p^\sigma - \eta^{\nu\sigma}p^2}
\right\}
\VEVOPITIL{\tilde{K}^\sigma_\kappa(p) \tilde{\omega}^\alpha(l)}^\OL=0.
\end{eqnarray}
The breaking $\Omega^{\phantom{\mu}\nu}_{\alpha\beta}(l,p)$ calculated directly
from the bare one-loop diagrams has the form
%\footnote{This result can be also obtained from Eqs.
%\refer{Eq:DeltaZ:AA:AplusBplusC}, \refer{Eq:DeltaZ:PhiA:A} and
%\refer{Eq:DeltaZ:KmuOmega}.}
\begin{eqnarray}
\Omega^{\phantom{\mu}\nu}_{\alpha\beta}(l,p)
=i\ \! l^\nu\frac{1}{(4\pi)^2} \mathbb{W}^{}_{\alpha\beta}(l,p),
\end{eqnarray}
where
\begin{eqnarray}\label{Eq:DeltaAA}
\mathbb{W}^{}_{\alpha\beta}(l,p)
&=&\left\{\Lambda^2+\frac{l^2}{3}\right\}\tr\big[\TF_\alpha \TF_\beta\big]
-\left\{\frac{1}{3}+\ln\frac{3}{4}\right\}
\tr\big[\!\left\{\TF_\alpha,\ \TF_\beta\right\}m_F^* m_F\big]
+\nn\\
&{}&\hspace*{-35 pt}-\left\{\frac{1}{3}-2\ln\frac{3}{4}\right\}
\tr\big[\TF_\alpha m_F^*  \TF_\beta^* m_F\big]
-\left\{\frac{\Lambda^2}{2}+\frac{5}{48}l^2\right\}
\tr\big[\TS_\alpha \TS_\beta\big]
+\frac{1}{2}\tr\big[m_S^2\TS_\alpha \TS_\beta\big]\!+\!\!\!\nn\\
&{}&\hspace*{-35 pt}-\frac{3}{4}\ln\frac{3}{4}\ \!
\varphi^{\rm T}\!\!\left\{\TS_\beta,\ \TS^\kappa\right\}\!\TS_\alpha\TS_\kappa
\varphi-\left\{{\Lambda^2}+\frac{5}{24}l^2\right\}
\tr\big[\TV_\alpha \TV_\beta\big]+\nn\\
&{}&\hspace*{-35 pt}+\frac{3}{4}\left\{2-\ln\frac{3}{4}\right\}
\tr\big[m_V^2\TV_\alpha \TV_\beta\big]\,.
\end{eqnarray}
Taking into account minimal counterterms (i.e. setting $\Lambda^2$ to zero
in $\Omega^{\phantom{\mu}\nu}_{\alpha\beta}(l,p)$) yields
$\tilde\Omega^{\phantom{\mu}\nu}_{\alpha\beta}(l,p)$.
Comparison of \refer{Eq:WT-A-A} with \refer{Eq:DodKontr:PostacSymb}
reveals that two non-minimal counterterms can contribute to \refer{Eq:WT-A-A}:
the already determined counterterm \refer{Eq:FunDeltaFlat_PhiA}
(contributing to the $\VEV{\phi A}$ function) and the one for the vacuum
polarization which must be of the general form
\begin{eqnarray}
\label{Eq:KontrFlat:Fun:AA}
\delta\tig^{\nu\rho}_{\!\beta\kappa}(p,-p)
=\eta^{\nu\rho}(\delta^\flat\!m^2_{V})_{\beta\kappa}
+p^{\nu}p^{\rho}(\delta^\flat\!z_A)_{\beta\kappa},
\end{eqnarray}
with symmetric matrices $\delta^\flat\!m^2_{V}$ and $\delta^\flat\!z_A$.
Fulfilling the identity \refer{Eq:WT-A-A} requires that
\begin{eqnarray}
\tilde\Omega^{\phantom{\mu}\nu}_{\alpha\beta}(l,-l)
%\big|_{\Lambda^2=0}
+i l_\mu \delta\tig^{\mu\nu}_{\alpha\beta}(l,-l)
+(\TS_\alpha \varphi)^i \delta\tig^{\phantom{i}\nu}_{i\beta}(l,-l)=0.
\end{eqnarray}
Using the explicit form \refer{Eq:KontrFlat:c} of
$\delta\tig^{\phantom{i}\nu}_{i\beta}(l,-l)$  one finds the unique solution:
\begin{eqnarray}\label{Eq:KontrFlat:z_A}
(\delta^\flat\!z_A)_{\alpha\beta}
=\frac{1}{(4\pi)^2}\left\{-\frac{1}{3} \tr\big[\TF_\alpha \TF_\beta\big]
+\frac{5}{48} \tr\big[\TS_\alpha \TS_\beta\big]
+\frac{5}{24} \tr\big[e_\alpha e_\beta\big]\right\},
\end{eqnarray}
and
\begin{eqnarray}\label{Eq:KontrFlat:m_V}
(4\pi)^2(\delta^\flat\!m_V^2)_{\alpha\beta}
&=&-\left\{\frac{1}{4}+\ln2\right\}
\tr\big[\!\left\{\TF_\alpha,\ \TF_\beta\right\}m_F^* m_F\big]+\\
&{}&+\left\{\frac{3}{2}+2\ln2\right\}
\tr\big[\TF_\alpha m_F^*  \TF_\beta^* m_F\big]
-\frac{1}{2}\tr\big[m_S^2\TS_\alpha \TS_\beta\big]\!+\!\!\!\nn\\
&{}&
%\hspace*{-35 pt}
-\frac{3}{4}\left\{2-\ln\frac{3}{4}\right\}
\tr\big[m_V^2\TV_\alpha \TV_\beta\big]
+\frac{3}{4}\ln\frac{3}{4}\ \!
\varphi^{\rm T}\!\!\left\{\TS_\beta,\ \TS^\kappa\right\}\!\TS_\alpha\TS_\kappa
\varphi+\nn\\\nn
&{}&%\hspace*{-35 pt}
+\frac{3}{4}\ln\frac{3}{4}\ \!
\varphi^{\rm T}\TS^\kappa\TS_\beta\TS_\kappa \TS_\alpha\varphi
-\frac{3}{4}\ln\frac{3}{2}\ \!
\varphi^{\rm T}\TS_\beta\TS^\kappa\TS_\kappa \TS_\alpha\varphi.
\end{eqnarray}

%%%%%%%%%%%%%%%%%%%%%%%%%%%%%%%%%%%%%%%%%%%%%%%%%%%%%%%%%%%%%%%%%%%%%%%%%%%%%%%%
\subsection{Identity involving the $\VEV{\phi\phi A}$ function}

\vspace*{-0.2 cm}
The ST identity involving the  $\tig^{\phantom{ij}\mu}_{ij\alpha}$ vertex has
the form
\begin{eqnarray}\label{Eq:WT-phi-phi-A}
{\phantom{a}}&{\phantom{=}}&\hspace*{-20 pt}
i l_\mu \tig^{\phantom{ij}\mu}_{ij\alpha}(p,p',l)^\OL
+\TS^n_{\alpha\,  i} \tig_{nj}(p+l,p')^\OL
+\tig_{in}(p,p'+l)^\OL \TS^n_{\alpha\, j}+\nn\\
{}&{}&\!\!\hspace*{-10 pt}
+(\TS_\alpha \varphi)^n \tig_{ijn}(p,p',l)^\OL-\mathcal{V}'_n(\varphi)
\VEVOPITIL{\tilde{K_n}(0)\tilde{\phi}^i(p)
\tilde{\phi}^j(p') \tilde{\omega}^\alpha(l)}^\OL+\nn\\
{}&{}&\!\!\hspace*{-10 pt}
-i p'{}^\mu (\TS_\gamma \varphi)_j
\VEVOPITIL{\tilde{\phi}^i(p)\tilde{K}^\mu_\gamma(p') \tilde{\omega}^\alpha(l)}^\OL
-i p^\mu (\TS_\gamma \varphi)_i
\VEVOPITIL{\tilde{\phi}^j(p')\tilde{K}^\mu_\gamma(p) \tilde{\omega}^\alpha(l)}^\OL
+\nn\\
{}&{}&\!\!\hspace*{-10 pt}
+(p^2\!-m_S^2)_{in}\VEVOPITIL{\tilde{K}_n(p)\tilde{\phi}^j(p')\tilde{\omega}^\alpha(l)}^\OL
\!+\!(p'{}^2\!-m_S^2)_{jn}\VEVOPITIL{\tilde{K}_n(p')\tilde{\phi}^i(p)\tilde{\omega}^\alpha(l)}^\OL
\!+\!\nn\\
{}&{}&\!\!\hspace*{-10 pt}
+i (p-p')^\mu  \TS_{\gamma\,ji}
\VEVOPITIL{\tilde{K}^\mu_\gamma(p+p') \tilde{\omega}^\alpha(l)}^\OL=0.\qquad
\end{eqnarray}
The contribution of purely one-loop diagrams to the LHS of
\refer{Eq:WT-phi-phi-A} in the limit $\Lambda\to\infty$ is finite and reads
%\footnote{This result can be also obtained from Eqs.
%\refer{Eq:DeltaZ:PhiPhiA}, \refer{Eq:DeltaZ:PhiPhi:A+B} and
%\refer{Eq:DeltaZ:KmuOmega}.}
\begin{eqnarray}
\tilde\Omega^{\phantom{ij}\phantom{\mu}}_{ij{\alpha}}(p,p',l)
&=&
\frac{1}{(4\pi)^2}(p^2-p'{}^2)
\left\{
\left(\frac{7}{12}+\ln\frac{3}{2}\right)
\tr\big[\YF_i\TF_\alpha Y_j^*-\YF^*_i Y_j \TF_\alpha\big]
+
\right.\\&{}&\left.%\hspace*{-15 pt}
\qquad\qquad\qquad\quad\
+
\frac{3}{4}
\left[
\ln2\, (\TS^\kappa\TS_\alpha\TS_\kappa)_{ij}
-\ln\frac{3}{2}\, \mind{e}{\kappa\delta}{\!\alpha}(\TS_\kappa\TS_\delta)_{ij}
\right]\right\},\nn
\end{eqnarray}
In agreement with the expectation
$\tilde\Omega^{\phantom{ij}\phantom{\mu}}_{ij{\alpha}}(p,p',l)$ is related to
\refer{Eq:WT-br-phi-A--jaw} by
\begin{eqnarray}\label{Eq:Del-Del-shift}
\tilde\Omega^{\phantom{ij}\phantom{\mu}}_{ij{\alpha}}(p,0,-p)
=\derp{}{\varphi^j}  ~\!\tilde\Omega_{i\alpha}(p,-p) \, .
\end{eqnarray}
According to \refer{Eq:DodKontr:PostacSymb} only the function
$\VEV{\phi\phi A}$ can have a non-minimal counterterm. Its form
\begin{eqnarray}
\label{Eq:FunDeltaFlat_PhiPhiA}
\delta\tig^{\phantom{ij}\mu}_{ij\alpha}(p,p',l)
=i \left(p^\mu \delta^\flat\! \mathscr{X}_{\alpha\, ji}
+p'{}^\mu \delta^\flat\! \mathscr{X}_{\alpha\, ij}\right),
\end{eqnarray}
with an arbitrary constant tensor $\delta^\flat\! \mathscr{X}_{\alpha\, ji}$
is dictated by the requirement of the Bose-Einstein statistics.
Fulfillment of \refer{Eq:WT-phi-phi-A} imposes the condition
\begin{eqnarray}
\tilde\Omega^{\phantom{ij}\phantom{\mu}}_{ij{\alpha}}(p,p',l)
+i\,l_\mu \delta\tig^{\phantom{ij}\mu}_{ij\alpha}(p,p',l)=0,
\end{eqnarray}
and the unique solution is  the  tensor
\begin{eqnarray}\label{Eq:KontrFlat:X}
\delta^\flat\! \mathscr{X}_{\alpha\, ij}
&=&
\frac{1}{(4\pi)^2}\left\{
\left(\frac{7}{12}+\ln\frac{3}{2}\right)
\tr\big[\YF_i\TF_\alpha Y_j^*-\YF^*_i Y_j \TF_\alpha\big]
+
\right.\\&{}&\left.%\hspace*{-15 pt}
\qquad\ \ \ \
+
\frac{3}{4}
\left[
\ln2\, (\TS^\kappa\TS_\alpha\TS_\kappa)_{ij}
-
\ln\frac{3}{2}\, \mind{e}{\kappa\delta}{\!\alpha}(\TS_\kappa\TS_\delta)_{ij}
\right]
\right\},
\nn
\end{eqnarray}
which is antisymmetric in the $ij$ indices. It is related to the
counterterm \refer{Eq:KontrFlat:c} by
\begin{eqnarray}
\label{Eq:RownGFGSD}
\delta^\flat\! \mathscr{X}_{\alpha\, ji}
=\derp{}{\varphi^j}\, \delta^\flat\!c_{i\alpha}.
\end{eqnarray}

\subsection{Identities involving the $\VEV{\phi A A}$ and
$\VEV{\phi\phi AA }$ functions}

The relations \refer{Eq:Del-Del-shift} and \refer{Eq:RownGFGSD}
reflect two facts: the
preservation by our regularization prescription of the shift symmetry
\refer{Eq:shift-sym} and that the same requirement has been imposed
on non-minimal counterterms in Section \ref{sec:Subtractions}
(see also Appendix \ref{App:Ren-Det:IndStep}). Therefore
the non-minimal counterterms
for the $\VEV{\phi AA}$ and $\VEV{\phi\phi AA}$ 1PI functions must be
given by (power counting implies they are momentum independent)
\begin{eqnarray}\label{Eq:Delta-phi-A-A}
\delta\tig^{\,\phantom{i}\nu\rho}_{i\beta\kappa}(l,p,p')
=\delta\tig^{\,\phantom{i}\nu\rho}_{i\beta\kappa}(0,0,0)
=\derp{}{\varphi^i}\,\delta\tig^{\nu\rho}_{\!\beta\kappa}(0,0)
=\eta^{\nu\rho}\derp{}{\varphi^i}\,(\delta^\flat\! m^2_{V})_{\beta\kappa},
\end{eqnarray}
and
\begin{eqnarray}
\delta\tig^{\,\phantom{ij}\nu\rho}_{ij\beta\kappa}(l,l',p,p')
%=
%\delta^\flat_{ct}\!\tig^{\,\phantom{ij}\nu\rho}_{ij\beta\kappa}(0,0,0,0)
=\frac{\partial^2}{\partial\varphi^i\partial\varphi^j}\,
\delta\tig^{\nu\rho}_{\!\beta\kappa}(0,0)=\eta^{\nu\rho}
\frac{\partial^2}{\partial\varphi^i\partial\varphi^j}\,
(\delta^\flat\! m^2_{V})_{\beta\kappa}.
\end{eqnarray}
The matrix $(\delta^\flat\! m^2_{V})_{\beta\kappa}$ is
given in Eq. \refer{Eq:KontrFlat:m_V}.

\subsection{Identity involving the $\VEV{A A A}$ function}\label{Sec:WT:AAA}
\vspace*{-0.3 cm}
Because of its relation to anomalies, one of the most interesting
is the ST identity involving the triple vector boson vertex
$\tig^{\mu\nu\rho}_{\alpha\beta\gamma}(l,p,p')^\OL$
\begin{eqnarray}%\label{Eq:WT-AAA}
{\phantom{a}}&{\phantom{=}}& \!\!\!\!\!\!
i l_\mu \tig^{\mu\nu\rho}_{\alpha\beta\gamma}(l,p,p')^\OL
+\eSC{\kappa}{\alpha}{\beta}\tig^{\nu\rho}_{\kappa\gamma}(p+l,p')^\OL
+\eSC{\kappa}{\alpha}{\gamma}\tig^{\rho\nu}_{\kappa\beta}(p'+l,p)^\OL+\nn\\
{}&{}&\!\!+(\TS_\alpha \varphi)^i \tig^{\phantom{i}\nu\rho}_{i\beta\gamma}(l,p,p')^\OL
-\mathcal{V}^\prime_i(\varphi)\VEVOPITIL{\tilde{K_i}(0)
\tilde{\omega}^\alpha(l) \tilde{A}^\beta_\nu(p) \tilde{A}^\gamma_\rho(p')}^\OL+\nn\\
{}&{}&\!\!+i p'^\rho (\TS_\gamma \varphi)_i
\VEVOPITIL{\tilde{K_i}(p') \tilde{\omega}^\alpha(l) \tilde{A}^\beta_\nu(p)}^\OL
+i p^\nu (\TS_\beta \varphi)_i
\VEVOPITIL{\tilde{K_i}(p) \tilde{\omega}^\alpha(l) \tilde{A}^\gamma_\rho(p')}^\OL
+\nn\\
{}&{}&\!\!
+\left\{{m_V^2}_{\gamma\kappa}\eta^{\rho\sigma}
+{\delta}_{\gamma\kappa}\bracket{p'^\rho p'^\sigma - \eta^{\rho\sigma}p'^2}\right\}
\VEVOPITIL{\tilde{K}^\sigma_\kappa(p')
\tilde{\omega}^\alpha(l) \tilde{A}^\beta_\nu(p)}^\OL
+\label{eqn:AAAidentity}\\
{}&{}&\!\!
+\left\{{m_V^2}_{\beta\kappa}\eta^{\nu\sigma}
+{\delta}_{\beta\kappa}\bracket{p^\nu p^\sigma - \eta^{\nu\sigma}p^2}\right\}
\VEVOPITIL{\tilde{K}^\sigma_\kappa(p)
\tilde{\omega}^\alpha(l) \tilde{A}^\gamma_\rho(p')}^\OL
+\nn\\
{}&{}&\!\!
+ie_{\gamma\beta\kappa}\left\{
(p-p')^\sigma\eta^{\rho\nu}+(2p'+p)^\nu\eta^{\rho\sigma}-(2p+p')^\rho\eta^{\nu\sigma}
\right\}
%\times
%\nn\\{}&{}&\!\!
%\phantom{aaaaaaaaaaaaaaaaaaaaaaaaaaaaaaaa}
%\times
\VEVOPITIL{\tilde{K}^\sigma_\kappa(p+p') \tilde{\omega}^\alpha(l)}^\OL=0\, .\nn
\end{eqnarray}
Contribution $\Omega^{\phantom{\mu|}\nu\rho}_{\alpha\beta\gamma}(l,p,p')$ of bare
one-loop diagrams to the left hand side of \refer{eqn:AAAidentity} obtained
by direct manipulation of regularized integrals is
\begin{eqnarray}\label{Eq:Delta-AAA}
\Omega^{\phantom{\mu|}\nu\rho}_{\alpha\beta\gamma}(l,p,p')&=&
\Omega^{\phantom{\mu|}\nu\rho}_{\alpha\beta\gamma}(l,p,p')^{\rm anom}+
\\
&{}&
+\frac{1}{6(4\pi)^2}\ \!
\tr\big(\!\big[\TS_\gamma, \ \TS_\beta\big] \TS_\alpha \!\big)
\times
\nn\\&{}&\qquad
\!\times
\left\{
(p^2-p'^2)\eta^{\nu\rho}\!\bracket{\frac{19}{24}+\ln\frac{3}{2}}
-(p^\rho p^\nu\! -\! p'^\rho p'^\nu)\bracket{\frac{17}{12}+\ln\frac{3}{2}}
\right\}
\nn\\&{}&\nn
+
\frac{1}{36(4\pi)^2}\tr(\TV_\gamma \TV_\alpha\TV_\beta)
\!\times
\Big\{
\eta^{\nu\rho}(p^2\!-\!p'^2)\!
\matri{24\ln{2}\!-\!6\ln{3}\!-\!19}
+
\nn\\&{}&{}\qquad\qquad\qquad\qquad\qquad\
\!-\!(p^\rho p^\nu\!\!-\!\! p'^\rho p'^\nu)\!
\matri{24\ln{2}\!-\!6\ln{3}\!-\!34}
\!\Big\}
+\nn\\&{}&\nn
+
\frac{1}{18}\!\cdot\!\frac{1}{(4\pi)^2}\ \!
\tr\!\left(\phantom{\big\{}\!\!\! \TF_\alpha \!
\left[\phantom{m_F^2}\!\!\!\!\!\!\!\!\TF_\beta,\ \!\TF_\gamma\right]\right)
\times\\
&{}&\quad\!\!
\times\left[
\left(\!1\!-\!12\ln\frac{3}{2}\right)\left(p'^\nu p'^\rho-p^\nu p^\rho\right)
-\eta^{\nu\rho}\left(\!5\!+\!12\ln\frac{3}{2}\right)\left(p^2- p'^2\right)
\right]\, .\nn
\end{eqnarray}
Once again $\tilde\Omega^{\phantom{\mu|}\nu\rho}_{\alpha\beta\gamma}(l,p,p')
=\Omega^{\phantom{\mu|}\nu\rho}_{\alpha\beta\gamma}(l,p,p')$.
The first term of \refer{Eq:Delta-AAA} is the true anomaly which in our
regularization has the form
\begin{eqnarray}\label{Eq:Delta-AAA-true-anomaly-only}
\Omega^{\phantom{\mu|}\nu\rho}_{\alpha\beta\gamma}(l,p,p')^{\rm anom}
&=&
\frac{2i}{3(4\pi)^2}\ \!
\tr\!\left(\phantom{\big\{}\!\!\!
\TF_\alpha\{\TF_\beta,\ \TF_\gamma\}
\right)\!\cdot\!
p_\sigma p'_\tau\epsilon^{\sigma\tau\nu\rho},\qquad\quad\epsilon^{0123}=-1.
\end{eqnarray}
Except for this one, all the remaining terms of \refer{Eq:Delta-AAA}
can be also obtained (as already explained) by inserting in
\refer{eqn:AAAidentity} the appropriate differences \refer{Eq:DeltaReg:def}.
The part of the $\VEV{AAA}$ vertex that involves the Levi-Civita tensor 
is ambiguous \footnote{Ambiguous terms are multiplied by $\tr\!\left(\phantom{\big\{}\!\!\!
\TF_\alpha\{\TF_\beta,\ \TF_\gamma\}
\right)$ and thus vanish if the
gauge group representation furnished by fermions  is non-anomalous.} in the 
DimReg with naive (anticommuting) $\gamma^5$ and therefore the term \refer{Eq:Delta-AAA-true-anomaly-only} can be obtained
only directly in $\La$Reg; the calculation is similar to the one for the $\VEV{A A A A}$
vertex which is outlined in \ref{Sec:ApB-AAAA} (we show there that
the anomalies are independent of the shape of regularizing function in
\refer{eqn:PrescriptionDef} as long as it satisfies the appropriate boundary
conditions).
%\footnote{More specifically, from Eqs. \refer{Eq:DeltaZ:AAA:X},
%\refer{Eq:DeltaZ:AA:AplusBplusC}, \refer{Eq:RelXYZ} and
%\refer{Eq:DeltaZ:KmuOmega}.}

According to \refer{Eq:DodKontr:PostacSymb} non-minimal counterterms are
allowed for the $\VEV{AAA}$,  $\VEV{AA}$ and $\VEV{\phi AA}$ vertices. The
last two have already been determined (the formulae
\refer{Eq:KontrFlat:Fun:AA} and \refer{Eq:Delta-phi-A-A}, respectively).
It is well known that in general the metric-independent part of the
counterterm to the $\VEV{AAA}$ vertex converts only one form of the
anomaly into another one but cannot remove it - the anomaly
is cohomologically nontrivial. Therefore, we seek only a metric-dependent
non-minimal counterterm.
The most general form of such a counterterm  (which takes into account
the requirements of the Bose-Einstein statistics) reads
\begin{eqnarray}\label{Eq:Kontr-Flat-AAA}
\delta\tig^{\mu_1\mu_2\mu_3}_{\alpha_1\alpha_2\alpha_3}(l_1,l_2,l_3)
&=&
-i\Big\{
\eta^{\mu_1\mu_2}\big[
l_1^{\mu_3}\delta^\flat\!\mathrm{a}_{\alpha_1\alpha_2\alpha_3}
+
l_2^{\mu_3}\delta^\flat\!\mathrm{a}_{\alpha_2\alpha_1\alpha_3}
\big]
+
\\&{}&\hspace*{-80 pt}
+\eta^{\mu_1\mu_3}\big[
l_1^{\mu_2}\delta^\flat\!\mathrm{a}_{\alpha_1\alpha_3\alpha_2}
+
l_3^{\mu_2}\delta^\flat\!\mathrm{a}_{\alpha_3\alpha_1\alpha_2}
\big]
+\eta^{\mu_2\mu_3}\big[
l_2^{\mu_1}\delta^\flat\!\mathrm{a}_{\alpha_2\alpha_3\alpha_1}
+
l_3^{\mu_1}\delta^\flat\!\mathrm{a}_{\alpha_3\alpha_2\alpha_1}
\big]
\Big\},
\nn
\end{eqnarray}
with an arbitrary constant tensor
$\delta^\flat\!\mathrm{a}_{\alpha_1\alpha_2\alpha_3}$.
The condition
%\refer{Eq:ZJ-for-delta-Gamma}
\begin{eqnarray}\label{Eq:OPIOK}
\tilde\Omega^{\phantom{\mu|}\nu\rho}_{\alpha\beta\gamma}(l,p,p')
&+& i l_\mu \delta\tig^{\mu\nu\rho}_{\alpha\beta\gamma}(l,p,p')
+\\&{}& \hspace*{-40 pt}
\ +\eSC{\kappa}{\alpha}{\beta}\delta\tig^{\nu\rho}_{\kappa\gamma}(p+l,p')
\!+\!
\eSC{\kappa}{\alpha}{\gamma}\delta\tig^{\rho\nu}_{\kappa\beta}(p'+l,p)
\!+\!
(\TS_\alpha \varphi)^i \delta\tig^{\phantom{i}\nu\rho}_{i\beta\gamma}(l,p,p')
=0,
\nn
\end{eqnarray}
has the unique solution (provided the condition
\refer{Eq:Non-anomaly-cond} is fulfilled):
\begin{eqnarray}\label{Eq:KontrFlat:a_AAA}
(4\pi)^2\delta^\flat\!\mathrm{a}_{\alpha\beta\gamma}
&=&
\mind{e}{\kappa}{\!\alpha\beta}
\Big\{
-\frac{1}{18}\left(\!5\!+\!12\ln\frac{3}{2}\right)
\tr(\TF_\gamma\TF_\kappa)
-\frac{1}{6}\bracket{\frac{19}{24}\!+\!\ln\frac{3}{2}}
\tr(\TS_\gamma\TS_\kappa)
\!+
\nn\\&{}&\qquad\ \
+
\frac{1}{72}\matri{24\ln{2}\!-\!6\ln{3}\!-\!19}
\tr(\TV_\gamma\TV_\kappa)
\Big\}.
\end{eqnarray}
Owing to the total antisymmetry of $\delta^\flat\!\mathrm{a}$, the counterterm
\refer{Eq:Kontr-Flat-AAA} differs from the tree-level vertex
\refer{Eq:RegFey:AAA-vertex} only by the replacement of structure constants
with $\delta^\flat\!\mathrm{a}$.

%%%%%%%%%%%%%%%%%%%%%%%%%%%%%%%%%%%%%%%%%%%%%%%%%%%%%%%%%%%%%%%%%%%%%%%%%%%%%%%%

\subsection{Identity involving the $\VEV{AAAA}$ function}
\label{Sec:AAAA}

The last non-minimal counterterm from the list \refer{Eq:DodKontr:PostacSymb}
to be determined is the one for the four-vector boson vertex. To make the
formulae simpler it is convenient to introduce the following notation
\begin{eqnarray}\label{Eq:Not-AAAA-1}
\imath_n\equiv(\alpha_n,\mu_n,l_n),
\phantom{aaaa}
\tilde{A}_{\imath_n}\equiv\tilde{A}^{\alpha_n}_{\mu_n}(l_n),
\end{eqnarray}
and to define the operator $\mathbb{S}$ which symmetrizes expressions
w.r.t. $(\imath_2,\imath_3,\imath_4)$:
\begin{eqnarray}\label{Eq:Not-AAAA-2}
\mathbb{S}\!\left\{F(\imath_1,\imath_2,\imath_3,\imath_4)\right\}
\equiv \frac{1}{3!}\sum_{\sigma\in {S(\{2,3,4\})}}
F(\imath_1,\imath_{\sigma(2)},\imath_{\sigma(3)},\imath_{\sigma(4)}).
\end{eqnarray}
In this notation the relevant ST identity takes the form
\begin{eqnarray}
\label{Eq:WT-AAAA}
{\phantom{a}}&{\phantom{=}}& \!\!\!\!\!\!
i\, (l_1)_{\mu_1}
\tig^{\mu_1\mu_2\mu_3\mu_4}_{\alpha_1\alpha_2\alpha_3\alpha_4}(l_1,l_2,l_3,l_4)^\OL
+(\TS_{\alpha_1} \varphi)^j\!\VEVOPITIL{\tilde{\phi}^j(l_1)
\tilde{A}_{\imath_2}\tilde{A}_{\imath_3}\tilde{A}_{\imath_4}}^\OL+\nn\\
{}&{}&\!\! \hspace{20pt}
+3\,\mathbb{S}\!\left\{\eSC{\kappa}{\alpha_1}{\alpha_2}
\VEVOPITIL{\tilde{A}^\kappa_{\mu_2}(l_1\!+\!l_2)
\tilde{A}_{\imath_3}\tilde{A}_{\imath_4}}^\OL\right\}+\nn\\
{}&{}&\!\!\hspace{20pt}
+\VEVOPITIL{\tilde{A}^\kappa_{\rho}(l_1)\tilde{A}_{\imath_2}
\tilde{A}_{\imath_3}\tilde{A}_{\imath_4}}^{(0)}
\VEVOPITIL{\tilde{K}^\rho_\kappa(-l_1) \tilde{\omega}^{\alpha_1}(l_1)}^\OL+\nn\\
{}&{}&\!\!\hspace{20pt}
+3\,\mathbb{S}\!\left\{
\VEVOPITIL{\tilde{A}^\kappa_{\rho}(l_1+l_4)\tilde{A}_{\imath_2}\tilde{A}_{\imath_3}}^{(0)}
\VEVOPITIL{\tilde{K}^\rho_\kappa(-l_1-l_4) \tilde{\omega}^{\alpha_1}(l_1)
\tilde{A}_{\imath_4}}^\OL\right\}+\nn\\
{}&{}&\!\!\hspace{20pt}+3\,\mathbb{S}\!
\left\{\VEVOPITIL{\tilde{A}^\kappa_{\rho}(-l_2)\tilde{A}_{\imath_2}}^{(0)}
\VEVOPITIL{\tilde{K}^\rho_\kappa(l_2) \tilde{\omega}^{\alpha_1}(l_1)
\tilde{A}_{\imath_3}\tilde{A}_{\imath_4}}^\OL\right\}+\nn\\
{}&{}&\!\!\hspace{20pt}
+3\,\mathbb{S}\!\left\{\VEVOPITIL{\tilde{\phi}^j(-l_2)\tilde{A}_{\imath_2}}^{(0)}
\VEVOPITIL{\tilde{K}_j(l_2) \tilde{\omega}^{\alpha_1}(l_1)\tilde{A}_{\imath_3}
\tilde{A}_{\imath_4}}^\OL\right\}+\nn\\
{}&{}&\!\!\hspace{20pt}
+3\,\mathbb{S}\!\left\{
\VEVOPITIL{\tilde{\phi}^j(-l_2-l_3)\tilde{A}_{\imath_2}\tilde{A}_{\imath_3}}^{(0)}
\VEVOPITIL{\tilde{K}_j(l_2+l_3) \tilde{\omega}^{\alpha_1}(l_1)
\tilde{A}_{\imath_4}}^\OL\right\}+\nn\\
{}&{}&\!\!\hspace{20pt}
-\mathcal{V}^\prime_j(\varphi)
\VEVOPITIL{\tilde{K}_j(0) \tilde{\omega}^{\alpha_1}(l_1)
\tilde{A}_{\imath_2}\tilde{A}_{\imath_3}\tilde{A}_{\imath_4} }^\OL=0.
\end{eqnarray}
Power counting, Lorentz properties and the antighost equation \refer{Eq:AntiGhost} imply that
in \refer{Eq:WT-AAAA} only the functions $\left<AAAA\right>$,
$\left<AAA\right>$, and $\langle{K}^\rho_\kappa {\omega}^\alpha\rangle$ can
be different
in $\Lambda$Reg and DimReg. Therefore,
$\Omega^{\phantom{\mu_1}\mu_2\mu_3\mu_4}_{\alpha_1\alpha_2\alpha_3\alpha_4}(l_1,l_2,l_3,l_4)$
which is
identical with $\tilde\Omega^{\phantom{\mu_1}\mu_2\mu_3\mu_4}_{\alpha_1\alpha_2\alpha_3\alpha_4}
(l_1,l_2,l_3,l_4)$ can be obtained using the differences \refer{Eq:Delta:AAAA},
\refer{Eq:DeltaZ:AAA:X} and \refer{Eq:DeltaZ:KmuOmega}.\footnote{
Unlike the previously considered identities, only the (potentially
anomalous) contribution of fermions to \refer{Eq:WT-AAAA} has been worked
out directly in $\La$Reg (the calculation is outlined in \ref{Sec:ApB-AAAA}).  
This contribution is correctly reproduced by the differences \refer{Eq:Delta:AAAA},
\refer{Eq:DeltaZ:AAA:X} and \refer{Eq:DeltaZ:KmuOmega} 
if \refer{Eq:Non-anomaly-cond} is satisfied.}
As follows from \refer{Eq:DodKontr:PostacSymb}, only the vertices
$\langle AAAA\rangle$ and $\left<AAA\right>$  have non-minimal counterterms;
\refer{Eq:WT-AAAA} requires therefore that
\begin{eqnarray}\label{Eq:OPIOK4}
&{}&
\tilde\Omega^{\phantom{\mu_1}\mu_2\mu_3\mu_4}_{\alpha_1\alpha_2\alpha_3\alpha_4}(l_1,l_2,l_3,l_4)
+3\,\mathbb{S}\!\left\{\eSC{\kappa}{\alpha_1}{\alpha_2}
\delta\tig^{\mu_2\mu_3\mu_4}_{\kappa_{\phantom{2}}\alpha_3\alpha_4}
(l_1\!+\!l_2,l_3,l_4)\right\}+\nn\\
{}&{}&\!\! \hspace{93pt}+\,i\, (l_1)_{\mu_1}
\delta\tig^{\mu_1\mu_2\mu_3\mu_4}_{\alpha_1\alpha_2\alpha_3\alpha_4}(l_1,l_2,l_3,l_4)=0.
\end{eqnarray}
The explicit form of
$\tilde\Omega^{\phantom{\mu_1}\mu_2\mu_3\mu_4}_{\alpha_1\alpha_2\alpha_3\alpha_4}
(l_1,l_2,l_3,l_4)$ is rather complicated, however simplifications
occur after combining it with the second term in which
$\delta\tig^{\mu_2\mu_3\mu_4}_{\kappa_{\phantom{2}}\alpha_3\alpha_4}
(l_1\!+\!l_2,l_3,l_4)$ is given by  \refer{Eq:Kontr-Flat-AAA}.
The general form of the $\left<AAAA\right>$ counterterm
(again, neglecting a possible metric-independent part) is
\begin{eqnarray}
\label{Eq:KontrFlat:Fun:AAAA}
\delta\tig^{\mu_1\mu_2\mu_3\mu_4}_{\alpha_1\alpha_2\alpha_3\alpha_4}
(l_1,l_2,l_3,l_4)
&=&
\big\{
\eta^{\mu_1\mu_2}\eta^{\mu_3\mu_4}
\delta^\flat\!\mathrm{q}_{(\alpha_1,\alpha_2),(\alpha_3,\alpha_4)}
+
\\&{}&\hspace*{-40 pt}
+
\eta^{\mu_1\mu_3}\eta^{\mu_2\mu_4}
\delta^\flat\!\mathrm{q}_{(\alpha_1,\alpha_3),(\alpha_2,\alpha_4)}
+
\eta^{\mu_1\mu_4}\eta^{\mu_2\mu_3}
\delta^\flat\!\mathrm{q}_{(\alpha_1,\alpha_4),(\alpha_2,\alpha_3)}
\big\},
\nn
\end{eqnarray}
where the otherwise arbitrary constant tensor
$\delta^\flat\!\mathrm{q}_{(\alpha_1,\alpha_2),(\alpha_3,\alpha_4)}$
must be symmetric w.r.t. interchanges of the grouped pairs of indices
and w.r.t. interchanges of the indices within the pairs. The solution to
\refer{Eq:OPIOK4} exists (if \refer{Eq:Non-anomaly-cond} is satisfied) and
is unique:
\begin{eqnarray}\label{Eq:KontrFlat:q_AAAA}
&{}&\!\!\!\!\!\!\!\!\!\!-24(4\pi)^2
\delta^\flat\!\mathrm{q}_{(\alpha_1,\alpha_2),(\alpha_3,\alpha_4)}=\\
&=&\nn
(13+8\ln2)\,
\tr(\TS_{\alpha_1}\TS_{\alpha_2}\{\TS_{\alpha_3},\ \TS_{\alpha_4}\})
-2(9+8\ln2)\,
\tr(\TS_{\alpha_1}\TS_{\alpha_3}\TS_{\alpha_2} \TS_{\alpha_4})+\\
&\phantom{=}&\nn
+2(13-4\ln2)\,
\tr(\TV_{\alpha_1}\TV_{\alpha_2}\{\TV_{\alpha_3},\ \TV_{\alpha_4}\})
-4(9-4\ln2)\,
\tr(\TV_{\alpha_1}\TV_{\alpha_3}\TV_{\alpha_2} \TV_{\alpha_4})+\\
&\phantom{=}&\nn
-16(1+2\ln2)\, \tr(\TF_{\alpha_1}\TF_{\alpha_3}\TF_{\alpha_2}\TF_{\alpha_4}+
\TF_{\alpha_1}\TF_{\alpha_4}\TF_{\alpha_2}\TF_{\alpha_3})+\\
&\phantom{=}&\nn
+4(1+4\ln2)\,
\tr(\{\TF_{\alpha_1},\ \TF_{\alpha_2}\}\, \{\TF_{\alpha_3},\ \TF_{\alpha_4}\}).
\end{eqnarray}
\vskip0.2cm

This completes the  determination at the one-loop order of the non-minimal
counterterms listed in \refer{Eq:DodKontr:PostacSymb}.
Adding them to the action $\tilde I_0$ obtained from $I_0$ according to the
rules \refer{Eq:tilde-I-n} and \refer{Eq:RenCond}
one obtains the action $I_1$.
In agreement with the results of Section \ref {sec:Subtractions}, applying
the operator ${\cal S}$  given by \refer{Eq:ZinnJustinOperQuad}
to the effective action $\Gamma_1$, which is the
asymptotic part (in the sense explained in Section \ref{sec:Subtractions})
of\footnote{Recall that in $I_1^\Lambda$ the substitution
\refer{eqn:PrescriptionDef} is made also in the counterterms (both, minimal
and non-minimal); the (momentum space) form of a regularized counterterm can
be unambiguously fixed by the comparison with the corresponding regularized
tree-level vertex (see the formulae (\ref{Eq:RegFey:AAA-vertex})).}
$\Gamma[I_1^\Lambda]$ one gets in general (using
\refer{Eq:Delta-AAA-true-anomaly-only} and \refer{Eq:AnomAAAA-praw}) that
\begin{eqnarray}
{\cal S}(\Gamma_1)=\hbar{\cal S}_{I_0}\Gamma^{(1)}+{\cal O}(\hbar^2)
\end{eqnarray}
with (using the notation of differential forms in which
$A\equiv\TF_\alpha A^\alpha_\mu {\rm d}x^\mu$,
$\omega\equiv\TF_\alpha \omega^\alpha$)
\begin{eqnarray}
\label{Eq:FormaAnomalii}
\mathcal{S}_{{I}_0}\Gamma^{(1)}=-\frac{i}{24\pi^2}~\!\tr\!\int\!\omega~\!
{\rm d}\left\{A\wedge {\rm d}A+\frac{1}{2}A\wedge A\wedge A\right\},
\end{eqnarray}
(in our conventions
${\rm d}x^0\wedge{\rm d}x^1\wedge{\rm d}x^2\wedge{\rm d}x^3\simeq{\rm d}^4{x}$).
This means that, when minimal and non-minimal counterterms are taken into account, the Zinn-Justin identity is broken only by the true anomaly, which in our regularization and subtraction prescription
(part of which is the condition that 
non-minimal counterterms \refer{Eq:Kontr-Flat-AAA} and \refer{Eq:KontrFlat:Fun:AAAA} do not involve the Levi-Civita tensor) has the well known canonical form
(see e.g. \cite{Chu:1996fr, PiguetSorella}). In the rest of the paper
we assume that the condition \refer{Eq:Non-anomaly-cond} for absence of
anomalies is satisfied.

\section{Relation between $\Lambda$-$\overline{\rm MS}$ and
DimReg-$\overline{\rm MS}$}
\label{Sec:RelSch}

\renewcommand{\thesection}{\arabic{section}}
\renewcommand{\theequation}{\arabic{section}.\arabic{equation}}
\renewcommand{\thefigure}{\arabic{section}.\arabic{figure}}

\setcounter{equation}{0}
\setcounter{figure}{0}

Having determined all one-loop counterterms, minimal\footnote{These can be
obtained immediately from divergent parts of formulae listed in
\ref{App:Diag}.}
and non-minimal ones, we can prove the equivalence at this order of the
$\Lambda$-$\overline{\rm MS}$ scheme and the DimReg-$\overline{\rm MS}$ scheme
with the naive, i.e. fully anticommuting, prescription for the $\gamma^5$
matrix. Equivalence at one-loop of renormalizable YM theories without scalar
fields renormalized in the latter scheme and in a \emph{consistent}
DimReg-based scheme with the
't~Hooft-Veltman-Breitenlohner-Maison prescription for $\gamma^5$ \cite{BreiMei} 
has been
demonstrated in \cite{Martin:1999cc}. Our calculation can be therefore
treated as an extension of the result of \cite{Martin:1999cc}, i.e. as
a proof that at one-loop the naive DimReg-$\overline{\rm MS}$ scheme is
consistent for the most general renormalizable YM theories.\footnote{
In view of this, it is natural to expect that renormalizable
YM theories renormalized in the $\Lambda$-$\overline{\rm MS}$ scheme  and in
DimReg-based schemes with non-naive $\gamma^5$ are also equivalent
(at least at one-loop).}
This requires relating renormalized parameters and fields in both schemes
and constitutes a nontrivial check of the renormalization procedure
developed in Sections \ref{sec:Subtractions} and \ref{sec:cterms}: for
example, relations of the gauge couplings in the two schemes determined using
different vertices must come out the same.

To make the formulae simple we denote collectively all parameters
(masses and couplings) and fields (including antifields) in the
$\Lambda$-$\overline{\rm MS}$ scheme $g^C$, $C=1,\dots$ and $\Phi$,
respectively. Their counterparts in the DimReg-$\overline{\rm MS}$
scheme will be denoted $\check{g}$ and $\check{\Phi}$.
Equivalence of the two schemes means that the \emph{renormalized} effective
action $\Gamma_{\rm Dim}\!\matri{\check{\Phi},\check{g},\check{\mu}}$ which is
the asymptotic (in the sense explained in Section \ref{sec:Subtractions})
part of\footnote{In full analogy with the notation introduced in Section
\ref{sec:Subtractions}, $I^d_\infty$ denotes the dimensionally regularized
action with all order counterterms included.}
$\Gamma[I_\infty^d]$ in the naive DimReg-$\overline{\rm MS}$ can be
obtained from its $\Lambda$-$\overline{\rm MS}$ scheme counterpart
${\Gamma}_\Lambda\!\matri{{\Phi},{g},{\mu}}$ - the asymptotic part of
$\Gamma[I_\infty^\Lambda]$ - through a ``finite renormalization'' of fields and
couplings:
\begin{equation}
{\Gamma}_{\rm Dim}\!\matri{\check{\Phi},G(g),\check{\mu}}
={\Gamma}_\Lambda\!\matri{\zeta\check{\Phi},{g},\check{\mu}},\label{Eq:RelGamma}
\end{equation}
where $\zeta$ is a matrix field rescaling
\begin{equation}\label{Eq:Zeta-fun}
\zeta=\zeta(g)=\mathds{1}
-\frac{\hbar}{(4\pi)^{2}} \, \xi_{(1)}(g)+\mathcal{O}({\hbar}^2),
\end{equation}
and
\begin{equation}\label{Eq:G-fun}
\check{g}^C=G^C(g)=g^C
+\frac{\hbar}{(4\pi)^{2}}\,\theta^C_{(1)}(g)+\mathcal{O}(\hbar^2).
\end{equation}
The formula \refer{Eq:RelGamma} assumes that the two renormalization
scales: $\mu$ of the $\Lambda$-$\overline{\rm MS}$ scheme  and
$\check\mu$ of the DimReg-$\overline{\rm MS}$ are identified
(in other words, one seeks to relate fields and parameters of both
schemes taken at the same numerical value of the two respective
renormalization scales).

The first step in relating the two schemes is to determine the rescaling
factors (matrices) $\zeta$ \refer{Eq:Zeta-fun} for all the fields.
To this end we equate the terms quadratic in the fields $\check{\Phi}$
on both sides of the condition \refer{Eq:RelGamma}. Having determined
$\zeta$'s in this way one can proceed to finding relations between
the parameters. We consider first the matching
conditions which do not depend on non-minimal counterterms.

For the scalar fields (up to the one-loop accuracy) one has the relation:
\begin{equation}
\left[\zeta_\phi^{\rm T}(p^2-m_S^2)\zeta_\phi\right]_{ij}
-(p^2-\check{m}_S^2)_{ij}=-\hbar\Delta_R\tig_{ij}(p,-p)^{\OL}+\mathcal{O}(\hbar^2).
\label{Eq:RelGamma:phiphi}
\end{equation}
On the right hand side of \refer{Eq:RelGamma:phiphi} the factor
\begin{eqnarray}
\Delta_R\tig_{ij}(p,-p)^{\OL}
\equiv\tig_{ij}(p,-p)^{\OL}_{\Lambda}-\tig_{ij}(p,-p)^{\OL}_{{\rm Dim}},
\label{Eq:Delta_R_def}
\end{eqnarray}
is the difference of renormalized one-loop contributions in the two schemes.
Since in this case  the 1PI  function $\tig_{ij}(p,-p)^{\OL}_{\Lambda}$
(subtracted in the $\Lambda$-$\overline{\rm MS}$ scheme) is not affected
by non-minimal counterterms (cf. \refer{Eq:DodKontr:PostacSymb}), the
difference \refer{Eq:Delta_R_def} is obtained  by simply setting
$\delta_{\rm Div}=\Lambda^2=0$ in the corresponding ``bare difference'' of
the form \refer{Eq:DeltaReg:def} which is given explicitly by
\refer{Eq:DeltaZ:PhiPhi:A+B}. (The formula \refer{Eq:DeltaDiv:Def:Mod}
for $\delta_{\rm Div}$ implies that for $\mu=\check{\mu}$ setting
$\delta_{\rm Div}=0$ is just the minimal subtraction of logarithmic
divergences in both schemes).
Solving \refer{Eq:RelGamma:phiphi} for $\zeta_\phi=\zeta_\phi^{\rm T}$
(because we work with real scalar fields) one finds (here $\hbar=1$):
\begin{eqnarray}
(\zeta_{\phi})_{ij}\!&=&\!\delta_{ij}\!-\!\frac{1}{(4\pi)^2}\!\left[
\left\{\frac{3}{16}\!+\!\frac{1}{4}\ln2 \right\}\!
\tr\!\left[ \YF_i \YF_j^*\!+\!\YF_i^* \YF_j\right]
\!+\!\left\{\frac{1}{4}\!+\!\frac{3}{8}\ln\frac{32}{9}\right\}\!
(\TS^\alpha\TS_\alpha)_{ij}\right].\!\phantom{aaaa}\label{Eq:ZetaPhi}
\end{eqnarray}
The formula \refer{Eq:RelGamma:phiphi} yields also the relation between the
mass matrices $\check{m}_S^2$ and $m_S^2$ of the scalar fields in both schemes:
\begin{eqnarray}\label{Eq:ScalMasRel}
[\check{m}_S^2]_{ij}=\left[\zeta_\phi^{\rm T}{m_S^2}\zeta_\phi\right]_{ij}
-\Delta_R\tig_{ij}(0,0)^{\OL}.
\end{eqnarray}
We do not give the explicit form of this relation here, because it can be
also obtained from the general relation between the scalar potentials in
both schemes which we derive below.

In the analogous manner one finds the relation $\psi=\zeta_\psi\check\psi$
between the Majorana fields in the two schemes. Using the difference
\refer{Eq:DeltaZ:PsiPsi:AiB} with $\delta_{\rm Div}$ set to zero
and solving the analog of the condition
\refer{Eq:RelGamma:phiphi} for $\zeta_\psi=\zeta_F P_L+\zeta_F^* P_R$
with Hermitian $\zeta_F$ one gets
\begin{eqnarray}
\zeta_F=\mathds{1}+\frac{1}{(4\pi)^2}\left\{\frac{1}{2}\TF_\alpha \TF^\alpha
+\frac{1}{4}\left[\ln\frac{3}{4}-\frac{1}{6}\right]\YF_i^{*}\YF^i\right\}.
\label{Eq:ZetaF}
\end{eqnarray}
The mass matrices $\check{m}_F$ and $m_F$ of the left-chiral Weyl
fields in the two schemes are related by
\begin{eqnarray}
\label{Eq:RelPar:m_F}
\check{m}_F&=&m_F+\frac{1}{(4\pi)^2}\Big\{\varphi^i \YF^j
(\TS^\alpha \TS_\alpha)_{ji}-\frac{1}{2}\left[\TF_\alpha^{\rm T}\TF^{\alpha\,{\rm T}}m_F
+m_F\,\TF_\alpha\TF^\alpha\right]+\nn\\
&{}&\qquad\qquad\quad\ +\frac{1}{4}\left[\ln\frac{3}{4}-\frac{1}{6}\right]
\left[\YF^j\YF_j^{*} m_F +m_F \YF_j^{*}\YF^j\right]\Big\}.
\end{eqnarray}
The two mass matrices depend on the background scalar fields renormalized
in two different schemes:
$\check{m}_F=\check{\mathcal{M}}_F(0)+\check Y_i\check\varphi^i$
and $m_F=\mathcal{M}_F(0)+Y_i\varphi^i$ (cf. \refer{Eq:m_F-def}).
Since in both schemes the 1PI generating functional
depends only on the sum $\phi+\varphi$, it is natural to set
\begin{eqnarray}
\label{Eq:RelPar:Varphi}
\varphi=\zeta_\phi \check{\varphi},
\end{eqnarray}
(with $\zeta_\phi$ given in \refer{Eq:ZetaPhi}). This allows to rewrite
\refer{Eq:RelPar:m_F} in the form  (neglecting higher order terms)
\begin{eqnarray}\label{Eq:RelPar:CalM_F}
\check{\mathcal{M}}_F(\varphi)&{}&\!\!\!-{\mathcal{M}}_F(\varphi)
={Y}_i \mind{(\zeta_\phi-\mathds{1})}{i}{j}\varphi^j+\frac{1}{(4\pi)^2}
\bigg[\varphi^i \YF^j \left(\TS^\alpha \TS_\alpha\right)_{ji}+\\
&{}&\qquad-\frac{1}{2}\left\{\mathcal{M}_F(\varphi) \TF_\alpha\TF^\alpha+ tp.
\right\}+\frac{1}{4}\left\{\ln\frac{3}{4}-\frac{1}{6}\right\}
\left\{\mathcal{M}_F(\varphi) \YF_j^{*}\YF^j+ tp. \right\}
\bigg]\, ,\quad\!\nn
\end{eqnarray}
($tp.$ stands for the transposition of the preceding term). The advantage
of the relation \refer{Eq:RelPar:CalM_F} is that differentiating it w.r.t.
$\varphi^i$ yields the difference of
the Yukawa couplings $Y_i$ and $\check Y_i$ in both schemes. The result
agrees with the one obtained directly from the $\phi\psi\psi$ vertex using
the difference \refer{Eq:DeltaZ:PsiPsiPhi:Tot}.
(This confirms the relation \refer{Eq:RelPar:Varphi}).

Considering the terms linear in the scalar fields $\check\phi$ on both
sides of the condition \refer{Eq:RelGamma} one gets
(using \refer{Eq:RelPar:Varphi})  the relation
\begin{eqnarray}
\mind{(\zeta_\phi)}{j}{i}\mathcal{V}^\prime_j(\zeta_\phi\check\varphi)
-\check{\mathcal{V}} ^\prime_i(\check\varphi)
=\Delta_R\tig_{i}(p)^\OL \, .\label{Eq:RownHGKD}
\end{eqnarray}
Again, $\Delta_R\tig_{i}(p)^\OL$ is obtained from the difference
\refer{Eq:DeltaZ:Phi} by setting in it $\delta_{\rm Div}=\Lambda^2=0$.
Integrating both sides of \refer{Eq:RownHGKD} w.r.t. the background
field $\check\varphi$ and taking the difference of the resulting
potentials $\mathcal{V}$ and  $\check{\mathcal{V}}$ at the same ``point''
$\varphi$ one obtains the relation (neglecting higher order terms)
\begin{equation}\label{Eq:RelPar:CalV}
\check{\mathcal{V}}({\varphi})-\mathcal{V}({\varphi})
=-\frac{1}{2(4\pi)^2}\tr\!\left\{\mathcal{M}_V^2(\varphi)^2\right\}
+\mind{(\zeta_\phi-\mathds{1})}{i}{j}\varphi^j\derp{}{\varphi^i}
{\mathcal{V}}(\varphi).
\end{equation}
Differentiating it w.r.t. the background $\varphi$
one gets the formulae relating the mass matrices and self-couplings
of the scalar fields in the two schemes.
The relations obtained in this way agree with the
one obtained from \refer{Eq:ScalMasRel}  and the other relations obtained
by considering the matching conditions relating directly the $\phi^3$
vertices in the two schemes.
\vskip0.2cm

In comparing the terms bilinear in the gauge fields on both sides of
\refer{Eq:RelGamma} one has to take into account also the non-minimal
counterterm $\delta\tig^{\mu\nu}_{\!\alpha\beta}$ \refer{Eq:KontrFlat:Fun:AA}
(with $\delta^\flat\!z_A$ and $\delta^\flat\!m_V^2$ given by
\refer{Eq:KontrFlat:z_A} and \refer{Eq:KontrFlat:m_V}, respectively)
which affects the relevant ``renormalized difference'':
\begin{eqnarray}
\Delta_R\tig^{\mu\nu}_{\!\alpha\beta}(p,-p)^{\OL}=\delta\tig^{\mu\nu}_{\!\alpha\beta}(p,-p)
+\Delta\tig^{\mu\nu}_{\!\alpha\beta}(p,-p)^{\OLB}\bigg|_{\delta_{\rm Div}=\Lambda^2=0}\, ,
\end{eqnarray}
(the ``bare'' difference $\Delta\tig^{\mu\nu}_{\!\alpha\beta}(p,-p)^{\OLB}$ is given
by \refer{Eq:DeltaZ:AA:AplusBplusC}). The comparison gives
\begin{eqnarray}
\label{Eq:ZetaA}
(\zeta_{A})_{\alpha\beta}&=&\delta_{\alpha\beta}+\frac{1}{2(4\pi)^2}\!\left\{
\left(\frac{11}{18}+\frac{2}{3}\ln2 \right)\tr\big[\TF_\alpha \TF_\beta\big]
+\left(\frac{7}{144}+\frac{1}{6}\ln2 \right)\tr\big[\TS_\alpha \TS_\beta\big]
+\!\!\right.\nn\\
&{}&\left.%\hspace*{-50 pt}
\qquad\qquad\qquad-\left(\frac{23}{72}+\frac{19}{6}\ln2
-\frac{3}{2}\ln3\right)\tr\big[e_\alpha e_\beta\big]\right\},
\end{eqnarray}
and
\begin{eqnarray}\label{Eq:RelPar:m_V}
[\check{m}_V^2]_{\alpha\beta}=
\left[\zeta_A^{\rm T}{m_V^2}\zeta_A\right]_{\alpha\beta}
+\Delta_R\tig^{00}_{\!\alpha\beta}(0,0)^{\OL}.
\end{eqnarray}

In the similar way, matching the terms proportional to the product
$\phi A$ on both sides of \refer{Eq:RelGamma} (using the ``bare''
difference \refer{Eq:DeltaZ:PhiA:A}, the non-minimal counterterm
\refer{Eq:FunDeltaFlat_PhiA} and Eq. \refer{Eq:RelPar:Varphi}) one gets the relation
\begin{equation}\label{Eq:RelPar:T_Salpha}
\check{\TS}_{\alpha}=\TS_{\alpha'}\!\left\{\mind{(\zeta_A)}{\alpha'}{\!\!\alpha}
+\mind{\delta\Xi}{\alpha'}{\!\!\alpha}\right\},
\end{equation}
between the gauge group generators ${\cal T}_\al$ and $\check{\cal T}_\al$
in the two schemes (that is between the gauge coupling constants)
in which $\delta\Xi$ is given by
\begin{eqnarray}
\label{Eq:RelPar:DeltaXi}
\mind{\delta\Xi}{\alpha'}{\!\!\alpha}&=&\frac{1}{(4\pi)^2}
\left(\frac{3}{4}\ln\frac{3}{4}-\frac{1}{8}\right)
\tr\!\left[\TV^{\alpha'} \TV_{\alpha}\right],
\end{eqnarray}
($\zeta_A$ is given in \refer{Eq:ZetaA}) and the relation
\begin{equation}\label{Eq:RelPar:Stue}
\check{\bar{P}}_{\alpha_A}=(\zeta_A)^{\beta_A}_{\ \ \alpha_A}\bar{P}_{\beta_A}\,,
\end{equation}
between the Stueckelberg parameters (cf. \refer{Eq:CovDer}) in the two schemes.

We have verified that the formulae \refer{Eq:RelPar:T_Salpha},
\refer{Eq:RelPar:Stue} and \refer{Eq:RelPar:Varphi} in conjunction with the
explicit expression \refer{Eq:RelA12} for $m_V^2$ in the
$\Lambda$-$\overline{{\rm MS}}$ (and its DimReg counterpart) reproduce
the relation \refer{Eq:RelPar:m_V}. The same relation
\refer{Eq:RelPar:T_Salpha} follows also, upon using \refer{Eq:DeltaZ:PhiPhiA}
and \refer{Eq:FunDeltaFlat_PhiPhiA}, from matching the $\phi\phi A$ vertices
in the two schemes. Furthermore, using \refer{Eq:ZetaF} together with
\refer{Eq:DeltaZ:PsiPsiA:Tot} and
\refer{Eq:FunDeltaFlat_PsiPsiA}  we have verified that the relation
between the fermionic generators $t_\alpha$ and $\check t_\alpha$ in both
schemes obtained by considering the vertex $\psi\psi A$ is identical to
\refer{Eq:RelPar:T_Salpha}, as expected. Similarly, using Eqs.
\refer{Eq:DeltaZ:AAA:X} and \refer{Eq:Kontr-Flat-AAA} the same relation
for the adjoint generators $e_\alpha$ and $\check e_\alpha$ is obtained from
matching the corresponding $A A A$ vertices. Moreover, the relation
\refer{Eq:RelPar:T_Salpha} is also consistent with the form of the $AAAA$
vertices (cf. Eqs. \refer{Eq:Delta:AAAA} and \refer{Eq:KontrFlat:Fun:AAAA}).

To complete establishing the equivalence of the $\Lambda$-$\overline{\rm MS}$
and DimReg-$\overline{\rm MS}$ schemes at the one-loop order, it is
necessary to relate vertices involving antifields (these vertices do not
have non-minimal counterterms). Of these only the two-point function
$\VEV{{K}^\mu_\alpha{\omega}^\gamma}$ has a non-vanishing ``bare difference" (see
\ref{App:Diag}). Eq. 
\refer{Eq:DeltaZ:KmuOmega} after minimal renormalization yields
\begin{eqnarray}\label{Eq:RownXAAAA}
\zeta_{\mathbb{K}}^{\rm T}\zeta_{\omega}=\mathds{1}-\frac{1}{(4\pi)^2}
\left(\frac{1}{8}+\frac{3}{4}\ln\frac{4}{3}\right)\TV_\gamma\TV^\gamma\,,
\end{eqnarray}
where $\zeta_{\mathbb{K}}$ relates the vector antifields $K^\mu$ and
$\check K^\mu$. Introducing the notation $\Psi\equiv(\phi,\psi,A_\mu,\omega)$
and $\mathcal{K}\equiv(K,\bar{K},K^\mu,L)$ and matching the $\VEV{\mathcal{K}\,{\omega}\, \Psi}$  vertices
in the two schemes we get the relation
\eq{\label{Eq:Rel-Gen-AF}
\check{T}^{(\Psi)}_\gamma=
\zeta_{\mathcal{K}}^{\rm T}\, T^{(\Psi)}_{\gamma'}\zeta_{\Psi}\,
\mind{(\zeta_\omega)}{\gamma'}{\gamma}\,
}
with ${T}^{(\Psi)}_\gamma=(\TS_\gamma,\TM_\gamma,\TV_\gamma,\TV_\gamma)$. It
follows that the formulae \refer{Eq:RownXAAAA} and \refer{Eq:Rel-Gen-AF}
are consistent with \refer{Eq:RelPar:T_Salpha} (and its counterparts for
the other kinds of generators) provided
\eq{\label{Eq:XX1}
\zeta_{\mathcal{K}}=(\zeta_{\Psi}^{-1})^{\rm T}\,,
}
and
\eq{\label{Eq:XX2}
\mind{(\zeta_\omega)}{\alpha'}{\!\!\alpha}=
\mind{(\zeta_A)}{\alpha'}{\!\!\alpha}
+\mind{\delta\Xi}{\alpha'}{\!\!\alpha}
+\mathcal{O}(\hbar^2)\,,
}
with $\mind{\delta\Xi}{\alpha'}{\!\!\alpha}$ introduced in
\refer{Eq:RelPar:DeltaXi}. The  matrix $\zeta_{\anti{\omega}}$ relating the
antighost fields in both schemes is equal
to $\zeta_{\mathbb{K}}$, because of the ghost equation \refer{Eq:Ghost}.
Similarly, the corresponding Nakanishi-Lautrup multipliers are related by
$\zeta_h=(\zeta^{\rm T}_A)^{-1}$. Finally, the block  of the $\zeta_\phi$ matrix
\refer{Eq:ZetaPhi} corresponding to the Stueckelberg fields
is the unit matrix. By comparison of the two point functions
$\VEV{{K}_i\,{\omega}^\gamma}$ in both schemes one concludes that the same
statement holds for Stueckelberg antifields, so that \refer{Eq:XX1} is
correct in this case as well. This establishes the equivalence of the two
considered schemes  at least with the one-loop accuracy.

The relations between quantities defined in the $\Lambda$-$\overline{\rm MS}$
and DimReg-$\overline{\rm MS}$ schemes found in this section, apart from
providing a useful consistency test of the entire subtraction procedure
defined in Section \ref{sec:Subtractions}, will allow us to obtain the
two-loop RGEs satisfied by the running parameters of the former scheme using
the known RGEs in the latter one. Moreover, since usually the parameters
that are extracted by fitting the SM to the data are the gauge (and other)
couplings in the DimReg-$\overline{\rm MS}$ scheme (at $\check\mu=M_Z$ or $M_t$), the
relations established here will allow us (in Section \ref{sec:HP}) to give
the proper numerical input to the  RGEs in the $\Lambda$-$\overline{\rm MS}$
scheme when analyzing the hierarchy problem.

\section{Renormalization group equations}
\label{sec:rg}

\renewcommand{\thesection}{\arabic{section}}
\renewcommand{\theequation}{\arabic{section}.\arabic{equation}}
\renewcommand{\thefigure}{\arabic{section}.\arabic{figure}}

\setcounter{equation}{0}
\setcounter{figure}{0}

The relations \refer{Eq:G-fun} and \refer{Eq:Zeta-fun} imply that the
one-loop RG equations in the $\Lambda$-$\overline{\rm MS}$ and the ordinary
DimReg-$\overline{\rm MS}$ schemes are identical. Moreover, having the one-loop
relations between renormalized parameters and fields in the two schemes, it
is possible, using the known two-loop
RG equations in the DimReg-$\overline{\rm MS}$ scheme \cite{JO,MV}, to
obtain also the two-loop RG equations for the parameters in the
$\Lambda$-$\overline{\rm MS}$ one. From the point of view of the RGE it
is more convenient to treat the background $\varphi$ as a part of the
scalar field $\Phi=\phi+\varphi$. The renormalized parameters of the
$\Lambda$-$\overline{\rm MS}$ scheme, collectively denoted $g^A$, whose
two-loop $\beta$ functions  are derived in this section, are therefore the
gauge couplings (one per each independent gauge group factor, at least in
the absence of the mixing of gauge fields corresponding to different $U(1)$
groups), derivatives of  the scalar potential $\mathcal{V}(\Phi)$ at
$\Phi=0$, the Yukawa matrices $Y_i$, the mass matrices $\mathcal{M}_F(0)$
of the fermionic fields and the Stueckelberg parameters $\bar{P}^i_{\alpha_A}$.
Before deriving these equations it is instructive, however, to take a look
at how the RG arises in YM theories regularized with the help of a
BRST-symmetry breaking cutoff.
\vskip0.2cm

The subtraction procedure defined in Section \ref{sec:Subtractions}
introduces an arbitrary mass parameter $\mu$. As a result, the action
$I_\infty^\Lambda$ depends on this scale and on $\Lambda$ through the
counterterms (this dependence on $\Lambda$ comes on the top on the
dependence through the exponential factors (\ref{eqn:PrescriptionDef})).
The arbitrary scale $\mu$ is expected to play a similar role as in the
DimReg-$\overline{\rm MS}$ scheme. In particular, one expects that
observables computed in terms of renormalized parameters are, for fixed
value of the cutoff scale $\Lambda$, independent of $\mu$, if these
parameters vary appropriately with $\mu$ and that Green's functions
computed in terms of renormalized parameters satisfy the appropriate
differential renormalization group equations.

In the case of non-gauge theories, or if the regularization does not
break the BRST invariance, the RG equations follow from the observation that
$I_\infty^\Lambda$ can be written in the form of the bare action $I^\Lambda_{\rm B}$
which has the same form as the starting action $I_0^\Lambda$, but with the
renormalized parameters $g^A$ replaced by the ``bare'' ones, $g^A_{\rm B}$,
and with each type of field multiplied by its $Z^{1/2}$ factor. The bare
parameters $g^A_{\rm B}$ and the $Z^{1/2}$ factors are constructed as power
series in renormalized couplings $g^A$ with coefficients formally
divergent in the limit of removed cutoff. The important fact (actually,
more important than the precise form of $I^\Lambda_{\rm B}$) is that bare and
renormalized parameters, $g^A_{\rm B}$ and $g^A$, are in the strict one-to-one
correspondence and that to each field corresponds a unique $Z^{1/2}$ factor.
Thus, in this case $I_\infty^\Lambda$ depends on $\mu$ only through the $Z^{1/2}$
factors and the bare parameters. The formal equivalence of the perturbative
expansions in renormalized parameters and in bare ones (the latter with a
non-perturbative treatment of the $Z^{1/2}$ factors):
\begin{eqnarray}
\Gamma[I^\Lambda_\infty[\phi,g,\mu,\Lambda]]
=\Gamma[I^\Lambda_0[Z^{1/2}\phi,g_{\rm B}]]~\!,\label{eqn:ExpansionEquiv}
\end{eqnarray}
then firstly implies, that observables computed in terms of renormalized
parameters and depending explicitly on $\mu$ are in fact $\mu$-independent
(if bare parameters are treated as $\mu$-independent, which is ensured by
giving the renormalized parameters an implicit $\mu$-dependence, which in
turn is unambiguous owing to the one-to-one correspondence of bare and
renormalized parameters and uniqueness of the field renormalization
$Z$ factors) and, secondly, allows, by applying to \refer{eqn:ExpansionEquiv}
the chain differentiation rule, to show that the effective
action $\Gamma[I^\Lambda_\infty]$ satisfies the standard RG equation with beta
functions which express the independence
of $\mu$ (for fixed value of the UV regulator) of the bare parameters.
Moreover, the equality\footnote{In the regularization of Section
\ref{Sec:UV-cutoff} the relation
$I^\Lambda_\infty[\phi,g,\mu,\Lambda]=I_0^\Lambda[Z^{1/2}\phi,g_{\rm B}]$ is
ensured (in theories without gauge symmetries) provided the substitution \refer{eqn:PrescriptionDef} is made
in all counterterm vertices.}
$I^\Lambda_\infty[\phi,g,\mu,\Lambda]=I_0^\Lambda[Z^{1/2}\phi,g_{\rm B}]$
implies also that the same RG equation is satisfied by
$I^\Lambda_\infty[\phi,g,\mu,\Lambda]$.

As emphasized in \cite{Martin:1999cc}, this standard reasoning cannot be
directly extended to the BRST symmetry breaking regularizations, because the
action $I_\infty^\Lambda$ constructed in the process of removing divergences
and restoring the BRST invariance of the effective action does not have
the form which allows for immediate identification of the $Z^{1/2}$ factors
and bare couplings: trivially speaking, as illustrated by the explicit
one-loop calculations presented in Section \ref{sec:cterms}, to each gauge
field there correspond in fact two different $Z$ factors - one multiplying
the structure $\partial_\mu A_\nu\partial^\mu A^\nu$ and another one (affected
by non-minimal counterterms) multiplying $\partial_\mu A^\mu\partial_\nu A^\nu$.
Furthermore, because of the non-minimal counterterms, different operator
structures involving gauge fields in the interaction part of $I_\infty^\Lambda$
are multiplied by different power series (with divergent, as
$\Lambda\rightarrow\infty$ and $\mu$-dependent coefficients) in
renormalized couplings, so that even if it were possible to extract
in each vertex the appropriate combination of field renormalization
constants $Z^{1/2}$, one would end up with several ``bare'' gauge couplings
$g^{A(i)}_{\rm B}$ (here $i$ labels different bare couplings corresponding
to an independent gauge group factor $A$). It would not be then obvious
that all the bare couplings
$g^{A(i)}_{\rm B}$ yield the same beta function $\beta^{A}\equiv\beta^{A(i)}$
for the renormalized coupling $g^A$ (in other words, that requiring $\mu$
independence of one of these bare gauge couplings will automatically make
$\mu$ independent also the remaining ones).

On the other hand, it is well known that the concept of bare couplings
is not indispensable to prove that observables and Green's functions
do satisfy the standard RG equations. Indeed, QAP allows to derive
\cite{PiguetSorella,Martin:1999cc} the RG equation directly in terms of
the $\Gamma$ functional. However, since from our point view the action
$I^\Lambda_\infty[\phi,g,\mu,\Lambda]$ should have the physical
interpretation of a bare action, it is important to show that
$I^\Lambda_\infty[\phi,g,\mu,\Lambda]$ and $\Gamma$ obey \emph{the same}
RGE. Therefore below, (modifying the reasoning of \cite{PiguetSorella})
we present a recursive proof of this important fact.

We first notice that\footnote{Recall (see Section \ref{sec:Subtractions}),
that $\Gamma_n$ is obtained from the 1PI effective action $\Gamma[I_n^\Lambda]$
by neglecting terms that vanish in the infinite cutoff limit.}
$I_0$, $I^\Lambda_0$ and $\Gamma_0$ trivially satisfy the following relations
\begin{eqnarray}
R_0I_0=R_0I_0^\Lambda=R_0\Gamma_0=0~\!,\phantom{aaaa}{\rm where}\phantom{aaa}
R_0\equiv\mu{\partial\over\partial\mu}~\!.
\end{eqnarray}
In the next step, defining the differential operator
\begin{eqnarray}
R_n=\mu{\partial\over\partial\mu}+\beta^C_n{\partial\over\partial g^C}
-{\cal N}_\phi(\gamma_n^\phi)
-{\cal N}_\psi(\gamma_n^\psi)
-{\cal N}_\omega(\gamma_n^\omega)
-{\cal N}_A(\gamma_n^A)~\!,\label{eqn:RnDef}
\end{eqnarray}
in which the ``counting operators'' ${\cal N}_X$, $X=\phi,\psi,\omega,A$
are given by \refer{Eq:cN-operators}, while $\beta^C_n$ and $\gamma_n^X$,
are some $\Lambda$-independent coefficients, we prove that if
\begin{eqnarray}\label{Eq:RGE-I-n=r-bar-def}
R_nI_n=\bar r_n \equiv \hbar^{n+1}~\!r_n +{\cal O}(\hbar^{n+2}),
\label{eqn:RnIn}
\end{eqnarray}
then also
\begin{eqnarray}\label{Eq:RGE-Gamma-n}
R_n\Gamma_n= \hbar^{n+1}\,  r_n +{\cal O}(\hbar^{n+2})~\!,
\label{eqn:RnGamman}
\end{eqnarray}
with precisely the same local functional $r_n$. The proof, relegated to
\ref{app:rge}, relies on the fact that the regularization
\refer{eqn:PrescriptionDef} is such that \refer{Eq:RGE-I-n=r-bar-def}
implies that the regularized functional $I^\Lambda_n$ automatically obeys a
similar equation\footnote{Since it is $I^\Lambda_n$ that generates
Feynman rules, in the reasoning of \ref{app:rge} it is crucial
that $I^\Lambda_n$ (rather than $I_n$) obeys the RGE \refer{eqn:RnInLambda}.
For this it is crucial that the derivatives in counterterms have to be also
replaced according to the rule \refer{eqn:PrescriptionDef}; otherwise there
would be no coefficients $\be_1$ and $\ga_1$ for which the condition
$R_1 I_1^\La=\cO(\hbar^2)$ would be satisfied.
}
\begin{eqnarray}
R_nI^\Lambda_n = \bar r^\Lambda_n = \hbar^{n+1}\, r^\Lambda_n
+{\cal O}(\hbar^{n+2})~\!,\label{eqn:RnInLambda}
\end{eqnarray}
where  $\bar r^\Lambda_n$ is obtained from $\bar r_n$ by the replacement
\refer{eqn:PrescriptionDef}, so that
\begin{eqnarray}\label{Eq:rnlam}
\bar r^\Lambda_n=\bar r_n+\cO(\Lambda^{-1})\,,
\end{eqnarray}
because quadratically divergent terms of $I_n$ are momentum-independent.

To argue that \refer{eqn:RnIn} can be extended to the next order
we notice that the functional
\begin{eqnarray}
J_{n+1}\equiv r_n+\mu{\partial\over\partial\mu}~\!
\delta\Gamma^{(n+1)}_{\rm tot} \,,
\end{eqnarray}
where the complete counterterm $\delta\Gamma^{(n+1)}_{\rm tot}=
-\Gamma^{(n+1){\rm div}}_{n}+\delta_{\flat}\!\Gamma^{(n+1)}_n$ is constructed as
in Section \ref{sec:Subtractions}, belongs to the kernel of ${\cal S}_{I_0}$.
This follows from the fact that, owing to the structure of the counting
operators \refer{Eq:cN-operators}, $R_n$ given by (\ref{eqn:RnDef})
satisfies the identity
\begin{eqnarray}
R_n{\cal S}(F) ={\cal S}_FR_nF~\!,
\end{eqnarray}
in which $F$ is an arbitrary functional. This allows to write
$R_n{\cal S}\left(\Gamma_n\right)$ in two different ways:
\begin{eqnarray}
R_n{\cal S}\left(\Gamma_n\right)
={\cal S}_{\Gamma_n}R_n\Gamma_n
=\hbar^{n+1}\,{\cal S}_{I_0}r_n+{\cal O}(\hbar^{n+2})~\!,\nonumber
\end{eqnarray}
and, using \refer{eqn:Omega_n},
\begin{eqnarray}
R_n{\cal S}\left(\Gamma_n\right)=\hbar^{n+1}\, R_n\Omega_n +{\cal O}(\hbar^{n+2})
 =\hbar^{n+1}\, \mu{\partial\over\partial\mu}~\!\Omega_n
+{\cal O}(\hbar^{n+2})~\!.\nonumber
\end{eqnarray}
Combining both results and recalling that
$\Omega_n=-{\cal S}_{I_0} \delta\Gamma^{(n+1)}_{\rm tot}$ we find that indeed
${\cal S}_{I_0}J_{n+1}=0$.

In the similar way one can show recursively that $r_n$ (and hence $J_{n+1}$)
satisfies the homogeneous versions of the auxiliary conditions listed in
Appendix \ref{App:Ren-Det:Part1-AuxCond}. As an element
of $\mathpzc{V}\cap \ker\cS_{I_0}$ (for the definition of the space
$\mathpzc{V}$ see the text above Eq. \refer{Eq:K-ker-0}), $J_{n+1}$
can be represented in the form (cf. \refer{Eq:K-ker=Der-I-0}):
\begin{eqnarray}
J_{n+1}=-\delta R\, I_0\equiv
-\left\{\delta\beta^A~\!{\partial\over\partial g^A}
-{\cal N}_\phi(\delta\gamma^\phi)
-{\cal N}_\psi(\delta\gamma^\psi)
-{\cal N}_\omega(\delta\gamma^\omega)
-{\cal N}_A(\delta\gamma^A)\right\}I_0~\!,\nn
%\label{eqn:Jnplus1Form}
\end{eqnarray}
with some coefficients\footnote{Note that the conditions
\refer{Eq:bE}-\refer{Eq:zeta-A-rel-2-000} impose some constraints on these
coefficients; the most interesting one of them relates the beta functions
of gauge coupling to the anomalous dimension
(in the Landau gauge) of the corresponding ghost field
${\de\be}^C {\pa} T_\alpha / {\pa g^C}=[\de\ga^\om]^\ka_{\ \al}\, T_\ka$.
%\,,\qquad\quad \text{for}
%\qquad\quad T_\gamma=\TS_\gamma,\,\TM_\gamma,\,\TV_\gamma\,,$.
}
$\delta\beta^A$, $\delta\gamma^X$. Defining then
$\beta^A_{n+1}=\beta^A_n+\hbar^{n+1}\,\delta\beta^A$ etc. it is easy to see that
($R_{n+1}\equiv R_n+\hbar^{n+1}\,\delta R$)
\begin{eqnarray}
R_{n+1}I_{n+1}=R_nI_n+\hbar^{n+1}\, R_n\delta\Gamma_{\rm tot}^{(n+1)}
+\hbar^{n+1}\,\delta R\, I_0+{\cal O}(\hbar^{n+2})\nonumber\\
=\hbar^{n+1}\left\{r_n+\mu~\!{\partial\over\partial\mu}
\delta\Gamma_{\rm tot}^{(n+1)}
+\delta R\, I_0\right\}+{\cal O}(\hbar^{n+2})~\!.\phantom{a}\!
\end{eqnarray}
Since $J_{n+1}=-\delta R I_0$, the curly bracket vanishes and we get
$R_{n+1}I_{n+1}={\cal O}(\hbar^{n+2})$. The reasoning presented in \ref{app:rge}
then shows that also $R_{n+1}\Gamma_{n+1}={\cal O}(\hbar^{n+2})$. This in turn
implies that the coefficients in $\delta R$ are $\Lambda$-independent.
On the other hand, the relation $\delta R I_0=-J_{n+1}$ tells us that
coefficients of $\delta R$ are polynomials in dimensional parameters of
$I_0$; this (in conjunction with their $\Lambda$-independence) ensures that
they do not depend explicitly on $\mu$. This completes the inductive step.
\vskip0.2cm

The above result shows that $R_\infty I_\infty=R_\infty I^\Lambda_\infty=0$ and,
therefore (\ref{app:rge}), $R_\infty \Gamma_\infty=0$. The solution of the
first of these equations by the method of characteristics tells us 
in general
\cite{PCHQFT} that the value of $I_\infty$ at a ``point'' $(\Phi,~\!g,~\!\mu)$
is equal to the value assumed by $I_\infty$ at a particular
point $(\Phi_\Sigma,~\!g_\Sigma,~\!\mu_\Sigma)$ on an arbitrarily chosen
hypersurface $\Sigma$ of codimension one, connected to the point
$(\Phi,~\!g,~\!\mu)$ by the characteristic curves specified by the equations
\begin{eqnarray}
&&{\dd\over \dd t}~\!\bar\mu(t,\mu)=\bar\mu(t,\mu)~\!,
\qquad\bar\mu(0,\mu)=\mu~\!, \nonumber\\
&&{\dd\over \dd t}~\!\bar g^A(t,g)=\beta^A(\bar g(t,g))~\!,
\qquad\bar g^A(0,g)=g^A~\!, \label{eqn:characteristics}\\
&&{\dd\over \dd t}~\!\bar\Phi^i(t,\Phi,g)=-[\gamma(\bar g(t,g))]^i_{\phantom{a}j}
\bar\Phi^j(t,\Phi,g)~\!,\qquad\bar\Phi^i(0,\Phi,g)=\Phi^i~\!.\nonumber
\end{eqnarray}
In the case of the $\Lambda$-$\overline{\rm MS}$ scheme distinguished is
the hypersurface $\Sigma$ defined by the condition ($\bar\Lambda$ is
defined in \refer{Eq:delta_Lambda})
\begin{eqnarray}
f(\Phi_\Sigma,~\!g_\Sigma,~\!\mu_\Sigma)\equiv\mu_\Sigma-\bar\Lambda=0~\!,
\end{eqnarray}
on which $I_\infty$ takes the simplest form because all minimal
logarithmically divergent as $\Lambda\rightarrow\infty$ counterterms
vanish there (non-vanishing are only the minimal counterterms proportional
to $\Lambda^2$ and the non-minimal ones). Thus,
\begin{eqnarray}
I_\infty[\Phi,~\!g,~\!\mu,~\!\Lambda]
=I_\infty[\Phi_\Sigma,~\!g_\Sigma,~\!\mu_\Sigma,~\!\Lambda]
=I_{\infty}[\bar\Phi(t_\mu^\Lambda,\Phi,g),~\!
\bar g(t_\mu^\Lambda,g),~\!\bar\Lambda,~\!\Lambda]~\!.\label{eqn:BareAction}
\end{eqnarray}
where (cf. Eq. \refer{Eq:delta_Lambda})
\begin{eqnarray}
t_\mu^\Lambda=\ln{\bar\Lambda\over\mu}=\frac{1}{2}\delta_\Lambda\,.
\end{eqnarray}

The formula (\ref{eqn:BareAction}) together with the identification
$\Phi_{\rm B}\equiv \bar\Phi(t_\mu^\Lambda,\Phi,g)$, provides the definition
of the bare action $I_{\rm B}$ as the action defined on $\Sigma$:
\begin{eqnarray}
I_{\rm B}[\Phi_{\rm B},g_{\rm B}]\equiv
I_\infty[\Phi_\Sigma,~\!g_\Sigma,~\!\mu_\Sigma,~\!\Lambda]~\!.
\label{eqn:BareIdef}
\end{eqnarray}
The bare couplings (cubic and quartic couplings in the scalar field potential,
Yukawa couplings, gauge couplings as well as the explicit mass parameters of
fermions) are then naturally defined as
\begin{eqnarray}
g_{\rm B}^A\equiv\bar g^A(t_\mu^\Lambda,g)~\!.\label{eqn:BareCouplingsDef}
\end{eqnarray}
Independence of $g_{\rm B}^A$ of $\mu$, that is $\mu (\dd/\dd\mu) g_{\rm B}^A=0$,
determines then, as usual, the $\mu$ dependence of the running couplings
$g^A(\mu)$. Since the autonomous ordinary differential equations
\refer{eqn:characteristics} imply automatically that \cite{PCHQFT}
\begin{eqnarray}
{\partial\over\partial t}~\!\bar g^A(t,g)=\beta^C(g)~\!
{\partial\over\partial g^C}~\!\bar g^A(t,g)~\!,
\end{eqnarray}
one obtains $\mu (\dd/\dd\mu) g^A(\mu)=\beta^A(g(\mu))$ as the RG equations
allowing to relate  $g^A(\mu)$ to $g^A(\mu^\prime)$. (Inverting the relations
\refer{eqn:BareCouplingsDef} expresses, of course, $g^A(\mu)$
through the bare couplings $g_{\rm B}^A$.)

According to this definition of the bare couplings,
in the bare action the coefficients of the various gauge field dependent
interaction vertices (affected by non-minimal counterterms) are given by
different infinite power series in the bare couplings.\footnote{The
exception are the terms coupling the ghost and gauge fields which, having
no non-minimal counterterms, are simply proportional to the bare gauge
couplings \refer{eqn:BareCouplingsDef}.}
Furthermore, the $Z_A=Z_A(g,\mu,\Lambda)$ factor of a gauge field $A$ is in
this way uniquely defined (it is the coefficient of the
$\partial_\mu A_\nu\partial^\mu A^\nu$ structure in $I_\infty$ which is not
affected by non-minimal counterterms), whereas the coefficient of the
structure $\partial_\mu A^\mu\partial_\nu A^\nu$ must be of the form
$Z_A\times F(\bar g(t_\mu^\Lambda,g))\equiv Z_A\times F(g_{\rm B})$
with $F(g_{\rm B})$ being an infinite powers series in the (dimensionless)
bare couplings. Finally, the bare masses squared of the scalar fields
are uniquely defined by \refer{eqn:BareIdef} as the coefficients
of the terms quadratic in bare scalar fields and
have the form (notice that on the left hand side of \refer{eqn:BareIdef}
there in no explicit $\Lambda$ dependence!)
\begin{eqnarray}
(m^2_{\rm B})^C=(4\pi)^{-2}\Lambda^2 f^C(\bar\lambda(t_\mu^\Lambda,\lambda))
+(\bar m^2)^C(t_\mu^\Lambda,m^2,\rho,\lambda)\nonumber\\
=(4\pi)^{-2}\Lambda^2 f^C(\lambda_{\rm B})
+(\bar m^2)^C(t_\mu^\Lambda,m^2,\rho,\lambda)\,,\phantom{aaa}
\label{eqn:barebosonmassesdef}
\end{eqnarray}
where  $\lambda$ denote generically parameters of dimension 0
(gauge, Yukawa and quartic scalar couplings) and $\rho$ stands for
generic cubic scalar couplings or explicit fermionic mass parameters.
It is also important to notice that because of the minimal
counterterms proportional to $\Lambda^2$ (as well as due to the presence
of non-minimal counterterms) the bare action $I_{\rm B}$ includes also bare
vector boson masses squared
\begin{eqnarray}
(M^2_V)_{\rm B}
&=&
\Lambda^2 H(\bar \lambda(t_\mu^\Lambda,\lambda))+
K(
\bar m^2(t_\mu^\Lambda,m^2,\rho,\lambda),\,
\bar \rho(t_\mu^\Lambda,\rho,\lambda),\,
\bar \la(t_\mu^\Lambda,\lambda)
)
\nonumber\\
&=&
\Lambda^2 H(\lambda_{\rm B})+
K(
\bar m^2(t_\mu^\Lambda,m^2,\rho,\lambda),\,
\rho_{\rm B},\,
\la_{\rm B}
)\,,
\label{eqn:barevectormassesdef}
\end{eqnarray}
- in the cutoff regularization there are unavoidably
quadratically divergent corrections also to the vector boson
two-point functions (see Section \ref{sec:cterms}).

Summarizing, we have shown, that the action $I_\infty$ obtained in the
process of constructing minimal and non-minimal counterterms indeed winds
up to a ``bare'' action $I_{\rm B}$ which has no explicit dependence on $\mu$:
the entire dependence on $\mu$ enters through the bare parameters
and the field renormalization factors $Z$. In particular, the
result \refer{eqn:barebosonmassesdef} provides
the general justification of the conjecture first formulated
in \cite{Fujikawa} and used in \cite{HAKAOD,CHLEMENI}, namely that
coefficients of quadratic divergences are $\Lambda$-independent functions
of bare couplings. It should be stressed once again, that this result
relies on the consistent application of the regularization prescription
of Section \ref{Sec:UV-cutoff} (that is, on making the substitution
\refer{eqn:PrescriptionDef} also in the counterterms).
\vskip0.2cm

After these considerations we return to the derivation of the two-loop beta
functions in the $\Lambda$-$\overline{\rm MS}$ scheme. The relation
\refer{Eq:RelGamma} allows us to express the beta functions and the
field anomalous dimensions
in $\Lambda$-$\overline{\rm MS}$ in terms of their DimReg-$\overline{\rm MS}$
counterparts
\begin{eqnarray}
{\beta}^A(g)=\matrixind{\Omega(g)^{-1}}{A}{C}\ \!\check{\beta}^C\!(G(g))~\!,
\phantom{aaaaaaaaaaaaaaaaa}~\label{Eq:UIUI}\\
{\gamma}(g)=\zeta(g){\check{\gamma}}(G(g))[\zeta(g)]^{-1}
+{\beta}^A(g)\zeta(g){\partial\over\partial g^A}~\![\zeta(g)]^{-1}~\!,\nn
\end{eqnarray}
(matrix multiplications in the second line are implicit),
where $\matrixind{\Omega(g)}{C}{A}\equiv\partial G^C(g)/\partial g^A$.
Expanding\footnote{We use the obvious notation
\begin{equation}
%\label{Eq:BetaExp}
\check{\beta}(\check{g})=\sum_{\ell=1}^{\infty}
\frac{\hbar^\ell}{(4\pi)^{2\ell}}\,\check{\beta}_{(\ell)}(\check{g}),
\qquad\qquad\check{\gamma}(\check{g})=\sum_{\ell=1}^{\infty}
\frac{\hbar^\ell}{(4\pi)^{2\ell}}\,\check{\gamma}_{(\ell)}(\check{g}).\nn
\end{equation}
}
the relations \refer{Eq:UIUI} in powers of $\hbar$
and using the differential operators 
(cf. Eqs. \refer{Eq:Zeta-fun}-\refer{Eq:G-fun})
\begin{equation}
\mathcal{B}=\check{\beta}^C_{(1)}(g)~\!\derp{}{g^C}~\!,\qquad\qquad
\Theta=\theta^C_{(1)}(g)~\!\derp{}{g^C}~\!,
\end{equation}
we get
\begin{equation}
\label{Eq:Beta-Diff-2loop}
{\beta}^A(g)=\check{\beta}^A(g)+\frac{\hbar^2}{(4\pi)^{4}}
\left\{\Theta\, \check{\beta}^A_{(1)}(g)
-\mathcal{B}\, \theta^A_{(1)}(g)\right\}+\mathcal{O}(\hbar^3)~\!,
\end{equation}
\begin{equation}
\label{Eq:Diff-Gamma-ogolne}
{\gamma}(g)=\check{\gamma}(g)+\frac{\hbar^2}{(4\pi)^{4}}
\left\{\Theta\, \check{\gamma}_{(1)}(g)
+\mathcal{B}\, \xi_{(1)}(g)
-\left[\xi_{(1)}(g),\ \check{\gamma}_{(1)}(g)\right]
\right\}+\mathcal{O}(\hbar^3)~\!.
\end{equation}
Instead of listing the beta functions for various couplings $g^C$ we follow
Jack and Osborn \cite{JO} and give formulae for the quantities
\begin{equation}\label{Eq:Beta-1-loop}
\beta^{T_\alpha}_{(1)}\equiv\mathcal{B}\,T_\alpha,
\qquad
\beta_{(1)}^{\mathcal{M}_F}(\varphi)\equiv\mathcal{B}\,{\mathcal{M}_F}(\varphi),
\qquad
\beta_{(1)}^{\mathcal{V}}(\varphi)\equiv\mathcal{B}\,\mathcal{V}(\varphi).
\end{equation}
Since the scalar background $\varphi$ is \emph{not} one of the couplings
$g^C$,  $\beta_{(1)}^{\mathcal{V}}(\varphi)$ is simply the scalar potential
$\mathcal{V}$ present in the Lagrangian \refer{Eq:LagrTreeGI} but
with each coupling replaced by its one-loop beta function. Similarly, the
beta function of the Yukawa matrices can be immediately obtained
as the derivatives
\begin{equation}
\beta_{(1)}^{Y_i}=\derp{}{\varphi^i}\beta_{(1)}^{\mathcal{M}_F}(\varphi).
\end{equation}
The explicit forms of \refer{Eq:Beta-1-loop} read\footnote{The one-loop
functions given below can also be obtained from the formulae listed in
\ref{App:Diag}, or, more specifically, from their parts proportional
to $\delta_{\rm Div}$.}
\cite{JO}
\begin{eqnarray}
\label{Eq:BetaCalMF-1loop}
\beta_{(1)}^{\mathcal{M}_F}(\varphi)
&=&3\left\{\mathcal{M}_F(\varphi) \TF_\alpha\TF^\alpha+ tp. \right\}
+2\YF_j \mathcal{M}_F(\varphi)^* \YF^j+
\\
&{}&+\frac{1}{2}\left\{\mathcal{M}_F(\varphi) \YF_j^{*}\YF^j+ tp. \right\}
+\frac{1}{2}\varphi^i \YF^j \tr\!\left\{\YF_i \YF_j^* +cc. \right\},\nn
\end{eqnarray}
\begin{eqnarray}
\label{Eq:BetaV-1-loop}
\beta_{(1)}^{\mathcal{V}}(\varphi)&=&
\frac{1}{2} \tr\!\left\{\mathcal{M}_S^2(\varphi)^2\right\}
-\tr\!\left\{\big[\mathcal{M}_F(\varphi)\mathcal{M}_F(\varphi)^*\big]^2\right\}
+\frac{3}{2} \tr\!\left\{\mathcal{M}_V^2(\varphi)^2\right\}+\qquad\nn\\
&{}&+\gamma^{\phi\, \,i}_{(1)\,j}\varphi^j\derp{}{\varphi^i}{\mathcal{V}}(\varphi),
\end{eqnarray}
where
\begin{eqnarray}
\gamma^\phi_{(1)\, ij}(g)=\check{\gamma}^\phi_{(1)\, ij}(g)
=\frac{1}{2}\tr\!\left\{\YF_i\YF_j^*+cc.\right\}
+3\left(\TS_\alpha\TS^\alpha\right)_{ij},
\end{eqnarray}
is the the one-loop anomalous dimension of the scalar fields in the Landau
gauge (see e.g. \cite{MV}). Finally, the well-known expression for the
beta functions of the gauge couplings has the form
\begin{equation}
\beta^{T_\alpha}_{(1)}=
\mathcal{B}\,T_\alpha=T_\kappa \mind{\mathcal{A}}{\kappa}{\alpha},
\qquad\quad
T_\alpha=\TF_\alpha,\TS_\alpha,\TV_\alpha,
\end{equation}
with
\begin{equation}
{{\mathcal{A}}}_{\kappa\alpha}=
\frac{11}{3}\ \!\tr\!\left\{\TV_{\kappa}\TV_{\alpha}\right\}
-\frac{1}{6}\ \!\tr\!\left\{\TS_{\kappa}\TS_{\alpha}\right\}
-\frac{2}{3}\ \! \tr\!\left\{\TF_{\kappa}\TF_{\alpha}\right\}.
\end{equation}
For completeness we give here also the one-loop anomalous dimensions of
the vector fields and the left-chiral Weyl fermions (in the Landau gauge)
\begin{equation}
\gamma_{(1)\kappa\alpha}^A(g)=
\check{\gamma}_{(1)\kappa\alpha}^A(g)
=\frac{13}{6}\ \!\tr\!\left\{\TV_{\kappa}\TV_{\alpha}\right\}
-\frac{1}{6}\ \!\tr\!\left\{\TS_{\kappa}\TS_{\alpha}\right\}
-\frac{2}{3}\ \! \tr\!\left\{\TF_{\kappa}\TF_{\alpha}\right\}.
\end{equation}
\begin{eqnarray}
\gamma_{(1)}^F(g)&=&\check{\gamma}_{(1)}^F(g)=\frac{1}{2} \YF_i^*\YF^i.
\end{eqnarray}
In complete analogy with \refer{Eq:Beta-1-loop} we define also
\begin{equation}
\theta_{(1)}^{\mathcal{M}_F}(\varphi)\equiv\Theta\,{\mathcal{M}_F}(\varphi)=
(4\pi)^2\left({\check{\mathcal{M}}_F}(\varphi)-{\mathcal{M}_F}(\varphi)\right),
\end{equation}
etc.; their explicit forms follow immediately from \refer{Eq:RelPar:CalM_F},
\refer{Eq:RelPar:CalV} and \refer{Eq:RelPar:T_Salpha}; e.g.
\begin{equation}
\theta^{T_\alpha}_{(1)}={\Theta}\,T_\alpha=(4\pi)^2
\left(\check{T}_\alpha- {T_\alpha}\right)=
T_\kappa \mind{{\Omega}}{\kappa}{\alpha},
\end{equation}
where
\begin{eqnarray}
{\Omega}_{\alpha\beta}&=&\left(\frac{11}{36}
+\frac{1}{3}\ln2 \right)\tr\!\left\{\TF_\alpha \TF_\beta\right\}
%+\\&{}&%\hspace*{-40 pt}\nn
+\!\left(\frac{7}{288}\!+\!\frac{1}{12}\ln2 \right)
\tr\!\left\{\TS_\alpha \TS_\beta\right\}+\qquad\nn\\
&{}&%\hspace*{-50 pt}
-\
\left(\frac{41}{144}\!+\!\frac{19}{12}\ln2
-\frac{3}{4}\ln\frac{9}{4} \right)\tr\!\left\{\TV_\alpha \TV_\beta\right\}.
\end{eqnarray}

We are now in a position to compute the differences of the beta
functions in the two schemes. The formula \refer{Eq:Beta-Diff-2loop} allows
us to obtain these differences by means of simple algebraic
manipulations\footnote{Note that $\beta^{\mathcal{X}}_{(1)}$ and
$\theta^{\mathcal{X}}_{(1)}$ are linear combinations of $\beta^{A}_{(1)}(g)$ and
$\theta^{A}_{(1)}(g)$, respectively, with $g$-independent coefficients.}
on objects $\beta^{\mathcal{X}}_{(1)}$ and $\theta^{\mathcal{X}}_{(1)}$.
The results can be conveniently expressed in terms of the two-loop
counterparts of \refer{Eq:Beta-1-loop}, i.e.
\begin{eqnarray}
\beta_{(2)}^{\mathcal{M}_F}(\varphi)\equiv\beta_{(2)}^A(g)
\derp{}{g^A}\,\mathcal{M}_F(\varphi),\\
\check{\beta}_{(2)}^{\mathcal{M}_F}(\varphi)\equiv
\check{\beta}_{(2)}^A(g)\derp{}{g^A}\,\mathcal{M}_F(\varphi),
\end{eqnarray}
etc. In the case of the gauge couplings we get
\begin{eqnarray}
\label{Eq:Beta-Diff-2loop-Gauge-Jawna}
\beta^{T_\alpha}_{(2)}(g)=\check{\beta}_{(2)}^{T_\alpha}(g)+
T_\kappa\mind{\big[\Omega,\ \mathcal{A}\big]}{\kappa}{\alpha}\, .
\end{eqnarray}
Both $\Omega$ and $\mathcal{A}$ are matrices of invariant forms on a Lie
algebra, hence the commutator in \refer{Eq:Beta-Diff-2loop-Gauge-Jawna}
vanishes if the gauge group contains at most a single $U(1)$ factor. In such
a case the functions $\beta_{(1)}$ and $\theta_{(1)}$ for a given gauge coupling
depend only on this coupling and the two-loop beta function is the same
in both schemes, similarly as in theories with a single coupling. In
theories with multiple $U(1)$ factors there are more Abelian gauge
couplings than independent Abelian generators and all of them can mix
with each other (see e.g. \cite{ChPoWa} and references therein).
The two-loop beta functions for Abelian couplings are then in general
different in both schemes.

The beta functions of the couplings parametrizing the
potential ${\cal V}$ can be obtained from
\begin{eqnarray}
\label{Eq:Beta-V-2loop-LamReg}
\beta_{(2)}^{\mathcal{V}}(\varphi)
&=&\check{\beta}_{(2)}^{\mathcal{V}}(\varphi)
+\mind{\left[\gamma^\phi_{(2)}(g)
-\check{\gamma}_{(2)}^\phi(g)\right]}{i}{j}\,\varphi^j\,
\derp{\mathcal{V}(\varphi)}{\varphi^i}
+\qquad\qquad\qquad\quad\!\!\!\nn\\
&{}&\hspace*{-40 pt}
+2\,\tr\!\left\{
\mathcal{M}_V^2(\varphi)^2\left[{\mathcal{A}}+3\,\Omega\right]\right\}
-2\mathcal{M}_V^2(\varphi)_{\alpha\beta}\,\varphi^{\rm T}\!\left[3{\xi}^\phi_{(1)}
-{\gamma}^\phi_{(1)}\right]\!\TS^\alpha\TS^\beta\varphi+\nn\\
&{}&\hspace*{-40 pt}
-2\,\tr\!\left\{\mathcal{M}_S^2(\varphi)^2{\xi}^\phi_{(1)}\right\}
+2\mathcal{M}_V^2(\varphi)_{\alpha\beta}\,
\tr\!\left\{\mathcal{M}_S^2(\varphi)\TS^\alpha\TS^\beta\right\}+\nn\\
&{}&\hspace*{-40 pt}
-\varphi^{\rm T}\!\left\{\TS^\alpha,\ \TS^\beta\right\}\!
\mathcal{M}_S^2(\varphi)\!\left\{\TS_\alpha,\
\TS_\beta^{\phantom{\beta}}\!\!\right\}\!\varphi
+2\,\tr\!\left\{\TF^{\alpha*}\TF_\alpha^*
\left(\mathcal{M}_F(\varphi)\mathcal{M}_F(\varphi)^*\right)^2+cc.\right\}
+\nn\\&{}&\hspace*{-40 pt}
-2\left[\TS^\alpha\TS_\alpha\varphi\right]^j
\tr\!\left\{\YF_j\mathcal{M}_F(\varphi)^*\!\mathcal{M}_F(\varphi)
\mathcal{M}_F(\varphi)^*+cc.\right\}
+\nn\\&{}&\hspace*{-40 pt}
-\!\left[\ln\frac{3}{4}-\frac{1}{6}\right]
\tr\!\left\{\YF^j\YF_j^*\left(\mathcal{M}_F(\varphi)
\mathcal{M}_F(\varphi)^*\right)^2+cc.\right\},
\end{eqnarray}
where $\xi^\phi_{(1)}$ is the one-loop contribution to $\zeta_\phi$ given in
\refer{Eq:ZetaPhi} taken with the opposite sign (in agreement with the
definition \refer{Eq:Zeta-fun}). The two-loop anomalous dimension of the
scalar fields in the Landau gauge reads
\begin{eqnarray}
\label{Eq:Gamma-Phi-2loop-Lam}
\gamma^\phi_{(2)\,i j}(g)&=&\check{\gamma}^\phi_{(2)\,i j}(g)+\left\{\left[
\frac{1}{2}+\frac{3}{4}\ln\frac{32}{9}
\right]{\mathcal{A}}^{\kappa\lambda}+6\,\Omega^{\kappa\lambda}\right\}
(\TS_\kappa\TS_\lambda)_{ij}
+\\&{}&\hspace*{-30 pt}
+\left[\frac{5}{4}+3\ln2\right]\tr\!\left\{\YF_i^*\YF_j\TF^\alpha\TF_\alpha
+cc.\right\}
+\left[\frac{3}{4}+\ln2\right]\tr\!\left\{\YF_i^*\YF^\ell\YF_j^*\YF_\ell
+cc.\right\}+\nn\\
&{}&\hspace*{-30 pt}
+\left[\frac{7}{24}+\frac{1}{2}\ln\frac{3}{2}\right]\!
\tr\!\left\{\left(\YF_i\YF_j^*+\YF_j\YF_i^*\right)\YF^\ell\YF^*_\ell\right\}
%+
%\nn\\
%&{}&\hspace*{-30 pt}
+\left[\frac{3}{4}\!-\!\frac{3}{8}\ln\frac{32}{9}\right]\!
(\TS^\kappa\TS_\kappa)_{i\ell}\,\mathbb{Y}^\ell_{\,\,j}
\, ,\nn
\end{eqnarray}
where
\begin{equation}\label{Eq:bbY-def}
\mathbb{Y}_{ij}\equiv\tr\left\{Y_i Y_j^*+cc.\right\}.
\end{equation}
For the beta function of the Weyl fermions mass matrices  one obtains
\begin{eqnarray}
\beta_{(2)}^{\mathcal{M}_F}(\varphi)
&=&
\check{\beta}_{(2)}^{\mathcal{M}_F}(\varphi)
+
4\left[\TS^\alpha\TS_\alpha-{\xi}^\phi_{(1)}\right]_{ij}
\YF^i\mathcal{M}_F(\varphi)^*\YF^j
+\\&{}&\hspace*{-45pt}
+\!
\left[\TS^\alpha\TS_\alpha-{\xi}^\phi_{(1)}-
\frac{1}{4}\left(\ln\frac{3}{4}-\frac{1}{6}\right)\mathbb{Y}\right]_{ij}
\left\{\YF^i\YF^{j*}\mathcal{M}_F(\varphi)+tp.\right\}
+\nn\\&{}&\hspace*{-45pt}
+
\varphi^i\YF^j\Big\{\!
\!\left[\frac{3}{4}\!-\!\frac{3}{8}\ln\frac{32}{9}\right]\!\!
\left[\mathbb{Y}\TS^\kappa\TS_\kappa\!
-\!2{\mathcal{A}}^{\lambda\kappa}\TS_\lambda\TS_\kappa\right]_{ij}
\!\!+\!\!\left[\frac{5}{4}\!+\!3\ln{2}\right]\!
\tr\!\matri{\YF_i\YF_j^*\TF_\alpha^*\TF^{\alpha*}\!\!+\!cc.}
\!\!\nn\\&{}&\hspace*{-45pt}
\phantom{+\varphi^i\YF^j\Big\{}\!
\!+\!\!\left[\frac{7}{24}\!+\!\frac{1}{2}\ln\frac{3}{2}\right]\!
\tr\!\matri{\YF_i\YF_j^*\YF^\ell\YF_\ell^*\!+\!cc.}\!
+\!\!
\left[\frac{3}{4}\!+\!\ln{2}\right]\!
\tr\!\matri{\YF_i\YF_\ell^*\YF_j\YF^{\ell*}\!+\!cc.}\!\!
\Big\}
\nn\\&{}&\hspace*{-45pt}
+\left[\ln\frac{3}{4}\!-\!\frac{1}{6}\right]\!
\left\{
\left(\YF^j\!\mathcal{M}_F(\varphi)^*\YF^\ell\YF_{\ell}^*\YF_j\!+\!tp.\right)
-
\left(\YF^j\YF_{\ell}^*\YF_j\YF^{\ell*}\!\mathcal{M}_F(\varphi)\!+\!tp.\right)
\right\}
+\nn\\&{}&\hspace*{-45pt}
-\left[\frac{3}{2}\ln\frac{3}{4}\!+\!\frac{1}{4}\right]\!
\left\{
\left(\mathcal{M}_F(\varphi)\YF_{\ell}^*\TF_\alpha^*\TF^{\alpha*}\YF^\ell\!
+\!tp.\right)
+
\left(\YF^\ell\YF_{\ell}^*\TF_\alpha^*\TF^{\alpha*}\!\mathcal{M}_F(\varphi)\!
+\!tp.\right)
\right\}
+\nn\\&{}&\hspace*{-45pt}
-\,\,2\left(\YF^\ell\TF_\alpha\TF^{\alpha}\!\mathcal{M}_F(\varphi)^*\YF_{\ell}\!
+\!tp.\right)
+\left({\mathcal{A}}^{\kappa\lambda}+6\,{{\Omega}}^{\kappa\lambda}\right)\!
\left(\mathcal{M}_F(\varphi)\TF_\kappa\TF_{\lambda}\!+\!tp.\right).
\nn
\end{eqnarray}
The two-loop anomalous dimension of the left-chiral Weyl
fields (in the Landau gauge) reads
\begin{eqnarray}
\gamma_{(2)}^F(g)
&=&
\check{\gamma}_{(2)}^F(g)
-{\mathcal{A}}^{\kappa\lambda}
\TF_\kappa\TF_{\lambda}
+\!
\left[\TS^\alpha\TS_\alpha-{\xi}^\phi_{(1)}-
\frac{1}{4}\left(\ln\frac{3}{4}-\frac{1}{6}\right)\mathbb{Y}\right]_{ij}
\YF^{i*}\YF^{j}
+
\nn\\&{}&\hspace*{-35pt}\qquad\ \ \
-\left[\frac{3}{2}\ln\frac{3}{4}\!+\!\frac{1}{4}\right]\!
\left\{
\YF_{\ell}^*\TF_\alpha^*\TF^{\alpha*}\YF^\ell
+
\TF^{\alpha}\TF_\alpha\YF_{\ell}^*\YF^\ell
\right\}
-
\left[\ln\frac{3}{4}\!-\!\frac{1}{6}\right]
\YF^{\ell*}\YF_j\YF_{\ell}^*\YF^j\, ,%\qquad
\nn
\end{eqnarray}
while that of  the vector fields has the form
\begin{equation}
\label{Eq:Gamma-Diff-2loop-Gauge-Jawna}
\gamma_{(2)\alpha\beta}^A(g)
=\check{\gamma}_{(2)\alpha\beta}^A(g)
+\,
{
\big[
\Omega,\ \mathcal{A}
\big]}_{\alpha\beta}+
\left\{
\left[\frac{3}{2}\ln\frac{3}{4}-\frac{1}{4}\right]\mathcal{A}_{\alpha\gamma}
-3\Omega_{\alpha\gamma}
\right\}\tr(\TV^\gamma\TV_\beta)\, .
\end{equation}
One should expect that the relation $\beta^{T_{\alpha_{\!A}}}=
T_{\kappa_{\!A}} (\gamma^A)^{\kappa_{\!A}}_{\ \ \alpha_{\!A}}\,$ holds in both
schemes,\footnote{For non-Abelian indices the relation $\beta^{T_\alpha}=
T_\kappa(\gamma^A)^\kappa_{\ \alpha}\,$ holds only in the background field gauge
and provided $\gamma^A$ is the anomalous dimension of the background vector
fields.}
so that \refer{Eq:Gamma-Diff-2loop-Gauge-Jawna} agrees with
\refer{Eq:Beta-Diff-2loop-Gauge-Jawna}.
Similarly, the beta functions for Stueckelberg parameters in \refer{Eq:CovDer}
are determined by the anomalous dimensions of the Abelian vector fields
\begin{eqnarray}
\beta^C\derp{}{g^C}{\bar{P}}_{\alpha_A}
={\bar{P}}_{\beta_A}(\gamma^A)^{\beta_A}_{\ \ \alpha_A}\,.
\end{eqnarray}

The above formulae have to be supplemented with the Jack-Osborn expressions
\cite{JO} for $\check{\beta}_{(2)}$ functions in the DimReg-$\overline{\rm MS}$
scheme (to be distinguished from DimRed results, which
are also given in \cite{JO}) and with the Machacek-Vaughn formulae \cite{MV} for $\check{\gamma}_{(2)}$ matrices. 
For completeness we list them (using our
conventions) in \ref{App:JO}. The explicit expressions for the $\beta$
and $\theta$ functions in the SM are given in \ref{App:SM}.

\section{The ``bare" scalar potential}
\label{Vac:Diag}

\renewcommand{\thesection}{\arabic{section}}
\renewcommand{\theequation}{\arabic{section}.\arabic{equation}}
\renewcommand{\thefigure}{\arabic{section}.\arabic{figure}}

\setcounter{equation}{0}
\setcounter{figure}{0}

%{\bf  The 2-loop diagrams are below -- remove nothing}
%\nothing
{
\begin{figure}[t]
\centering
\includegraphics[width=0.9\textwidth]{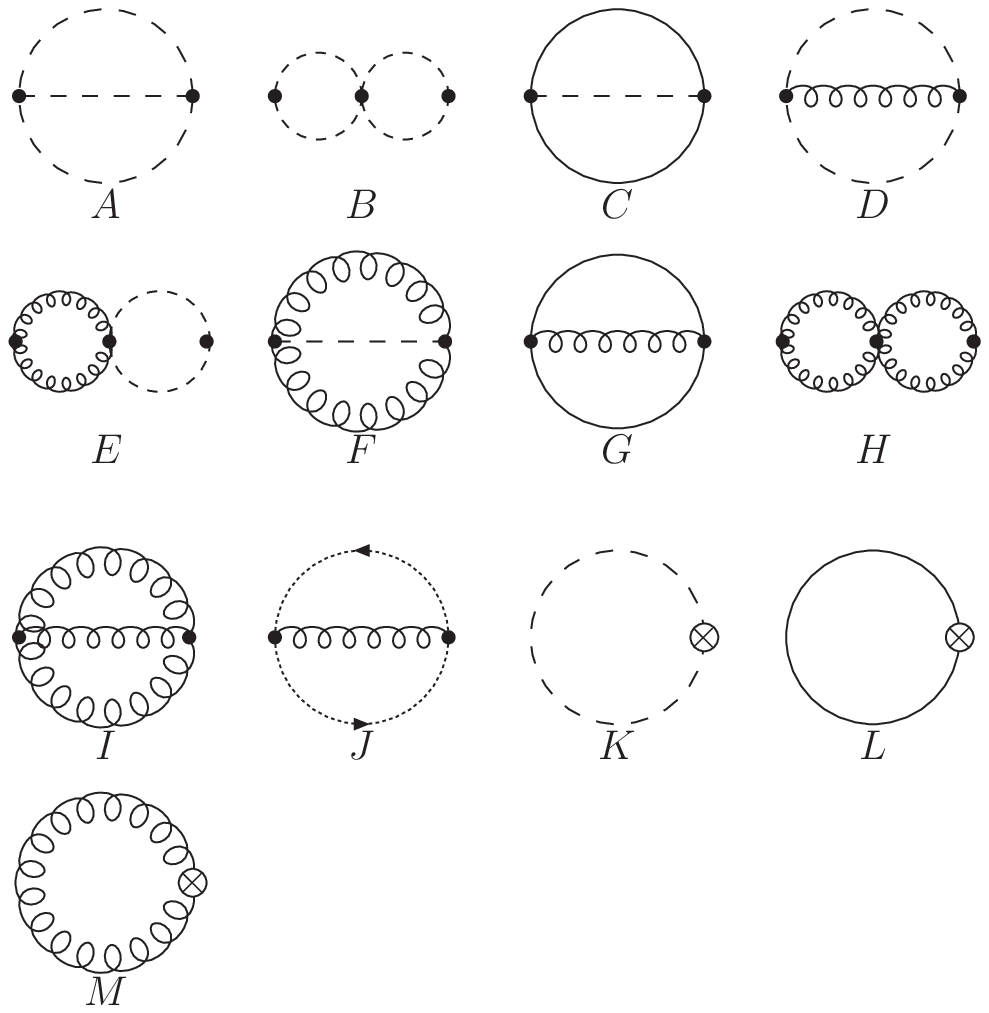}
\caption{Order $\hbar^2$ vacuum graphs.}
\label{Rys:DiVac}
\end{figure}
}

As a further consistency check of the renormalization scheme defined in
Section \ref{sec:Subtractions} and as an example of dealing with the
regularization \refer{eqn:MomSpacePrescription} in higher orders we consider
in this section the order $\hbar^2$ contribution to the constant term
$\Gamma[0]$ of the effective action $\Gamma[\phi,A,\ldots]$, i.e. to the
background field dependent zero-point 1PI function. Owing to the ``shift''
symmetry \refer{Eq:shift-sym} which is preserved by the cutoff
regularization of Section \ref{Sec:UV-cutoff},
calculating $\Gamma[0]$ in order $\hbar^2$ is equivalent to the
determination of the two-loop contribution to the effective potential
$\cV_{\rm eff}(\varphi)$
\begin{eqnarray}
\Gamma[\phi,~\!{{\rm other\ (anti)fields}} =0]
\! =\! -\!\int\!\dd^4x\left\{{\cal V}_{\rm eff}(\phi+\varphi)
+{\rm derivative~terms}\right\}.
\nonumber
\end{eqnarray}
However because calculation of the complete two-loop Feynman integrals in the
regularization of Section \ref{Sec:UV-cutoff} is quite cumbersome, here we will
content ourselves\footnote{The complete two-loop ${\cal V}_{\rm eff}(\vp)$ of an arbitrary
renormalizable gauge theory in
the DimReg-$\overline{\rm MS}$ and DimRed schemes is given in \cite{MartinVeff}.}
with calculating only the divergent part of $\Gamma_1[0]$ (in agreement
with the notation introduced in Section \ref{sec:Subtractions} the subscript
$1$ indicates that the calculation proceeds from the
action $I_1^\Lambda$).
In other words, we want to find the $\varphi$-dependent counterterm that ensures the finiteness of $\Gamma[0]$ in the $\cO(\hbar^2)$ order.
This will provide some nontrivial consistency checks and will also allow
to determine the two-loop coefficients of quadratic divergences in the bare
action $I_{\rm B}$ introduced in Section \ref{sec:rg}.
\vskip0.2cm

Diagrams relevant for calculating the two-loop contribution to the zero-point
function $\Gamma[0]$ are shown in Fig. \ref{Rys:DiVac}. By an
appropriate change of the basis in the field space the background field
dependent mass matrices $\mathcal{M}_S^2(\varphi)$, $\mathcal{M}^2_V(\varphi)$
and $\mathcal{M}_F(\varphi)$ can be made diagonal. In this special basis the
integrals corresponding to the genuine two-loop diagrams $A$-$J$ of Fig.
\ref{Rys:DiVac}, which can be written down using the rules for propagators
and vertices given in Section \ref{Sec:UV-cutoff}, reduce to the nine integrals
listed in \ref{App:2LoopInt}. All these integrals are fully regularized by the
prescription \refer{eqn:PrescriptionDef} and can, in turn, be reduced to the
four basic integrals \refer{Eq:Itot-def}-\refer{Eq:M1tot-def} whose divergent
parts we are here interested in, are determined in \ref{App:2LoopInt}. The
results for divergent parts can be then written back in the initial field basis.

Diagrams $K$, $L$ and $M$ of Fig. \ref{Rys:DiVac} are the one-loop diagrams with
insertions of the one-loop counterterms corresponding to the ``wave function''
and mass renormalization. We discuss them in more detail here in order to
illustrate the working of our regularization scheme. As explained in Section
\ref{sec:Subtractions}, the momenta in the counterterms must also be replaced
according to \refer{eqn:MomSpacePrescription}; for example, the counterterm
for the $\VEVOPI{A_\mu^\alpha A_\nu^\beta}$ function must have the form
\begin{equation}\label{Eq:Vect-2-point-contr}
\delta_{tot}\tig^{\mu\nu}_{\!\alpha\beta}(p,-p)
=\eta^{\mu\nu}(\tilde{\delta} m^2_{V})_{\alpha\beta}
+\KR^{\mu}\!(p)\KR^{\nu}\!(p)(\delta Z_{A,L})_{\alpha\beta}
-\eta^{\mu\nu}\KR(p)^2(\delta Z_{A,T})_{\alpha\beta}.
\end{equation}
As stressed, this is necessary for consistency of the
$\Lambda$-$\overline{\rm MS}$ scheme based on the regularization
\refer{eqn:MomSpacePrescription}: as revealed by the analysis of Section
\ref{sec:rg} (and \ref{app:rge}) only then it is possible to derive the
RG equations and give the action $I_\infty$ the meaning of the
bare action $I_{\rm B}$.
Because of this rule the integrals corresponding to the diagrams $K$, $L$
and $M$ of Fig. \ref{Rys:DiVac} are not completely regularized by the
prescription \refer{eqn:PrescriptionDef}. As found in Section
\ref{Sec:UV-cutoff}, one-loop vacuum graphs are  the only ones for which
such a situation can occur. However, unregularized parts of these diagrams
are background-independent and can be omitted in the calculation of
the effective potential ${\cal V}_{\rm eff}(\vp)$. Indeed,
${\cal V}_{\rm eff}(\vp)$ can be also determined by computing the
background-dependent contributions to the scalar one-point 1PI functions (i.e. to the scalar field tadpoles,
which according to the analysis of Section \ref{Sec:UV-cutoff} get
completely regularized by the prescription \refer{eqn:MomSpacePrescription}),
and integrating them with respect to $\varphi$.
Similarly, we will omit also all other $\varphi$-independent terms
proportional to $\Lambda^4$ in Eqs. \refer{Eq:V2L:A}-\refer{Eq:V2L:HIJ} below (in particular, the contribution of the ghost analog of the
diagram $K$ which is background independent in the Landau gauge).

The background-dependent contributions of the diagrams $K$, $L$, $M$ reduce to the single integral
\begin{eqnarray}\label{Eq:FunAexp-dokl}
(4\pi)^2\int\volfour{k} \frac{\,i}{\KR(k)^2-m^2}
&=&
\Lambda^2+m^2\left\{\ln\frac{m^2}{\mu^2}-1-\delta_\Lambda\right\}+\\
&{}&
+\frac{m^4}{\Lambda^2}\left\{3\ln\frac{m^2}{\mu^2}+\ln\frac{27}{8}
-5-3\,\delta_\Lambda\right\}+\mathcal{O}\!\bracket{\Lambda^{-3}},\nn
\end{eqnarray}
and read (tildes on $\Gamma$ indicate that the factors
$(2\pi)^4\delta^{(4)}_{\rm mom}(0)\equiv\int\!d^4x$ have been removed)
\begin{equation}
\tig^{}(|K)=\frac{1}{2}
\int\volfour{k}\ \! i\ \! \tr\!\left\{
\left[\KR(k)^2-m_S^2\right]^{-1}
\left[m_S^2\delta Z_\phi-\tilde{\delta}m_S^2\right]
\right\},\phantom{aaaaaaaa}~
\end{equation}
\begin{equation}%\label{Diag:Phi:S}
\tig^{}(|M)=\frac{3}{2}
\int\volfour{k}\ \! i\ \! \tr\!\left\{
\left[\KR(k)^2-m_V^2\right]^{-1}
\left[m_V^2\delta Z_{A,T}-\tilde{\delta}m_V^2\right]
\right\},\phantom{aaaaaa}~
\end{equation}
\begin{eqnarray}
%\label{Diag:Phi:S}
\tig^{}(|L)&=&-\int\volfour{k}\ \! i\ \! \tr\Big\{\!
\left[m_F^* m_F \delta Z_{F}\!+\!\delta Z_{F}m_F^* m_F
\!-\!\left(m_F^*\, \tilde{\delta}m_F\!+\!\tilde{\delta}m_F^*\, m_F
\right)\right]\times\nn\\
&{}&\qquad\qquad\qquad
\times\left[\KR(k)^2-m_F^* m_F\right]^{-1}\Big\}.
\end{eqnarray}
(The traces reduce to simple sums over mass eigenvalues in the basis
in which the mass matrices are diagonal). Here
\begin{equation}\label{Eq:KontrDiv-Z_AT-1loop}
(\delta Z_{A,T} )_{\alpha\beta}=\frac{\delta_\Lambda}{(4\pi)^2}\left\{
-\frac{13}{6}\tr\big[e_\alpha e_\beta\big]
+\frac{1}{6} \tr\big[\TS_\alpha \TS_\beta\big]
+\frac{2}{3}\tr\big[\TF_\alpha \TF_\beta\big]\right\},
\end{equation}
\begin{equation}\label{Eq:DeltaZ-Phi-1loop}
(\delta Z_\phi)_{ij}=\frac{\delta_\Lambda}{(4\pi)^2}\Big\{
-\frac{1}{2}\tr\!\left\{\YF_i\YF_j^*+cc.\right\}
-3\left(\TS_\alpha\TS^\alpha\right)_{ij}\Big\},
\end{equation}
\begin{equation}\label{Eq:KontrDiv:Z_F}
\delta Z_F=\frac{\delta_\Lambda}{(4\pi)^2}
\left\{ -\frac{1}{2}Y_i^* Y^i\right\},
\end{equation}
as well as (cf. Eq. \refer{Eq:RegFey:LamRhoMs})
\begin{eqnarray}%\label{Eq:KontrFlat:m_V}
(\tilde{\delta}m_S^2)_{ij}
&=&
\frac{\delta_\Lambda}{(4\pi)^2}\Big\{
-\frac{3}{2}(m_V^2)_{\alpha\beta}\!\left\{\TS^\alpha,\ \TS^\beta\right\}_{ij}
+\frac{3}{2}(\varphi^{\rm T}\!\!\left\{\TS_\alpha,\ \TS_\kappa\right\})_i\,
(\left\{\TS^\alpha,\ \TS^\kappa\right\}\!\varphi)_j\nn\\
&{}&%\hspace*{-35 pt}
\qquad\quad\
+\frac{1}{2}\lambda_{ijkl}(m_S^2)^{kl}
+\frac{1}{2}\tr\big[\rho_i\rho_j\big]
-2\,\tr\big[\YF_j \YF_i^* m_F m_F^*+cc.\big]+\nn
\\&{}&%\hspace*{-35 pt}
\qquad\quad\ -\tr\big[\YF_im_F^* \YF_j m_F^*+cc. \big]\Big\}+\nn\\
&{}&%\hspace*{-35 pt}
+\frac{\Lambda^2}{(4\pi)^2}
\left\{3(\TS^\alpha \TS_\alpha)_{ij}-\frac{1}{2}\lambda_{ijkl}\delta^{kl}
+\tr\big[\YF_i \YF_j^*+cc.\big]\right\},
\end{eqnarray}
\begin{equation}\label{Eq:KontrDiv:m_F}
\tilde{\delta} m_F=\frac{\delta_\Lambda}{(4\pi)^2}
\left\{-3 \TF_\gamma^{\rm T} m_F \TF^\gamma+Y_i m_F^* Y^i\right\},
\end{equation}
are minimal counterterms extracted from the expressions
\refer{Eq:DeltaZ:AA:AplusBplusC}, \refer{Eq:DeltaZ:PhiPhi:A+B} and
\refer{Eq:DeltaZ:PsiPsi:AiB}.
Non-minimal counterterms enter only through the diagram $M$ in which
\begin{equation}
\tilde{\delta}m^2_{V}=\tilde{\delta}^{0}\!m^2_{V}+{\delta}^{\flat}\! m^2_{V}\, ,
\end{equation}
consists of the minimal part
\begin{eqnarray}%\label{Eq:KontrFlat:m_V}
(\tilde{\delta}^0\!m_V^2)_{\alpha\beta}
&=&\frac{\delta_\Lambda}{(4\pi)^2}\Big\{
\frac{3}{4}\tr\big[m_V^2\TV_\alpha \TV_\beta\big]
-2\,\tr\big[\TF_\alpha m_F^*  \TF_\beta^* m_F\big]
+\tr\big[\!\left\{\TF_\alpha,\ \TF_\beta\right\}m_F^* m_F\big]+\nn\\
&{}&%\hspace*{-35 pt}
\qquad\quad\
+\frac{3}{4}\varphi^{\rm T}\!\!\left\{\TS_\alpha,\ \TS^\kappa\right\}\!
\left\{\TS_\beta,\ \TS_\kappa\right\}\!\varphi \Big\}+\nn\\
&{}&%\hspace*{-35 pt}
+\frac{\Lambda^2}{(4\pi)^2}\left\{\tr\big[e_\alpha e_\beta\big]
+\frac{1}{2} \tr\big[\TS_\alpha \TS_\beta\big]
-\tr\big[\TF_\alpha \TF_\beta\big]\right\},
\end{eqnarray}
and the non-minimal one, ${\delta}^{\flat}\! m^2_{V}$, given by
\refer{Eq:KontrFlat:m_V}. The counterterm $\delta Z_{A,L}$
in \refer{Eq:Vect-2-point-contr}
in which $\delta Z_{A,L}=\delta Z_{A,T}+\delta^\flat\!z_A$, where the
non-minimal $\delta^\flat\!z_A$ part is given by \refer{Eq:KontrFlat:z_A},
does not contribute because the vector propagator is
transverse in the Landau gauge (cf. \refer{Eq:FermProp}).

In combining the contributions of the genuine two-loop diagrams $A$-$J$
with those of the counterterm diagrams $K$, $M$, $L$, it is convenient to
decompose the diagrams $K$-$M$ into pieces proportional to different types
of couplings; schematically:
\begin{equation}
L=L_{\YF}\oplus L_{\TF},\quad
M=M_{\TS}\oplus M_{\TF}\oplus M_{\TV} \oplus M_{\varphi},\quad
K=K_{\YF}\oplus K_{\TS}\oplus K_{\lambda} \oplus K_{\rho}\oplus K_{\varphi}.\quad
\end{equation}
Similarly, it is convenient to decompose contributions of the fermionic
two-loop diagrams $C$ and $G$ into pieces $I$ and $II$
\begin{equation}
C=C_I\oplus C_{II},
\qquad
G=G_I\oplus G_{II}.
\end{equation}
corresponding to the product of, respectively, two masses and two momenta
arising from numerators of propagators of the Majorana fields. As usually,
combining contributions of genuine two-loop diagrams
with those of the counterterm diagrams should remove all divergences
non-polynomial in the background field dependent mass matrices
providing thereby a nontrivial check of the consistency of the
whole computation.

Having computed the divergent ($\varphi$-dependent) contributions to the
zero-point 1PI function one can determine those counterterms of $I_2$ which
are necessary to renormalize up to the order $\hbar^2$ the effective
potential ${\cal V}_{\rm eff}$. In other words, one can determine the
counterterm $\mathcal{V}_\infty^{(2)}(\varphi)$ in
\begin{equation}\label{Eq:V_inf_expanded}
I_2=-\left\{\mathcal{V}(\phi+\varphi)
+\sum_{\ell=1}^{2}\frac{\hbar^\ell}{(4\pi)^{2\ell}}\,
\mathcal{V}_\infty^{(\ell)}(\phi+\varphi)\right\}+\ldots\,,
\end{equation}
(the ellipsis stand for derivative terms, and terms involving fields other than $\phi$). In the $\Lambda$-$\overline{\rm MS}$ scheme defined
in Section \ref{sec:Subtractions} the functions $\mathcal{V}_\infty^{(\ell)}$
are pure divergences, that is, vanish if one sets first $\delta_\Lambda=0$
and then $\Lambda^2=0$.
The one-loop counterterm $\mathcal{V}_\infty^{(1)}$ can be read off from
\refer{Eq:DeltaZ:Phi} and reads
\begin{eqnarray}\label{Eq:V-inf-1loop}
\mathcal{V}_\infty^{(1)}(\varphi)
&=&
-\frac{\Lambda^2}{2}\Big[
\tr\!\left\{\mathcal{M}_S^2(\varphi)\right\}
-2\tr\!\left\{\mathcal{M}_F(\varphi)\mathcal{M}_F(\varphi)^*\right\}
\!+3\tr\!\left\{\mathcal{M}_V^2(\varphi)\right\}
\!\Big]\phantom{aaaa}
\\&{}&%\!\!\!\!\!\!\!\!\!\!\!\!\!\!\!
+\frac{\delta_\Lambda}{2}\Big[
\frac{1}{2} \tr\!\left\{\mathcal{M}_S^2(\varphi)^2\right\}
-\tr\!\left\{\big[\mathcal{M}_F(\varphi)\mathcal{M}_F(\varphi)^*\big]^2\right\}
\!+\!\frac{3}{2} \tr\!\left\{\mathcal{M}_V^2(\varphi)^2\right\}\!
\Big].\nn
\end{eqnarray}
We present the result for
$\mathcal{V}_\infty^{(2)}$ dividing it (using an obvious notation, e.g. writing
${\cal M}_X$ for ${\cal M}_X(\varphi)$ and ${\cal M}_X^4$ for
$[{\cal M}^2_X(\varphi)]^2$ etc., see
also the definitions \refer{Eq:RegFey:LamRhoMs} and  \refer{Eq:bbY-def})
into pieces which remove divergences from the sums of genuine
two-loop diagrams and the counterterm diagrams of Fig.
\ref{Rys:DiVac}, in which cancellations of nonlocal divergences occur:
\begin{eqnarray}\label{Eq:V2L:A}
\mathcal{V}^{(2)}_{\infty}(\mathcal{\varphi}|A\oplus K_\rho)
&=&
\frac{1}{4}\Lambda^2\left(-\delta_\Lambda+\ln\frac{4}{3}\right)
          \delta^{ij}\delta^{k m}\delta^{l n}
                  \mathcal{V}'''_{ikl}(\varphi)\mathcal{V}'''_{jmn}(\varphi)
+\nn\qquad\quad\\&{}&
+\frac{1}{8}\left(\delta_\Lambda^2-2\delta_\Lambda\right)
          \mathcal{M}_S^{2}(\varphi)^{ij}\delta^{k m}\delta^{l n}
          \mathcal{V}'''_{ikl}(\varphi)\mathcal{V}'''_{jmn}(\varphi),
\end{eqnarray}
\begin{eqnarray}
\mathcal{V}^{(2)}_{\infty}(\mathcal{\varphi}|B\oplus K_{\lambda})
&=&
-\frac{1}{4}\Lambda^2\delta_\Lambda \,
\lambda_{ijkl}\mathcal{M}_S^{2}{}^{ij}\delta^{kl}
%+\nn\qquad\quad\\&{}&
+\frac{1}{8}\delta^2_\Lambda\, \lambda_{ijkl}
\mathcal{M}_S^{2}{}^{ij}\mathcal{M}_S^{2}{}^{kl},
\end{eqnarray}
\begin{eqnarray}
\mathcal{V}^{(2)}_{\infty}(\mathcal{\varphi}|C_I\oplus K_{Y,I}\oplus L_{Y,I})
&=&
\frac{3}{2}\Lambda^2\left(\delta_\Lambda-\ln\frac{4}{3}\right)
          \tr\!\left\{\YF^i\!\mathcal{M}_F^*\YF_i\mathcal{M}_F^*+cc.\right\}
+\qquad\quad\\&{}&\hspace*{-140 pt}
-\frac{1}{4}\!\!\left(\delta_\Lambda^2\!-\!2\delta_\Lambda\right)\!\!
   \left\{
   \mathcal{M}^2_{S\, ij}\,
   \tr\!\left[\YF^i\!\mathcal{M}_F^*\YF^j\!\mathcal{M}_F^*\!+\!cc.\!\right]
\!+\!2\,\tr\!\left[\YF^j\!\mathcal{M}_F^*\YF_j\mathcal{M}_F^*
\mathcal{M}_F\mathcal{M}_F^*\!+\!cc.\!\right]\right\},\nn
\end{eqnarray}
\begin{eqnarray}
\mathcal{V}^{(2)}_{\infty}(\mathcal{\varphi}|C_{II}\oplus K_{Y,II}\oplus L_{Y,II})
&=&
\Lambda^2\delta_\Lambda\left\{
\frac{3}{2} \tr\!\left[\YF^i\!\mathcal{M}_F^*\mathcal{M}_F\YF_i^*+cc.\right]
-\frac{1}{4}\tr\!\left[\mathcal{M}_S^2\mathbb{Y}\right]
\right\}\!\!
\nn\\&{}&\hspace*{-150 pt}
+\Lambda^2\Big\{
\frac{1}{12}(64\ln2+15\ln3-25\ln5-11\ln11)\tr\!\left[\YF^i\!
\mathcal{M}_F^*\mathcal{M}_F\YF_i^*+cc.\right]
\nn\\&{}&\hspace*{-150 pt}
\qquad\ \  +
\frac{1}{2}\left[ -1 -\frac{45}{2}\ln3+\frac{25}{2}\ln5+9\ln2 \right]
\tr\!\left[\mathcal{M}_S^2\mathbb{Y}\right]
\Big\}
+\nn\\&{}&\hspace*{-150 pt}
-\frac{\delta_\Lambda^2}{8}\Big\{
2\tr\!\left[\YF_i^*\!(\mathcal{M}_F\mathcal{M}_F^*)^2\YF^i\!+\!cc.\right]
\!+\!4\mathcal{M}_{S\, ij}^2 \tr\!\left[\YF^{i*}\!\mathcal{M}_F\mathcal{M}_F^*\YF^j\!+\!cc.\right]
\!-\!\tr\!\left[\mathcal{M}_S^4\mathbb{Y}\right]
\!\Big\}\!\!\!
\nn\\&{}&\hspace*{-150 pt}
+\delta_\Lambda\Big\{
\frac{1}{2}\tr\!\left[\YF_i^*\!\mathcal{M}_F\mathcal{M}_F^*\YF^i\mathcal{M}_F^*\mathcal{M}_F\!+\!cc.\right]
\!-\frac{1}{16}\left(7+4\ln2\right)\!\tr\!\left[\mathcal{M}_S^4\mathbb{Y}\right]
\!+\!
\nn\\&{}&\hspace*{-150 pt}
\qquad\
+\frac{1}{12}\left[7+6\ln\frac{4}{3}\right]\tr\!
\left[\YF_i^*\!(\mathcal{M}_F\mathcal{M}_F^*)^2\YF^i\!+\!cc.\right]\!\Big\}\, ,
\end{eqnarray}
\begin{eqnarray}
\mathcal{V}^{(2)}_{\infty}(\mathcal{\varphi}|D\oplus E \oplus K_{\TS}\oplus M_{\TS})
&=&
\Lambda^2\delta_\Lambda\left\{
-\frac{3}{2} \tr\!\left[\mathcal{M}_S^2\TS_\alpha\TS^\alpha\right]
+\frac{7}{4}\tr\!\left[\TS^\alpha\TS^\beta\right]\mathcal{M}_{V\, \alpha\beta}^2
\right\}\!\!
\nn\\&{}&\hspace*{-150 pt}
+\Lambda^2\Big\{
\left[\frac{17}{8}-\frac{51}{4}\ln2+\frac{459}{16}\ln3-\frac{125}{8}\ln5\right]
\tr\!\left[\TS^\alpha\TS^\beta\right]\mathcal{M}_{V\, \alpha\beta}^2
+\nn\\&{}&\hspace*{-150 pt}
\qquad\
+\!
\left[\frac{29}{16}-\frac{1901}{72}\ln2-\frac{369}{32}\ln3
\!+\!\frac{125}{72}\ln5+\frac{847}{72}\ln11\right]
\!\tr\!\left[\mathcal{M}_S^2\TS_\alpha\TS^\alpha\right]\!
\Big\}
\nn\\&{}&\hspace*{-150 pt}
+\frac{\delta_\Lambda^2}{8}\Big\{
6\tr\!\left[\mathcal{M}_S^4\TS_\alpha\TS^\alpha\right]
-6\tr\!\left[\mathcal{M}_S^2\TS^\beta\TS^\alpha\right]\mathcal{M}_{V\,\alpha\beta}^2
-\tr\!\left[\TS^\alpha\TS^\beta\right]\mathcal{M}_{V\, \alpha\beta}^4
\Big\}+
\nn\\&{}&\hspace*{-150 pt}
+\delta_\Lambda\Big\{
\frac{1}{96}\left[47+24\ln2\right]
\tr\!\left[\TS^\alpha\TS^\beta\right]\mathcal{M}_{V\, \alpha\beta}^4
-\frac{3}{8}\left[2+\ln\frac{32}{9}\right]
\tr\!\left[\mathcal{M}_S^4\TS_\alpha\TS^\alpha\right]
+\nn\\&{}&\hspace*{-150 pt}
\qquad\ -\frac{3}{2}\tr\!\left[\TS_\alpha\mathcal{M}_S^2\TS^\alpha
\mathcal{M}_S^2\right]
-\frac{3}{2}\tr\!\left[\mathcal{M}_S^2\TS^\beta\TS^\alpha\right]
\mathcal{M}_{V\,\alpha\beta}^2\Big\}\, ,
\end{eqnarray}
\begin{eqnarray}
\mathcal{V}^{(2)}_{\infty}(\mathcal{\varphi}|F \oplus K_{\varphi}\oplus M_{\varphi})
&=&
\left[-\frac{15}{8}\Lambda^2\delta_\Lambda\,
+\Lambda^2
\left(-\frac{1}{8}+\frac{189}{32}\ln2-\frac{45}{16}\ln3\right)\right]\times
\nn\\&{}&\hspace*{-110 pt}
\times\varphi^{\rm T}\!\big\{\TS_\alpha,\ \TS_\beta\big\}
\big\{\TS^\alpha,\ \TS^\beta\big\}\varphi
+
\nn\\&{}&\hspace*{-110 pt}
+{\delta_\Lambda^2}\Big[
\frac{3}{8}
\varphi^{\rm T}\!\big\{\TS_\alpha,\ \TS_\beta\big\}\mathcal{M}_S^2
\big\{\TS^\alpha,\ \TS^\beta\big\}\varphi+
\frac{9}{16}
\varphi^{\rm T}\!\big\{\TS_\alpha,\ \TS^\gamma\big\}
\big\{\TS^\alpha,\ \TS^\delta\big\}\varphi\, \mathcal{M}_{V\, \gamma\delta}^2
\Big]
\nn\\&{}&\hspace*{-110 pt}
-{\delta_\Lambda}\Big[
\frac{3}{8}
\varphi^{\rm T}\!\big\{\TS_\alpha,\ \TS_\beta\big\}\mathcal{M}_S^2
\big\{\TS^\alpha,\ \TS^\beta\big\}\varphi
+
\nn\\&{}&\hspace*{-110 pt}
\qquad\
+\frac{3}{32}\left(14+3\ln\frac{32}{9}\right)
\varphi^{\rm T}\!\big\{\TS_\alpha,\ \TS^\gamma\big\}
\big\{\TS^\alpha,\ \TS^\delta\big\}\varphi\, \mathcal{M}_{V\, \gamma\delta}^2\Big],
\end{eqnarray}
\begin{eqnarray}
\mathcal{V}^{(2)}_{\infty}(\mathcal{\varphi}|G_{II}\oplus M_{\TF,II})
&=&
\Lambda^2\delta_\Lambda\left\{
\tr\!\left[\TF^\alpha\TF^\beta\right]\mathcal{M}_{V\, \alpha\beta}^2
-\frac{3}{2}
 \tr\!\left[\mathcal{M}_F^*\mathcal{M}_F\TF_\alpha\TF^\alpha\!+\!cc.\right]
\right\}
+
\nn\\&{}&\hspace*{-110 pt}
+\Lambda^2\Big\{
\left[-\frac{13}{4}+\frac{61}{2}\ln2-{54}\ln3+{25}\ln5\right]
\tr\!\left[\TF^\alpha\TF^\beta\right]\mathcal{M}_{V\, \alpha\beta}^2
+\nn\\&{}&\hspace*{-110 pt}
\qquad\
+\!
\left[\frac{667}{18}\ln2\!+\!21\ln3
\!+\!\frac{25}{9}\ln5\!-\!\frac{187}{9}\ln11\!-\!\frac{13}{8}\right]
\!\tr\!\left[\mathcal{M}_F^*\mathcal{M}_F\TF_\alpha\TF^\alpha\!+\!cc.\right]
\!\Big\}
\nn\\&{}&\hspace*{-110 pt}
+{\delta_\Lambda^2}\Big\{
\frac{3}{4}\tr\!\left[\mathcal{M}_F^*\mathcal{M}_F\TF^\alpha\TF^\beta\!+\!cc.\right]
\mathcal{M}_{V\, \alpha\beta}^2
-\frac{1}{2}\tr\!\left[\TF^\alpha\TF^\beta\right]\mathcal{M}_{V\, \alpha\beta}^4
\Big\}+
\nn\\&{}&\hspace*{-110 pt}
+{\delta_\Lambda}\Big\{
\!-\!
\tr\!\left[(\mathcal{M}_F^*\mathcal{M}_F)^2\TF^\alpha\TF_\alpha\!+\!cc.\right]
+
\frac{3}{8}(1-4\ln2)
\tr\!\left[\mathcal{M}_F^*\mathcal{M}_F\TF^\alpha\TF^\beta\!+\!cc.\right]
\mathcal{M}_{V\, \alpha\beta}^2
\nn\\&{}&\hspace*{-110 pt}
\qquad\ \ \!
+\frac{1}{12}(19+6\ln4)
\tr\!\left[\TF^\alpha\TF^\beta\right]\mathcal{M}_{V\, \alpha\beta}^4
\Big\},
\end{eqnarray}
\begin{eqnarray}
\mathcal{V}^{(2)}_{\infty}(\mathcal{\varphi}|G_{I}\oplus M_{\TF,I}\oplus L_{\TF})
&=&
\Lambda^2\!\left[\frac{9}{2}\delta_\Lambda
\!+\!\frac{9}{2}\ln{3}\!-\!\frac{21}{2}\ln2\!-\!\frac{9}{8}\right]
\!\tr\!\left[\mathcal{M}_F\TF^\alpha\!\mathcal{M}_F^*\TF^*_\alpha\!+\!cc.\right]
\nn\\&{}&\hspace*{-110 pt}
-
\delta^2_\Lambda
\Big\{
\frac{3}{2}
\tr\!\left[\mathcal{M}_F\mathcal{M}_F^*\mathcal{M}_F\TF^\alpha\!\mathcal{M}_F^*\TF^*_\alpha\!+\!cc.\right]
+
\frac{3}{4}
\tr\!\left[\mathcal{M}_F\TF^\alpha\!\mathcal{M}_F^*\TF^{\beta*}\!+\!cc.\right]
\mathcal{M}_{V\, \alpha\beta}^2
\Big\}
+
\nn\\&{}&\hspace*{-110 pt}
+\delta_\Lambda
\Big\{
3
\tr\!\left[\mathcal{M}_F\mathcal{M}_F^*\mathcal{M}_F\TF^\alpha\!
\mathcal{M}_F^*\TF^*_\alpha\!+\!cc.\right]
+\nn\\&{}&\hspace*{-110 pt}
\qquad\
+
\frac{3}{8}(7+4\ln2)
\tr\!\left[\mathcal{M}_F\TF^\alpha\!\mathcal{M}_F^*\TF^{\beta*}\!+\!cc.\right]
\mathcal{M}_{V\, \alpha\beta}^2\Big\},
\end{eqnarray}
\begin{eqnarray}\label{Eq:V2L:HIJ}
\mathcal{V}^{(2)}_{\infty}(\mathcal{\varphi}|H\oplus I \oplus J\oplus M_{\TV})
&=&
\Lambda^2\delta_\Lambda\left\{
-\frac{35}{8}
\tr\!\left[\TV^\beta\TV^\alpha\right]\mathcal{M}_{V\, \beta\alpha}^2
\right\}+
\nn\\&{}&\hspace*{-130 pt}
+\Lambda^2
\left[
\frac{39}{4}-\frac{9695}{96}\ln2+\frac{81}{4}\ln3
-\frac{625}{24}\ln5+\frac{847}{24}\ln11\right]
\tr\!\left[\TV^\beta\TV^\alpha\right]\mathcal{M}_{V\, \beta\alpha}^2
\nn\\&{}&\hspace*{-130 pt}
+{\delta_\Lambda^2}\Big\{
\frac{13}{8}\tr\!\left[\TV^\beta\TV^\alpha\right]\mathcal{M}_{V\, \beta\alpha}^4
+\frac{9}{16}
\tr\!\left[\TV_\alpha\mathcal{M}_V^2\TV^\alpha\mathcal{M}_V^2\right]
\Big\}
+
\nn\\&{}&\hspace*{-130 pt}
-
{\delta_\Lambda}\Big\{
\left[\frac{61}{48}+\frac{103}{16}\ln2-\frac{27}{8}\ln3\right]
\tr\!\left[\TV^\beta\TV^\alpha\right]\mathcal{M}_{V\, \beta\alpha}^4+
\nn\\&{}&\hspace*{-130 pt}
\qquad\
+\left[\frac{129}{16}+\frac{3}{32}\ln\frac{2^{15}}{3^6}\right]
\tr\!\left[\TV_\alpha\mathcal{M}_V^2\TV^\alpha\mathcal{M}_V^2\right]
\Big\}.
\end{eqnarray}

The function ${\cal V}_\infty^{(2)}$ given by the sum of the expressions
\refer{Eq:V2L:A}-\refer{Eq:V2L:HIJ} is indeed polynomial in the
$\varphi$-dependent masses, in agreement with the expectations. Furthermore,
it has been established in Section \ref{sec:rg} that the (local) action
\refer{Eq:V_inf_expanded} with the two-loop counterterms included should
satisfy  the RGE of the form $R_2I_2=\mathcal{O}(\hbar^3)$
(cf. Eq. \refer{eqn:RnDef}); this, in particular, implies the
following relation
\begin{equation}\label{Eq:Kombinacja}
{\beta}_{(2)}^\mathcal{V}(\varphi)-\matrixind{\gamma_{(2)}^\phi}{i}{j}
{\varphi}^j\derp{}{{\varphi}^i}\mathcal{V}(\varphi)=v^{(2)}(\varphi)\,,
\end{equation}
where $v^{(2)}(\varphi)$ is the coefficient of ${1\over2}\de_\La$ (i.e.
of $\ln(\Lambda/\mu)$) in $\cV_\infty^{(2)}(\vp)$.
We have verified that $v^{(2)}(\varphi)$ extracted from the formulae
\refer{Eq:V2L:A}-\refer{Eq:V2L:HIJ} agrees with the left hand side of
\refer{Eq:Kombinacja} computed using the result \refer{Eq:Beta-V-2loop-LamReg}
combined with the DimReg result \refer{Eq:BetaVCheck2L}.
(Notice that \refer{Eq:Beta-V-2loop-LamReg} gives precisely the difference
appearing on the left hand side of \refer{Eq:Kombinacja}.)
Moreover, the RGE $R_2 I_2=\cO(\hbar^3)$ implies that the coefficients
of the $\Lambda^2\times \delta_\Lambda$ and $\delta^2_\Lambda$ terms in
\refer{Eq:V2L:A}-\refer{Eq:V2L:HIJ} should be entirely fixed by the 1-loop
divergences
\refer{Eq:V-inf-1loop} and the 1-loop $\beta$ and $\gamma$ functions; we have
verified that this is indeed the case. In particular, up to the (background)
field renormalization, in $\mathcal{V}_\infty^{(2)}(\varphi)$ the terms
proportional to $\Lambda^2\times \delta_\Lambda$ can be obtained from quadratic
divergences in \refer{Eq:V-inf-1loop} by replacing there the renormalized
couplings with the bare ones.

Finally, in the results \refer{Eq:V2L:A}-\refer{Eq:V2L:HIJ} there is a new
information, which is not a mere consistency check of our earlier results:
this is the 2-loop coefficients of the quadratic divergence (of $\Lambda^2$)
which is important for the hierarchy problem (Section \ref{sec:HP}). The
explicit form of this coefficient for the SM is given by \refer{Eq:DelFH}.
It differs from the one derived in \cite{HAKAOD} where superficially a
similar regularization was used.

A possible explanation of the discrepancy of our result \refer{Eq:DelFH}
and that of \cite{HAKAOD} follows from the observation that the latter one
is reproduced if: {\it i)} after the reduction to the basic integrals
\refer{Eq:Itot-def}-\refer{Eq:M1tot-def} only the ``sunset'' integrals
\refer{Eq:Itot-def} contribute to the two-loop coefficients of the quadratic
divergences (in other words, contributions of the remaining basic integrals
are assumed to cancel exactly with the contributions of the counterterm
diagrams $K$, $L$ and $M$, {\it ii)} all the sunset integrals
\refer{Eq:Itot-def} occur in the same version $n_1=n_2=n_3=1$.
These assumptions are satisfied in the DimRed scheme because the coefficient
of the quadratic divergence of the sunset integral \refer{Eq:Itot-def} is in
DimRed given by the residue of the pole at $d=3$ \cite{Jones,JonesSTARY},
while the quadratic divergences of the remaining $\cO(\hbar^2)$ contributions
correspond to residues of the poles at $d=2$. The fact that the result of
\cite{HAKAOD} agrees with the one of \cite{Jones,JonesSTARY} suggests that
the cutoff on the integrals was in \cite{HAKAOD} imposed after their
reduction to the basic integrals. In contrast, our result
\refer{Eq:DelFH} is obtained using the consistent implementation of the
cutoff procedure of Section \ref{Sec:UV-cutoff}
(which requires making the substitution \refer{eqn:PrescriptionDef} at the
level of the complete action, including the counterterms, that is
\emph{before} reducing Feynman integrals to the basic ones, so that
no operations on divergent integrals are performed) which violates
the above assumptions (for example  {\it ii)} is violated by the
fermionic loop of the diagram $C$). It is this
consistent implementation which allows to
prove the RGE and is therefore the one in which the conjecture of
\cite{Fujikawa,HAKAOD} (proved in Section \ref{sec:rg}) is valid.

\section{Bare parameters and the Hierarchy problem}
\label{sec:HP}

\renewcommand{\thesection}{\arabic{section}}
\renewcommand{\theequation}{\arabic{section}.\arabic{equation}}
\renewcommand{\thefigure}{\arabic{section}.\arabic{figure}}

\setcounter{equation}{0}
\setcounter{figure}{0}

In its most applications the role of QFT is to establish relations
between various low energy data. In this context renormalization allows
to parametrize predictions of a concrete model, like the SM, in terms
of a small set of finite parameters. Regularization is then only an
auxiliary procedure which is chosen following the requirements of
calculational convenience and counterterms implementing subtractions
are not treated as carrying any physical information - they become infinite
when, at the end, the regularization is removed. In such applications of
QFT the origin and magnitude of finite parameters, like masses
of physical particles, are not an issue.

With an explicit UV cutoff, like the one introduced in Section
\ref{Sec:UV-cutoff}, one can, however, take another point of view
(ubiquitous in applications of field theory to critical phenomena) and,
keeping the UV cutoff finite and fixed, treat the action $I^\Lambda_\infty$
with the counterterms constructed in the process of renormalization as
the fundamental object - the ``bare'' action expressed in terms of ``bare''
parameters and ``bare'' fields. That such a bare action can be defined
when the cutoff breaks the BRST symmetry\footnote{If there is a physical
regulator preserving all symmetries necessary for quantum consistency (or as
in the $\phi^4$ model, there is simply no continuous symmetries) there
is no need to construct counterterms: it is possible to start directly
from the bare action which takes then the same form as $I^\Lambda_0$.}
has been shown in Section \ref{sec:rg}. Once such a bare action
$I^\Lambda_{\rm B}$ is obtained, there is in fact no need to split  bare
parameters (and fields) into renormalized ones and counterterms: it is
perfectly possible to compute Green's functions directly in terms of bare
parameters (keeping the regularizing cutoff finite)
- when they are used to express, order by order in the perturbative
expansion, physical quantities in terms of a selected
set of other physical quantities (like $M_Z$, $G_F^{-1/2}$, $\alpha_{\rm EM}$, etc. 
in the SM) all potential infinities  disappear leaving relations which
would remain finite
in the limit of removed UV cutoff (in practical application it is then
convenient to remove the cutoff entirely to simplify the results;
nothing however prevents in principle keeping the regulator finite, at
least when no gauge fields are present - see the remarks at the end of
Section \ref{sec:Subtractions}).
In such an approach the UV cutoff can be given a physical meaning
e.g. of the inverse of the lattice spacing of a statistical model
underlying the considered field theory model or, as we want to treat it
here, the characteristic scale
of a more fundamental finite theory. The question why the measured
masses of physical particles described by the model, like $W^\pm$, $Z^0$
or the Higgs boson are orders of magnitude smaller than the value of the
physical UV cutoff $\Lambda$, which should be comparable to the
Planck scale\footnote{Each physical
intermediate scale between the electroweak scale and the Planck one
potentially generates a hierarchy problem, if the effective quantum field
theory valid below the intermediate scale involves scalar fields; in our
considerations  we assume absence of such intermediate scales.}, 
becomes then important and is known as
the hierarchy problem.

To study the hierarchy problem as described above 
one has to assume that at the most fundamental level physics of  all
interactions, including the gravitational ones, is described by some
(most probably finite) theory, which may be not a QFT, and (like
Loop Quantum Gravity) may even
give a completely different view on space and time, which predicts all
measured quantities in terms of a single dimensionful parameter, to be
identified with $\Lambda$, which is its intrinsic scale.
It is then quite natural to expect that all predictions of this
hypothetical fundamental theory pertaining to low energy physics (low with
respect to $\Lambda$), in the limit in which  departures of the
space-time from the flat Minkowski space-time are neglected and coupling
to the gravitational sector ignored, can be obtained from an
effective finite field theory whose bare action $I_{\rm B}^\Lambda$
and bare parameters are fixed by the fundamental theory. Moreover, taking
into account the putative finiteness of the fundamental theory, it is natural
to assume that it is the intrinsic scale of the latter that acts in the
effective theory as the UV cutoff. It is also conceivable that the complete 
effective field theory action $I^\Lambda_{\rm B}$ contains also terms 
suppressed by $\Lambda$ whose effect is such that amplitudes
computed in the effective theory eventually do satisfy for finite $\Lambda$
all the necessary ST identities, even though $I^\Lambda_{\rm B}$ 
is not BRST invariant. 
In such a scenario the underlying
hypothetical fundamental theory of all interactions must by itself solve the
fundamental aspect of the hierarchy problem, that is predict the ratio
$M_W/M_{\rm Pl}\sim M_W/\Lambda$. But even if it does,
the  hierarchy problem generically manifests itself at the level
of the effective low energy field theory as the fine cancellation between
the bare mass square parameters $m^2_{\rm B}$
(like \refer{eqn:barebosonmassesdef}) of the scalar fields (if
such fields are present in the low energy theory) and, as is clear from
\refer{eqn:barevectormassesdef}, also of bare masses squared of the vector
fields $(M^2_V)_{\rm B}$ (if
the built-in cutoff violates explicitly the gauge symmetry) and the
order $\Lambda^2$ contributions in the perturbative calculation
of the physical $W^\pm$, $Z^0$ and the Higgs boson masses.

Of course, if it is assumed, as it must, that the fundamental theory
predicts correctly the ratio $M_W/M_{\rm Pl}$ (and $M_h/M_{\rm Pl}$), the above
cancellation is an artifact of using the effective field theory.
Nevertheless it is precisely this cancellation (which can be termed
the ``technical'' aspect of the hierarchy problem),
which from the point of view of the low energy effective theory
is perceived as the  main hierarchy problem and attempts at solving it
entirely within the effective theory, undertaken over years, have
led to many ideas such as technicolor or low energy supersymmetry,
extra dimensions, etc.

If one adopts this attitude toward the hierarchy problem, it is just the
cutoff dependent bare action of the effective theory which is of special
interest. Of course the fundamental theory is unknown and, therefore, neither
the corresponding bare action  nor the way the intrinsic scale $\Lambda$
of the fundamental theory acts in it as a cutoff are known. Nevertheless, it
may by enlightening, using the bare action $I_{\rm B}^\Lambda$ of Section
\ref{sec:rg} (which, with the cutoff $\Lambda$ implemented as in Section
\ref{Sec:UV-cutoff}, has many features expected from the realistic
effective theory - after expanding the regularizing exponential functions
\refer{eqn:PrescriptionDef} it consists of infinite set of operators of
growing dimensions, coefficients of operators containing gauge fields are
given by infinite series in bare couplings) and assuming a concrete form
of the action $I_0$ - be it the SM or some of its extensions
- to pursue a kind of a ``bottom-up'' approach
and investigate the resulting structure of the bare effective action
(as a function of the unknown scale $\Lambda$)
implied  by the low energy data. In particular it can be interesting
within such an approach to see, using the RG equations of Section \ref{sec:rg}
to evolve the renormalized parameters from the electroweak scale up to 
 the high scale (of order $M_{\rm Pl}$) where they become bare parameters of $I_{\rm B}^\Lambda$
(see \refer{eqn:BareCouplingsDef}),  whether one can get some clues  to the
technical aspect of the hierarchy problem, at least as far as the cancellation
between bare masses squared of scalar fields and the $\Lambda^2$
contributions are concerned.\footnote{It seems, however, that the problem of
a similar cancellation for vector fields must be taken care of by some other
mechanism operating at the level of the fundamental theory, possibly
related to the one which is necessary to restore the BRST invariance
for finite $\Lambda$ - see the remarks at the end of Section
\ref{sec:Subtractions}.}

In \cite{CHLEMENI}, inspired by the study
\cite{HAKAOD}, we have envisaged a possibility which, if realized in Nature,
would in fact imply absence of such a cancellation. This possibility - viewed
from the perspective of the bottom-up approach - is the potential existence
of a particular cutoff scale\footnote{It is more convenient to work with the rescaled cutoff $\bar\Lambda\approx 0.32 \La$ introduced in Eq. \refer{Eq:delta_Lambda}.} 
$\bar\Lambda=\bar\Lambda_{\star}$ at which all  
contributions proportional to $\Lambda^2$ to the counterterms to the scalar
field masses squared, that is all coefficients $f^C(\lambda_{\rm B}(\bar\La))$ in
\refer{eqn:barebosonmassesdef}, simultaneously vanish. If such a value of
$\bar\La$ exists and is reasonably close to Planck scale,
one can take the position that $\Lambda_{\star}\approx 3\bar\Lambda_{\star}$ is perhaps the
intrinsic scale of the fundamental theory and that the obtained
bare action $I^\Lambda_{\rm B}$ defined as in \refer{eqn:BareIdef}
is the bare action of the corresponding
effective field theory. Absence of terms proportional to $\Lambda^2$
in \refer{eqn:barebosonmassesdef}, i.e. the vanishing of all coefficients $f^C$,
would of course
mean absence of the technical hierarchy problem in the effective theory.
While this bears some resemblance to the well known Veltman
condition \cite{VE}, the important differences should be noted: the Veltman
condition was imposed on the \emph{renormalized} couplings at the
electroweak scale. Moreover, if the DimRed is used as the regularization
(as advocated by Veltman), the leading quadratic divergences correspond, at
$L$-th loop, to simple poles at $d=4-2/L$; therefore vanishing
of quadratic divergences requires in DimRed an infinite number of
constraints on coupling constants (which can be simultaneously satisfied
only if there is a special symmetry, like e.g. the supersymmetry).
In contrasts, in the consistent regularization based on a physical UV
momentum cutoff $\Lambda$, like the one of Section \ref{Sec:UV-cutoff},
coefficients of quadratic divergences arising from
consecutive loops combine (as shown in Section \ref{sec:rg}) to
cutoff independent functions of \emph{bare}  couplings,
and the number of constraints coincides with the (finite and small) number
of scalar field multiplets and, therefore, their vanishing does not
require any additional (super)symmetry.

If all coefficients $f^C(\lambda_{\rm B})$ in \refer{eqn:barebosonmassesdef}
do vanish simultaneously, the smallness of the electroweak scale $G_F^{-1/2}$
and Higgs boson mass(es) compared to the Planck scale must be ensured by
the smallness compared to the scale $\Lambda$ of the functions
$(\bar m^2)^C$ in \refer{eqn:barebosonmassesdef}; this, in turn, must
be ensured by the fundamental theory,
much in the same way as the smallness of soft supersymmetry breaking
scalar masses in supersymmetric  low energy effective theories
must be ensured by a supersymmetry breaking mechanism operating in the
underlying more fundamental theory.
\vskip0.2cm

In \cite{CHLEMENI} using the one-loop RG equations (which are identical in
$\Lambda$-$\overline{{\rm MS}}$ and in DimReg-$\overline{{\rm MS}}$ schemes)
and one-loop approximation to the functions $f^C$ in
\refer{eqn:barebosonmassesdef} we have shown that the scenario described
above can be realized in the extension of the SM considered earlier
in \cite{MeNi} and consisting of an extra complex singlet scalar field
and three right-chiral gauge singlet neutrino fields. Here, using the
two-loop RG  equations derived in Section \ref{sec:rg} and the two-loop
approximation to the functions $f^C$ we analyze this possibility taking
for $I_0$ the SM action. One of the reasons for doing this exercise is to
get an estimate on the simplest possible example of changes brought in by
the systematic inclusion of all two-loop effects.

In the SM there is only one $SU(2)$ doublet of scalar fields and,
consequently, only one function $f$ defined by \refer{eqn:barebosonmassesdef}
(in the model analyzed in \cite{CHLEMENI} there were two such functions).
The one-loop contribution $f^{(1)}$ to
\begin{eqnarray}\label{Eq:ffff}
f=f^{(1)}+(4\pi)^{-2} f^{(2)}+\ldots
\end{eqnarray}
can be read off from \refer{Eq:V-inf-1loop} and reads (for the normalization
of the couplings - see Eqs. \refer{Eq:Pot-SM}-\refer{Eq:Killing}; all Yukawa
couplings other than the top one, $y_t$, are neglected):
\eqs{\label{Eq:QuadDiv1L_SM}
f^{(1)}&\equiv& -6\lambda_1-\frac{3}{4}(3 g_w^2+g_y^2)+6 y_t^2\,.
}
$f^{(2)}$  is given in \refer{Eq:DelFH}. All couplings in
\refer{Eq:QuadDiv1L_SM} and  \refer{Eq:DelFH} are the bare couplings
(the subscripts ${\rm B}$ are omitted for simplicity).

To find the dependence of the SM function $f$ on the rescaled cutoff
scale $\bar\Lambda\approx0.32\Lambda$ (cf. \refer{Eq:delta_Lambda})
we evolve the SM couplings in the $\La$-$\anti{{\rm MS}}$ scheme using the
two-loop RG equations of \ref{App:SM} from the scale $\mu=M_t$ up to
some high scale. (Recall that, in agreement with
\refer{eqn:BareCouplingsDef}, the cutoff-dependent dimensionless bare coupling
$g_{\rm B}(\bar\La)$ is simply given by the running  one $g(\mu)$ extrapolated to high scales, i.e. $g_{\rm B}(\bar\La)=g(\mu=\bar\La)$.)  As the initial conditions for the RG evolution we take
the known values of the SM DimReg-$\anti{{\rm MS}}$ scheme couplings 
\cite{Buttazzo:2013uya}
\eqs{\nn
\check{\lambda}(M_t)&=&0.12604
+ 0.00206\left\{\frac{M_h}{\rm GeV}-125.15\right\}+\\
&{}&
- 0.00004\left\{\frac{M_t}{\rm GeV}-173.34\right\}
\pm 0.0003_{\rm th},
\nn}
\eqs{\nn
\check{y}_t(M_t)&=&0.93690
+ 0.00556\left\{\frac{M_t}{\rm GeV}-173.34\right\}
\pm 0.0005_{\rm th},
}
\eqs{\label{Eq:RG-ini-cond}
\check{g}_w(M_t)&=&0.64779
+ 0.00004\left\{\frac{M_t}{\rm GeV}-173.34\right\},
}
\eqs{\nn
\check{g}_y(M_t)&=&0.35830
+ 0.00011\left\{\frac{M_t}{\rm GeV}-173.34\right\},
}
\eqs{\nn
\check{g}_s(M_t)&=&1.1666
- 0.00046\left\{\frac{M_t}{\rm GeV}-173.34\right\},
}
(the central value of $g_s$ corresponds to $\al_s(M_Z)= 0.1184$) 
in which $M_t=(173.34\pm 0.75)$ GeV and
$M_h=(125.15\pm 0.24 )$ GeV are the pole top quark and Higgs boson masses,
and convert them with the help of the relation \refer{Eq:G-fun}
which takes here the form
\begin{equation}\label{Eq:RelAtTop}
g^C({M_t}) = \check{g}^C({M_t}) -\frac{1}{(4\pi)^{2}}\,
\theta^C_{(1)}(\check{g}({M_t}))\,,
\end{equation}
with the one-loop $\theta$ functions given in
\refer{Eq:!!!ThetaMNJaw0}-\refer{Eq:!!!ThetaMNJaw1}, into the values
appropriate for the $\La$-$\anti{{\rm MS}}$ scheme.\footnote{Up to
the two-loop accuracy we could alternatively evolve the  couplings
\refer{Eq:RG-ini-cond} using the two-loop RG equations of the
DimReg-$\anti{{\rm MS}}$ scheme and convert them
at the scale $\mu=\bar\Lambda$ into the $\La$-$\anti{{\rm MS}}$ scheme
couplings using \refer{Eq:RelAtTop} with $M_t$ replaced by $\bar\Lambda$.}

%%%%%%%%%%%%%%%  PLOTS -- Remove NOTHING  %%%%%%%%%%%%%%%%%%%%%%%%%%%%%%
%\nothing
{
\begin{figure}[t]
\centering
\includegraphics[width=0.80\textwidth]{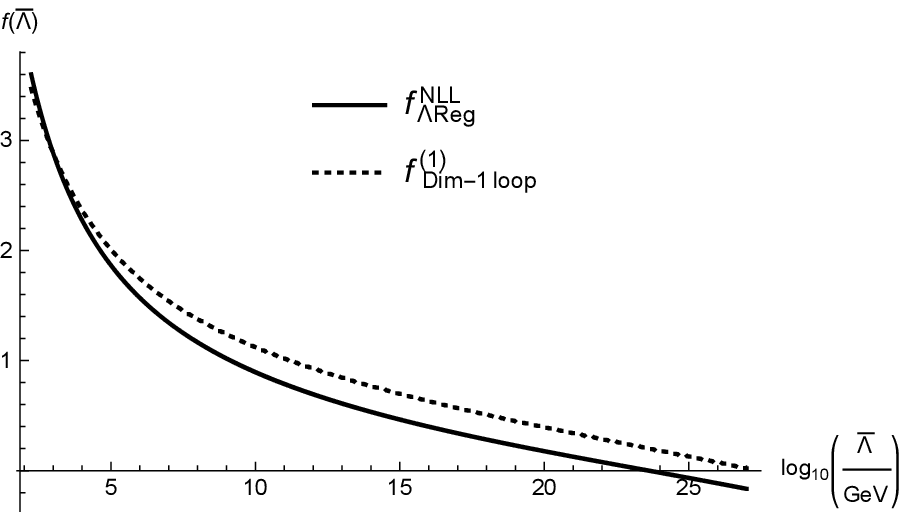}
\caption{Coefficient $f$ of the quadratic divergence (of the term proportional
to $\Lambda^2$ in \refer{eqn:barebosonmassesdef}) in the SM. The solid line
shows the results of the full two-loop (NLL) calculation (i.e. the full two-loop coefficient $f$ with $\La$-$\anti{{\rm MS}}$
couplings running according to two-loop beta functions). The short-dashed line
shows the one-loop coefficient $f^{(1)}$ with DimReg-$\anti{{\rm MS}}$
couplings running according to one-loop beta functions. Both curves correspond
to the central values of the DimReg-$\anti{{\rm MS}}$ initial data given by
\refer{Eq:RG-ini-cond}.}
\label{Fig:Plot1Lvs2L}
\end{figure}

\begin{figure}[t]
\centering
\includegraphics[width=0.95\textwidth]{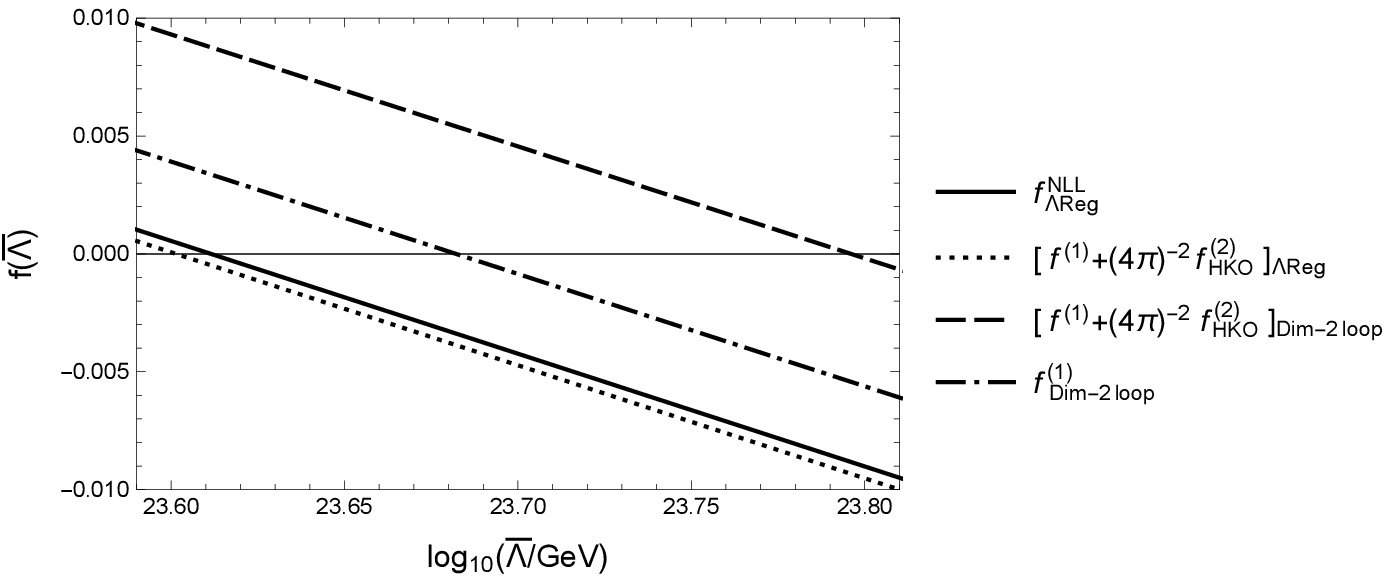}
\caption{Comparison of the result of the consistent two-loop calculation
of $f$ (solid line) with other approaches: as indicated, the dotted
line shows the result of replacing $f^{(2)}$ given in \refer{Eq:DelFH} by
$f^{(2)}$ of \cite{HAKAOD}, the dashed line corresponds to using in
addition the two-loop running couplings of the DimReg-${\anti{\rm MS}}$
(instead of $\La$-${\anti{\rm MS}}$) scheme.
Finally, the dot-dashed line shows the result of approximating $f$
by $f^{(1)}$ and using the two-loop running couplings of the DimReg-${\anti{\rm MS}}$ scheme. In all cases central values of the initial
values of \refer{Eq:RG-ini-cond} are used.}
\label{Fig:PlotCompare}
\end{figure}

\begin{figure}[t]
\centering
\includegraphics[width=0.80\textwidth]{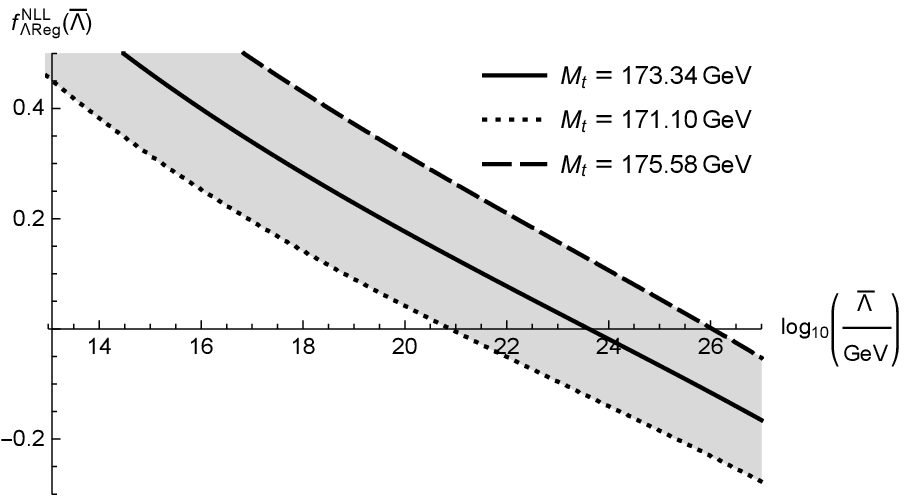}
\caption{Uncertainty of the SM function $f^{\rm NLL}_{\Lambda{\rm Reg}}$
corresponding to the uncertainty in the value of the top mass.
The band corresponds to $3\sigma$ deviations of $M_t$ from the central
value $M_t=(173.34\pm 0.75)$ GeV. Central value of $M_h$ is used.}
\label{Fig:Plot3sigma}
\end{figure}
}%END OF NOTHING

The dependence of the SM function $f$ of \refer{eqn:barebosonmassesdef}
on the rescaled cutoff $\bar\La$ for the central values of the
couplings \refer{Eq:RG-ini-cond} is shown in Figure \ref{Fig:Plot1Lvs2L}.
It is seen that the two-loop effects lower the scale $\bar\Lambda$ at which
$f$ vanishes by about 3 orders of magnitude. Nevertheless, this scale
remains too high to  reasonably identify $\Lambda\approx 3\bar\Lambda$
with the intrinsic scale of a fundamental theory which, as argued, should
be related to the Planck scale $M_{\rm Pl}=1.8\times10^{18}$ GeV.

In Figure \ref{Fig:PlotCompare} we compare the results of various approaches.
It is clear that replacing only $f^{(2)}$ given in \cite{HAKAOD} by the result
\refer{Eq:DelFH} of  the systematic calculation in the consistent
regularization scheme of Section \ref{Sec:UV-cutoff} is not very significant
numerically. The difference is larger if the actual approach taken in
\cite{HAKAOD} (dashed line) is compared with our result (solid line).
Still, this comparison shows that the estimate of the scale
$\bar\Lambda$ at which $f$ vanishes is not very sensitive to the details
(nor to the consistency) of the approach taken to estimate the two-loop
effects. This is important for the interpretation of the hierarchy problem
proposed in this section. Since the one-loop beta functions are
(for mass independent schemes) universal and the function $f^{(1)}$ in
\refer{Eq:ffff} is (up to a multiplicative constant) independent of the
precise form of the momentum cutoff\footnote{At least if the cutoff
does not differentiate between fields of different spins - but this
seems a reasonable assumption in view of the universality of gravity.}
 it should be possible, unless large values of some couplings come into play,
and if the uncertainty in the top mass is reduced - see below - to reliably
test whether a given extension of the SM involving elementary scalar fields
is consistent with the proposed solution
to the hierarchy problem, that is whether it predicts (with the uncertainty
of one-two orders of magnitude) $\bar\Lambda$
sufficiently close to the Planck scale.

As illustrated in Figure \ref{Fig:Plot3sigma}
the value of scale $\bar\Lambda$ at which the SM function
$f$ vanishes strongly depends on the actual value of top mass.
This is, however, not surprising since
the instability scale of the SM is also strongly dependent on the value
of $M_t$ \cite{Buttazzo:2013uya}.

\section{Conclusions}
\label{sec:Final}

In this paper we have considered renormalization of a general renormalizable 
YM theory 
with scalar and spinor fields in the regularization based on a physical
UV momentum cutoff which explicitly breaks the BRST symmetry.
In this connection we have recalled the general renormalization procedure
based on QAP. We have proposed a concrete consistent realization of such
a regularization and formulated a mass-independent
renormalization procedure in terms
of counterterms to the action which implement the necessary subtractions.
Using our scheme we have performed a systematic one-loop renormalization
of a general YM theory obtaining explicitly the one-loop counterterms
(minimal and non-minimal ones). The proposed renormalization scheme,
similarly to the conventional $\overline{\rm MS}$ scheme, introduces an
arbitrary
renormalization scale $\mu$. Therefore, we have proved that the parameters
and Green's functions computed in this scheme satisfy the appropriate
RG equations ensuring independence of $\mu$ of physical quantities.
We have also established the
relations between parameters of the theory renormalized in our
scheme and those of the ordinary $\overline{\rm MS}$ scheme. This allowed us
to obtain explicit expressions for the two-loop RG equations satisfied by
the parameters renormalized in our scheme. Their correctness has been partly
checked by the direct calculation in our scheme of divergences of two-loop
vacuum graphs.

The established RG invariance of physics allowed to define the $\mu$
independent bare couplings and formulate the theory in terms of the
bare action dependent on the cutoff scale $\Lambda$ only through the
regularizing exponential function. The structure of this bare action
has been elucidated.

Finally, the concept of the bare action
allowed us to speculate in the last part of the paper on
the hierarchy problem. We have formulated a condition which, if realized
in Nature, could be considered  a solution of this problem, at least
as far as it concerns scalar fields only.
It should be stressed that this solution does not require any additional
(super)symmetry. Using the results of the paper we have analyzed
whether the SM itself can be consistent with this possibility.
While it turns out that the renormalization scale at which
the parameters of the SM satisfy the necessary condition is too high
to be accepted, from comparing on the example
of the SM different approximations
we have gained
some useful insight into the reliability of the similar checks based
on simple one-loop calculations for potential extensions of the SM.
\vspace{0.5cm}

\noindent{\bf Acknowledgments:}
We thank Hermann Nicolai for discussions. K.A.M. thanks the Albert Einstein
Institute in Potsdam for hospitality and support. A.L. and K.A.M. were
supported by the Polish NCN grant DEC-2013/11/B/ST2/04046.

\setcounter{section}{0}

\renewcommand{\thesection}{Appendix~\Alph{section}}
\renewcommand{\theequation}{\Alph{section}.\arabic{equation}}
\renewcommand{\thefigure}{\Alph{section}.\arabic{figure}}

\setcounter{equation}{0}
\setcounter{footnote}{0}
\setcounter{figure}{0}

\section{Auxiliary conditions in the subtraction procedure }

\renewcommand{\thesection}{\Alph{section}}
\renewcommand{\theequation}{\Alph{section}.\arabic{equation}}
\renewcommand{\thefigure}{\Alph{section}.\arabic{figure}}

\subsection{Auxiliary conditions}
\label{App:Ren-Det:Part1-AuxCond}

Here we list the auxiliary conditions which together with the ZJ identity
$\mathcal{S}(\Gamma)=0$ specify the 1PI effective action $\Gamma$.
For convenience we write these conditions for an arbitrary functional
$G$. These are (see \cite{PiguetSorella} and references therein):
\begin{itemize}
\item The ``translational Ward identity" \cite{Pilaf}
\eq{\label{Eq:TransWT}
\tau_i\, G=0\, ,\qquad\quad
\tau_i\equiv-\derp{}{\varphi^i}+\int\!{{\rm d}^4}{x}\, \derf{}{\phi^i(x)}\,.
}
\item Symmetry w.r.t. global gauge transformations (cf. formulae
\refer{Eq:RelKom-ZJ--AntiGh}-\refer{Eq:Glob-loc} below for the definition
of $\cW_\alpha$)
\eq{\label{Eq:GlobalSym}
\cW_\alpha G=0\,,
}
\item The ghost equation
\eq{\label{Eq:Ghost}
\cG^\al(x)G=0\,,
\qquad\quad
\mathcal{G}^\alpha(x)
\equiv
\derf{}{\anti{\omega}_{\alpha}\!\bracket{x}}
-\derp{}{x_\mu}\derf{}{{K}^\mu_\alpha(x)} ~\!,
}
\item The (Landau) gauge condition
\eq{\label{Eq:Gauge}
\derf{G}{h_\beta(x)}=\bar{\Delta}^\beta_h(x)\,,
\qquad\quad \bar{\Delta}^\beta_h(x)\equiv-\partial^\nu\!\!A_\nu^\beta(x)~\!,
}
\item The antighost equation \cite{BPS:anti-ghost-eq}
\eq{\label{Eq:AntiGhost}
\anti{\cG}_\alpha G=\Delta^{\anti{\cG}}_\alpha\,,
\qquad\quad\anti{\cG}_\alpha\equiv\int\!{{\rm d}^4}{x}
\bigg\{\derf{}{{\omega}^\alpha\!\bracket{x}}
-\anti{\omega}_\gamma(x)\mind{e}{\gamma}{\alpha\beta}
\derf{}{h_\beta\!\bracket{x}}\bigg\},
}
in which
\eqs{\nn
\Delta_\alpha^{\anti{\cG}}&\equiv&
\int\!{{\rm d}^4}{x}\bigg\{L_\beta\mind{e}{\beta}{\alpha\gamma}\omega^\gamma
-K^\mu_\beta\mind{e}{\beta}{\alpha\gamma}A^\gamma_\mu
-K_i\matri{\TS_\alpha(\phi+\varphi)+\bar{P}_\alpha}^i
+\bar{K}_a\matri{\TM_\alpha\psi}^a\bigg\},
\nn}
\end{itemize}

If the gauge Lie algebra has an Abelian ideal, an additional condition,
the {\it local} Abelian antighost equation (AAE) \cite{Grassi}
\eq{\label{Eq:AbelAntiGhost}
\derf{G}{{\omega}^{\alpha_A}\!\bracket{x}}
=\bar{\Delta}_{\alpha_A}^{\anti{\cG}}(x)\, ,
}
with
\eqs{\nn
\bar{\Delta}_{\alpha_A}^{\anti{\cG}}
&\equiv&-\partial_\mu\{K^\mu_{\alpha_A}-\partial^\mu\anti{\omega}_{\alpha_A}\}
-K_i\matri{\TS_{\alpha_A}(\phi+\varphi)+\bar{P}_{\alpha_A}}^i
+\bar{K}_a\matri{\TM_{\alpha_A}\psi}^a,
\nn}
is imposed.

If Stueckelberg fields are present in the model (i.e. at least one vector
$\bar{P}_{\beta_A}$ in (\ref{Eq:CovDer})
is nonzero) one can ensure (by performing, if necessary,
an orthogonal rotation in the space of scalars $\phi^i$) that only the last
$N_{St}$ rows $\bar{p}^s_{\ \alpha_A}$ with $s=1,\ldots, N_{St}$ of the matrix
$[\bar{P}^i_{\ \alpha_A}]$ are non-vanishing and that they are linearly
independent. The corresponding components of the (rotated) scalar field
$\phi$ are the Stueckelberg fields $\xi^{s}$. In such a case
two further conditions (the Stueckelberg equations)
\begin{eqnarray}
\derf{G}{{\xi}^{s'}\!\bracket{x}}=\bar{\Delta}_{s'}^{\Xi}(x)\,,
\qquad\bar{\Delta}_{s'}^{\Xi}\equiv-\delta_{s' s}\,\partial^\mu_{}
\{\partial_\mu \xi^{s}+\bar{p}^{s}_{\ \gamma_A}A_\mu^{\gamma_A}\}~\!,\label{Eq:Stue-1}
\end{eqnarray}
and
\eq{\label{Eq:Stue-2}
\mathscr{T}^{s}_\mu(x)G=0\,,
\qquad\qquad\mathscr{T}^{s}_\mu(x)\equiv\derp{}{x^\mu}\derf{}{K_{s}(x)}
+\bar{p}^{s}_{\ \gamma_A}\derf{}{K^\mu_{\gamma_A}(x)}\, ,
}
are imposed. Further conditions on $I_0$ and on $\Gamma$ may result from
imposing other continuous or discrete global (non-gauge) symmetries.

All the conditions \refer{Eq:TransWT}-\refer{Eq:Stue-2} are satisfied by the
tree-level action $I_0$ \refer{Eq:I_0=I_0^GI+I_0^Rest} in the Landau gauge
\refer{Eq:Gauge_Fix_Funct_falpha_Gen_Gauge} expressing their
``accidental symmetries" or specifying their breaking (factors $\Delta$).
It the analysis it is important that because all the $\Delta$ factors are
linear in the quantum (propagating) fields they do not affect quantum
corrections.

Most of the conditions \refer{Eq:TransWT}-\refer{Eq:Stue-2} play only a
simplifying role in our analysis: imposed on $\Gamma$, they enforce
the Landau gauge as
a particular choice in the class of $R_\xi$ gauges. An important exception
is the Abelian antighost equation \refer{Eq:AbelAntiGhost} which specifies
Abelian gauge currents beyond the tree-level: if the theory has continuous
(non-gauge) symmetries, in ker${\cal S}_{I_0}$ there are terms corresponding
to couplings of Abelian gauge fields to conserved currents of these symmetries
\cite{BBBC2,Barnich:1994ve}. Such terms, which unlike other elements of
$\ker\SI$, do not correspond to infinitesimal changes of parametrization of
the tree-level action $I_0$, are excluded by the AAE \cite{Grassi,BBBC2}. That
this is indeed so can be seen by noticing that for an arbitrary functional
$F$ \cite{Grassi} the following ``anti-commutation relation" holds:
\begin{eqnarray}
\derf{}{{\omega}^{\alpha_A}\!\bracket{x}}\cS(F)+
\cS_F \left(\derf{F}{{\omega}^{\alpha_A}\!\bracket{x}}
-\bar{\Delta}_{\alpha_A}^{\anti{\cG}}(x)\right)
=\fW_{\alpha_A}(x)F-\partial_\mu\partial^\mu h_{\alpha_A}(x)~\!.
\label{Eq:RelKom-ZJ--abel-AntiGh}
\end{eqnarray}
Here $\fW_{\alpha_A}(x)$ is the infinitesimal generator of Abelian gauge
transformations
\eqs{\label{Eq:Def:WT-ab-loc}
\fW_{\alpha_A}&=&\partial_\mu\derf{}{A^{\alpha_A}_\mu}+
\matri{\TS_{\alpha_A}(\phi\!+\!\varphi)\!+\!\bar{P}_{\alpha_A}}^i\!
\derf{}{\phi^i}
+\matri{{\TM_{\alpha_A} \psi}}^a\derf{}{\psi^a}
+\nn\\&{}&
-K_j\matri{{\TS}_{\alpha_A}}^j_{\  i}\derf{}{K_i}
-\anti{K}_b\matri{{\TM}_{\alpha_A}}^b_{\  a} \derf{}{\anti{K}_a}\,.
}
From \refer{Eq:RelKom-ZJ--abel-AntiGh} one learns that if $\Gamma$ satisfies
the ZJ identity and the AAE then it also  obeys Abelian Ward-Takahashi
(WT) identities which ensure that Abelian gauge bosons couple only to
gauge currents.

(Anti)commutation relations, similar to \refer{Eq:RelKom-ZJ--abel-AntiGh},
hold also for all other differential operators
in \refer{Eq:TransWT}-\refer{Eq:Stue-2} (some of them can be found in
\cite{PiguetSorella}). Here we show only the one satisfied by
$\anti{\cG}_\alpha$ in order to specify the $\cW_\alpha$ operator in
\refer{Eq:GlobalSym}
\begin{eqnarray}
\anti{\cG}_\alpha\cS(F)+\cS_F\left(\anti{\cG}_\alpha F
-\Delta^{\anti{\cG}}_\alpha\right)=\cW_\alpha F~\!.\label{Eq:RelKom-ZJ--AntiGh}
%,\qquad\qquad \forall F\, .
\end{eqnarray}
In particular, comparing \refer{Eq:RelKom-ZJ--AntiGh} with
\refer{Eq:RelKom-ZJ--abel-AntiGh} we get the relation between
$\cW_{\alpha_A}$ and  \refer{Eq:Def:WT-ab-loc}
\eq{\label{Eq:Glob-loc}
\cW_{\alpha_A}=\int{\rd^4 x\,\, \fW_{\alpha_A}}(x)\,.
}
Relations like \refer{Eq:RelKom-ZJ--AntiGh} mean that for functionals $G$
which satisfy the ZJ identity (\ref{eqn:ZJidBasic}) not all conditions
\refer{Eq:TransWT}-\refer{Eq:Stue-2} are independent. They are, however, all
necessary to specify the actions $I_n$ which do not satisfy this identity.

Finally, we remark that, as can be seen from \refer{Eq:ZinnJustinOperQuad},
gauge singlet fields are in our formalism treated on an equal footing  with
non-singlet ones. In particular, we do not exclude the possibility that
antifields corresponding to gauge singlets (i.e. $L_{\al_A}$ and, say,
$K_{i_s}$ and $\bar{K}_{a_s}$) appear in counterterms even though they are
absent in the tree-level action $I_0$.
Assigning to them the same ghost numbers and power-counting dimensions as to
their non-singlet counterparts one concludes that the conditions
\eq{\nn %\label{Eq:other-cond}
\derf{I_n}{K^\mu_{\alpha_A}}=-\partial_\mu\omega^{\alpha_A}\,,
\qquad
\derf{I_n}{L_{\alpha_A}}=0\,,
\qquad
\derf{I_n}{K_{i_s}}=0\,,
\qquad
\derf{I_n}{\bar{K}_{a_s}}=0\,,
}
follow already from the conditions \refer{Eq:TransWT}-\refer{Eq:Stue-2}
imposed on the $I_n$ functional - they do not have to be imposed
separately.\footnote{Alternatively, these new constraints can be imposed
by appropriately restricting the form of the $\cS(\cdot)$ operation
\cite{BBBC1,BBBC2}.}

\subsection{Completion of the inductive step}
\label{App:Ren-Det:IndStep}

To complete the inductive step discussed in Section \ref{sec:Subtractions} we have
to show that the auxiliary conditions \refer{Eq:TransWT}-\refer{Eq:Stue-2}
are satisfied by $I_{n+1}$. To this end, we first notice that the identities
\refer{Eq:TransWT}, \refer{Eq:GlobalSym} and \refer{Eq:AntiGhost} are
preserved by the regularization prescription (\ref{eqn:PrescriptionDef}),
i.e. if $I_n$ obeys them, then so does $I^\Lambda_n$. By using the
well-known arguments \cite{Z-J:74} one concludes that $\Gamma^\Lambda_n$ and
its ``asymptotic part" $\Gamma_n$ satisfy these identities. In particular,
this means that $\Gamma_n^{(n+1){\rm div}}$ obeys their homogeneous counterparts.
The same arguments show that $\Gamma_n^{(n+1){\rm div}}$ possesses all global
(non-gauge) symmetries of $I_0$. In fact, $\Gamma_n^{(n+1){\rm div}}$
satisfies also homogeneous versions of all the  remaining conditions
listed in Appendix \ref{App:Ren-Det:Part1-AuxCond},
even though, due to their dependence on derivatives, these identities are
not (exactly) preserved by the regularization (\ref{eqn:PrescriptionDef}).
Let us consider first
\refer{Eq:Ghost} with  $G=I_n$. This identity implies that $I_n$ depends on
the antighost $\anti\omega_\alpha$ only through the difference
$K^\mu_\alpha-\partial^\mu \anti\omega_\alpha$. That the same is true also for
$\Gamma_n$ is obvious from Feynman diagrams, the only subtlety being that
the derivative acting on the antighost field in $I_n^\Lambda$ is replaced
according to (\ref{eqn:PrescriptionDef}). However, the additional
exponential that could spoil this identity for $\Gamma_n^\Lambda$ necessarily
contains an \emph{external} momentum. Thus, the
breaking of identity \refer{Eq:Ghost} for $\Gamma_n^\Lambda$ tends to zero
in the infinite cutoff limit (1PI functions with external
antighost lines are at most linearly divergent, due to the derivative on
each antighost field in the action $I_n$). Thus, the entire
$\Gamma_n^{(n+1)}$ and in particular,
$\Gamma_n^{(n+1){\rm div}}$ satisfy the identity \refer{Eq:Ghost}.
The same arguments show that $\Gamma_n^{(n+1){\rm div}}$ obeys
\refer{Eq:Stue-2}. Finally, the conditions, \refer{Eq:Gauge},
\refer{Eq:AbelAntiGhost} and \refer{Eq:Stue-1} applied to $I_n$ restrict its
vertices  in such a way that it is impossible to construct 1PI loop diagrams
which would contribute to functions with external
lines of, respectively,
$h_\alpha$, $\omega^{\alpha_A}$ and $\xi^{\ep_G}$. Thus, $\Gamma_n^{(n+1){\rm div}}$
obeys homogeneous versions of these identities as well.
\vskip0.2cm

To prove that $I_{n+1}$
obeys \refer{Eq:TransWT}-\refer{Eq:Stue-2} we still have to show that the
non-minimal counterterm $\delta_\flat\!\Gamma_n^{(n+1)}$ in Eq.
\refer{Eq:Def:In+1} satisfies their homogeneous versions. This is an
important point since \emph{a priori} these identities could be in conflict
with the condition \refer{Eq:DeltaGammaCond} of restoration of the ZJ
identity. That this is not the case follows from (anti)commutation
relations like  \refer{Eq:RelKom-ZJ--AntiGh}. More precisely, the above
arguments show that the $\tilde{{I}}_n$ functional, cf. \refer{Eq:tilde-I-n},
obeys all the conditions \refer{Eq:TransWT}-\refer{Eq:Stue-2}. Similarly
as above one concludes that $\tilde{\Gamma}_n$ obeys them as well. Thus,
the relation \refer{Eq:RelKom-ZJ--AntiGh} applied to $F=\tilde{\Gamma}_n$ in
conjunction with \refer{Eq:Def:tilde-Delta} tells us that
$\tilde{\Omega}_n$ obeys the homogeneous version of the antighost
equation \refer{Eq:AntiGhost}
\begin{eqnarray}
\anti\cG_\alpha\,\tilde{\Omega}_n=0~\!.
\end{eqnarray}

Using the counterparts of \refer{Eq:RelKom-ZJ--AntiGh} for the other
functional differential operators appearing in
\refer{Eq:TransWT}-\refer{Eq:Stue-2} one concludes that $\tilde{\Omega}_n$
satisfies also homogeneous versions of all the remaining conditions
\refer{Eq:TransWT}-\refer{Eq:Stue-2} except for the
Abelian antighost equation \refer{Eq:AbelAntiGhost}.\footnote{
The difference between the Abelian antighost equation and other auxiliary
identities is caused by the fact that the Abelian WT identity is badly broken
by our regularization prescription and thus $\tilde{\Gamma}_n$ does not
satisfy it. From \refer{Eq:RelKom-ZJ--abel-AntiGh} we get
\eq{\nn
\hbar^{n+1}\derf{\tilde{\Omega}_n}{{\omega}^{\alpha_A}\!\bracket{x}}
+\cO(\hbar^{n+2})
=\fW_{\alpha_A}(x)\tilde{\Gamma}_n-\partial_\mu\partial^\mu h_{\alpha_A}(x)\,,
}
with $\fW_{\alpha_A}(x)$ defined in \refer{Eq:Def:WT-ab-loc}. Incidentally
this relation shows that the counterterms which remove the breakings
$\tilde{\Omega}_n$ \refer{Eq:Def:tilde-Delta} automatically restore
also Abelian WT identities.
}
Exploiting these constraints and assuming that $\tilde{\Omega}_n$ is
cohomologically trivial  (cf. the remarks below Eq. \refer{Eq:Om0=}),
\eq{\nn
%\label{Eq:K-Cond}
\tilde{\Omega}_n=\mathcal{S}_{{I}_0}\tilde\cC_n \, ,
}
we have verified that $\tilde\cC_n$ must be the sum of two terms
\eq{\label{Eq:K-decomp}
\tilde\cC_n=\tilde\cC^{0}_n+\tilde\cC^{1}_n~\!,
}
of which $\tilde\cC^{0}_n$ satisfies the homogeneous versions of all the
conditions \refer{Eq:TransWT}-\refer{Eq:Stue-2} 
(including \refer{Eq:AbelAntiGhost}) and can be assumed
to be invariant under global and discrete symmetries of $I_0$, while
$\tilde\cC^{1}_n$ belongs to $\ker\cS_{I_0}$. Let us consider
\refer{Eq:TransWT} as an example. For
$\tilde\cC_n=\tilde\cC_n[\phi,\ldots;\varphi]\,$ one can define
$\tilde\cC_n^0$ as
\begin{eqnarray}\nn
\tilde\cC_n^0=\tilde\cC_n^0[\phi,\ldots;\varphi]
\equiv
\tilde\cC_n[\phi+\varphi,\ldots;0]\,,
\end{eqnarray}
so that
\eqs{
\tilde\cC_n^1=\tilde\cC_n-\tilde\cC_n^0
&\equiv&\int^1_0{\rm d}t\,\,
\deru{}{t}~\!\tilde\cC_n[\phi+(1-t)\varphi,\ldots;t\varphi]\nn\\
&=&-\varphi^i\int^1_0{\rm d}t
\left\{(\tau_i~\!\tilde\cC_n)[\underline\phi,\ldots;\underline\varphi]
\right\}
\bigg|_{{}^{\scriptsize
\begin{array}{l}
\underline\phi=\phi+(1-t)\varphi \\
\underline\varphi=t\,\varphi\end{array}}}~\, .
\nn}
Using the relation $\tau_i \cS_{I_0} \tilde\cC_n =0$ and the fact that
$[\tau_i,\ \cS_{I_0}]=0$, one concludes that the above difference belongs
to the kernel of $\cS_{I_0}$. In order to arrive at similar conclusions for
the identities \refer{Eq:Gauge}, \refer{Eq:Ghost} and \refer{Eq:GlobalSym}
we have used arguments of \cite{PiguetSorella}. For the remaining ones we have
performed a ``brute force" analysis of all possible terms in $\tilde\cC_n$
consistent with the power-counting.  Finally, for continuous global symmetries
of $I_0$ Ward identities can be used, in parallel with \refer{Eq:GlobalSym}
while for discrete symmetries
$\tilde\cC^{0}_n$ can be averaged over the group of discrete symmetries to
obtain ``new" $\tilde\cC^{0}_n$ possessing discrete symmetries in question.

Obviously, $\tilde\cC^{1}_n\in\ker\cS_{I_0}$ can be discarded
as far as restoration of the ZJ identity is concerned, cf.
\refer{Eq:DeltaGammaCond}. In other words, for the counterterm restoring
the BRST symmetry in the order $\hbar^{n+1}$ one can take
\eq{
\delta_\flat\!\Gamma_n^{(n+1)}=-\tilde\cC^{0}_n\,
}
preserving in this way the
additional symmetries \refer{Eq:TransWT}-\refer{Eq:Stue-2} of the next order
local action $I_{n+1}$. This completes the inductive
step.

\renewcommand{\thesection}{Appendix~\Alph{section}}
\renewcommand{\theequation}{\Alph{section}.\arabic{equation}}
\renewcommand{\thefigure}{\Alph{section}.\arabic{figure}}

\setcounter{equation}{0}
\setcounter{footnote}{0}
\setcounter{figure}{0}

\section{One-loop diagrams}
\label{App:Diag}

Here we list the differences, defined in Eq. \refer{Eq:DeltaReg:def},
between the values of the one-loop diagrams in
$\Lambda$Reg and DimReg.
They have been generated by a dedicated Mathematica based package in
which the steps explained in Section  \ref{Sec:UV-cutoff} have
been implemented. These are
\begin{itemize}
\item expansion of regularized propagators according
      to \refer{eqn:PropagatorExpansion},
\item introduction of the Feynman parameters (at the level of
      \emph{tensor} integrals),
\item shift of the integration variable producing ``spherically" symmetric
      denominators,
\item expansion of the exponential factors in powers of external momenta,
\item carrying out the integrations over angular variables in $d$ dimensions
      (making the standard replacements
      $k^\mu k^\nu\mapsto k^2\,\eta^{\mu\nu}/d$, etc.),
\item transition to the Euclidean space (i.e. formal Wick rotation),
\item contractions of tensor structures  in $d$ dimensions.
\end{itemize}
%%%%%%%%%%%%%%%%%%%%%%%
%%%%%% DIAGRAMS %%%%%%%
%%%%%%%%%%%%%%%%%%%%%%%
%{\bf REMOVE nothing TO TEX THE DIAGRAMS!}
%\nothing
{
\begin{figure}[pth]
\centering
\includegraphics[width=0.5\textwidth]{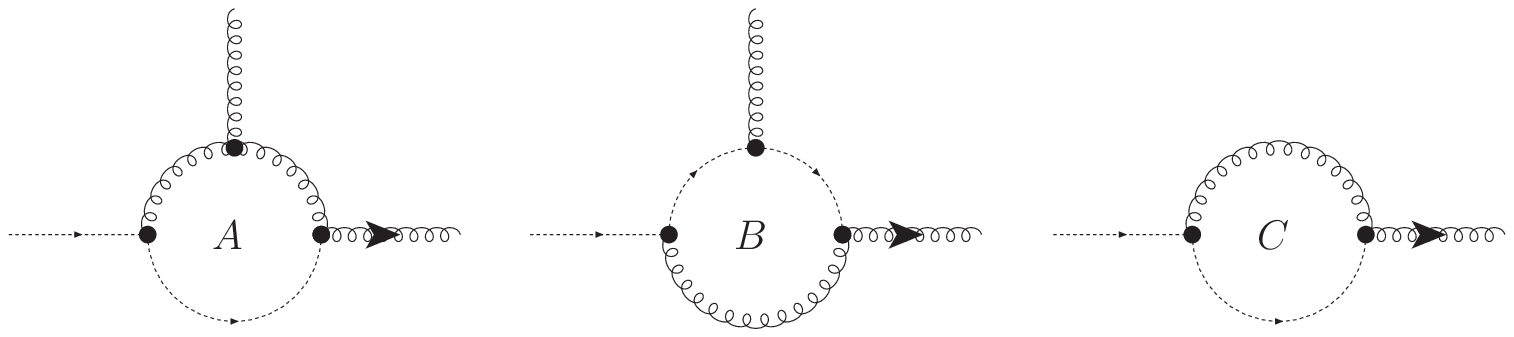}
\caption{Corrections to the BRST transformation of vector fields.}
\label{Rys:KmuOmA_i_KmuOm}
\end{figure}

\begin{figure}[pth]
\centering
\includegraphics[width=0.6\textwidth]{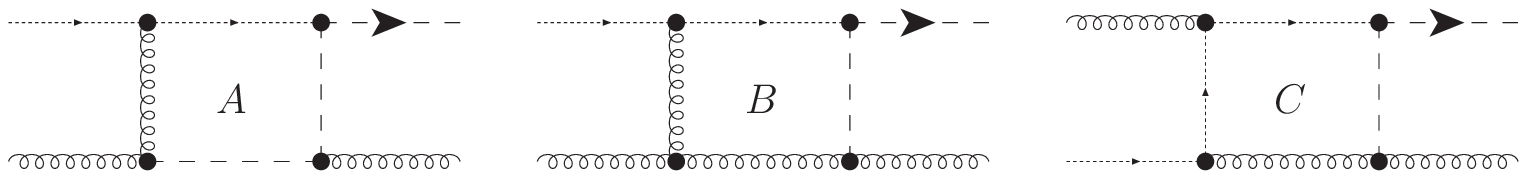}
\caption{Function $\VEVOPITIL{\tilde{K_i}(q) \tilde{\omega}^\alpha(l)
\tilde{A}^\beta_\nu(p)\tilde{A}^\gamma_\rho(p')}^\OLB$.}
\label{Rys:KiOmAA}
\end{figure}

\begin{figure}[pth]
\centering
\includegraphics[width=0.2\textwidth]{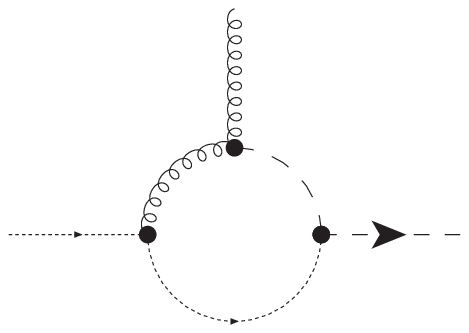}
\caption{Function $\VEVOPITIL{\tilde{K_i}(q) \tilde{\omega}^\alpha(l)
\tilde{A}^\beta_\nu(p)}^\OLB$.}
\label{Rys:KiOmA}
\end{figure}
\clearpage

\begin{figure}[pth]
\centering
\includegraphics[width=0.2\textwidth]{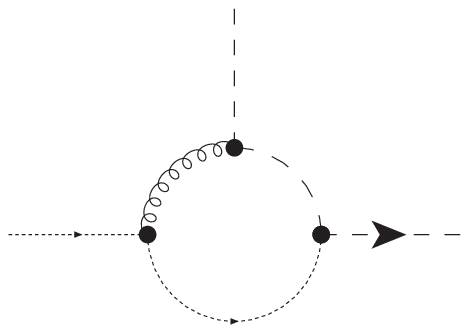}
\caption{Corrections to the BRST transformation of scalars.}
\label{Rys:KiOmPhi}
\end{figure}

\begin{figure}[pth]
\centering
\includegraphics[width=0.2\textwidth]{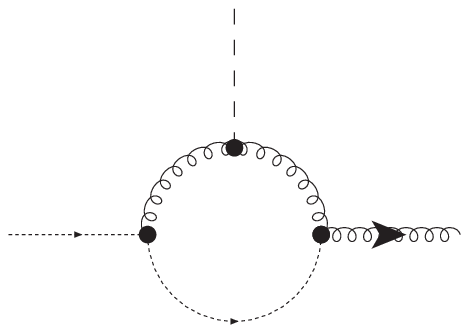}
\caption{Function $\VEVOPITIL{
\tilde{K}^\mu_\gamma(q)\tilde{\omega}^\alpha(l)\tilde{\phi}^j\!(p)  }^\OLB$.}
\label{Rys:KmuOmPhi}
\end{figure}

\begin{figure}[pth]
\centering
\includegraphics[width=0.4\textwidth]{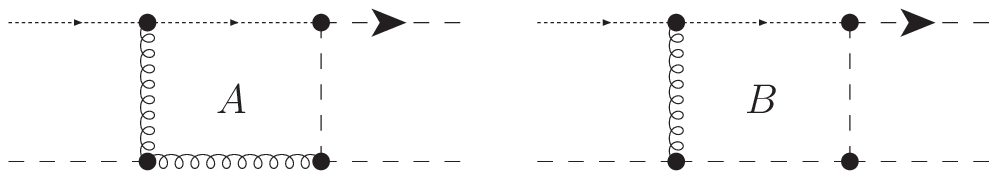}
\caption{Function
$\VEVOPITIL{\tilde{K}_n(q)\tilde{\omega}^\alpha(l)
\tilde{\phi}^i\!(p)\tilde{\phi}^j\!(p')}^\OL$.}
\label{Rys:KiOmPhiPhi}
\end{figure}

\begin{figure}[t]
\centering
\includegraphics[width=0.2\textwidth]{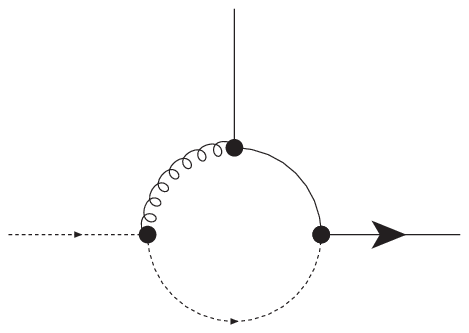}
\caption{Corrections to the BRST transformation of fermions.}
\label{Rys:KaBarOmPsi}
\end{figure}

\begin{figure}[pth]
\centering
\includegraphics[width=0.2\textwidth]{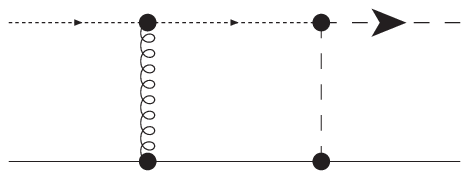}
\caption{Function
$\VEVOPITIL{\tilde{K}_i(l)\tilde{\omega}^\gamma(q)
\tilde{\psi}^{b_1}\!(p_1)\tilde{\psi}^{b_2}\!(p_2)}^\OL$.}
\label{Rys:KiOmPsiPsi}
\end{figure}

\begin{figure}[pth]
\centering
\includegraphics[width=0.35\textwidth]{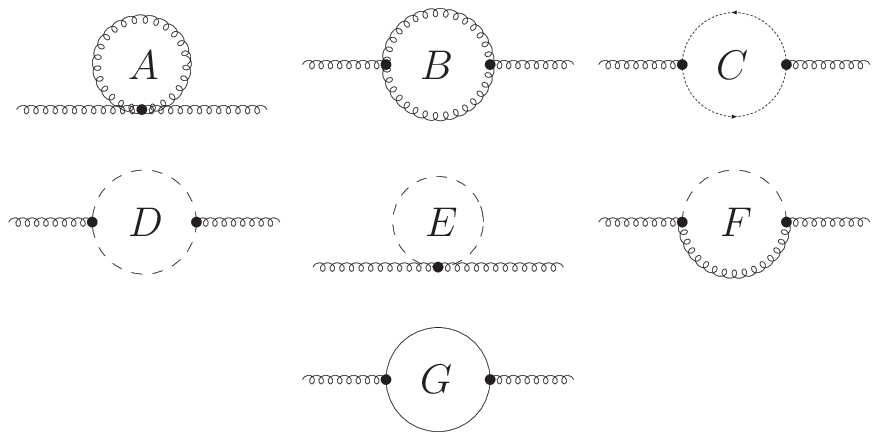}
\caption{One-loop contributions to the vacuum polarization
$\tig^{\mu\nu}_{\alpha\beta}(l,-l)$.}
\label{Rys:AA}
\end{figure}

\begin{figure}[pth]
\centering
\includegraphics[width=0.5\textwidth]{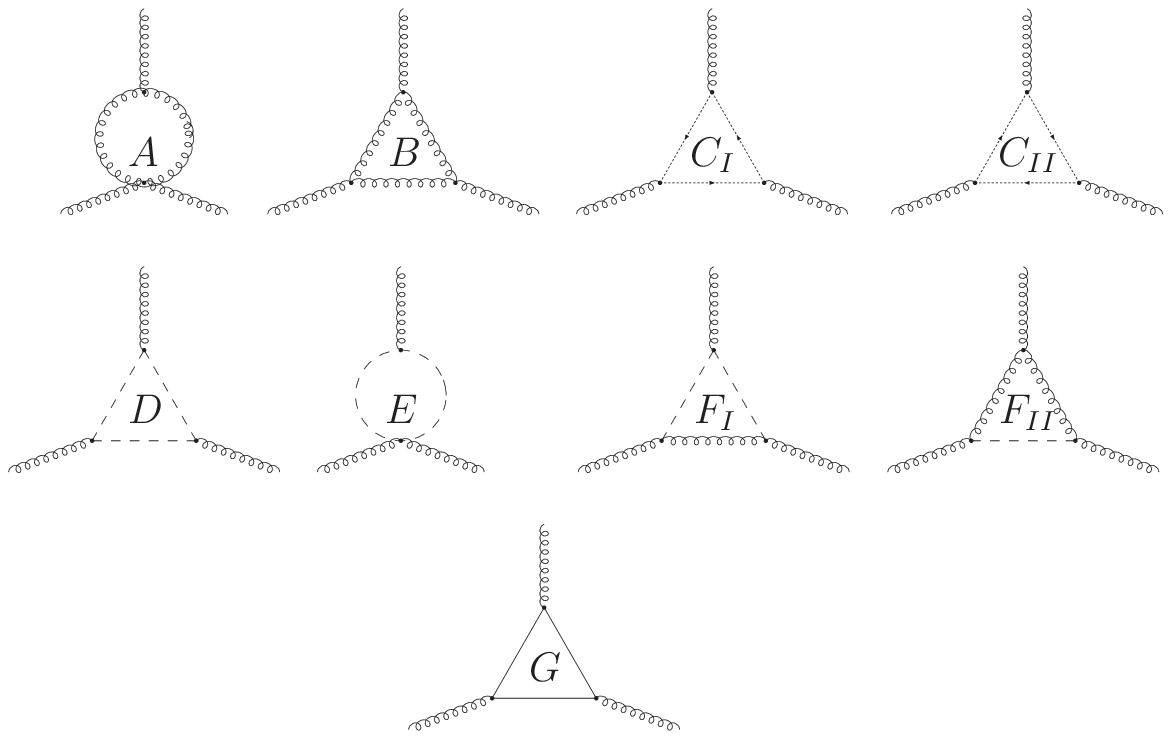}
\caption{One-loop contributions to $\tig^{\mu\nu\rho}_{\alpha\beta\gamma}(l,p,p')$.}
\label{Rys:AAA}
\end{figure}

\begin{figure}[pth]
\centering
\includegraphics[width=0.5\textwidth]{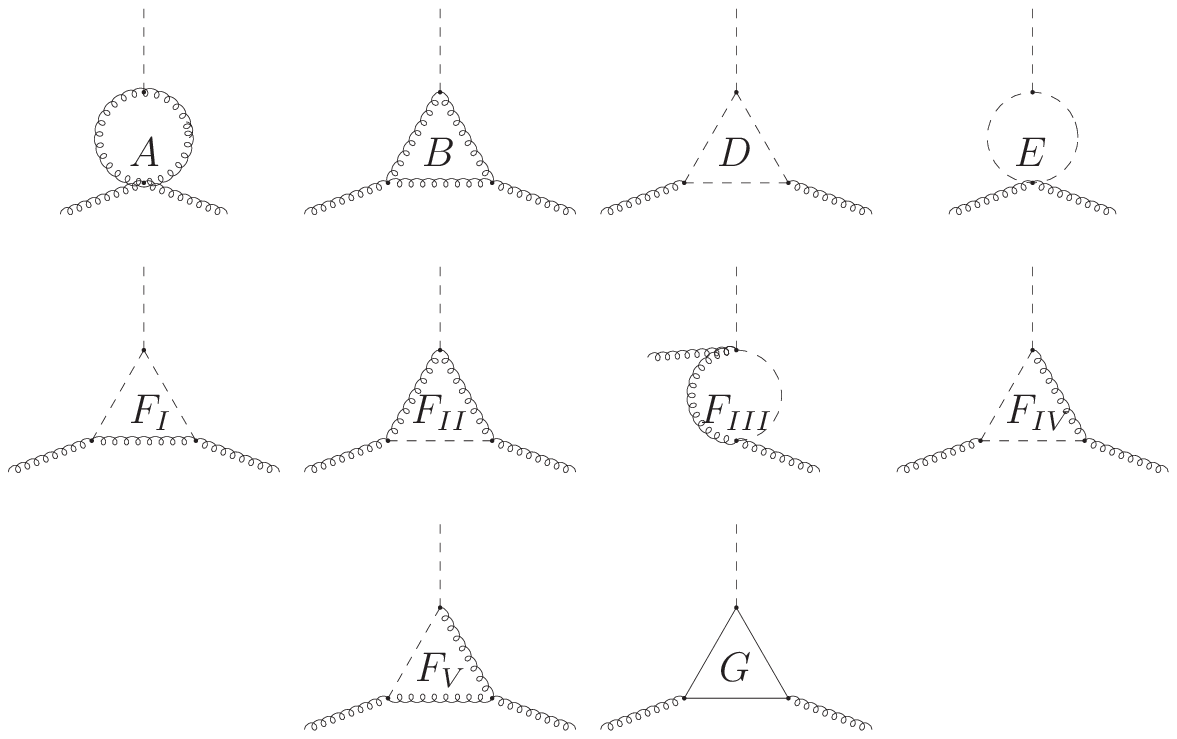}
\caption{One-loop contributions to $\tig^{\ \nu\rho}_{i\beta\gamma}(l,p,p')$.}
\label{Rys:PhiAA}
\end{figure}

%EPS:
\begin{figure}[pth]
\centering
\includegraphics[width=0.3\textwidth]{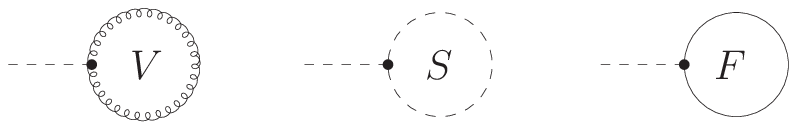}
\caption{One-loop contributions to $\tig^{}_{i}$.}
\label{Rys:Phi}
\end{figure}

\begin{figure}[pth]
\centering
\includegraphics[width=0.45\textwidth]{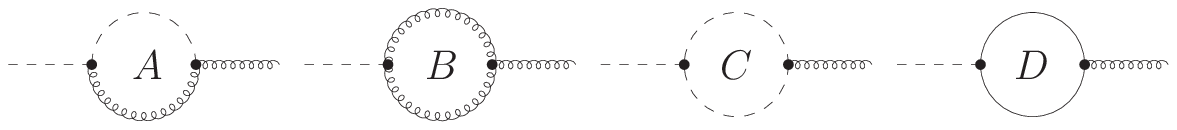}
\caption{One-loop contributions to $\tig^{\ \!\nu}_{i\beta}(l,-l)$.}
\label{Rys:PhiA}
\end{figure}

\begin{figure}[pth]
\centering
\includegraphics[width=0.35\textwidth]{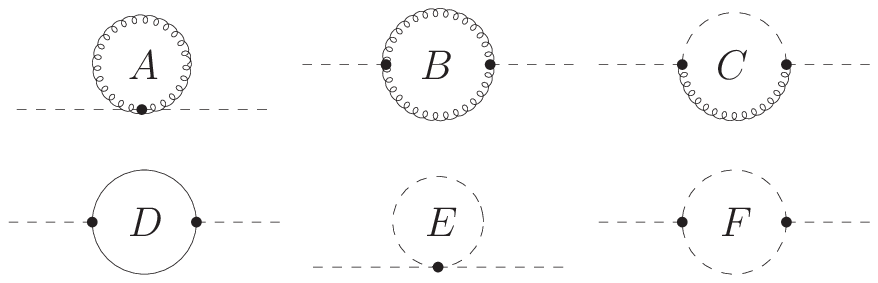}
\caption{One-loop contribution to $\tig_{i_1 i_2}(p,-p)$.}
\label{Rys:PhiPhi}
\end{figure}

\begin{figure}[pth]
\centering
\includegraphics[width=0.5\textwidth]{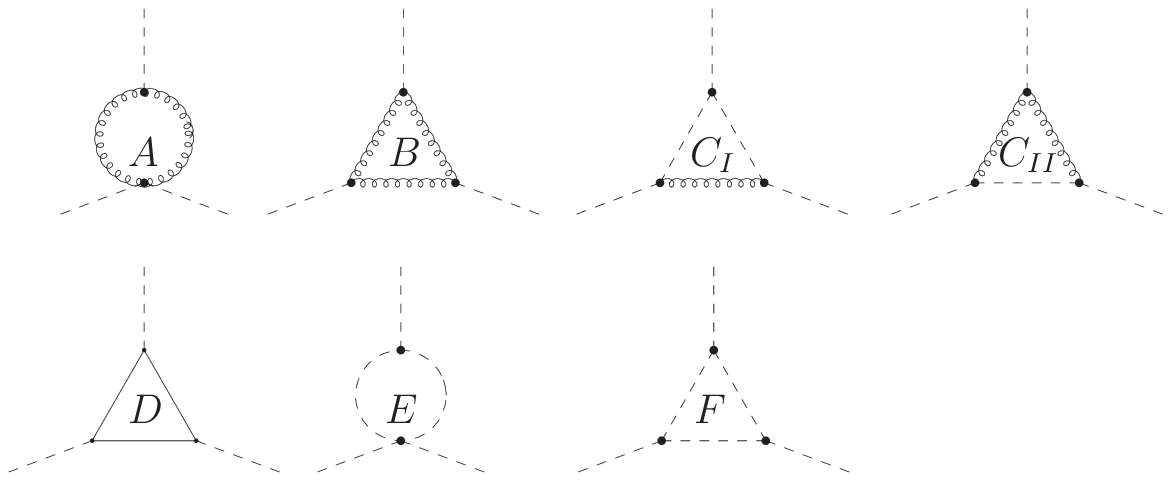}
\caption{One loop contributions to $\tig_{ijn}(p,p',p'')$.}
\label{Rys:PhiPhiPhi}
\end{figure}

\begin{figure}[pth]
\centering
\includegraphics[width=0.5\textwidth]{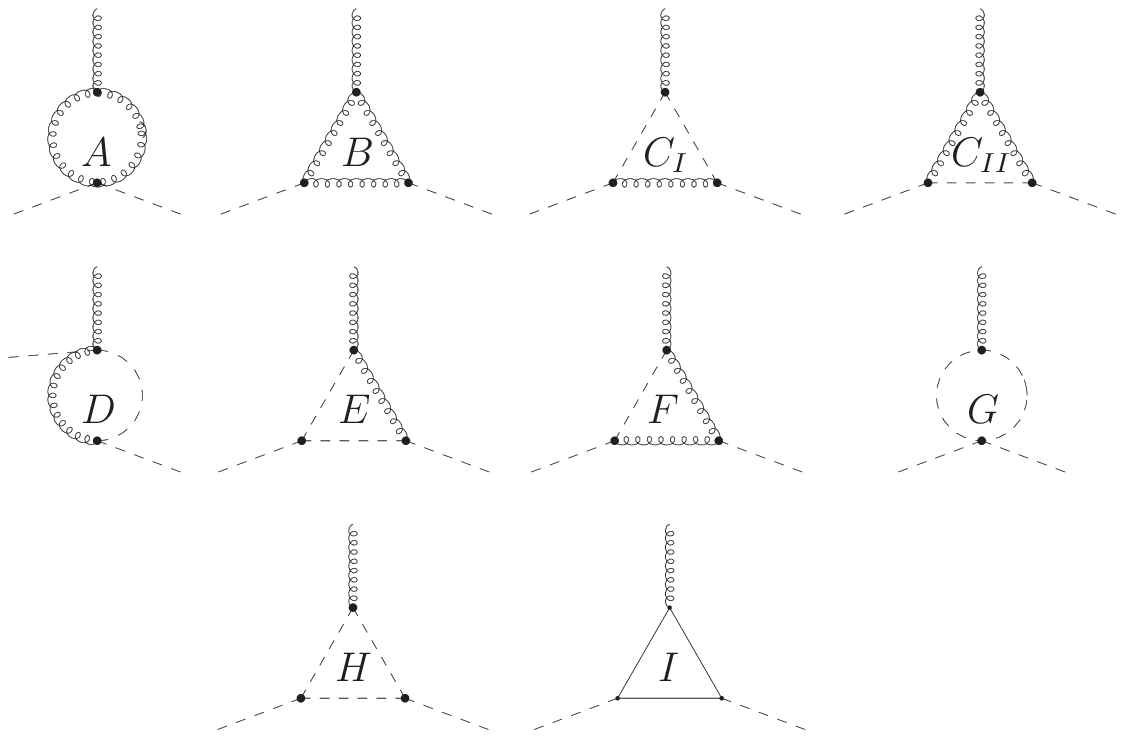}
\caption{One-loop contributions to
$\tig^{\ \ \!\mu}_{ij\alpha}(p,p',l)$.
}
\label{Rys:PhiPhiA}
\end{figure}

\begin{figure}[pth]
\centering
\includegraphics[width=0.3\textwidth]{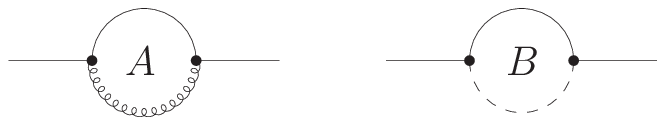}
\caption{One-loop contributions to  $\tig_{a_1 a_2}(p,-p)$.}
\label{Rys:PsiPsi}
\end{figure}

\begin{figure}[pth]
\centering
\includegraphics[width=0.4\textwidth]{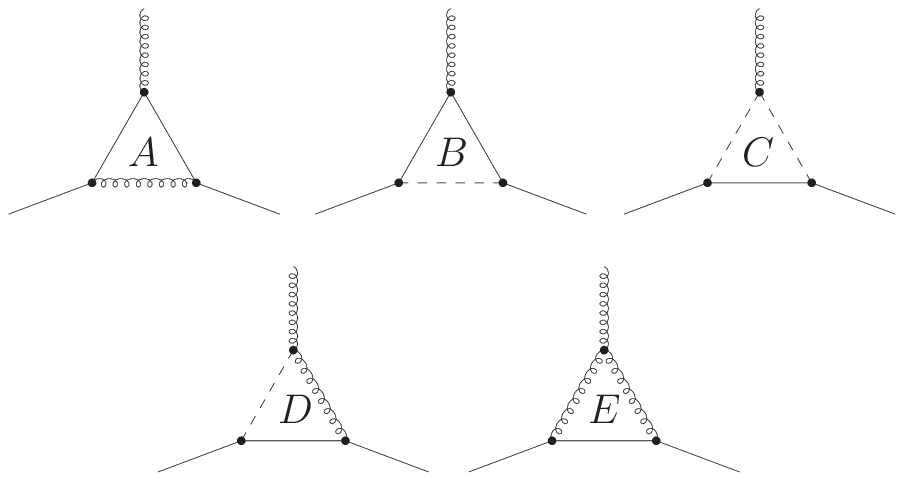}
\caption{One-loop contributions to
$\tig^{\  \ \ \ \ \!\mu}_{a_1a_2\alpha}(p_1,p_2,q)$.}%
\label{Rys:PsiPsiA}
\end{figure}

\begin{figure}[pth]
\centering
\includegraphics[width=0.4\textwidth]{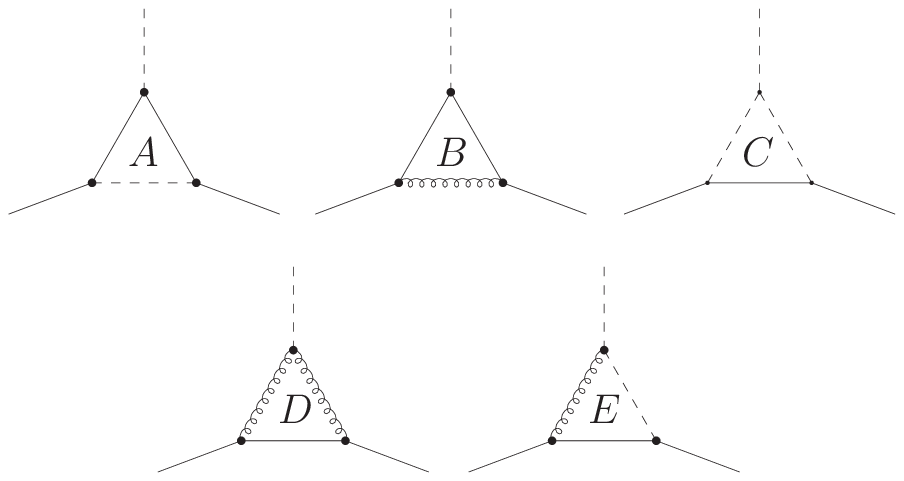}
\caption{One-loop contributions to
 $\tig_{a_1a_2 i}(p_1,p_2,q)$.
}
\label{Rys:PsiPsiPhi}
\end{figure}

\begin{figure}[pth]
\centering
\includegraphics[width=0.5\textwidth]{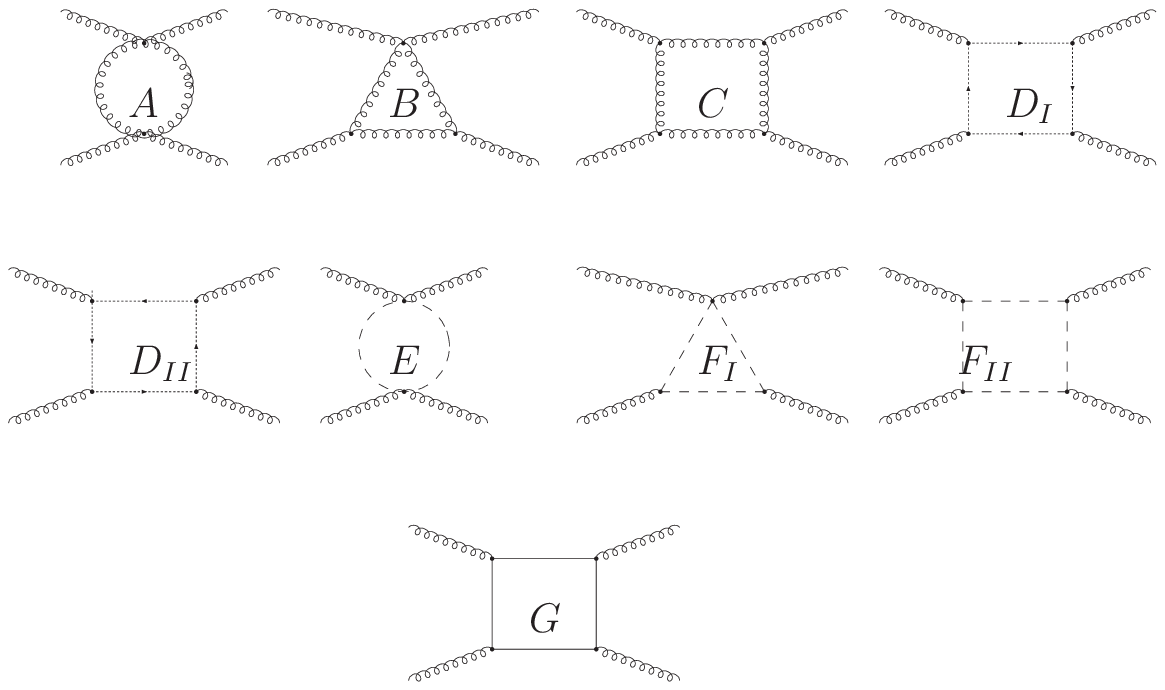}
\caption{\emph{Regularization-dependent} one-loop contributions to
$\tig^{\mu_1\mu_2\mu_3\mu_4}_{\alpha_1\alpha_2\alpha_3\alpha_4}(l_1,l_2,l_3,l_4)$.}
\label{Rys:AAAA}
\end{figure}

}%END OF NOTHING

The expression corresponding to a one-loop diagram, obtained according to the
above prescription, has the form of an integral over the Feynman parameters
and over the length of the Euclidean momentum $k_E$. For $d\to4$ it gives
the value of the diagram in the $\Lambda$Reg, while for $\Lambda\to\infty$
- in the DimReg. Starting from this point the package treats both cases
separately. The expression corresponding to $\Lambda$Reg is integrated
over $k_E$ ``algebraically'', that is by exploiting the definition
\refer{eqn:ConflHypGeom} of the confluent hypergeometric function.
Analogous ``algebraic'' integration in the DimReg case exploits, instead
of \refer{eqn:ConflHypGeom}, the standard representation of the Euler beta
function. Both are the Mathematica built-in functions
($\texttt{HypergeometricU}$ and $\texttt{Beta}$, respectively) and their
asymptotic forms can be found by calling the $\texttt{Series}$ procedure.
After the expansion nonlocal parts of both expressions manifestly cancel out
in the difference, which becomes, therefore, a polynomial in the Feynman
parameters which can be integrated over by using the Mathematica
$\texttt{Integrate}$ function.
The package has been tested on many examples. In particular, we have
verified that violations of the ST identities analyzed in Sections
\ref{Sec:WT:psipsiA}-\ref{Sec:WT:AAA}, which were obtained by the direct
calculation in $\Lambda$Reg (of the type presented in \ref{Sec:ApB-AAAA}),
are reproduced by employing the trick discussed around
Eq. \refer{Eq:DeltaReg:def} using the formulae listed below.

Two remarks are in order. Firstly, the algorithm is simple, because the
Feynman parameters are introduced at the level of tensor integrals. While
this methods does yield also expressions for the nonlocal parts of
diagrams, their comparison with the result obtained with the help of the
standard Passarino-Veltman reduction usually requires lengthy integrations
by parts. Secondly, the package assumes that the $\varphi$-dependent mass
matrices $m_F$, $m_S^2$ and $m_V^2$ are real and diagonal. These assumptions
are satisfied only in a special basis in the field space, but results for
the general case can always be unambiguously recovered. In particular, all
the formulae given in this Appendix and in the main text are correct
for arbitrary mass matrices.

In the differences of the 1PI functions generated by the package one-loop
logarithmic divergences always appear in the combination\footnote{This
reflects the universality of one-loop logarithmic divergences which are
related to the structure of non-local terms in the 1PI effective
action.}
\eq{\label{Eq:DeltaDiv:Def}
\delta_{\rm Div}=\ln{\Lambda^2\over\mu_{H}^2}-{1\over\epsilon}-1-\ln{8\pi},
}
in which $\epsilon\equiv (4-d)/2$ and $\mu_H$ is the 't Hooft mass - the natural
mass unit of the DimReg (see e.g. the expression \refer{Eq:FeynIntEx} below),
which is also the
renormalization scale of the ordinary DimReg-MS scheme. Since we are
interested in renormalized parameters of the DimReg-$\overline{\rm MS}$
scheme, it is more convenient to express $\delta_{\rm Div}$ through the
renormalization scale $\check{\mu}$ of the latter scheme which is related to $\mu_H$ by
\begin{equation}
\check{\mu}\equiv\mu_H \sqrt{4\pi}\, e^{-{\gamma_E}/{2}},
\end{equation}
and the ``fundamental divergence'' $\delta_\Lambda$ of the
$\Lambda$-$\overline{\rm MS}$ scheme, defined in \refer{Eq:delta_Lambda}:
\begin{equation}\label{Eq:DeltaDiv:Def:Mod}
\delta_{\rm Div}=\delta_{\rm \Lambda}-{1\over\epsilon}-2\ln{\check{\mu}\over\mu}.
\end{equation}
It is clear that divergent parts of bare 1PI functions (in either
regularization) can be easily recovered from the formulae listed below.
\vskip0.2cm

To illustrate the method described above we quote here the explicit expression
for the one-loop correction (diagram $C$ of Fig. \ref{Rys:KmuOmA_i_KmuOm})
to the function $\VEV{K^\mu_\alpha \omega^\gamma}$
entering all the ST identities analyzed in Section \ref{sec:cterms}
\eqs{\label{Eq:FeynIntEx}
\VEVOPITIL{\tilde{K}^\mu_\alpha(-q) \tilde{\omega}^\gamma(q)}^\OLB
&=&-i \mind{\bracket{\TV_\beta \TV_\kappa}}{\alpha}{\gamma}\ \times\\
&{}& \times\
\mu_H^{4-d}\!\int\vol{k}\left[\frac{i}{\KR(k)^2-m_V^2}\right]^{\beta\kappa}
\left[\eta_{\mu\nu}-\frac{k_\mu k_\nu}{k^2}\right]
\frac{\KR^\nu(k+q)}{  \KR(k+q)^2}\,.
\nn}
(In diagrams external lines of antifields are marked in the same way
as those of the
corresponding fields but carry the extra arrow pointing the direction of
the flow of the ghost number). For $d=4$ the integral in \refer{Eq:FeynIntEx}
is regularized according to the prescription \refer{eqn:MomSpacePrescription},
whereas for $\La=\infty$ it is regularized dimensionally. For the difference
defined in \refer{Eq:DeltaReg:def}, omitting terms which vanish in the
limits $\Lambda\rightarrow\infty$, $\epsilon\rightarrow0$, one gets
\eqs{
\label{Eq:DeltaZ:KmuOmega}
&{}&\Delta\!\VEVOPITIL{\tilde{K^{\sigma}_\kappa}(-l)\tilde{\omega}^\alpha(l)}^\OLB
=\frac{i\ \! l_\sigma}{(4\pi)^2}\ \!
\mind{\left(\TV_\gamma\TV^\gamma\right)}{\kappa}{\alpha}
\left(\frac{1}{8}+\frac{3}{4}\ln\frac{4}{3}+\frac{3}{4}\delta_{\rm{Div}}\right).
\qquad\quad
}
As to the other 1PI functions of antifields,
there are no one-loop diagrams contributing to the function
$\VEVOPITIL{\tilde{K_i}(q) \tilde{\omega}^\alpha(l)}^\OLB$.
To the 
function $\VEVOPITIL{\tilde{K}^\sigma_\kappa(p') \tilde{\omega}^\alpha(l)
\tilde{A}^\beta_\nu(p)}^\OLB $ contribute the diagrams $A$ and $B$ shown in Fig.
\ref{Rys:KmuOmA_i_KmuOm}. However, owing to the antighost equation
\refer{Eq:AntiGhost} both these
contributions are independent of the regularization (despite being
superficially logarithmically divergent). Likewise  diagrams
with external lines of antifields, shown in
Figs. \ref{Rys:KiOmAA}--\ref{Rys:KiOmPsiPsi}, which contribute
to the following functions of the antifields
\eqs{
&{}&
\VEVOPITIL{\tilde{K_i}(q) \tilde{\omega}^\alpha(l)}^\OLB,
\quad
\VEVOPITIL{\tilde{K^{\sigma}_\kappa}(p') \tilde{\omega}^\alpha(l)
\tilde{A}^\beta_\nu(p)}^\OLB,
\quad
\VEVOPITIL{\tilde{K_i}(p') \tilde{\omega}^\alpha(l) \tilde{A}^\beta_\nu(p)
}^\OLB,
\nn\\&{}&\nn\\
&{}&
\VEVOPITIL{\tilde{K_i}(q) \tilde{\omega}^\alpha(l) \tilde{A}^\beta_\nu(p)
\tilde{A}^\gamma_\rho(p')}^\OLB,
\quad
\VEVOPITIL{\tilde{K}_n(r)\tilde{\omega}^\alpha(l)\tilde{\phi}^i\!(p)}^\OLB,
\quad
\VEVOPITIL{\tilde{K}^\mu_\gamma(q)\tilde{\omega}^\alpha(l)\tilde{\phi}^j\!(p)}^\OLB,
\nn\\&{}&\nn\\
&{}&
\VEVOPITIL{\tilde{K}_n(q)\tilde{\omega}^\alpha(l)\tilde{\phi}^i\!(p)
\tilde{\phi}^j\!(p')}^\OL,
\quad
\VEVOPITIL{\tilde{K}_n(l)\tilde{\omega}^\gamma(q)\tilde{\psi}^{b_1}\!(p_1)
\tilde{\psi}^{b_2}\!(p_2)}^\OL,
\nn\\&{}&\nn\\
&{}&
\VEVOPITIL{
\tilde{\bar{K}}_{a_1}(p_1)\tilde{\omega}^\gamma(q)\tilde{\psi}^{a_2}\!(p_2)}^\OLB
\nn
}
are independent of regularization. Therefore, the
differences \refer{Eq:DeltaReg:def} corresponding to these function vanish.
\vskip0.2cm

One-loop diagrams contributing to the two-point function
$\tig^{\mu\nu}_{\alpha\beta}$ are shown in Figure \ref{Rys:AA}.
The corresponding difference \refer{Eq:DeltaReg:def} reads
\eqs{\label{Eq:DeltaZ:AA:AplusBplusC}
(4\pi)^2 \Delta\tig^{\mu\nu}_{\alpha\beta}(p,-p)^\OLB
&=&\tr\!\left(\TV_\alpha\TV_\beta m_V^2\right)\eta^{\mu\nu}
\left(\frac{11}{8}-\frac{3}{4}\delta_{\rm{Div}}\!\right)
+\\&{}&\hspace*{-80 pt}
+\tr(\TV_\alpha \TV_\beta) \left[
{p^\mu}{p^\nu}
\left(\frac{1}{9}+\frac{13}{6}\delta_{\rm{Div}}+\frac{19}{6}\ln2
- \frac{3}{2}\ln3\right)+\right.
\nn\\&{}&\hspace*{-80 pt} \qquad\qquad\ \ \
+\left.\!\eta^{\mu\nu}\!\left(-\frac{23}{72}p^2-\frac{13}{6}p^2\delta_{\rm{Div}}
-\frac{19}{6}p^2\ln2+\frac{3}{2}p^2\ln{3}-{\Lambda^2}\right)
\right]+\nn\\
&{}&\hspace*{-80 pt}
%+\\&{}&
%\hspace*{-50 pt}
+\tr(\TS_\alpha \TS_\beta) \left[
{p^\mu}{p^\nu}
\left(-\frac{11}{72}-\frac{1}{6}\delta_{\rm{Div}}-\frac{1}{6}\ln2\right)+
\right.
\nn\\&{}&\hspace*{-80 pt} \qquad\qquad \ \ \ \
+\left.
\!\eta^{\mu\nu}\!\left(\frac{7}{144}p^2
+\frac{1}{6}p^2\delta_{\rm{Div}}+\frac{1}{12}p^2\ln4-\frac{\Lambda^2}{2}\right)
\right]+\nn\\
&{}&\hspace*{-80 pt} %\qquad\qquad\ \ \
+\frac{1}{4}\eta^{\mu\nu}
\tr\left(\{\TS_\alpha, \  \TS_\beta\}m_S^2\right)
-\left\{\frac{1}{8}+\frac{3}{4}\delta_{\rm{Div}}\right\}\eta^{\mu\nu}
\varphi^{\rm T}\!\{\TS_\alpha,\  \TS_\epsilon\}\{\TS^\epsilon,\  \TS_\beta\}\varphi
+\nn
\\&{}&\hspace*{-80 pt}\nn
-\tr(\TF_\alpha \TF_\beta) \left[
{p^\mu}{p^\nu}\left(\frac{5}{18}+\frac{2}{3}\delta_{\rm{Div}}
+\frac{2}{3}\ln2\right)\right.
+\\&{}& \hspace*{-80 pt}\qquad\qquad\ \ \
-\left.\eta^{\mu\nu}\left(\Lambda^2\!+\!\frac{11}{18}p^2\!
+\!\frac{2}{3}p^2\delta_{\rm{Div}}\!+\!\frac{2}{3}p^2\ln2\right)\right]+\nn\\
&{}& \hspace*{-80 pt}
+\eta^{\mu\nu}\left[
2\delta_{\rm{Div}}\tr\left(\TF_\alpha m_F^{*} \TF_\beta^{*} m_F\right)
\!-\!\left(\delta_{\rm{Div}}\!+\!\frac{1}{2}\right)
\tr\left(\{\TF_\alpha,\ \TF_\beta\} m^{*}_F m_F\right)
\right].
\nn}

One-loop corrections to the three-point  function
$\tig^{\mu\nu\rho}_{\alpha\beta\gamma}(l,p,p')$ are displayed in Fig. \ref{Rys:AAA}.
For fermions in a non-anomalous representation the corresponding difference
\refer{Eq:DeltaReg:def}
(diagrams $E$, $F_I$ and $F_{II}$ do not contribute to it) reads
\eqs{\label{Eq:DeltaZ:AAA:X}
(4\pi)^2 \Delta\tig^{\mu\nu\rho}_{\alpha\beta\gamma}(l,p,p')^\OLB
&=&\big[
\eta^{\mu\rho}(l-p')^\nu+\eta^{\mu\nu}(p-l)^\rho+\eta^{\nu\rho}(p'-p)^\mu
\big]\times
\nn\\&{}&\hspace*{-120 pt}
\times\bigg\{
{i}\,\tr\left(\TV_\alpha\TV_\beta \TV_\gamma\right)
\left(\frac{11}{12}\!+\!\frac{17}{6}\delta_{\rm{Div}}
\!+\!\frac{4}{3}\ln\frac{4}{3}\right)-
\frac{i}{6}\ \!\tr\left(\TS_\alpha\TS_\beta \TS_\gamma\right)
\left(-1\!+\!2\delta_{\rm{Div}}\!+\!2\ln\frac{4}{3}\right)
+\nn\\&{}&\hspace*{-120 pt}
\phantom{\times\bigg\{}
-
\frac{i}{3}\ \!\tr\left(\TF_\alpha\big[\TF_\beta,\ \TF_\gamma\big]\right)
\left(1+2\delta_{\rm{Div}}+2\,\ln\frac{4}{3}\right)
\bigg\}.\qquad
}

Diagrams contributing to the function $\VEV{\phi AA}$  are shown in
Fig. \ref{Rys:PhiAA} (the diagram $C$ is not shown, because ghosts do not
couple to scalars in the Landau gauge). Power-counting, the  Lorentz symmetry
and the translational invariance \refer{Eq:shift-sym} of the effective action
imply the relation
\eq{\label{Eq:RelXYZ}
\Delta\tig^{\phantom{i}\nu\rho}_{i\beta\gamma}(l,p,p')^\OLB=
\Delta\tig^{\phantom{i}\nu\rho}_{i\beta\gamma}(0,p,-p)^\OLB=
\derp{}{\varphi^i}\Delta\tig^{\nu\rho}_{\beta\gamma}(p,-p)^\OLB\,,
}
between \refer{Eq:DeltaZ:AA:AplusBplusC} and
$\Delta\tig^{\phantom{i}\nu\rho}_{i\beta\gamma}(l,p,p')^\OLB$.
By computing the latter difference directly we have checked that the
relation \refer{Eq:RelXYZ} does indeed hold.

The difference \refer{Eq:DeltaReg:def} corresponding to the
one-loop tadpole diagrams shown in Fig. \ref{Rys:Phi} is
\begin{eqnarray}
(4\pi)^2\Delta\tig^{}_{i}(p)^\OLB=
\Lambda^2\tr\!\left\{Y_i m_F^{*}+cc.\right\}
-\delta_{\rm  Div}\tr\!\left\{Y_i m_F^{*} m_F m_F^{*}+cc.\right\}
-\frac{1}{2}\Lambda^2\tr\!\left\{\rho_i\right\}+\nonumber\\
+\frac{1}{2}\delta_{\rm  Div}\tr\!\left\{\rho_i m^2_S\right\}
+\frac{1}{2}
\left[3\Lambda^2\delta_{\alpha\beta}
-(3\delta_{\rm  Div}+2)m_{V\,\alpha\beta}^2\right]
\left(\varphi^{\rm T}\!\left\{\TS^\alpha,\ \TS^\beta\right\}\right)_i~\!.
\phantom{aaa}\label{Eq:DeltaZ:Phi}
\end{eqnarray}
(The couplings $\rho_{ijk}$ and the scalar fields mass matrix
$m^2_S$ are defined in \refer{Eq:RegFey:LamRhoMs}).

Diagrams contributing to the $\VEV{\phi A}$ function are presented in
Fig. \ref{Rys:PhiA} (diagrams $B$ and $C$ have the same value in both
regularizations). The corresponding difference \refer{Eq:DeltaReg:def} reads
\eqs{\label{Eq:DeltaZ:PhiA:A}
(4\pi)^2 \Delta\tig^{\ \nu}_{i\beta}(l,-l)^\OLB
&=&-i\,l^\nu
\left(\frac{1}{4}+\frac{3}{2}\delta_{\rm{Div}}-\frac{3}{4}\ln\frac{3}{4}\right)
\left(\TS^\kappa\!\left\{\TS_\kappa,\ \TS_\beta\right\}\!\varphi\right)_i
+\\&{}&\hspace*{-80 pt}
+
i\,l^\nu
\left(\frac{1}{12}+\frac{1}{2}\delta_{\rm{Div}}-\frac{1}{2}\ln\frac{3}{4}\right)
\tr\big(\YF_i m_F^{*} \TF_\beta^* -\YF_i \TF_\beta  m_F^{*} + \YF_i^*  m_F \TF_\beta
- \YF_i^* \TF_\beta^* m_F \big).
\nn}

Diagrams contributing to the function $\VEV{\phi\phi}$ are shown in
Fig. \ref{Rys:PhiPhi}. They give
\eqs{\label{Eq:DeltaZ:PhiPhi:A+B}
(4\pi)^2 \Delta\tig_{i_1 i_2}(l,-l)^\OLB
&=&3\Lambda^2(\TS^\alpha\TS_\alpha)_{i_1 i_2}+
\\&{}&\hspace*{-100 pt}\nn
-\!\left[\frac{3}{2}\delta_{\rm{Div}}\!+\!1\right]\!
\Big[(m_V^2)_{\alpha\beta}\left\{\TS^\alpha,\ \TS^\beta\right\}_{i_1 i_2}
-\left(\left\{\TS^\alpha,\ \TS^\beta\right\}\!\varphi\right)_{i_1}
\left(\big\{\TS_\alpha,\ \TS_\beta\big\}\varphi\right)_{i_2}\Big]+\\
&{}&\hspace*{-100 pt}\nn
+l^2\left(3\delta_{\rm{Div}}+\frac{1}{2}+\frac{3}{4}\ln\frac{32}{9}\right)
(\TS^\alpha\TS_\alpha)_{i_1 i_2}+\\
&{}&\hspace*{-100 pt}\nn+\left[\Lambda^2+l^2\left(\frac{1}{2}\delta_{\rm{Div}}
+\frac{3}{8}+\frac{1}{2}\ln2\right)\right]\!\tr\!\left[\YF_{i_1}\YF_{i_2}^*
\!+\!cc.\right]+\!\\
&{}&\hspace*{-100 pt}\nn
-\delta_{\rm{Div}}\tr\!\left[\left(Y_{i_1}m_F^{*}m_F Y_{i_2}^*
+Y_{i_1}Y_{i_2}^*m_F m_F^{*}
+Y_{i_1} m_F^{*} Y_{i_2}m_F^{*} \right)\!+\!cc.\right]+\\
&{}&\hspace*{-100 pt}\nn
-\frac{1}{2}\lambda_{i_1 i_2 j_1 j_2}
\left[\delta^{j_1 j_2} \Lambda^2 -\delta_{\rm{Div}} (m_S^2)^{j_1 j_2}
\right]+
\frac{1}{2}\delta_{\rm{Div}} \tr\!\left(\rho_{i_1} \rho_{i_2}\right).
}

Contributions to the $\VEV{\phi\phi\phi}$ vertex are displayed in
Fig. \ref{Rys:PhiPhiPhi} (only diagrams $A$, $D$ and $E$ are different
in $\Lambda$Reg and DimReg). Since both regularizations preserve
(\ref{Eq:shift-sym}), $\Delta\tilde\Gamma_{i_1i_2i_3}^{(1B)}$ obtained by
direct calculation coincides with the result of differentiating
$\Delta\tig_{i_1 i_2}(l,-l)^\OLB$ given in \refer{Eq:DeltaZ:PhiPhi:A+B}
with respect to the background $\varphi$.

Only diagrams $D$ and $I$ of Fig. \ref{Rys:PhiPhiA}
contribute to the difference
\eqs{\label{Eq:DeltaZ:PhiPhiA}
(4\pi)^2 \Delta\tig^{\ \ \!\mu}_{ij\alpha}(p,p',l)^\OLB
&=&i\,(p^\mu\!-\!p'{}^\mu)\times\\
&{}&\hspace*{-110 pt}
\times\left\{\left[\frac{1}{6}\!+\!\delta_{\rm{Div}}\!-\!\ln\frac{3}{4}\right]
\tr\big[\YF_i^* \YF_j \TF_\alpha\!+\!cc.\big]
-\left[\frac{1}{4}\!+\!\frac{3}{2}\delta_{\rm{Div}}\!
-\!\frac{3}{4}\ln\frac{3}{4}\right]
\left(\TS_\beta\!\left\{\TS_\alpha,\ \TS^\beta\right\}\right)_{ij}
\right\}.\nn
}
corresponding to the $\VEV{\phi \phi A }$ vertex.

For 1PI functions with two external fermionic lines
(recall that we work with Majorana fermions) we use the matrix
notation in which spinor indices are omitted; we write for example
\eq{\label{Eq:FunPsiPsiA:Matrix-Conv}
\tig^{ \phantom{\psi\otimes\psi}}_{\psi\otimes\psi}(p_1,p_2)
\equiv\matri{\tig^{\phantom{a_1} \phantom{a_2}}_{a_1a_2}(p_1,p_2,q)},
\qquad\tig^{ \phantom{\psi\otimes\psi}~\!\mu}_{\psi\otimes\psi~\!\alpha}(p_1,p_2,q)
\equiv\matri{\tig^{\phantom{a_1} \phantom{a_2}~\!\mu}_{a_1a_2~\!\alpha}(p_1,p_2,q)},\nn
}
etc. In this notation the diagrams of Fig. \ref{Rys:PsiPsi} give
\eqs{\label{Eq:DeltaZ:PsiPsi:AiB}
(4\pi)^2 \Delta\tig_{\psi\otimes\psi}(p,-p)^{\OLB}
&=&C\ds{p}\left\{\TM_\gamma \TM^\gamma
+\frac{1}{2}\left[\ln\frac{3}{4}-\frac{1}{6}-\delta_{\rm Div}\right]
\YM_i^{*}\YM^i\right\}+\nn\\
&{}&\,-\!\left(3\,\delta_{\rm Div}\!+\!2\right)
C\TM_\gamma^{\rm T}\widehat{m}_F \TM^\gamma
\!+\!\delta_{\rm Div}\,C\YM_i \widehat{m}_F^{*} \YM^i.
}
for the difference \refer{Eq:DeltaReg:def}
of the corresponding  $\VEV{\psi\psi}$ functions in the two schemes,
\eqs{\label{Eq:DeltaZ:PsiPsiA:Tot}
(4\pi)^2 \Delta \tig^{ \phantom{\psi\otimes\psi}\mu}_{\psi\otimes\psi\alpha}(p_1,p_2,q)^\OLB
&=&i\,C\Big\{\!\!
-\!\frac{3}{2}\gamma^\mu\TM_\kappa \TM_\alpha \TM^\kappa
\!+\!\frac{1}{4}\left(2\delta_{\rm Div}\!+\!1\right)
\YM^i\TM_\alpha\YM_i^{*}\gamma^\mu
\!\!+\!\!\!\!\nn\\&{}&\hspace*{-120 pt}
+\frac{1}{4}\left(2\delta_{\rm Div}-1\right)\gamma^\mu\YM_i^*\YM_j\TS_\alpha^{ij}
+\frac{3}{2}\left[\delta_{\rm Div}\!+\!\frac{1}{6}\right]\gamma^\mu
\TM_\kappa \TM_\beta e^{\kappa\beta}_{\phantom{\kappa\beta}\!\alpha}
\Big\}.
}
for the  difference of the $\VEV{\psi\psi A}$ vertices
(diagrams of Fig. \ref{Rys:PsiPsiA}) and
\eq{\label{Eq:DeltaZ:PsiPsiPhi:Tot}
(4\pi)^2 \Delta\tig_{\psi\otimes\psi i}(p_1,p_2,q)^\OLB
=C\big\{-\!\left(3\,\delta_{\rm Div}\!+\!2\right)
\TM_\gamma^{\rm T}\YM_i \TM^\gamma
\!+\!\delta_{\rm Div}\,\YM_j \YM_i^* \YM^j
\big\}.\quad
}
for the difference of the  $\VEV{\psi\psi\phi}$ vertices (diagrams of
Fig. \ref{Rys:PsiPsiPhi}).
As expected, (\ref{Eq:DeltaZ:PsiPsiPhi:Tot}) is just the derivative of
(\ref{Eq:DeltaZ:PsiPsi:AiB}) w.r.t $\varphi$.

Regularization-dependent contributions to the four-point function
$\VEV{AAAA}$ are shown in Fig. \ref{Rys:AAAA}. The corresponding difference
\refer{Eq:DeltaReg:def} has (for non-anomalous
representations $\TF_\alpha$) the unambiguous form
\eqs{\label{Eq:Delta:AAAA}
(4\pi)^2
\Delta\tig^{\mu_1\mu_2\mu_3\mu_4}_{\alpha_1\alpha_2\alpha_3\alpha_4}(l_1,l_2,l_3,l_4)^\OLB
&=&\\&{}&\hspace*{-95 pt}
=\eta^{\mu_1\mu_2}\eta^{\mu_3\mu_4}
\mathbb{V}_{\alpha_1\alpha_2\alpha_3\alpha_4}+
\eta^{\mu_1\mu_4}\eta^{\mu_3\mu_2}
\mathbb{V}_{\alpha_1\alpha_4\alpha_3\alpha_2}+
\eta^{\mu_1\mu_3}\eta^{\mu_2\mu_4}
\mathbb{V}_{\alpha_1\alpha_3\alpha_2\alpha_4},
\nn}
where
\eqs{
\mathbb{V}_{\alpha_1\alpha_2\alpha_3\alpha_4}
&=&
\frac{1}{9}\Big\{\!
\left(11\!+\!12\,\delta_{\rm{Div}}\right)
\tr(\TV_{\alpha_1}\TV_{\alpha_2}\{\TV_{\alpha_3},\ \TV_{\alpha_4}\})
\!-\!
8\left(2\!+\!3\,\delta_{\rm{Div}}\right)
\tr(\TV_{\alpha_1}\TV_{\alpha_3}\TV_{\alpha_2}\TV_{\alpha_4})
\Big\}
+\nn\\
&{}&
+\frac{1}{6}\Big\{\!
\left(\delta_{\rm{Div}}-1\right)
\tr\left(\left[\TS_{\alpha_2},\ \TS_{\alpha_3}\right]
\left[\TS_{\alpha_1},\ \TS_{\alpha_4}\right]
+\left[\TS_{\alpha_1},\ \TS_{\alpha_3}\right]
\left[\TS_{\alpha_2},\ \TS_{\alpha_4}\right]
\right)
+\nn\\
&{}&\phantom{+\frac{1}{6}\Big\{\!}
+ 2\, \tr\left(\TS_{(\alpha_1}\TS_{\alpha_2}\TS_{\alpha_3}\TS_{\alpha_4)}\right)
\Big\}
+\nn\\
&{}&
-\frac{2}{3}\,\delta_{\rm{Div}}
\tr\left(\left[\TF_{\alpha_4},\ \TF_{\alpha_1}\right]
\left[\TF_{\alpha_2},\ \TF_{\alpha_3}\right]
+\left[\TF_{\alpha_1},\ \TF_{\alpha_3}\right]
\left[\TF_{\alpha_4},\ \TF_{\alpha_2}\right]
\right)
+\nn\\
&{}&%\phantom{+\frac{1}{6}\Big\{\!}
-\frac{4}{9}\,
\tr\left(\left\{\TF_{\alpha_1},\ \TF_{\alpha_2}\right\}
\left\{\TF_{\alpha_3},\ \TF_{\alpha_4}\right\}\right)
+\frac{5}{9}\, \tr\left(\TF_{\alpha_1}\TF_{\alpha_3}\TF_{\alpha_2}\TF_{\alpha_4}
+\TF_{\alpha_3}\TF_{\alpha_1}\TF_{\alpha_4}\TF_{\alpha_2}\right),
}
(notice the symmetrization of the indices $\alpha_1,\dots,\alpha_4$
in the third line).
Divergences are associated only with the structure constants as expected.
In general (i.e. not assuming \refer{Eq:Non-anomaly-cond})
the difference \refer{Eq:Delta:AAAA} would contain also terms proportional
to the Levi-Civita tensor $\epsilon^{\alpha_1\alpha_2\alpha_3\alpha_4}$ which
cannot be determined uniquely because of the ambiguities of the DimReg
scheme with the
naive $\gamma^5$. Such terms are multiplied by tensors of the form
\eq{
\tr(\TF_{\alpha_1}\TF_{\alpha_2}\TF_{\alpha_3}\TF_{\alpha_4}-cc.)
\equiv\frac{1}{2}
\tr(\{\TF_{\alpha_1},\ \TF_{\alpha_2}\}\,[\,\TF_{\alpha_3},\ \TF_{\alpha_4}])
+\frac{1}{2}
\tr([\,\TF_{\alpha_1},\ \TF_{\alpha_2}]\,\{\TF_{\alpha_3},\ \TF_{\alpha_4}\}),
}
and vanish for a non-anomalous fermionic representation.

\renewcommand{\thesection}{Appendix~\Alph{section}}
\renewcommand{\theequation}{\Alph{section}.\arabic{equation}}
\renewcommand{\thefigure}{\Alph{section}.\arabic{figure}}

\setcounter{equation}{0}
\setcounter{footnote}{0}
\setcounter{figure}{0}

\section{Chiral anomaly}
\label{Sec:ApB-AAAA}
In this appendix we determine directly the contribution of fermionic loops
to the ST identity \refer{Eq:WT-AAAA} involving the $\VEV{AAAA}$ vertex.
We use the notation introduced in Subsection \ref{Sec:AAAA}
(Eqs. \refer{Eq:Not-AAAA-1} and \refer{Eq:Not-AAAA-2}).

Fermionic loops contribute only to the first three terms of the LHS of the
identity \refer{Eq:WT-AAAA}. Since it potentially can have
a \emph{true} anomaly, it will be instructive
to generalize the regularization \refer{eqn:MomSpacePrescription} by
not specifying explicitly the profile $g(\cdot)$ of the function
entering the regularization prescription
\begin{eqnarray}
k^\mu\rightarrow\KR^\mu(k) =  \frac{1}{\tilde{g}(k)}\times k^\mu,
\qquad\quad
\tilde{g}(k)\equiv g\big({k^2}/{\Lambda^2}\big),
\end{eqnarray}
but assuming only that $g$ is an analytic function satisfying the
boundary conditions
\begin{eqnarray}\label{Eq:Reg:BoundCond}
g(0)=1,\qquad
g(x)\stackrel{x\to-\infty}{\longrightarrow} 0.
\end{eqnarray}
We also use the following notation for fermionic
propagators (cf.  Eq. \refer{Eq:FermProp}):
\begin{eqnarray}
\FeMaR{k}\equiv S(k;\, \La) C,\qquad\quad
\FeMa{k}\equiv \lim_{\Lambda\to\infty}\FeMaR{k}.
\end{eqnarray}
The contribution of fermions to the $\VEV{AAAA}$ vertex (diagram $G$ of Fig.
\ref{Rys:AAAA}) reads
\begin{eqnarray}\label{Diag:AAAA:G}
\tig^{\mu_1\mu_2\mu_3\mu_4}_{\alpha_1\alpha_2\alpha_3\alpha_4}(l_1,l_2,l_3,l_4|G)\equiv
\mathscr{G}_{\imath_1\imath_2\imath_3\imath_4}=
\mathbb{G}_{\imath_1\imath_2\imath_3\imath_4}+
\mathbb{G}_{\imath_1\imath_2\imath_4\imath_3}+
\mathbb{G}_{\imath_1\imath_3\imath_2\imath_4},\quad
\end{eqnarray}
where
\begin{eqnarray}
\mathbb{G}_{\imath_1\imath_2\imath_3\imath_4}
\phantom{aaaaaaaaaaaaaaaaaaaaaaaaaaaaaaaaaaaaaaaaaaaaaaaaaaaaaaaaaaaaa}\\
%&=&\\\nn
%{}&{}& \hspace*{-53 pt}
=\!\tr\!\!\int\volfour{k}~\!
i\gamma^{\mu_1} \TM_{\alpha_1}  \FeMaR{k}
\gamma^{\mu_2} \TM_{\alpha_2}  \FeMaR{k\!-\!l_2}
\gamma^{\mu_3} \TM_{\alpha_3}  \FeMaR{k\!-\!l_2\!-\!l_3}
\gamma^{\mu_4} \TM_{\alpha_4} \FeMaR{k\!+\!l_1},\nonumber
\end{eqnarray}
while the analogous contribution to the $\left<\phi A A A\right>$ vertex has the form
\begin{eqnarray}\label{Diag:PhiAAA:G}
\tig^{\phantom{j}\,\mu_2\mu_3\mu_4}_{j\,\alpha_2\alpha_3\alpha_4}(l_1,l_2,l_3,l_4|G)
\equiv
\anti{\mathbb{G}}_{\mathbbm{j}\imath_2\imath_3\imath_4}+
\anti{\mathbb{G}}_{\mathbbm{j}\imath_2\imath_4\imath_3}+
\anti{\mathbb{G}}_{\mathbbm{j}\imath_3\imath_2\imath_4},
\end{eqnarray}
with
\begin{eqnarray}
\anti{\mathbb{G}}_{\mathbbm{j}\imath_2\imath_3\imath_4}
\phantom{aaaaaaaaaaaaaaaaaaaaaaaaaaaaaaaaaaaaaaaaaaaaaaaaaaaaaaaaaaaaa}\\
%&=&\\\nn
%{}&{}& \hspace*{-45 pt}
=i\,\tr\!\int\volfour{k}\,
i\, \YM_j  \FeMaR{k}
\gamma^{\mu_2} \TM_{\alpha_2}  \FeMaR{k\!-\!l_2}
\gamma^{\mu_3} \TM_{\alpha_3}  \FeMaR{k\!-\!l_2\!-\!l_3}
\gamma^{\mu_4} \TM_{\alpha_4} \FeMaR{k\!+\!l_1}.\nonumber
\end{eqnarray}
Finally the diagram $G$ of Fig. \ref{Rys:AAA} gives
\begin{eqnarray}\label{Diag:AAA:G}
\tig^{\mu\nu\rho}_{\alpha\beta\gamma}(l,p,p'|G)
=i~\!\tr\!\int\volfour{k}~\!
i\gamma^\nu \TM_\beta  \FeMaR{k}
\gamma^\rho \TM_\gamma  \FeMaR{k-p'}
\gamma^\mu \TM_\alpha  \FeMaR{k+p}.
%\nn\\
\end{eqnarray}

The total contribution of these three functions to the LHS of
\refer{Eq:WT-AAAA} will be denoted ${\bf\Omega}_{4F}$. To simplify it, we
decompose the contribution of \refer{Diag:AAAA:G}
contracted with the momentum $l_1^{\mu_1}$ using the identity
\eq{\label{Eq:RoownXXX}
\ds{l}_1 \TM_{\alpha_1}=\{\ds{k}+\ds{l}_1-\widehat{m}_F\}\TM_{\alpha_1}
+\TM^{\rm T}_{\alpha_1}\{\ds{k}-\widehat{m}_F\}-(\TS_{\alpha_1}\varphi)^j \YM_j,
}
(following from the gauge symmetry of $I_0^{GI}$)
into three parts. The one that originates from the last term of
\refer{Eq:RoownXXX} cancels exactly in ${\bf\Omega}_{4F}$ the contribution
of \refer{Diag:PhiAAA:G}. Furthermore, after expanding the propagators in the
remaining terms of
${\bf\Omega}_{4F}$ according to \refer{eqn:PropagatorExpansion} and retaining
only those integrals that do not vanish in the limit $\Lambda\to\infty$,
``similar'' terms\footnote{Some of these
``similarities'' become visible only after shifting the
integration momentum and replacing the matrix under the trace by its
transposition, with the aid of standard relations
\begin{eqnarray}
%\label{Eq:EpsilonMatr}
C{\gamma^\mu}^{\rm T} C={\gamma^\mu},
\qquad
C{\gamma^5}={\gamma^5}C,
\qquad
C^{\rm T}=C^{-1}=-C.\nn
\end{eqnarray}
}
can be combined. As a result of these operations ${\bf\Omega}_{4F}$ can be
represented in the form
\eq{%\label{Eq:deltaF-AAAA-ver0}
{\bf\Omega}_{4F}
=(\mathbb{X}^{I}\!+\!\mathbb{X}^{II})_{\imath_1\imath_2\imath_3\imath_4}+
(\mathbb{X}^{I}\!+\!\mathbb{X}^{II})_{\imath_1\imath_4\imath_3\imath_2}+
(\mathbb{X}^{I}\!+\!\mathbb{X}^{II})_{\imath_1\imath_2\imath_4\imath_3}
+\mathcal{O}(\Lambda^{-1}),
\nn}
(notice some reordering of the symbols $\imath_n$ as compared to
\refer{Diag:AAAA:G}!)
where the integrands of the integrals
\eqs{
\mathbb{X}^{I}_{\imath_1\imath_2\imath_3\imath_4}
&=&
\frac{i}{2}\,
\tr\!\! \int\volfour{k}
i \FeMa{k}
\gamma^{\mu_2} \TM_{\alpha_2}  \FeMa{k\!-\!l_2}
\gamma^{\mu_3} \TM_{\alpha_3}  \FeMa{k\!-\!l_2\!-\!l_3}
\gamma^{\mu_4} \{\TM_{\alpha_4},\ \TM_{\alpha_1}\}
\times
\nn\\&{}&
\qquad\qquad\quad\times
\left\{\tilde{g}(k+l_1)-\tilde{g}(k+l_4)\right\}~\!
\tilde{g}(k)~\!\tilde{g}(k-l_2)~\!\tilde{g}(k-l_2-l_3),
\nn}
and
\eqs{
\mathbb{X}^{II}_{\imath_1\imath_2\imath_3\imath_4}
&=&
\frac{i}{2}\eSC{\kappa\!}{\alpha_4}{\alpha_1}~\!
\tr\!\int\volfour{k}
i \FeMa{k}
\gamma^{\mu_2} \TM_{\alpha_2}  \FeMa{k\!-\!l_2}
\gamma^{\mu_3} \TM_{\alpha_3}  \FeMa{k\!-\!l_2\!-\!l_3}
\gamma^{\mu_4} \TM_{\kappa}
\times
\nn\\&{}&
\qquad\qquad\times
\left\{\tilde{g}(k+l_1)+\tilde{g}(k+l_4)-2\right\}
\tilde{g}(k)~\!\tilde{g}(k-l_2)~\!\tilde{g}(k-l_2-l_3).
\nn}
vanish in the limit $\Lambda\to\infty$. (The term corresponding to $-2$ in
the curly brackets of $\mathbb{X}^{II}$ originates from the contribution of
the $\VEV{AAA}$ vertex to the considered ST identity.)
The factors $\tilde{g}(k+l)$ can be now expanded in powers of the external
momentum $l$. Performing next the integrals over the angular variables one
finds that in the limit $\Lambda\to\infty$ the fermionic contribution can
be written in the form
\begin{eqnarray}
\label{Eq:deltaF-AAAA}
{\bf\Omega}_{4F}\equiv
{\bf\Omega}_{4F}^{\mathbb{A}}+
{\bf\Omega}_{4F}^{\mathbb{N}}
=(\mathbb{A}\!+\!\mathbb{N})_{\imath_1\imath_2\imath_3\imath_4}+
(\mathbb{A}\!+\!\mathbb{N})_{\imath_1\imath_4\imath_3\imath_2}+
(\mathbb{A}\!+\!\mathbb{N})_{\imath_1\imath_2\imath_4\imath_3}+
\mathcal{O}(\Lambda^{-1})\,,\ \ \ \ \ \  ~
\end{eqnarray}
in which
\begin{eqnarray}
\label{Eq:AnomAAAA-praw}
\mathbb{A}_{\imath_1\imath_2\imath_3\imath_4}
&=&
\frac{\mathscr{C}_0}{(4\pi)^2}\ \!
\epsilon^{\nu\mu_2\mu_3\mu_4}
\bigg\{\frac{1}{24}\,
(l_3-l_2)_\nu\,\,\eSC{\kappa\!}{\alpha_4}{\alpha_1}
\tr\!\left(\phantom{\big\{}\!\!\!\TF_\kappa
\{\TF_{\alpha_3},\ \TF_{\alpha_2}\}\right)
+\nn\\&{}&\hspace*{77 pt}
+\frac{1}{8}\,(l_1-l_4)_\nu\,\,\eSC{\kappa\!}{\alpha_3}{\alpha_2}
\tr\!\left(\phantom{\big\{}\!\!\!\TF_\kappa
\{\TF_{\alpha_4},\ \TF_{\alpha_1}\}\right)
\bigg\}\, ,\qquad
\end{eqnarray}
is the true anomaly, while
\begin{eqnarray}
\mathbb{N}_{\imath_1\imath_2\imath_3\imath_4}
&=&\nn\\&{}&\hspace*{-50 pt}
=\frac{i\, \mathscr{C}_0}{(4\pi)^2}\times \frac{1}{24}
\left\{
\eta^{\nu\mu_2}\eta^{\mu_3\mu_4}+\eta^{\nu\mu_3}\eta^{\mu_2\mu_4}+
\eta^{\nu\mu_4}\eta^{\mu_2\mu_3}\right\}
\times\qquad\qquad\qquad\qquad
\nn\\&{}&\hspace*{-50 pt}
\quad\ \ \times
\bigg\{\frac{1}{3}\,
(l_3\!-\!l_2)_\nu\,\eSC{\kappa\!}{\alpha_4}{\alpha_1}
\tr\!\left(\phantom{\big\{}\!\!\!\TF_\kappa
[\TF_{\alpha_2},\ \TF_{\alpha_3}]\right)
%+\nn\\&{}&\hspace*{77 pt}
\!+\!(l_4\!-\!l_1)_\nu\,
\tr\!\left(\phantom{\big\{}\!\!\!\!
\{\TF_{\alpha_2},\ \TF_{\alpha_3}\}\,
\{\TF_{\alpha_4},\ \TF_{\alpha_1}\}\!\right)\!\!
\bigg\}\!+
\nn\\&{}&\hspace*{-50 pt}
\phantom{=}
+\frac{i\, \mathscr{C}_1}{(4\pi)^2}\times \frac{2}{3}
\left\{
-(l_2+2l_3)^{\mu_2}\eta^{\mu_3\mu_4}
+(2l_2+l_3)^{\mu_3}\eta^{\mu_2\mu_4}
-(l_2-l_3)^{\mu_4}\eta^{\mu_2\mu_3}
\right\}\times
\nn\\&{}&\hspace*{-50 pt}
\phantom{+}\quad\ \
\times
\eSC{\kappa\!}{\alpha_4}{\alpha_1}
\tr\!\left(\phantom{\big\{}\!\!\!\TF_\kappa
[\TF_{\alpha_2},\ \TF_{\alpha_3}]\right).
\end{eqnarray}
is the cohomologically trivial breaking
(the ``\emph{spurious} anomaly'') of the ST identity.
Thus, the complete result for ${\bf\Omega}_{4F}$ depends
(in the $\Lambda\to\infty$ limit) on only two integrals:
\begin{eqnarray}
\mathscr{C}_0=4\!\int_{-\infty}^0\!{\rm d}t~\! g(t)^3 g^\prime(t)=1,
\phantom{aaaa}{\rm and}\phantom{aaaa}
\mathscr{C}_1=\int_{-\infty}^0\!\frac{{\rm d}t}{t}\left\{g(t)^4-g(t)^3\right\},
\end{eqnarray}
of which only $\mathscr{C}_1$, entering the spurious part of the anomaly
depends on the shape of the regularizing function $g$
($\mathscr{C}_1=\ln(4/3)$ for $g(x)=\exp(x/2)$, which corresponds to the prescription \refer{eqn:PrescriptionDef}), while $\mathscr{C}_0$, which multiplies
the true anomaly, is independent of the specific shape of $g$ and
depends only on the boundary conditions \refer{Eq:Reg:BoundCond}.
\vskip0.2cm

To close the analysis of the anomalies we will argue that  in the
regularization (\ref{eqn:PrescriptionDef}) the ST identity involving
the $\langle AAAAA\rangle$ vertex is free of anomaly, that is that
${\bf\Omega}_{5F}$ vanishes in the limit $\Lambda\rightarrow\infty$.
To this end we first notice that all integrals that enter in this identity,
whose form is analogous to \refer{Eq:WT-AAAA}, are (at worst) logarithmically
divergent. As far as fermionic contributions are concerned relevant are only
the $\langle A^5\rangle$, $\VEV{A^4\,\phi}$ and $\VEV{A^4}$ functions.
In the logarithmically divergent contribution \refer{Diag:AAAA:G} to the
$\VEV{A^4}$ function one can make the replacement
\eqs{\label{Eq:QWERTY}
\FeMaR{p_i}\mapsto \gt(p_i)\FeMa{p_i},
}
because the terms of the integral \refer{Diag:AAAA:G} omitted in this way
all vanish in the limit $\Lambda\to\infty$. In the convergent fermionic
diagrams contributing to the $\VEV{A^5}$ and $\VEV{A^4 \phi}$ functions
$\Lambda$ can be sent to infinity. However, to make the
cancellations in the ST identity manifest, the fermionic contribution to
$\VEV{A^5}$ (contracted with the momentum $l_1^{\mu_1}$) has to be decomposed
using \refer{Eq:RoownXXX} into a combinations of integrals which
individually are logarithmically divergent.
Therefore, before making the decomposition \refer{Eq:RoownXXX}, we multiply
the \emph{unregulated} integrands (i.e. the ones in which $\Lambda$
has been sent to infinity) of the fermionic contributions to
the $\VEV{A^5}$ and $\VEV{A^4 \phi}$ functions by the factor
$\gt(k)^4$, where $k$ is the loop momentum.
(Making instead the replacement \refer{Eq:QWERTY} in the regulated
integrands would produce five factors of $\gt$.) Performing next,
just as before, appropriate shifts of the integration momentum gives
${\bf\Omega}_{5F}$ in the form of the integral whose
integrand is a homogeneous function of fourth degree in $\gt$ and vanishes
in the limit $\Lambda\to\infty$. The integral is therefore  similar
to  $\mathbb{X}^I$ given above.
However, because the present integral is
(unlike $\mathbb{X}^I$) only logarithmically
divergent when all the $\gt$ factors are omitted, the momenta
$q_i$ in $\gt(k+q_i)$ can give only a contribution of the order of
$\Lambda^{-2} \times \cO(\Lambda \ln \Lambda)$ which vanishes in the
infinite cutoff limit. Thus, ${\bf\Omega}_{5F}$ indeed vanishes for
$\Lambda\to\infty$.
To complete the argument, it is sufficient to notice that the  counterterm
\refer{Eq:KontrFlat:Fun:AAAA} does not break the considered ST identity
either, because $\delta^\flat\!\mathrm{q}$ is an invariant tensor of the
Lie algebra.

\setcounter{equation}{0}
\setcounter{footnote}{0}
\setcounter{figure}{0}

\setcounter{equation}{0}
\setcounter{footnote}{0}
\setcounter{figure}{0}

\section{RGE}
\label{app:rge}

Here we show that Eq. \refer{Eq:RGE-Gamma-n} indeed follows from
\refer{Eq:RGE-I-n=r-bar-def}. To this end we notice that the functional
$\bar r_n^\Lambda$ defined by \refer{eqn:RnInLambda} and 
\refer{eqn:RnDef} is local and of renormalizable form
when the regularization is neglected. The arguments of Section
\ref{Sec:UV-cutoff} then ensure that Feynman diagrams generated by the
auxiliary action ($\Phi$ and $\mathcal{K}$ stand for all fields and
antifields, respectively)
\begin{eqnarray}\label{Eq:I-bar-def}
\bar I^\Lambda_n[\Phi,\mathcal{K};g,c,\mu,\Lambda]
\equiv I^\Lambda_n[\Phi,\mathcal{K};g,\mu,\Lambda]
-c~\!\bar{r}^\Lambda_n[\Phi,\mathcal{K};g,\mu,\Lambda]~\!,
\end{eqnarray}
with the additional ``coupling constant"  $c$, are convergent.
The RGE \refer{eqn:RnInLambda} can be rewritten as
\begin{eqnarray}\label{Eq:R-bar-n-I-bar-n}
\left.
\bar{R}_n\bar I^\Lambda_n[\Phi,\mathcal{K};g,c,\mu,\Lambda]
\right|_{c=0}=0\, , \phantom{aaa}{\rm with}\phantom{aaa}
\bar{R}_n\equiv R_n+\derp{}{c}
\end{eqnarray}
In the following it is convenient to decompose the $R_n$ operator defined
in \refer{eqn:RnDef} (and analogously $\bar{R}_n$) in the following way
\begin{eqnarray}\label{Eq:R_n-decomp}
R_n=B_n - \mind{(\gamma^\Phi_n)}{i}{j}\Phi^j\!\cdot\! \derf{}{\Phi^i} \, .
\end{eqnarray}
${B}_n$ contains then only derivatives w.r.t. the parameters (including
the antifields):
\begin{eqnarray}\label{Eq:B_n:def}
B_n=\mu\derp{}{\mu}+\beta^A_n\derp{}{g^A}-{(\gamma^\mathcal{K}_n)}_{i}^{\phantom{i}j}
\mathcal{K}_j\!\cdot\!\derf{}{\mathcal{K}_i} \, .
\end{eqnarray}
Using Eq. \refer{Eq:R-bar-n-I-bar-n} and the functional
integration by parts one easily checks that the generating functional
\begin{eqnarray}
\bar{Z}_n[J,\mathcal{K};g,c,\mu,\Lambda]=\int[\mathcal{D}\Phi]~\!
e^{i(\bar I^\Lambda_n+J_k\cdotp\Phi^k)}\, ,
\end{eqnarray}
satisfies the following RGE:
\eq{
\left.\bar{\mathscr{R}}_n\bar{Z}_n[J,\mathcal{K};g,c,\mu,\Lambda]\right|_{c=0}
=\Theta\times\bar{Z}_n[J,\mathcal{K};g,0,\mu,\Lambda],
}
in which
\eq{
\bar{\mathscr{R}}_n\equiv
\bar{B}_n
+\mind{(\gamma^\Phi_n)}{i}{j}J_i\cdotp\! \derf{}{J_j}
=
{B}_n
+\derp{}{c}
+\mind{(\gamma^\Phi_n)}{i}{j}J_i\cdotp\! \derf{}{J_j}
\, .
}
Although $\Theta$ is a badly divergent factor
\eq{
\Theta=\sum_i\, (\mp\gamma^\Phi_n)^i_{\ i}\times
\int{{\rm d}^4 x\, \delta^{(4)}_{\rm position}(0)}\, ,
}
(upper/lower sign corresponds to bosonic/fermionic $\Phi^i$),
it drops out from the RGE satisfied by the  functional
$\bar W_n$ generating connected Green's functions defined by
\begin{eqnarray}
e^{i\bar{W}_n[J,\mathcal{K};g,c,\mu,\Lambda]}
=\frac{\bar{Z}_n[J,\mathcal{K};g,c,\mu,\Lambda]}
{\bar{Z}_n[0,\mathcal{K};g,c,\mu,\Lambda]}\, ,
\end{eqnarray}
which satisfies the simple relation
\begin{eqnarray}\label{Eq:RGE-dla-W-bar}
\left.
\bar{\mathscr{R}}_n\bar{W}_n[J,\mathcal{K};g,c,\mu,\Lambda]
\right|_{c=0}=0.
\end{eqnarray}
Passing next to the functional
$\bar{\Gamma}^\La_n\equiv\Gamma[\bar I_n^\Lambda]$ generating 1PI functions,
which is given as usually by the Legendre transform
\begin{eqnarray}
\bar{\Gamma}^\La_n[\Phi,\mathcal{K};g,c,\mu,\Lambda]\equiv
\bar{W}_n[\mathcal{J}^\Phi,\mathcal{K};g,c,\mu,\Lambda]
-\mathcal{J}^\Phi_k\,\cdotp\Phi^k\,,
\end{eqnarray}
with the source $\mathcal{J}^\Phi$ determined by the condition
\eq{
\left.\derf{ \bar{W}_n[J,\mathcal{K};g,c,\mu,\Lambda] }
{J_i(x)}\right|_{J=\mathcal{J}^\Phi}=\Phi^i(x)~\!,
}
and using the inverse relations
\eq{
\left.\derf{  \bar{\Gamma}^\La_n[\Phi,\mathcal{K};g,c,\mu,\Lambda] }
{\Phi^i(x)}\right.=\mp\mathcal{J}^\Phi_i(x),
}
one finds that
\eqs{
\bar{R}_n \bar{\Gamma}^\La_n[\Phi,\mathcal{K};g,c,\mu,\Lambda]
&=&
\left.\left\{
\bar{B}_n \bar{W}_n[J,\mathcal{K};g,c,\mu,\Lambda]
+\mind{(\gamma^\Phi_n)}{i}{j}J_i\cdotp\!
\derf{\bar{W}_n[J,\mathcal{K};g,c,\mu,\Lambda]}{J_j}
\right\}\right|_{J=\mathcal{J}^\Phi}
\nn\\
&=&
\bar{\mathscr{R}}_n\bar{W}_n[J,\mathcal{K};g,c,\mu,\Lambda]
\Big|_{J=\mathcal{J}^\Phi}\, .
}
This means that
\begin{eqnarray}
\bar{R}_n \bar{\Gamma}^\La_n[\Phi,\mathcal{K};g,c,\mu,\La]\big|_{c=0}=0~\!.
\label{eqn:G16}
\end{eqnarray}
Since $\Gamma_n^\Lambda\equiv{\Gamma}[I_n^\Lambda]
=\bar{\Gamma}_n^\La[\Phi,\mathcal{K};g,0,\mu,\La]$,
\refer{eqn:G16} implies that
\begin{eqnarray}
R_n\Gamma_n^\Lambda[\Phi,\mathcal{K};g,\mu,\La]
=-\left.\derp{\bar\Gamma_n^\La[\Phi,\mathcal{K};g,c,\mu,\La]}{c}\right|_{c=0}\, ,
\end{eqnarray}
However, from \refer{Eq:I-bar-def} it follows immediately that
\begin{eqnarray}\label{Eq:XYZa}
\left.\derp{\bar{\Gamma}^\La_n[\Phi,\mathcal{K};g,c,\mu,\Lambda]}{c}\right|_{c=0}
=-\bar{r}_n^\Lambda+\mathcal{O}({\hbar^{n+2}})\, .
\end{eqnarray}
To see that the quantum correction in \refer{Eq:XYZa} is indeed of order
$\mathcal{O}(\hbar^{n+2})$, it is sufficient to notice that $c$-dependent
contributions to $\bar{\Gamma}_n^\La$ (except for the tree-level one, i.e.
$-c\,\bar{r}^\Lambda_n$) are generated only by loop diagrams which
contain one or more $-c\,\bar{r}^\Lambda_n$ vertices. Since by the inductive
hypothesis $\bar r^\Lambda_n=\mathcal{O}(\hbar^{n+1})$, such diagrams are
necessarily at least of order of $\mathcal{O}(\hbar^{n+2})$. Finally, using
\refer{Eq:rnlam} and the $\La$-independence of the coefficients in $R_n$ one
concludes that Eq. \refer{Eq:RGE-Gamma-n} indeed follows from \refer{eqn:RnIn}.

\setcounter{equation}{0}
\setcounter{footnote}{0}
\setcounter{figure}{0}

\section{DimReg-$\overline{\rm MS}$ beta function $\check{\beta}$}
\label{App:JO}

In this appendix, for the reader's convenience, we recall in our notation
(see Section \ref{Sec:Dzialanie_S_I_itp})
the Jack-Osborn \cite{JO} expressions for the two-loop contributions to the
beta functions $\check{\beta}$ in the DimReg-$\overline{\rm MS}$ scheme.
We begin, however, with Machacek-Vaughn \cite{MV} formulae for two-loop
contributions to the anomalous dimensions $\check{\gamma}$ of the scalars
and left-chiral Weyl fermions in this scheme in the Landau gauge:
\begin{eqnarray}
\label{Eq:GammaCheck2L}
\check{\gamma}^\phi_{(2)\,i j}
&=&
\left\{
-\frac{35}{3}\,\tr\!\left[\TV_\alpha \TV_\beta\right]
+\frac{11}{12}\,\tr\!\left[\TS_\alpha \TS_\beta\right]
+\frac{20}{12}\,\tr\!\left[\TF_\alpha \TF_\beta\right]
\right\}\left(\TS^\alpha\TS^\beta\right)_{ij}
+\qquad\qquad\nn\\&{}&%\hspace*{-50 pt}
+\frac{1}{12}\lambda_{ikln}{\lambda}_j^{\phantom{j}kln}
+\frac{3}{2}\left[\left(\TS^\alpha\TS_\alpha\right)^2\right]_{ij}
-\frac{5}{2}\,\tr\big\{\TF^\alpha\TF_\alpha\YF_i^*\YF_j+cc.\big\}
+\nn\\&{}&%\hspace*{-50 pt}
-\frac{3}{4}\,\tr\!\left\{\YF_i\YF_j^*\YF^\ell\YF^*_\ell+cc.\right\}
-\frac{1}{2}\,\tr\!\left\{\YF_i^*\YF^\ell\YF_j^*\YF_\ell+cc.\right\},
\end{eqnarray}
\begin{eqnarray}
\check{\gamma}_{(2)}^F
&=&
%%%%%%
\left\{
\frac{25}{4}\,\tr\!\left[\TV_\alpha \TV_\beta\right]
-\frac{1}{4}\,\tr\!\left[\TS_\alpha \TS_\beta\right]
-\tr\!\left[\TF_\alpha \TF_\beta\right]\right\}\TF^\alpha\TF^{\beta}
-\frac{3}{2}\left(\TF_\kappa\TF^\kappa\right)^2
-\frac{3}{8}\mathbb{Y}^{ij}\YF_j^*\YF_i+
\nn\\&{}&%\hspace*{-45pt}
-\frac{1}{8}\YF_i^*\YF^{j}\YF_j^*\YF^{i}
-\frac{9}{2}\left[\TS^\alpha\TS_\alpha\right]^{ij}\YF_i^*\YF_{j}
+\frac{7}{4}\TF_\alpha\TF^{\alpha}\YF_\ell^*\YF^{\ell}
+\frac{1}{4}\YF_{\ell}^*\TF_\alpha^*\TF^{\alpha*}\YF^\ell\,.
\nn
\end{eqnarray}
We have not found in the literature the analogous Landau gauge two-loop
contributions to the anomalous dimensions of the vector fields.
(Of course, anomalous dimensions of Abelian gauge fields
can be read off from the expression for the beta function of the
corresponding Abelian gauge couplings given below.)

Two-loop contributions to the beta functions of the parameters of the
scalar fields potential can be extracted by taking the
appropriate derivatives with respect to $\varphi$ (to save the space
we write $\mathcal{M}_{X}$ for $\mathcal{M}_X(\varphi)$,
and $\mathcal{M}^4_X$ for $[\mathcal{M}^2_X(\varphi)]^2$) of
\begin{eqnarray}
\label{Eq:BetaVCheck2L}
\check{\beta}_{(2)}^\mathcal{V}(\varphi)
&=&\matrixind{\check{\gamma}_{(2)}^\phi(g)
   -\frac{1}{4}\,\tr\!\left[\TV_\alpha \TV_\beta\right]\TS^\alpha\TS^\beta
      }{i}{j}\,\varphi^j\,
\derp{\mathcal{V}(\varphi)}{\varphi^i}
+\varphi^{\rm T}
    \TS^\alpha\TS^\beta\!\mathcal{M}_S^2\TS_\beta\TS_\alpha\varphi\!+
\nn\\&{}&\hspace*{-40 pt}
+\left\{
-\frac{161}{12}\,\tr\!\left[\TV_\alpha \TV_\beta\right]
+\frac{7}{6}\,\tr\!\left[\TS_\alpha \TS_\beta\right]
+\frac{8}{3}\,\tr\!\left[\TF_\alpha \TF_\beta\right]
\right\}[\mathcal{M}_V^4]^{\alpha\beta}
-{27\over2}~\!\tr\left\{\TV_\alpha\mathcal{M}_V^2\TV^\alpha\mathcal{M}_V^2\right\}
+\nn\\
&{}&\hspace*{-40 pt}
-15[\mathcal{M}^2_V]_{\alpha\beta}\varphi^{\rm T}\TS^\alpha\TS^\beta
\TS^\kappa\TS_\kappa\varphi
-5[\mathcal{M}^2_V]_{\alpha\beta}\,\tr\!
\left\{\mathcal{M}_S^2\TS^\alpha\TS^\beta\right\}
-\tr\!\left\{\mathcal{M}_S^4\TS^\alpha\TS_\alpha\right\}
+\nn\\&{}&\hspace*{-40 pt}
-3\,\tr\!\left\{\mathcal{M}_S^2\TS^\alpha\mathcal{M}_S^2\TS_\alpha\right\}
-\frac{1}{2}[\mathcal{M}_S^2]^{ij}\delta^{k m}\delta^{l n}
                  \mathcal{V}'''_{ikl}(\varphi)\mathcal{V}'''_{jmn}(\varphi)
+\\&{}&\hspace*{-40 pt}
+6\mathcal{M}^2_{V\,{\alpha\beta}}\,\tr\!\left\{\TF^\alpha\mathcal{M}_F^*\,\TF^{\beta*}\mathcal{M}_F+cc.\right\}
+2\,\tr\!\left\{\TF^{\alpha*}\TF_\alpha^*\left(\mathcal{M}_F\mathcal{M}_F^*\right)^2+cc.\right\}
+\nn\\&{}&\hspace*{-40 pt}
-\left[\TS^\alpha\TS_\alpha\varphi\right]^j
\tr\!\left\{\YF_j\mathcal{M}_F^*\mathcal{M}_F\mathcal{M}_F^*+cc.\right\}
+\mathcal{M}^2_{S\, ij}\,
\tr\!\left\{\YF^i\!\mathcal{M}_F^*\YF^j\mathcal{M}_F^*+cc.\right\}
+\nn\\&{}&\hspace*{-40 pt}
-\frac{1}{2}\mathbb{Y}^{ij}\left\{
          \mathcal{M}^4_{S}-[\mathcal{M}^2_V]_{\alpha\beta}
          \TS^\alpha\varphi\varphi^{\rm T}\TS^\alpha \right\}_{i j}
+\tr\!\left\{\YF^j\YF_j^*\left(\mathcal{M}_F\mathcal{M}_F^*\right)^2+cc.\right\}
+\nn\\&{}&\hspace*{-40 pt}
+2\,\tr\!\left\{\YF^j\mathcal{M}_F^*\YF_j
\mathcal{M}_F^*\mathcal{M}_F\mathcal{M}_F^*+cc.\right\}
+\,\tr\!\left\{\YF^j\mathcal{M}_F^*\mathcal{M}_F\YF_j^*
\mathcal{M}_F\mathcal{M}_F^*+cc.\right\},\qquad\nn
\end{eqnarray}
where $\mathbb{Y}$ defined in \refer{Eq:bbY-def}.
The above formula is basically the sum of the expressions (3.46) and (3.47)
given in \cite{JO}. We have however explicitly rewritten the traces over
Dirac's (or Majorana's) indices to the traces over the Weyl's indices, 
expressing \refer{Eq:BetaVCheck2L} (and the formulae below) in terms of 
simpler matrices corresponding to the Weyl fermions. 
The second term in the bracket in the first line
``correcting'' the anomalous dimension
originates from the fact that the contribution to
$\check\beta_{(2)}^\cV(\varphi)$ naturally generated by the
Feynman rules has the form
\begin{equation}
\nn
\frac{1}{4}
\varphi^{\rm T}\!\{\TS^\alpha,\ \TS^\beta\}
\mathcal{M}_S^2\{\TS_\alpha,\ \TS_\beta^{\phantom{\beta}}\!\}\varphi
=\varphi^{\rm T}\TS^\alpha\TS^\beta\!\mathcal{M}_S^2\TS_\beta\TS_\alpha\varphi
-\frac{1}{4}
\tr\!\left[\TV_\alpha \TV_\beta\right]\!
            \left[\TS^\alpha\TS^\beta\varphi\right]^{i}\!\mathcal{V}'_i(\varphi),
\end{equation}
(the decomposition follows from the gauge invariance of the tree-level
potential $\mathcal{V}$).

The beta function for the $\mathcal{M}_F(\varphi)$ matrix reads \cite{JO}
\begin{eqnarray}
\check{\beta}_{(2)}^{\mathcal{M}_F}(\varphi)
&=&
\left\{
-\frac{97}{6}\,\tr\!\left[\TV_\alpha \TV_\beta\right]
+\frac{11}{12}\,\tr\!\left[\TS_\alpha \TS_\beta\right]
+\frac{5}{3}\,\tr\!\left[\TF_\alpha \TF_\beta\right]
\right\}\!
\left\{\mathcal{M}_F\TF^\alpha\TF^{\beta}\!+\!tp.\right\}
+\nn\\&{}&\hspace*{-45pt}
-\frac{3}{2}\left\{\mathcal{M}_F\left(\TF_\kappa\TF^\kappa\right)^2\!+\!tp.\right\}
+6\left[\TS^\alpha\TS_\alpha\varphi\right]^i\!
\left\{\YF_i\TF_\kappa\TF^\kappa\!+\!tp.\right\}
-3\left\{\YF_{\ell}\mathcal{M}_F^*\YF^\ell\TF_\alpha\TF^{\alpha}\!+\!tp.\right\}
+
\nn\\&{}&\hspace*{-45pt}
-5\left\{\YF^\ell\TF_\alpha\TF^{\alpha}\!\mathcal{M}_F^*\YF_{\ell}\!+\!tp.\right\}
-3\left\{\YF^{\ell}\YF_\ell^*\mathcal{M}_F\TF_\alpha\TF^{\alpha}\!+\!tp.\right\}
+\nn\\&{}&\hspace*{-45pt}
+\frac{7}{4}\left\{\YF^\ell\YF_{\ell}^*\TF_\alpha^*\TF^{\alpha*}\!\mathcal{M}_F\!+\!tp.\right\}
-\frac{11}{4}\left\{\mathcal{M}_F\YF_{\ell}^*\TF_\alpha^*\TF^{\alpha*}\YF^\ell\!+\!tp.\right\}
+\nn\\&{}&\hspace*{-45pt}
-6\left[\TS^\alpha\varphi\right]^i\TS_\alpha^{jk}\!\left\{\YF_j\YF_k^*\YF_i\!+\!tp.\right\}
-\frac{3}{2}\left[\TS^\alpha\TS_\alpha\right]^{ij}\!
\left\{
\left[\YF_i\YF_j^*\!\mathcal{M}_F\!+\!tp.\right]
+4\YF_i\mathcal{M}_F^*\YF_j\right\}
+\nn\\&{}&\hspace*{-45pt}
+3\left[\TS^\alpha\TS_\alpha\varphi\right]^i\!\left\{
\left[\YF^j\YF_j^*\YF_i\!+\!tp.\right]+4\YF^j\YF_i^*\YF_j\right\}
+2\YF^j\YF_i^*\mathcal{M}_F\YF_j^*Y^i
+\nn\\&{}&\hspace*{-45pt}
-2\YF^i\YF^{j*}\mathcal{M}_F\YF_j^*Y_i
-\left\{\YF^j\!\mathcal{M}_F^*\YF^\ell\YF_{\ell}^*\YF_j\!+\!tp.\right\}
-\frac{1}{8}\left\{\YF^i\YF_{j}^*\YF^j\YF_{i}^{*}\!\mathcal{M}_F\!+\!tp.\right\}
+\nn\\&{}&\hspace*{-45pt}
-2\mathcal{V}'''_{ijk}(\varphi)\YF^i\YF^{j*}\YF^k
-\!\mathbb{Y}^{ij}\left\{
\YF_i\mathcal{M}_F^*\YF_j+\frac{3}{8}\left[\YF_i\YF_j^*\!\mathcal{M}_F\!+\!tp.\right]
\right\}
+\nn\\&{}&\hspace*{-45pt}
+Y^i\varphi^j\Big\{\!
\left[\frac{49}{4}\,\tr\!\left[\TV_\alpha \TV_\beta\right]
-\frac{1}{4}\,\tr\!\left[\TS_\alpha \TS_\beta\right]
-\,\tr\!\left[\TF_\alpha \TF_\beta\right]
\right]\!\left[\TS^\alpha\TS^\beta\right]_{ij}
\!-\frac{21}{2}\left[\left(\TS^\alpha\TS_\alpha\right)^2\right]_{ij}
+\nn\\&{}&%\hspace*{-50 pt}
+\frac{1}{12}\lambda_{ikln}{\lambda}_j^{\phantom{j}kln}
-\frac{5}{2}\,\tr\big\{\TF^\alpha\TF_\alpha\YF_i^*\YF_j+cc.\big\}
-\frac{3}{4}\,\tr\!\left\{\YF_i\YF_j^*\YF^\ell\YF^*_\ell+cc.\right\}
+\nn\\&{}&%\hspace*{-50 pt}
-\frac{1}{2}\,\tr\!\left\{\YF_i^*\YF^\ell\YF_j^*\YF_\ell+cc.\right\}\Big\}.
\end{eqnarray}
Finally, those of the gauge couplings read \cite{JO,MV}
\begin{eqnarray}
\check{\beta}_{(2)}^{T_\alpha}=
T_\kappa&{}&\!\!\!\!
\Big\{\!
-\!\frac{1}{3}
\left\{34\,\tr\!\left[\TV^\kappa \TV^\beta\right]
\!-\!\tr\!\left[\TS^\kappa \TS^\beta\right]
\!-\!10\,\tr\!\left[\TF^\kappa \TF^\beta\right]\right\}
\tr\!\left[\TV_\beta\TV_\alpha\right]
+\\&{}&\,
\!+2\,\tr\!\left[\TS^\kappa \TS_\alpha\TS^\beta \TS_\beta\right]
\!+\tr\!\left[\TF^\kappa \TF_\alpha\TF^\beta \TF_\beta\!+\!cc.\right]
\!+\frac{1}{2}\,\tr\!\left[\TF^\kappa \TF_\alpha\YF_i^*\YF^i\!+\!cc.\right]
\!\Big\}\, .\nn
\end{eqnarray}

\setcounter{equation}{0}
\setcounter{footnote}{0}
\setcounter{figure}{0}

\section{The Standard Model case}
\label{App:SM}

In this Appendix we list the Standard Model two-loop $\beta$ functions
in the $\Lambda$-$\overline{\rm MS}$ scheme, the two-loop coefficient of
quadratic divergence of the scalar fields and the factors $\theta$
relating renormalized parameters in this scheme to the ones in the
DimReg-$\overline{\rm MS}$ scheme. We use the
notation in which the scalar potential has the form
\eqs{\label{Eq:Pot-SM}
\mathcal{V}(H)=m_H^2H^\dagger H+\lambda_1(H^\dagger H)^2.
}
The normalization of the $H$ field VEV is such that
$\langle H_i\rangle=\frac{1}{\sqrt2}v_H\delta_{i2}$.
The tree-level masses of the (usually most relevant) heavy states read
\eq{\label{Eq:MasySM}
M_t=\frac{1}{\sqrt{2}}y_t v_H,\qquad M_W=\frac{1}{2}g_w v_H,\qquad
M_Z=\frac{1}{2}\sqrt{g^2_w+g^2_y}~\!v_H~\!.
}
The strong coupling  constant $g_s$ is normalized so that
the adjoint representation generators $\TV_{\alpha_c}$ of the $SU(3)_C$
group satisfy the relation
\eq{\label{Eq:Killing}
\tr(\TV_{\alpha_c} \TV_{\beta_c})=-3 g_s^2 \delta_{\alpha_c \beta_c}~\!.
}
All the Yukawa couplings other than $y_t$ are neglected.
\vskip0.2cm

The beta functions of the gauge couplings are the same in both schemes and
read (see the explanation below \refer{Eq:Beta-Diff-2loop-Gauge-Jawna})
\eq{\nonumber
\beta_{(1)}^{g_w}=-\frac{19}{6}g_w^3,
\qquad\beta_{(2)}^{g_w} =\frac{1}{6}g_w^3
\left(72 g_s^2 + 35g_w^2+9g_y^2-9y_t^2\right),\quad\ \
}
\eq{
\beta_{(1)}^{g_y}=\frac{41}{6}g_y^3,
\qquad\beta_{(2)}^{g_y} =\frac{1}{18}g_y^3
\left(264 g_s^2 + 81g_w^2 + 199g_y^2 - 51y_t^2\right),
}
\eq{\nonumber
\beta_{(1)}^{g_s}=-7 g_s^3,
\qquad\beta_{(2)}^{g_s} =-\frac{1}{6}g_s^3
\left(156 g_s^2 -27 g_w^2 -11 g_y^2+12 y_t^2\right).
}
The one- and two-loop pieces of the beta functions
of the $\Lambda$-$\overline{\rm MS}$ scheme coupling $\lambda_1$ read
\eqs{\nonumber
\beta_{(1)}^{\lambda_1} &=& 24\lambda_1^2-3\lambda_1\left(3g_w^2+g_y^2-4y_t^2\right)
+\frac{9}{8}g_w^4+\frac{3}{4}g_w^2 g_y^2+\frac{3}{8}g_y^4- 6 y_t^4,
\phantom{aaaa}\nonumber\\
%}
%\eqs{\nonumber
\beta_{(2)}^{\lambda_1} &=& -312\lambda_1^3 +\lambda_1^2
\left(3(12+c_3)(3g_w^2 + g_y^2) - 36(4+c_5)y_t^2\right)+\nn\\
&{}&
+\lambda_1
\left\{
\frac{1}{8}\left(\frac{3}{2}c_{15}-73\right)g_w^4
+\left(\frac{39}{4}-36c_0\right)g_w^2g_y^2+
\right.\nonumber \\
&{}&
\left.
\qquad\
+\frac{1}{24}\left(\frac{3}{2}c_{10}+629\right)g_y^4
+(\frac{3}{2}c_2-3)y_t^4
+\right.\\
&{}&
\left.
\qquad\
+\left(8(10-c_6) g_s^2+\frac{9}{4}(10-c_8)g_w^2
+\frac{1}{12}(170-c_{12})g_y^2\right)y_t^2
\right\}+\nonumber\\
&{}&
-\frac{1}{16}\left(9(4-c_1) g_w^4 -6(28+c_1) g_w^2 g_y^2
+ (76-3c_1) g_y^4 \right)y_t^2+\nonumber\\
&{}&
+\left[-8(4+12c_0) g_s^2 +27c_0 g_w^2
+ (c_0-\frac{8}{3})g_y^2\right]y_t^4 + (c_4+30)y_t^6+\nonumber\\
&{}&
+\frac{1}{192}\left\{3(1220+3c_9) g_w^6 - (1156+3c_{16})g_w^4 g_y^2
- (2236-3c_{13})g_w^2 g_y^4+\right.\nn\\
&{}&\left.\qquad\quad
-(1516+3c_{11}) g_y^6\right\},\nonumber
}
The one- and two-loop pieces of the beta function of the
$\Lambda$-$\overline{\rm MS}$ scheme mass parameter $m^2_H$ are
\begin{eqnarray}
\beta^{m^2_H}_{(1)}\Big\slash m^2_H&=&12\lambda_{1}+ 6y^2_t
-\left(\frac{9}{2}g_w^2+\frac{3}{2}g_y^2\right),\nn\\
\beta^{m^2_H}_{(2)}\Big\slash m^2_H&=&
-60\la^2_{1}+\left(24+\frac{3}{2}c_3\right)\la_{1}\left(3g_w^2+g_y^2\right)
+\left(\frac{3}{32}c_{14}-\frac{145}{16}\right)g_w^4+\\
&{}&
+\left(\frac{15}{8}-9c_0\right)g_w^2g_y^2
+\left(\frac{c_{17}}{32}+\frac{557}{48}\right)g_y^4
-(72+18c_5)\la_1 y^2_t +{}\nn\\
{}&{}&
+\frac{1}{24}\left[96(10-c_6)g_s^2
\!+\!27(10-c_8) g_w^2\!+\!(170-c_{12})g_y^2\right]y^2_t
+\left[\frac{3}{4}c_2-\frac{27}{2}\right]y^4_t
\nonumber.
\end{eqnarray}
The beta function of the $\Lambda$-$\overline{\rm MS}$ scheme  top quark
Yukawa coupling reads
\begin{eqnarray}
\beta_{(1)}^{y_t}&=&y_t\left\{
\frac{9}{2}y_t^2  -8g_s^2  -\frac{9}{4}g_w^2
-\frac{17}{12}g_y^2\right\},\nonumber\\
\beta_{(2)}^{y_t} &=&y_t\left\{
y_t^2\left[(36-4c_7)g_s^2  +\frac{1}{32}(450-9c_{21})g_w^2
+\left(\frac{131}{16}-\frac{c_{19}}{96}\right)g_y^2 -12\lambda_1\right]+
\right.\nonumber\\
{}&{}&
\qquad\!
-12y_t^4+\left(\frac{c_{18}}{3}-108\right)g_s^4  +g_s^2\left(9 g_w^2
+\frac{19}{9}g_y^2\right)+\\
{}&{}&
\left.\qquad\!
+\left[\left(\frac{{c_{20}}}{64}-\frac{23}{4}\right)g_w^4
-\frac{3}{4}g_w^2g_y^2 +\left(\frac{{c_{22}}}{16}
+\frac{1187}{216}\right)g_y^4\right]
+6\lambda_1^2\right\}.\nonumber
\end{eqnarray}
The coefficients $c_n$ appearing in these beta functions  read

\vspace*{10 pt}
\noindent\begin{tabular}{lll}
%\hline
$c_0=\frac{1}{3},\qquad$ & $c_1=1+12 \ln 2,\qquad$ &
			$c_2=7+12 \ln \frac{3}{2},$ \\
$c_3=2+3 \ln\frac{32}{9},\qquad$&
          $c_4=3\left(1+6\ln\frac{4}{3}\right),\qquad$& $c_5=3+4 \ln 2,$\\
$c_6=5+12 \ln 2,\qquad$ & $c_7=2+3 \ln \frac{16}{3},\qquad$ &
          $c_8=11+3 \ln\frac{9}{2},\qquad$ \\
\end{tabular}\\
\vspace*{10 pt}

\noindent and \\

%DRUGA TABELA:
%
\noindent\begin{tabular}{ll}
$c_9=257 \ln 2-9 (19+26 \ln 3),\qquad$ & $c_{10}=145+82 \ln\frac{9}{2},\qquad$  \\
$c_{11}=59+373 \ln2-18\ln3,\qquad$ & $c_{12}=139+69\ln2+54 \ln3,\qquad$  \\
$c_{13}=13-553 \ln 2+90 \ln 3,\qquad$ & $c_{14}=101-2 (101 \ln2-106\ln3),\qquad$  \\
\end{tabular}\\
%
%TRZECIA TABELA
%
\noindent\begin{tabular}{ll}
$c_{15}=53-2 (101 \ln2-106 \ln3),\qquad\!\!\!\!\!\!$ & $c_{16}=13 \ln2+63 (1+\ln9),\qquad$  \\
$c_{17}=193+82 \ln\frac{9}{2},\qquad$ & $c_{18}=103-348 \ln2+216 \ln3,\qquad$  \\
$c_{19}=298+411 \ln2-42 \ln3,\qquad$ & $c_{20}=295-6 (101 \ln2-106 \ln 3),\qquad$  \\
$c_{21}=26+3 \ln 18,\qquad$ &
$c_{22}=\frac{13931}{108}+\frac{41}{18} (23 \ln 2+18 \ln 3).\qquad$  \\
\end{tabular}
\vspace*{0.3 cm}
\vspace*{3 pt}

\noindent Setting them to zero one recovers the beta functions
of the DimReg-$\overline{\rm MS}$ scheme.\\

The two-loop coefficient in front of the quadratic divergence, normalized
as in Eqs. \refer{eqn:barebosonmassesdef} and \refer{Eq:ffff},  has the form
\eqs{\label{Eq:DelFH}
f^{(2)}&=&
\frac{1}{2} \left(-3 \left(\bar{c}_3-\bar{c}_5+6 \bar{c}_6
+2 \bar{c}_9\right) g_w^4+6 \bar{c}_5 g_w^2 g_y^2
-\left(\bar{c}_3-\bar{c}_5+10 \bar{c}_6\right) g_y^4\right)
+\nn\\
&{}&
-\frac{1}{6} \left(96 \left(\bar{c}_7+\bar{c}_8\right) g_s^2
+27 \bar{c}_7 g_w^2+\left(17 \bar{c}_7+8 \bar{c}_8\right) g_y^2\right) y_t^2
+18 \bar{c}_1 y_t^4+\nn\\
&{}&
+72 \bar{c}_2 \lambda _1 y_t^2+36 \lambda _1^2\ln\frac{4}{3}
-3 \bar{c}_4 \lambda _1 \left(3 g_w^2+g_y^2\right),
}
where
\eq{
\bar{c}_1=\frac{1}{12} (64 \ln 2+15 \ln 3-25 \ln 5-11 \ln 11),
}
\eq{
\bar{c}_2=\frac{1}{2}\left(-1+9\ln2-\frac{45}{2}\ln3+\frac{25}{2}\ln5\right),
}
\eq{
\bar{c}_3=\frac{17}{8}-\frac{51}{4}\ln2+\frac{459}{16}\ln3-\frac{125}{8}\ln5,
}
\eq{
\bar{c}_4=\frac{29}{16}-\frac{1901}{72}\ln2-\frac{369}{32}\ln3
+\frac{125}{72}\ln5+\frac{847}{72}\ln11,
}
\eq{
\bar{c}_5=
-\frac{1}{8}+\frac{189}{32} \ln 2-\frac{45}{16} \ln 3,
}
\eq{
\bar{c}_6=
-\frac{13}{4}+\frac{61}{2}\ln2-54\ln3+25\ln5,
}
\eq{
\bar{c}_7=
-\frac{13}{8}+\frac{667}{18}\ln2+21\ln3+\frac{25}{9}\ln5-\frac{187}{9}\ln11,
}
\eq{
\bar{c}_8=
\frac{1}{2} \left(-\frac{9}{4}-21 \ln2+9 \ln3\right)
,}
\eq{
\bar{c}_9=
\frac{39}{4}-\frac{9695}{96} \ln 2+\frac{81}{4} \ln 3-\frac{625 }{24}\ln 5+\frac{847}{24} \ln 11.
}
\\

Finally, the factors $\theta$ appearing in the formula \refer{Eq:G-fun}
relating the renormalized parameters of the $\Lambda$-$\overline{\rm MS}$
and DimReg-$\overline{\rm MS}$ schemes read
\eq{\label{Eq:!!!ThetaMNJaw0}
\theta_{(1)}^{g_s}(g)=\left[\frac{29}{4}\ln 2-\frac{9 }{2}\ln3
-\frac{47}{48}\right] g_s^3,
}
\eq{
\theta_{(1)}^{g_w}(g)=\left[\frac{49}{12}\ln2-3\ln3
-\frac{371}{288}\right] g_w^3,
}
\eq{
\theta_{(1)}^{g_y}(g)=-\frac{1}{288} (887+984 \ln2) g_y^3,
}
\eqs{
\theta_{(1)}^{\lambda_1}(g)&=&-\frac{1}{8} \left(2 g_w^2 g_y^2+3 g_w^4+g_y^4\right)
+\frac{1}{8} \left[2+3 \ln \frac{32}{9}\right]
\lambda _1 (3 g_w^2+g_y^2) +\nn\\
&{}&
-\left[\frac{9}{2}+6 \ln 2\right]\lambda _1  y_t^2,
}
\eq{
\theta_{(1)}^{m_H^2}(g)=m_H^2 \left\{
\frac{1}{16} \left[2+3 \ln\frac{32}{9}\right](3 g_w^2+g_y^2)
-\frac{3}{4} (3+4 \ln2) y_t^2\right\},
}
\eqs{\label{Eq:!!!ThetaMNJaw1}
\theta_{(1)}^{y_t}(g)&=&y_t \left\{\frac{4}{3} g_s^2
+\frac{3}{32} \left[3 \ln \frac{32}{9}-2\right]g_w^2
+\left[\frac{7}{144}+\frac{3}{32} \ln \frac{32}{9}\right]g_y^2
\right.+\nn\\
&{}&\left.\quad\ \,-\left[\frac{5}{4}
+\frac{3}{4} \ln \frac{16}{3}\right] y_t^2\right\}.
}

\setcounter{equation}{0}
\setcounter{footnote}{0}
\setcounter{figure}{0}

\section{Basic 2-loop integrals}
\label{App:2LoopInt}

Here we list the nine basic 2-loop integrals to which the genuine 2-loop
vacuum graphs $A$-$J$ shown in Figure \ref{Rys:DiVac} (Section \ref{Vac:Diag})
regularized using the prescription of Section \ref{Sec:UV-cutoff}
can be reduced. We also calculate their divergent parts.

The integrals are
\begin{eqnarray}
I^{tot}_\Lambda\!\left(m_1,m_2,m_3|n_1,n_2,n_3\right)&\equiv&\label{Eq:Itot-def}\\
&{}&\hspace{-90 pt}
=i^2\!\int{\volfour{k}}\!\int{\volfour{q}}
\frac{e^{(n_1-1)\frac{k^2}{\Lambda^2}}}{\KR(k)^2-m_1^2}
\frac{e^{(n_2-1)\frac{q^2}{\Lambda^2}}}{\KR(q)^2-m_2^2}
\frac{e^{(n_3-1)\frac{(k+q)^2}{\Lambda^2}}}{\KR(k+q)^2-m_3^2},\qquad\nn
\end{eqnarray}

\begin{eqnarray}
J^{tot}_\Lambda\!\left(m_1,m_2|n_1,n_2,n_3\right)
&\equiv&\label{Eq:Jtot-def}\\
&{}&\hspace{-20 pt}=i^2\!\int{\volfour{k}}\!\int{\volfour{q}}\,
e^{n_3\frac{(k+q)^2}{\Lambda^2}}
\frac{e^{(n_1-1)\frac{k^2}{\Lambda^2}}}{\KR(k)^2-m_1^2}
\frac{e^{(n_2-1)\frac{q^2}{\Lambda^2}}}{\KR(q)^2-m_2^2},\nn
\end{eqnarray}

\begin{eqnarray}
N^{tot}_{2,\Lambda}\!\left(m_1,m_2|n_1,n_2,n_3\right)&\equiv&\label{Eq:N2tot-def}\\
&{}&\hspace{-90 pt}
=i^2\!\int{\volfour{k}}\!\int{\volfour{q}}\,
e^{n_3\frac{(k+q)^2}{\Lambda^2}}
\frac{e^{(n_1-1)\frac{k^2}{\Lambda^2}}}{\KR(k)^2-m_1^2}
\frac{e^{(n_2-1)\frac{q^2}{\Lambda^2}}}{\KR(q)^2-m_2^2}
\times
\frac{(k\!\cdot\!q)^2}{k^2 q^2},\qquad\nn
\end{eqnarray}

\begin{eqnarray}
M^{tot}_{1,\Lambda}\!\left(m_1,m_3|n_1,n_2,n_3\right)&\equiv&\label{Eq:M1tot-def}\\
&{}&\hspace{-90 pt}
=-i^2\!\int{\volfour{k}}\!\int{\volfour{q}}\,
e^{n_2\frac{(k+q)^2}{\Lambda^2}}
\frac{e^{(n_1-1)\frac{k^2}{\Lambda^2}}}{\KR(k)^2-m_1^2}
\frac{e^{(n_3-1)\frac{q^2}{\Lambda^2}}}{\KR(q)^2-m_3^2}
\times\frac{k\!\cdot\!q}{q^2},\qquad\nn
\end{eqnarray}

\begin{eqnarray}
I^{tot}_{1,\Lambda}\!\left(m_1,m_2,m_3|n_1,n_2,n_3\right)
&\equiv&\label{Eq:I1tot-def}\\
&{}&\hspace{-130 pt}
=i^2\!\int{\volfour{k}}\!\int{\volfour{q}}
\frac{e^{(n_1-1)\frac{k^2}{\Lambda^2}}}{\KR(k)^2-m_1^2}
\frac{e^{(n_2-1)\frac{q^2}{\Lambda^2}}}{\KR(q)^2-m_2^2}
\frac{e^{(n_3-1)\frac{(k+q)^2}{\Lambda^2}}}{\KR(k+q)^2-m_3^2}\times{1\over(k+q)^2},\nn
\end{eqnarray}
\begin{eqnarray}\label{Eq:I2tot-def}
I^{tot}_{2,\Lambda}\!\left(m_1,m_2,m_3|n_1,n_2,n_3\right)&\equiv&\\
&{}&\hspace{-130 pt}
=i^2\!\int{\volfour{k}}\!\int{\volfour{q}}
\frac{e^{(n_1-1)\frac{k^2}{\Lambda^2}}}{\KR(k)^2-m_1^2}
\frac{e^{(n_2-1)\frac{q^2}{\Lambda^2}}}{\KR(q)^2-m_2^2}
\frac{e^{(n_3-1)\frac{(k+q)^2}{\Lambda^2}}}{\KR(k+q)^2-m_3^2}\!\times\!
\frac{1}{(k\!+\!q)^2q^2},\nn
\end{eqnarray}

\begin{eqnarray}
J^{tot}_{1,\Lambda}\!\left(m_1,m_2|n_1,n_2,n_3\right)\!&\equiv&\!
\label{Eq:J1tot-def}\\
&{}&\hspace{-20 pt}
=i^2\!\!\int{\volfour{k}}\!\!\int{\volfour{q}}\,
e^{n_3\frac{(k+q)^2}{\Lambda^2}}\!
\frac{e^{(n_1-1)\frac{k^2}{\Lambda^2}}}{\KR(k)^2\!-\!m_1^2}
\frac{e^{(n_2-1)\frac{q^2}{\Lambda^2}}}{\KR(q)^2\!-\!m_2^2}\,\frac{1}{k^2},\nn
\end{eqnarray}
\begin{eqnarray}
J^{tot}_{2,\Lambda}\!\left(m_1,m_2|n_1,n_2,n_3\right)
\!&\equiv&\!\label{Eq:J2tot-def}\\
&{}&\hspace{-10 pt}=i^2\!\!\int{\volfour{k}}\!\!\int{\volfour{q}}\,
e^{n_3\frac{(k+q)^2}{\Lambda^2}}\!
\frac{e^{(n_1-1)\frac{k^2}{\Lambda^2}}}{\KR(k)^2\!-\!m_1^2}
\frac{e^{(n_2-1)\frac{q^2}{\Lambda^2}}}{\KR(q)^2\!-\!m_2^2}\frac{1}{k^2q^2},\!\nn
\end{eqnarray}

\begin{eqnarray}\label{Eq:M2tot-def}
M^{tot}_{2,\Lambda}\!\left(m_1,m_3|n_1,n_2,n_3\right)&\equiv&\\
&{}&\hspace{-60 pt}
=-i^2\!\int{\volfour{k}}\!\int{\volfour{q}}\,
e^{n_2\frac{(k+q)^2}{\Lambda^2}}
\frac{e^{(n_1-1)\frac{k^2}{\Lambda^2}}}{\KR(k)^2-m_1^2}
\frac{e^{(n_3-1)\frac{q^2}{\Lambda^2}}}{\KR(q)^2-m_3^2}
\times\frac{k\!\cdot\!q}{q^2\,k^2}.\nn\qquad
\end{eqnarray}
All these integrals are convergent provided all $n_i$'s in the exponents
are nonnegative and at least two of them are strictly positive. The
integrals arising in the decompositions of the genuine 2-loop vacuum diagrams
of Figure \ref{Rys:DiVac} all fulfill these conditions.

Strictly speaking, only the integrals
\refer{Eq:Itot-def}-\refer{Eq:M1tot-def} are independent; the remaining
ones are their linear combinations. For instance,
\begin{eqnarray}\label{Eq:I-dec-ex}
I^{tot}_{2,\Lambda}\!\left(m_1,m_2,m_3|n_1,n_2,n_3\right)
&=&\frac{1}{m_2^2\, m_3^2}
\big\{I^{tot}_{\Lambda}\!\left(m_1,m_2,m_3|n_1,n_2\!-\!1,n_3\!-\!1\right)+\qquad\!\!
\nn\\&{}&\hspace*{-125 pt}
-I^{tot}_{\Lambda}\!\left(m_1,m_2,0|n_1,n_2\!-\!1,n_3\!-\!1\right)
\!-\!I^{tot}_{\Lambda}\!\left(m_1,0,m_3|n_1,n_2\!-\!1,n_3\!-\!1\right)+
\nn\\&{}&\hspace*{-125 pt}
+I^{tot}_{\Lambda}\!\left(m_1,0,0|n_1,n_2\!-\!1,n_3\!-\!1\right)\big\}.\qquad
\end{eqnarray}
However, because the integrals on the right hand side of \refer{Eq:I-dec-ex}
have lower values of $n_i$'s, they can in principle be divergent even if
the one on the right hand side is not. In the case of the diagrams
$A$-$J$ of Figure \ref{Rys:DiVac} the integrals
\refer{Eq:Itot-def}-\refer{Eq:M1tot-def}
arising on right hand sides of decompositions analogous to \refer{Eq:I-dec-ex}
have still nonnegative $n_i$'s but
in some cases more than one $n_i$ vanishes. The decompositions like
\refer{Eq:I-dec-ex} are then justified if one makes the replacement
\begin{equation}
%\label{Eq:DDDodRegPozDiv}
n_i\to n_i + c_i\, \vep,\ \ \ \qquad c_i>0,\  \vep>0,\nn
\end{equation}
first. 
Singularities arising for $\vep\to 0$ cancel out in the sums like
\refer{Eq:I-dec-ex}. Below we give explicit expressions only
for the integrals \refer{Eq:Itot-def}-\refer{Eq:M1tot-def}.
\vskip0.2cm

It is convenient to start with the following auxiliary integral
\begin{equation}\label{Eq:I-nie-tot-def}
\mathbb{I}_\Lambda\equiv I_\Lambda\!\left(m_1,m_2,m_3|n_1,n_2,n_3\right)
\equiv
i^2\!\int{\volfour{k}}\!\int{\volfour{q}}
\frac{e^{n_1\frac{k^2}{\Lambda^2}}}{k^2-m_1^2}
\frac{e^{n_2\frac{q^2}{\Lambda^2}}}{q^2-m_2^2}
\frac{e^{n_3\frac{(k+q)^2}{\Lambda^2}}}{(k+q)^2-m_3^2}.
\end{equation}
In the Schwinger parametrization $\mathbb{I}_\Lambda$ takes the form
\begin{equation}
(4\pi)^4\mathbb{I}_\Lambda=-\int\limits_{\mathbb{R}_+^3}\!{\rm d}^3\xi\,
{e^{-\xi_1 m_1^2} e^{-\xi_2 m_2^2} e^{-\xi_3 m_3^2}}H_\Lambda(\xi)\, ,
\end{equation}
where ($\epsilon_i\equiv n_i/\Lambda^2$)
\begin{equation}
H_\Lambda(\xi)=
{\left[(\xi_1\!+\!\epsilon_1)(\xi_2\!+\!\epsilon_2)
\!+\!(\xi_1\!+\!\epsilon_1)(\xi_3\!+\!\epsilon_3)
\!+\!(\xi_2\!+\!\epsilon_2)(\xi_3\!+\!\epsilon_3)\right]^{-2}}\, .\nonumber
\end{equation}
With the help of the  identity
\begin{equation}
e_1 e_2 e_3 =\left\{-2+\sum_i e_i\right\}
+\!\sum_{i,j;\, i<j}\!(1-e_i)(1-e_j)-\prod_i (1-e_i)\, ,\nonumber
\end{equation}
$\mathbb{I}_\Lambda$ can be split into several pieces:
\begin{equation}
\mathbb{I}_\Lambda
=\mathbb{I}^{\rm div}_\Lambda
-\!\sum_{i,j;\, i<j}\!\mathbb{I}^{(2) ij}_\Lambda+\mathbb{I}^{(3)}_\Lambda,
\end{equation}
where
\begin{equation}\label{Eq:Idiv-def}
(4\pi)^4\mathbb{I}^{\rm div}_\Lambda
=-\int\limits_{\mathbb{R}_+^3}\!{\rm d}^3\xi
\left\{-2+\sum_{i=1}^3 e^{-\xi_i m_i^2}\right\}H_\Lambda(\xi)\, ,
\end{equation}
etc. Using the inequalities:
\begin{equation}
0\leq H_\Lambda(\xi)\leq H_\infty(\xi)=
{\left[\xi_1 \xi_2\!+\!\xi_1 \xi_3\!+\!\xi_2\xi_3\right]^{-2}}\, ,\nonumber
\end{equation}
and $1-e^{-x}\leq 2x/(1+x)$ (the latter valid for $x\geq0$)
it is easy to prove that the integrals $\mathbb{I}^{(2) ij}_\Lambda$ and
$\mathbb{I}^{(3)}_\Lambda$ have finite limits for $\Lambda\to\infty$.
Aiming at computing the divergent parts of the diagrams
$A$-$J$ in Fig. \ref{Rys:DiVac}
we focus, therefore, on $\mathbb{I}^{\rm div}_\Lambda$ only.

Two out of the three integrals over $\xi_i$ in $\mathbb{I}^{\rm div}_\Lambda$
are elementary. Taking them we get
\begin{eqnarray}
(4\pi)^4\mathbb{I}^{\rm div}_\Lambda=
-\mathcal{K}_\Lambda(m_1|n_1,n_2,n_3)
-\mathcal{K}_\Lambda(m_2|n_2,n_1,n_3)
-\mathcal{K}_\Lambda(m_3|n_3,n_1,n_2)\nn\\
+2\,\mathcal{K}_\Lambda(0|n_1,n_2,n_3)\, ,
\phantom{aaaaaaaaaaaaaaaaaaaaaaaaaaaaaaaaa}\label{Eq:Idiv-K-Lam}
\end{eqnarray}
where
\begin{equation}\label{Eq:K-Lam-LapForm}
\mathcal{K}_\Lambda\!\left(m|n_1,n_2,n_3\right)
=\Lambda^2\!
\int\limits_0^\infty{\rm d}t\,
\exp(-s_m t)\,f_{n_1 n_2 n_3}(t)\, ,
\qquad
s_m\equiv \frac{m^2}{\Lambda^2},
\end{equation}
with
\begin{eqnarray}
f_{n_1 n_2 n_3}(t)&=&
\frac{1}{(t+n_1)^2}\Big\{
\ln(t+n_1+n_3)+\ln(t+n_1+n_2)
+\nn\\&{}&\qquad\qquad\ \
-\ln\!\left[t\!+\!\frac{n_1n_2\!+\!n_1n_3\!+\!n_2n_3}{n_2\!+\!n_3}\right]
-\ln(n_2+n_3)\Big\}\, .\qquad
\end{eqnarray}
To find the required terms in the expansion of the Laplace transform
\refer{Eq:K-Lam-LapForm} the Handelsman-Lew theorem \cite{HANLEW}
can be employed. It gives
\begin{eqnarray}
\mathcal{K}_\Lambda\!\left(m|n_1,n_2,n_3\right)
=-\Lambda^2 \mathcal{Q}(n_1,n_2,n_3)-\mathscr{L}_\Lambda(m|n_2,n_3)
+\mathcal{O}(\Lambda^0),
\end{eqnarray}
where
\begin{equation}\nn
\mathcal{Q}(n_1,n_2,n_3)
=\left\{\sum_{l=1}^3\frac{1}{n_l}\right\}
\ln\left\{ \sum_{i=1}^2\sum_{j=i+1}^{3}\!\! n_in_j\!\right\}
\!-\!\sum_{i=1}^2\sum_{j=i+1}^{3}\!\!
\left\{\frac{1}{n_i}\!+\!\frac{1}{n_j}\right\}\ln(n_i\!+\!n_j),
\end{equation}
and
\begin{equation}
\mathscr{L}_\Lambda(m|n_2,n_3)=m^2\ln\frac{m^2}{\Lambda^2}\left\{
\frac{1}{2}\ln\frac{m^2}{\Lambda^2}+\gamma_E-1+\ln(n_2\!+\!n_3)\right\}\, .
\end{equation}
Thus, the integral \refer{Eq:I-nie-tot-def} has the form
\begin{eqnarray}\label{Eq:I-nie-tot-DivPart}
I_\Lambda\!\left(m_1,m_2,m_3|n_1,n_2,n_3\right)
&=&\frac{1}{(4\pi)^4}\big\{\Lambda^2 \mathcal{Q}(n_1,n_2,n_3)+\\
&{}&\hspace*{-70 pt}\nn
+\mathscr{L}_\Lambda(m_1|n_2,n_3)
+\mathscr{L}_\Lambda(m_2|n_1,n_3)
+\mathscr{L}_\Lambda(m_3|n_1,n_2)\big\}+\mathcal{O}(\Lambda^0).
\end{eqnarray}
Note that the integral $I_\Lambda\left(0,0,0|n_1,n_2,n_3\right)$ is
infrared convergent and elementary (in the Schwinger parametrization):
\begin{equation}\nn
I_\Lambda\!\left(0,0,0|n_1,n_2,n_3\right)
=\frac{1}{(4\pi)^4}~\!\Lambda^2 \mathcal{Q}(n_1,n_2,n_3)\,.
\end{equation}
Combining the above formula with the inequality
\begin{eqnarray}
\frac{e^{-(n-1)k_E^2/\Lambda^2}}{k_E^2\,e^{k_E^2/\Lambda^2}+m^2}
\leq\frac{e^{-n\,k_E^2/\Lambda^2}}{k_E^2+m^2}
+2\,\frac{m^2}{\Lambda^2}\,\frac{e^{-n\,k_E^2/\Lambda^2}}{k_E^2}\,,\nn
\end{eqnarray}
and monotonicity of the first term on the RHS in $m$, it is easy to show
that the difference between $I_\Lambda$ \refer{Eq:I-nie-tot-def} and
$I^{tot}_{\Lambda}$ \refer{Eq:Itot-def} is a bounded function of $\Lambda$ for
$\Lambda\rightarrow\infty$.
Thus, in this limit,
\begin{eqnarray}%\label{Eq:I-nie-tot-DivPart}
I^{tot}_\Lambda\!\left(m_1,m_2,m_3|n_1,n_2,n_3\right)
=I_\Lambda\!\left(m_1,m_2,m_3|n_1,n_2,n_3\right)+\mathcal{O}(\Lambda^0).
\end{eqnarray}
\vskip0.2cm

The integral
$\mathbb{J}_\Lambda=
J^{tot}_\Lambda\!\left(m_1,m_2|n_1,n_2,n_3\right)$ defined in
\refer{Eq:Jtot-def} can be conveniently calculated as a power series in $n_3$:
\begin{equation}\label{Eq:J_Lam-Series}
(4 \pi)^4\mathbb{J}_\Lambda
=\sum_{s=0}^{\infty}
\frac{(2 n_3)^{2s}}{(2s)!}\tilde{K}_s\,  F_s(\alpha_1,m_1^2)F_s(\alpha_2,m_2^2),
\qquad
\alpha_i\equiv n_i+n_3,
\end{equation}
where
\begin{equation}
\tilde{K}_n=
\frac{\Gamma(n+\frac{1}{2})}{\Gamma(n+2)\Gamma(\frac{1}{2})},
\end{equation}
and
\begin{equation}\label{Eq:F_s-Series}
F_s(\alpha,x)\equiv
2\int\limits_0^\infty\!{\rm d}k\, k^3\, \frac{k^{2 s}}{\Lambda^{2 s}}\,
\frac{e^{-(\alpha-1){k^2}/{\Lambda^2}}}{e^{{k^2}/{\Lambda^2}}k^2+x}
=\sum_{n=0}^\infty x^n G_n(\alpha,s,x)\,,\,
\end{equation}
with
\begin{equation}
G_n(\alpha,s,x)=
\int\limits_0^\infty{\rm d}t\,\, \frac{t^{s+1}}{\Lambda^{2 s}}\,
\frac{e^{-\alpha\, t/\Lambda^2}}{(t+x)^{n+1}}(1-e^{- t/\Lambda^2})^n   \, .
\end{equation}
Since $G_0=\mathcal{O}(\Lambda^2)$ and $G_3=\mathcal{O}(\Lambda^{-4})$, it
is easy to show that the series \refer{Eq:F_s-Series} can be replaced by
the sum of its first three terms at most
(the remaining terms give to the sum \refer{Eq:J_Lam-Series} contributions
of order $\mathcal{O}(\Lambda^{-1})$). It is also easy to determine with
required accuracy the asymptotic, for $\Lambda\to\infty$, forms of $G_0$,
$G_1$ and $G_2$ (the cases $s=0$, $s=1$ and $s\geq2$ have to be considered
separately). With these approximations the series \refer{Eq:J_Lam-Series}
can be summed. Thus,
\begin{eqnarray}
J^{tot}_\Lambda\!\left(m_1,m_2|n_1,n_2,n_3\right)
&=&\frac{1}{(4\pi)^4}~\!\frac{1}{2!}
\left\{\phantom{+}\!\!\!
\tilde{J}^{tot}_\Lambda\!\left(m_1^2,m_2^2|n_1\!+\!n_3,n_2\!+\!n_3,n_3\right)
+\right.\nn\\&{}&
\qquad\qquad\left.
+\!\tilde{J}^{tot}_\Lambda\!\left(m_2^2,m_1^2|n_2\!+\!n_3,n_1\!+\!n_3,n_3\right)
\right\}\!
,\qquad\ \
\end{eqnarray}
where the divergent part of $\tilde{J}^{tot}_\Lambda$ is
\begin{eqnarray}
\tilde{J}^{tot}_\Lambda\!\left(x_1,x_2|\alpha_1,\alpha_2,n_3\right)
&=&
-\frac{\Lambda^4}{{n_3^2}}\ln\!\left(1-\frac{n_3^2}{\alpha _1 \alpha_2}\right)
+\delta_\Lambda^2\, x_1 x_2+
\nn\\&{}&\hspace*{-70 pt}
+\delta_\Lambda  \left\{
\frac{n_3^2 x_1^2}{\alpha _2^2}-\frac{2\left(\alpha_1+2\right)x_1^2}{\alpha_2}
-2 x_1 x_2 \left[\ln \frac{1+\alpha_1}{2}+\ln\frac{x_1}{\mu^2}-1\right]\right\}
+\nn\\&{}&\hspace*{-70 pt}
+2\Lambda^2\, \frac{x_1}{{\alpha_2}}
\left\{\ln\frac{x_1}{\mu^2}-(2+\delta_\Lambda +\ln2)
       +\ln\!\left(1+\alpha _1-\frac{n_3^2}{\alpha _2}\right)
\right.
+\nn\\&{}&\hspace*{-70 pt}\qquad\qquad\
\left.
-\alpha_2\frac{1+\alpha _1}{{n_3^2}}
\ln\!\left(1-\frac{n_3^2}{\alpha _2+\alpha _1 \alpha_2}\right)\right\}
+\mathcal{O}(\Lambda^0).
\end{eqnarray}
(The dependence on $\mu$ is spurious - it cancels between logarithms and
$\delta_\Lambda$). Terms of order $\mathcal{O}(\Lambda^0)$ can be determined
in the same way (with the aid of expansion of the dilogarithm ${\rm Li}_2(z)$).
\vskip0.2cm

The remaining two integrals, \refer{Eq:N2tot-def} and \refer{Eq:M1tot-def},
can be calculated in precisely the same way as $J^{tot}_\Lambda$:
\begin{eqnarray}
N^{tot}_{2,\Lambda}\!\left(m_1,m_2|n_1,n_2,n_3\right)\!
&=&\!
\frac{1}{(4\pi)^4}\frac{1}{2!}
\!\left\{\phantom{+}\!\!\!
\tilde{N}^{tot}_{2,\Lambda}\!\left(m_1^2,m_2^2|n_1\!+\!n_3,n_2\!+\!n_3,n_3\right)
+\right.\nn\\&{}&
\qquad\qquad\left.
\!+\!\tilde{N}^{tot}_{2,\Lambda}\!\left(m_2^2,m_1^2|n_2\!+\!n_3,n_1\!
+\!n_3,n_3\right)
\right\}\!
,\qquad\ \
\end{eqnarray}
where
\begin{eqnarray}
\tilde{N}^{tot}_{2,\Lambda}\!\left(x_1,x_2|\alpha_1,\alpha_2,n_3\right)
&=&
\frac{\Lambda^4}{{2\,n_3^4}}\left\{-3\,n_3^2
+(n_3^2-3\alpha_1\alpha_2)\ln\!\left(1-\frac{n_3^2}{\alpha_1\alpha_2}\right)
\right\}+
\nn\\&{}&\hspace*{-120 pt}
-\frac{1}{2}\delta_\Lambda\left\{
x_1^2\frac{\alpha_1+2}{\alpha_2}
-
\frac{n_3^2 x_1^2}{\alpha _2^2}
+ x_1 x_2 \left[
\ln \frac{1+\alpha_1}{2}
+\ln\frac{x_1}{\mu^2}-1\right]
\right\}+\frac{1}{4}\delta_\Lambda^2\, x_1 x_2
+\nn\\&{}&\hspace*{-120 pt}
+
\Lambda^2\, \frac{x_1}{2\,{\alpha_2}}
\left\{
\ln\frac{x_1}{\mu^2}-\left[3\alpha_2\frac{1+\alpha_1}{n_3^2}
+\delta_\Lambda +\ln2+\frac{1}{2}\right]
       +\ln\!\left(1+\alpha _1-\frac{n_3^2}{\alpha _2}\right)
\right.
+\nn\\&{}&\hspace*{-120 pt}\qquad\qquad\
\left.
+\alpha_2\frac{1+\alpha _1}{{n_3^4}}
\left( 2n_3^2-3\alpha_2\left(1+\alpha_1\right)   \right)
\ln\!\left(1-\frac{n_3^2}{\alpha _2+\alpha _1 \alpha_2}\right)\right\}
+\mathcal{O}(\Lambda^0).
\nn
\end{eqnarray}
and
\begin{eqnarray}
M^{tot}_{1,\Lambda}\!\left(m_1,m_3|n_1,n_2,n_3\right)\!
=\!
\frac{1}{(4\pi)^4}
\tilde{M}^{tot}_{1,\Lambda}\!\left(m_1^2,m_3^2|n_1\!+\!n_2,n_2,n_3\!+\!n_2\right)
,\qquad
\end{eqnarray}
where
\begin{eqnarray}
\tilde{M}^{tot}_{1,\Lambda}\!\left(x_1,x_3|\alpha_1,n_2,\alpha_3\right)
&=&
-\frac{\Lambda^4}{{\alpha_1\,n_2^3}}\left\{
n_2^2+\alpha_1\alpha_3\ln\!\left(1-\frac{n_2^2}{\alpha_1\alpha_3}\right)
\right\}+
\nn\\&{}&\hspace*{-120 pt}
+\delta_\Lambda \frac{n_2}{6\alpha_1^3\alpha_3}\left\{
3x_1^2\alpha_1^3+3x_1 x_3\frac{\alpha_1^3\alpha_3}{1+\alpha_1}
+x_3^2\alpha_3(2n_2^2-3\alpha_1(2+\alpha_3))
\right\}
+\nn\\&{}&\hspace*{-120 pt}
+
\frac{\Lambda^2}{4\,n_2^3{\alpha_1^2}}
\bigg\{
2\,x_3\, n_2^4\ln\frac{x_3}{\mu^2}
\!-\!2n_2^2\alpha_1(x_3\!+\!2x_1\alpha_1\!+\!x_3\alpha_3)
\!-\!n_2^4x_3(3\!+\!2\delta_\Lambda\!+\!\ln 4 )\!
+\nn\\&{}&\hspace*{-120pt}\qquad\qquad\
+2\,x_3\,n_2^4\ln\!\left(\!1\!+\!\alpha_3\!-\!\frac{n_2^2}{\alpha_1}\right)
\!-\!2\,x_3\,\alpha_1^2(1\!+\!\alpha_3)^2
\ln\!\left(\!1\!-\!\frac{n_2^2}{\alpha_1\!+\!\alpha_1\alpha_3}\!\right)
+\nn\\&{}&\hspace*{-120 pt}\qquad\qquad\
\left.
+4x_1\alpha_1^2\left( n_2^2-\alpha_3\left(1+\alpha_1\right)\right)
\ln\!\left(1-\frac{n_2^2}{\alpha_3+\alpha_1\alpha_3}\right)\right\}
+\mathcal{O}(\Lambda^0).
\nn
\end{eqnarray}

\end{document}